\documentclass[aps,prx,twocolumn,floatfix,superscriptaddress]{revtex4-2}

\usepackage{physics}
\usepackage{graphicx}
\usepackage{lineno,hyperref}
\usepackage{amsmath,amssymb}
\usepackage{array}
\usepackage{graphicx}
\usepackage{dcolumn}
\usepackage{bm}

\usepackage{url}

\hypersetup{colorlinks=true,linkcolor=blue, filecolor=magenta, citecolor=blue,urlcolor=blue}

\begin{document}


\title{Emergent vortex Majorana zero mode in iron-based superconductors}

\author{Lingyuan Kong}
\affiliation{Beijing National Laboratory for Condensed Matter Physics and Institute
of Physics, Chinese Academy of Sciences, Beijing 100190, China}
\affiliation{School of Physical Sciences, University of Chinese Academy of Sciences, Beijing 100049, China}

\author{Hong Ding}%
  \email{dingh@iphy.ac.cn}
\affiliation{Beijing National Laboratory for Condensed Matter Physics and Institute
of Physics, Chinese Academy of Sciences, Beijing 100190, China}
\affiliation{School of Physical Sciences, University of Chinese Academy of Sciences, Beijing 100049, China}
\affiliation{CAS Center for Excellence in Topological Quantum Computation, University
of Chinese Academy of Sciences, Beijing 100190, China}
    
\date{May 12, 2020}

\begin{abstract}
The vortex of iron-based superconductors is emerging as a promising platform for Majorana zero mode, owing to a magic integration among intrinsic vortex winding, non-trivial band topology, strong electron-electron correlations, high-\(T_\mathrm{c}\) superconductivity and the simplification of single material. It overcomes many difficulties suffered in heterostructure-based Majorana platforms, including small topological gap, interfacial contamination, lattice imperfections, and etc. Isolated zero-bias peaks have been found in vortex of several iron-based superconductors. So far,  studies from both experimental and theoretical aspects strongly indicate the realization of vortex Majorana zero mode, with a potential to be applied to fault-tolerant quantum computation. By taking Fe(Te,Se) superconductor as an example, here we review the original idea and research progress of Majorana zero modes in this new platform. After introducing the identifications of topological band structure and real zero modes in vortex, we summarize the physics behaviors of vortex Majorana zero modes systematically. First, relying on the behavior of the zero mode wave function and evidence of quasiparticle poisoning, we analyze the mechanism of emergence of vortex Majorana zero modes. Secondly, assisted with some well-established theories, we elaborate the measurements on ``Majorana symmetry'' and topological nature of vortex Majorana zero modes. After that, we switch from quantum physics to quantum engineering, and analyze the performance of vortex Majorana zero modes under real circumstances, which may potentially benefit the exploration of practical applications in the future. This review follows the physics properties of vortex Majorana zero modes, especially emphasizes the link between phenomena and mechanisms. It provides a chance to bridge the gap between the well-established theories and the newly discovered ``iron home'' of Majoranas. 
\end{abstract}
\maketitle


\section{Introduction}\label{sec:1}

Majorana zero mode (MZM) is a zero-energy quasiparticle with exotic properties. In a condensed matter system, it usually appears in a topological superconductor as a bound state~\citep{1_nayak2008non,2_wilczek2009majorana,3_alicea2012new,4_beenakker2013search,5_elliott2015colloquium}. MZMs have three fundamental properties. First of all, the hole and particle components of MZM are equal. The so-called ``Majorana symmetry'' requires self-conjugation of the generation and annihilation operators, demonstrating the charge neutrality of MZMs. MZM is a solid-state analogue to the Majorana fermions in the universe, of which the antiparticle is the particle itself. Similar to the Majorana fermions, MZMs can only exist individually. Two MZMs fuse into a normal fermion at a finite energy. The deviation from the zero energy increases exponentially with the two MZMs approaching to each other. Secondly, MZMs are topological in nature. The appearance of MZM is accompanied by nontrivial topological invariants. MZM can be viewed as a topological boundary state of an effective topological superconductor. Thirdly, MZMs obey non-Abelian anyonic statistics, owing to their fractional quantum dimension in a value of \(\sqrt{2}\).  MZM is regarded as  ``half'' of a fermion in some literature. The braiding operations on these ``halves'' constitute a building block of fault-tolerant topological quantum computation~\citep{6_kitaev1997quantum,7_kitaev2003fault,8_kitaev2006anyons,9_aasen2016milestones}. These exotic properties propel the exploration of MZM on both fundamental physics research and potential practical application over the last decades. Majorana physics has become one of the most active topics in the frontier of condensed matter physics and quantum physics.

Owing to the particle-hole redundancy of superconductor, a superconducting quasiparticle is the quantum superposition of electron and hole. It provides an opportunity for emergence of MZM on some topological defects of a superconductor. MZM should be at the absolute zero energy, owing to the coexistence of Majorana symmetry and particle–hole symmetry (\(C_{E}^{\dagger}=C_{-E}\)). Vortex is a typical topological defect. It appears spontaneously in type-II superconductors under a sufficient larger magnetic field~\citep{10-abrikosov2004nobel,11-blatter1994vortices,12-suderow2014imaging}. In the vicinity of a vortex, the superconducting order parameter is spatially nonuniform. The amplitude of superconducting order parameter decreases when moving to the vortex center at which the amplitude vanishes, while the phase changes with the azimuth around the vortex. Note that the vorticity is defined as the ratio of the change of superconducting phase to the change of azimuth. Vortex can be regarded as a superconducting version of quantum well, in which superconducting quasiparticles form bound states. It can be described by Bogoliubov-de Genne (BdG) equation,
\begin{equation}\label{eq:1}
H_{\mathrm{BdG}}=H_{0}(\boldsymbol{r}) \boldsymbol{\tau}_{z}+\varDelta(\boldsymbol{r}) \mathrm{e}^{\mathrm{i} \theta(x, y)} \boldsymbol{\tau}_{x}
\end{equation}
where \(H_{0}(\boldsymbol{r})\) is the Fourier transformation of normal states Hamiltonian \(H_{0}(\boldsymbol{k})=(\hbar \boldsymbol{k})^{2} /(2 m)-\mu\). \(\varDelta(\boldsymbol{r})\) is the amplitude of superconducting gap. \(\theta(x,y)\) is the azimuth in the real space, describing the superconducting phase winding around the vortex. \(\tau_{x}\) and \(\tau_{z}\) are the Pauli matrices in particle-hole space. and \(m\) is the effective mass, and \(\mu\) is the chemical potential. When the superconducting gap is nodeless, the solution of BdG equation manifests the well-defined vortex bound states. Owing to in-plane rotational symmetry, the energy of the vortex bound states can be expressed by the eigenvalue of angular momentum (\(\nu\)). In conventional \(s\)-wave superconductors, the vortex bound state is known as Caroli-de Gennes-Matricon (CdGM) bound state~\citep{13-caroli1964bound,14-hess1989scanning,15-gygi1990electronic}. Zero energy mode is forbidden in the level sequence, which can be expressed as \(E_{n} \approx \varDelta / k_{\mathrm{F}} \zeta \approx(n+1 / 2) \varDelta^{2} / E_{\mathrm{F}}\), here \(n+1/2=\nu\), \(n\) is an integer, \(k_\mathrm{F}\) is the Fermi wavevector, \(E_\mathrm{F}\) is the Fermi energy, and \(\zeta\) is the coherence length of the superconductor. Ultimately, the absence of zero-energy vortex bound state is a requirement of quantum uncertainty principle. In a conventional superconductor, it is reflected in the antiperiodic boundary conditions of the quasiparticles, which is a natural result of the \(\pi\)-flux vortex. 

There are two prerequisites in order to create MZM in a superconducting vortex. First, satisfaction of the zero mode condition. It requires to introduce an additional geometric phase with odd times of \(\pi\) into the BdG equation, so that, the energy level of vortex bound states turns to be \(E_{n} \approx n\varDelta^{2}/E_{F}\). Second, stabilization of a single zero mode. It requires to elevate the spin degeneracy of the system. Following these clues, people first realized that the chiral \(p\)-wave superconductor (\(p_{x}+ip_{y}\)) is a candidate which satisfies the requirements. The  \(p_{x}+ip_{y}\) superconducting order parameter carries the intrinsic \(\pi\) phase, which counteracts the antiperiodic boundary condition of vortex, enables the zero-energy vortex bound state~\citep{16-read2000paired,17-ivanov2001non,18-volovik1999fermion,19-senthil2000quasiparticle,20-stone2004edge}. Theoretical analysis further manifests that the \(p_{x}+ip_{y}\) superconductors are intrinsic topological superconductors in the weak pairing phase~\citep{16-read2000paired}. In a spinless \(p_{x}+ip_{y}\) superconductor, a single MZM can be stabilized in a conventional vortex \((\phi = h/(2e))\). But for spinful \(p_{x}+ip_{y}\) superconductors, two MZMs emerges in a conventional vortex, they are not protected and fuse to be a complex fermion immediately. Further studies resolve that the spin degree of freedom can be eliminated in a half-quantum vortex \((\phi = h/(4e))\), in which the superconducting phase winding only couple to one of the spin components. Thus in the spinful case, a  single MZM can be stabilized in a half-quantum vortex~\citep{17-ivanov2001non,18-volovik1999fermion}. The idea of \(p_{x}+ip_{y}\) superconductors can be traced back to the studies of 5/2 fractional quantum Hall effect~\citep{21-willett2013quantum} in which a special ground state named as Paffian state (also known as Moore-Read state) was proposed. The Paffian state maps the spin-polarized electrons under a strong magnetic field into the spinless \(p_{x}+ip_{y}\) superfluid of composite fermions under the zero field. Majorana modes were proved to appear on topological defects (\(e.g.\) domain walls, vortex) of the Paffian state of 5/2 fractional quantum Hall~\citep{22-moore1991nonabelions}. Although the proposal of intrinsic topological superconductors is crystal clear in theory, both the \(p_{x}+ip_{y}\) superconductivity and the half-quantum vortices are extremely difficult to be realized in experiments~\citep{23-mackenzie2017even}. It poses huge constrictions on the research of MZM at that time.

This deadlock was not broken until the advent of the Fu-Kane model in 2008. Fu-Kane model studies the quasiparticle behavior of a superconducting Dirac surface state in two dimension. It proves that a Dirac surface state proximitized by a conventional superconductor has equivalent low-energy quasiparticle as a spinless \(p_{x}+ip_{y}\) superconductor. MZM can emerge in a conventional vortex of a \(s\)-wave superconductor/topological insulator heterostructure. Fu-Kane Hamiltonian is
\begin{equation}\label{eq:2}
H_{\text {Fu-Kane }}=\psi^{\dagger}\left[(\boldsymbol{\sigma} \cdot \boldsymbol{k}-\mu) \boldsymbol{\tau}_{z}+\Delta \boldsymbol{\tau}_{x}\right] \psi
\end{equation}
where \(\boldsymbol{\sigma}\) is the Pauli matrix in spin space. Under the spatial inversion transformation \((\boldsymbol{r} \rightarrow-\boldsymbol{r})\), the spin of the Dirac surface state is rotated by \(2\pi\) (give a minus sign) owing to spin-momentum locking (\(\boldsymbol{\sigma}\cdot\boldsymbol{k}\)), thus the superconducting pairing \(c^{\dagger}_{k\uparrow}c^{\dagger}_{-k\downarrow}\) remain unchanged, compatible with the proximitized \(s\)-wave pairing. The Fu-Kane model satisfies the two prerequisites mentioned above for the creation of MZM. But unlike the \(p_{x}+ip_{y}\) superconductor, the extra phase, which is required to enable the zero mode, is provided by the spin-momentum locking of Dirac surface state, and the spin degeneracy of electrons is elevated by the spin-orbital coupling~\citep{25-hasan2010colloquium,26-qi2011topological,27-hsieh2009tunable}. By taking the transformation \(c_{k}=\left(\psi_{\uparrow k}+\mathrm{e}^{\mathrm{i} \theta_{k}} \psi_{\downarrow k}\right) / \sqrt{2}\), where \(\theta_{k}\) is the azimuth in \(k\)-space, Fu-Kane Hamiltonian becomes,
\begin{equation}\label{eq:3}
H_{\mathrm{eff}}=\sum_{k}(v|\boldsymbol{k}|-\mu) c_{k}^{\dagger} c_{k}+\left(\Delta \mathrm{e}^{\mathrm{i} \theta_{k}} c_{k}^{\dagger} c_{-k}^{\dagger}+\text { h.c. }\right) / 2
\end{equation}
Eq.(\ref{eq:3}) describes an isotropic electronic band taking spinless \(p_{x}+ip_{y}\) superconducting pairing. Thus it proves the equivalence between the Fu-Kane model and a spinless \(p_{x}+ip_{y}\) superconductor. A stable single MZM can emerge in a conventional vortex of the Fu-Kane model.

\begin{figure*} 
\begin{centering}
\includegraphics[width=2\columnwidth]{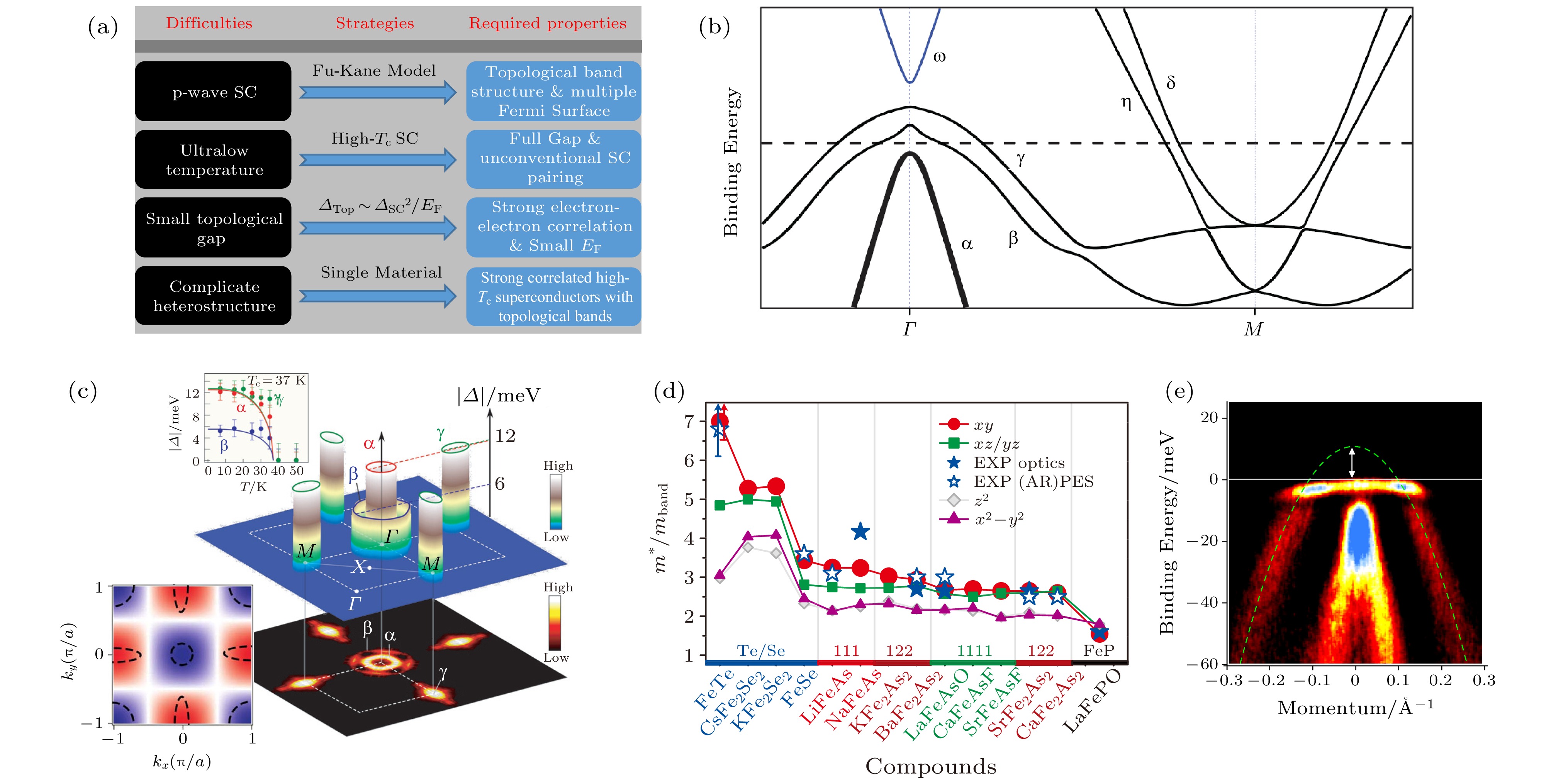}
\par\end{centering}
\caption{\textbf{Iron-based superconductors as a promising Majorana platform}. (a) the original idea for searching for Majorana zero mode (MZM) in iron-based superconductors; (b) typical band structure of iron-based superconductors, the orbital characters of each band are, \(\alpha~(d_{xz})\); \(\beta~(d_{yz})\); \(\gamma~(d_{xy})\); \(\eta~(d_{xy})\); \(\delta~(d_{xz})\); \(\omega~(p_{z})\) ; (c) typical Fermi surfaces and superconducting order parameters of iron-based superconductors~\citep{74-ding2008observation,76-richard2015arpes}; (d) mass renormalization among different compounds, indicating strong electron-electron correlations in iron-based superconductors~\citep{81-yin2011kinetic}; (e) band structure near the \(\Gamma\) point of Fe(Te,Se) single crystals measured by ARPES~\citep{82-rinott2017tuning}.}
\label{fig1}
\end{figure*}

Owing to the successful circumvention of \(p_{x}+ip_{y}\) superconductor, Fu-Kane model starts a new phase of Majorana research by facilitating the advances from theoretical hypothesis to experimental realization. Enlightened by the idea of Fu and Kane, various platforms were designed to realize Majorana modes in condensed matter systems, and some experimental evidence of MZM has been reported. These Majorana platforms can be categorized according to the physics mechanism as follows, 1) Zeeman gap assisted Rashba nanowires~\citep{28-lutchyn2010majorana,29-oreg2010helical,30-potter2010multichannel}, including semiconductor nanowires~\citep{31-mourik2012signatures,32-gazibegovic2017epitaxy,33-lutchyn2018majorana,34-prada2019andreev} and Au nanowires~\citep{35-potter2012topological,36-manna2020signature,37-xie2020topological}; 2) spin chains with strong exchange coupling, including magnetic atom chains~\citep{38-braunecker2010spin,39-nadj2013proposal,40-nadj2014observation,41-jeon2017distinguishing} and magnetic carbon nanotubes~\citep{42-desjardins2019synthetic}; 3) chiral topological superconductivity on a Yu-Shiba lattice~\citep{43-rontynen2015topological,44-li2016two,45-rachel2017quantized,46-menard2017two,47-menard2019yu,48-palacio2019atomic}; 4) superconducting proximitized quantum anomalous Hall~\citep{49-qi2010chiral,50-chen2018emergent,51-lian2018topological,52-he2017chiral,53-shen2020spectroscopic,54-kayyalha2020absence}; 5) Fu-Kane magnetic boundary states~\citep{24-fu2008superconducting}, including Majorana end mode in one dimension~\citep{55-jack2019observation} and chiral Majorana modes in two dimension~\citep{56-fu2009probing,57-akhmerov2009electrically}; 6) helical Majorana modes in Fu-Kane Josephson junctions~\citep{24-fu2008superconducting,58-williams2012unconventional}; 7) MZM in Fu-Kane vortex~\citep{24-fu2008superconducting,59-wang2012coexistence,60-xu2014artificial,61-xu2015experimental,62-sun2016majorana}. In these platforms, superconductivity is provided by an \(s\)-wave superconductor; and the realization of MZM relies on combing another topological material with the superconductor to form a heterostructure. Due to the low critical temperature (\(T_\mathrm{c}\)) and small superconducting gap of the \(s\)-wave superconductor, the topological gap (\(\varDelta_{\text {Top }} \approx \varDelta^{2} / E_{\mathrm{F}} \ll 0.1~\mathrm{meV}\)), which separates the MZMs from the lowest excitations, is also very small. It requires a very low experimental temperature (\(T_{\exp }<100~\mathrm{mK}\))  to make a reliable observation of MZMs. It not only increases the experimental cost, but also makes MZM more vulnerable to other trivial quasiparticles, thus hinders the further studies of MZMs in experiments. In addition, the heterostructure induces uncontrollable disorders and fabrication complexity~\citep{63-takei2013soft}, and may induce other trivial mechanisms in the experiments which mimic Majorana-like signals, making the unequivocal confirmation of MZM difficult~\citep{34-prada2019andreev,54-kayyalha2020absence}. Therefore, a new leap in the Majorana research needs a paradigm innovation of material platform of MZMs.

In order to explore a better material platform of MZMs, here we review the difficulties suffered in the previous Majorana platforms systematically. Symptomatic solutions are found through the one-by-one analysis (Fig.~\ref{fig1}(a)). Firstly, the intrinsic \(p\)-wave superconductivity is rare in nature and difficult to be realized. This difficulty is solved by the Fu-Kane model, by using the spin Berry phase of the topological surface state to replace the superconducting phase winding of intrinsic topological superconductor. This requires the coexistence of topological band structure and superconductivity. Instead of making a Fu-Kane heterostructure, we realized that in a single material such the coexistence requires a multiband structure to possess both topology and superconductivity. Secondly, the required experimental temperature is extremely low. A MZM is a superconducting quasiparticle, the survival temperature of the MZM is positively correlated to the value of  \(T_\mathrm{c}\). Incorporating a high-\(T_\mathrm{c}\) superconductor can effectively increase the survival temperature of the MZM. Furthermore, considering the necessary conditions of the existence of a well-defined vortex bound state, the superconductor should be full-gapped. So that the \(s\pm\) wave unconventional superconductors~\citep{64-mazin2008unconventional,65-kuroki2008unconventional,66-seo2008pairing,67-chen2009strong}, and the nodeless \(d\)-wave unconventional superconductors~\citep{68-maier2011d,69-khodas2012interpocket,70-agterberg2017resilient,71-lee2018routes} may be promising choices. Thirdly, the tiny topological gap, separating the MZM from the lowest quasiparticles, causes severe quasiparticle poisoning. In vortices, the topological gap is proportional to \(\varDelta^{2}/E_\mathrm{F}\). A large topological gap requires a small Fermi energy, which can be realized in a material with strong electron-electron correlations. Fourthly, the heterostructure induces difficulties. The most direct way to remove those difficulties is to use a single material which integrates all of the three properties mentioned above. To sum up, a better Majorana platform could be potentially found in a material combining the properties of multiple bands, strong electron-electron correlations, high-\(T_\mathrm{c}\) superconductivity and topological band structure.

Our research experiences on both superconductors and topological matters inspired the idea that the iron-based superconductors could be such kind of candidates. First of all, the low-energy electronic states of iron-based superconductors~\citep{72-paglione2010high,73-chen2014iron} are mainly composed of the \(t_{2g}\) orbitals of Fe atom (\(d_{x z}, d_{y z}, d_{x y} \) orbitals) and the \(p\) orbitals of chalcogenides. Near the Fermi level, three hole bands appear in the center of the Brillouin zone, and two electron bands appear at the corners of the Brillouin zone (Fig.~\ref{fig1}(b)). These bands pass through the Fermi level and form multiple Fermi pockets. It is observed experimentally that a full gap opens on the Fermi surfaces below \(T_\mathrm{c}\)~\citep{74-ding2008observation}. Although a universally-acknowledged superconducting mechanism is not reached for the iron-based superconductors, a large number of theoretical and experimental results support the \(s\pm\)-wave pairing related to spin fluctuations (Fig.~\ref{fig1}(c))~\citep{74-ding2008observation,75-hirschfeld2011gap,76-richard2015arpes,77-hanaguri2010unconventional,78-liu2019spectroscopic,79-liu2019detection,80-chen2020bosonic}. Secondly, many iron-based superconductors have strong electron-electron correlations~\citep{81-yin2011kinetic}, which leads to strong mass renormalization of their bands. For example, the mass renormalization factor of Fe(Te,Se) is close to that of cuprate superconductors (about 7) (Fig.~\ref{fig1}(d)). Therefore, the bulk bands of Fe(Te,Se) acquire very small Fermi energies (about 10 meV) (Fig.~\ref{fig1}(e)), which were clearly observed by angle-resolved photoelectron spectroscopy (ARPES) experiments~\citep{82-rinott2017tuning}. Finally, some theories showed that the iron-based superconductors have topologically non-trivial band structures~\citep{83-hnwlxb,84-hao2019topological}. Topological band inversion was predict in multiple compounds, such as FeSe/STO monolayer~\citep{85-hao2014topological}, Fe(Te,Se) monolayer~\citep{86-wu2016topological}, Fe(Te,Se) single crystals~\citep{87-wang2015topological}, CaFeAs\(_{2}\)~\citep{88-wu2015cafeas}, and etc. Theories also predicted the appearance of topological band structure owing to electronic orders~\citep{89-ran2009nodal,90-morinari2010topological,91-richard2010observation,92-huynh2011both,93-hao2017topological,94-wu2016nematic,95-tan2016observation,96-watson2016evidence,97-phan2017effects}. For example, the evidence of topological bands has been found in the SDW state of BaFe\(_{2}\)As\(_{2}\)~\citep{89-ran2009nodal,90-morinari2010topological,91-richard2010observation,92-huynh2011both} and the nematic state of FeSe single crystals~\citep{94-wu2016nematic,95-tan2016observation,96-watson2016evidence,97-phan2017effects}. 

Most importantly, some early experimental evidence of topological bands was found in Fe(Te,Se) single crystals that started the notion of the iron-based Majorana platform~\citep{87-wang2015topological,98-zhang2014observation,99-yin2015observation}. In 2014, by implementing \(in\)-\(situ\) surface dosing on FeTe\(_{0.55}\)Se\(_{0.45}\) single crystals, an electron band was found above the original Fermi level~\citep{98-zhang2014observation}. Later, the signature of band inversion was observed by synchrotron-based ARPES experiments with variable photon energy and polarization~\citep{87-wang2015topological}. At the same time, a scanning tunneling microscopy/spectroscopy (STM/S) experiment found a robust zero-energy bound state on the interstitial Fe atom~\citep{99-yin2015observation}. All these pieces of evidence suggest non-trivial topology in FeTe\(_{0.55}\)Se\(_{0.45}\) single crystals. These clues attracted the attentions on FeTe\(_{0.55}\)Se\(_{0.45}\) single crystals and inspired more systematic experiments to identify the superconducting topological surface states~\citep{100-zhang2018observation,101-zhang2019multiple} and vortex MZM~\citep{102-wang2018evidence,103-kong2019majorana,104-kong2019half,105-zhu2020nearly,106-liu2019new}.

In this review, we systematically summarize the original ideas and research progress of the emergent vortex MZM in Fe(Te,Se) single crystals with rich details. We aim to bridge the gap between the classical Majorana theories and the emerging iron-based Majorana platform. The rest of the paper is organized as follows. In Section~\ref{sec:2}, we introduce the topological band structure and the experimental observation of the surface states~\citep{87-wang2015topological,100-zhang2018observation,101-zhang2019multiple}. In Section~\ref{sec:3}, the zero-bias conductance peak (ZBCP) was clearly observed in the vortices. By introducing a careful validation of vortex MZMs~\citep{102-wang2018evidence}, we discuss the issue of ``what'' is the ZBCPs. In Section~\ref{sec:4}, we discuss ``how'' the vortex MZM emerges in a real material, \(e.g.\) Fe(Te,Se) single crystals. The wavefunction and quasiparticle poisoning behavior of the vortex MZM are carefully introduced. The effects of various realistic issues on emergence of vortex MZM are identified~\citep{102-wang2018evidence}. In Section~\ref{sec:5}, we review the Majorana nature of the vortex MZM~\citep{105-zhu2020nearly}. The experimental evidence of particle-hole equivalence is introduced in combination with the theory of Majorana induced resonant Andreev reflection. In Section~\ref{sec:6}, we review the topological nature of the vortex MZM~\citep{104-kong2019half}. Based on the energy eigenvalue and wavefunction distribution of the vortex bound states, we introduce the half-integer level shift of bound states, which is concomitant with emergence of vortex MZM. In Section~\ref{sec:7}, we switch our angle to focus the realistic details, which is an engineering concern. It has been observed that MZM is absent in some vortices on Fe(Te,Se). We analyze the possible mechanisms governing this phenomenon~\citep{104-kong2019half}, which may benefit the exploration of practical applications in the future. In Section~\ref{sec:8}, we present the conclusions and outlooks. The research progresses of the iron-based Majorana platform are briefly reviewed, and the main obstacles in the way to topological quantum computation are briefly analyzed.

\begin{figure*} 
\begin{centering}
\includegraphics[width=2\columnwidth]{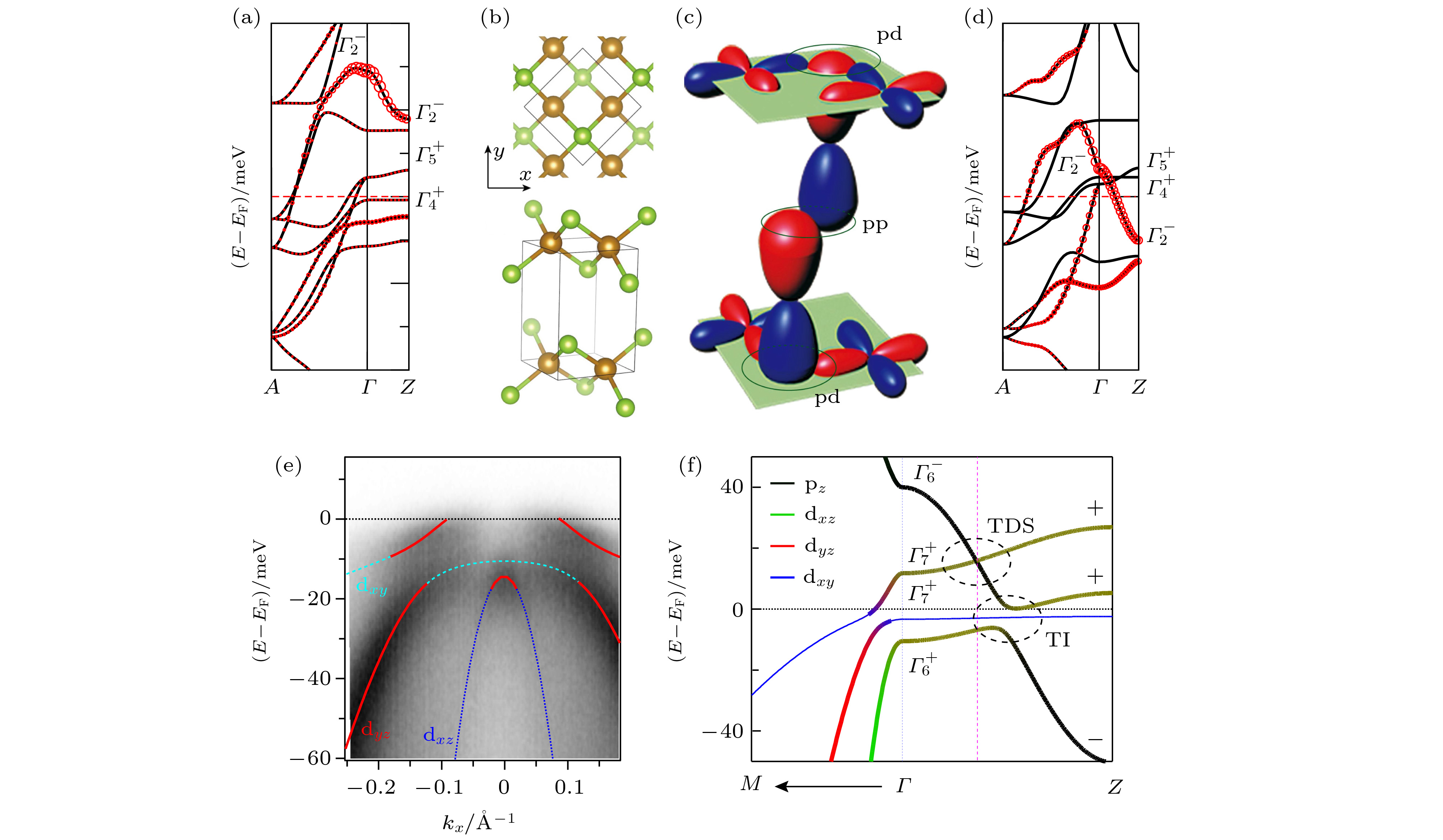}
\par\end{centering}
\caption{\textbf{Mechanism of topological band structure and band inversion of Fe(Te,Se)}. (a) first-principle calculation of band structure of FeSe (without SOC), the size of red circles represents the components of the \(p_{z}\) orbital~\citep{87-wang2015topological}; (b) crystal structure of Fe(Te,Se)~\citep{87-wang2015topological}; (c) band inversion mechanism and orbital overlapping in Fe(Te,Se)~\citep{87-wang2015topological}; (d) first-principle calculation of band structure of FeTe\(_{0.5}\)Se\(_{0.5}\) (without SOC)~\citep{87-wang2015topological}; (e) experimental band structure around \(\Gamma\) in FeTe\(_{0.55}\)Se\(_{0.45}\) measured by ultra-high resolution laser ARPES~\citep{101-zhang2019multiple}; (f) realistic topological band structure in FeTe\(_{0.55}\)Se\(_{0.45}\) (with SOC). TDS stands for topological Dirac semimetal, and TI stands for topological insulator~\citep{101-zhang2019multiple}.
}
\label{fig2}
\end{figure*}

\section{Topological band structure}\label{sec:2}
Iron-based superconductors can have multiple topological states, produced by the concurrence of strong interlayer coupling and spin-orbital coupling (SOC)~\citep{83-hnwlxb,84-hao2019topological,87-wang2015topological,101-zhang2019multiple}. Topological band inversion can be engineered by adjusting the \(k_{z}\) dispersion of the \(p_{z}\) orbital which belongs to the anions. Under appropriate conditions, the odd-parity \(p_{z}\) orbital intersects with the even-parity \(d\) orbitals (\(d_{xz}\), \(d_{yz}\), \(d_{xy}\)) along \(\Gamma-Z\). Those nontrivial band crosses form topological insulator state or topological Dirac semimetal state, contingent upon the same or different angular momentums of the crossing bands respectively. We note that the band inversion along \(k_{z}\) is not unique to iron-based superconductors, it occurs ubiquitously in the 1T phase of transition metal dichalcogenides (such as PtTe\(_{2}\), PdTe\(_{2}\), and NiTe\(_{2}\))~\citep{107-bahramy2018ubiquitous,108-clark2018fermiology,109-mukherjee2019fermi}. In principle, topological band structure can be produced in all the iron-based superconductors under sufficient tuning. But some compounds are naturally so under ideal conditions, topological band inversions have been verified in Fe(Te,Se), LiFeAs and CaKFe\(_{4}\)As\(_{4}\) by first-principles calculations and ARPES measurements. Taking Fe(Te,Se) single crystals as an example, in this section we introduce the mechanism of band inversion and experimental progress on the studies of topological band structure.

\subsection{Mechanism of band inversion}\label{sec:2.1}
A first-principles calculation of FeSe single crystals is shown in Fig.~\ref{fig2}(a)~\citep{87-wang2015topological}. Three bands appear near the Fermi level in the absence of SOC. The corresponding irreducible representations are marked as \(\Gamma^{-}_{2}\) (\(p_{z}\)/\(d_{xy}\) antibonding orbital), \(\Gamma^{+}_{4}\) (\(d_{xy}\) orbital), and \(\Gamma^{+}_{5}\) (\(d_{xz}\)/\(d_{yz}\) orbital), respectively. The odd-parity \(\Gamma^{-}_{2}\) band is always above the even-parity bands, so there is no topological band inversion in FeSe. Theoretically, the energy position of \(\Gamma^{-}_{2}\) band can be adjusted by the lattice constant. As shown in Fig.~\ref{fig2}(c), the \(d_{xy}\) orbital of Fe atoms lies on the Fe plane (shown as the cyan plane), its electron cloud spreads a little bit along the c-direction. On the contrary, the \(p_{z}\) orbital of Se atoms is distributed vertically, with one end coupling with the \(d_{xy}\) orbital on the Fe plane, and the other end crossing the van der Waals (vdW) gap and coupling with the \(p_{z}\) orbital of the adjacent Se-Fe-Se layer. By changing the distance between the anion and the Fe plane, the relative strength of the intralayer p-d coupling and the interlayer p-p coupling can be effectively controlled. The competition between them determines the behavior of the \(\Gamma^{-}_{2}\) band. The tellurium substitution on the anionic site can increase the distance between the anion and the Fe plane, thus weakens the intralayer p-d coupling. So the \(\Gamma^{-}_{2}\) band moves to lower energy to adapt to this change. On the other hand, the tellurium substitution enhances the interlayer p-p coupling, which leads to a larger \(k_{z}\) dispersion of the \(\Gamma^{-}_{2}\) band. As a consequence, the odd-parity \(\Gamma^{-}_{2}\) crosses the even-parity \(\Gamma^{+}_{4}\) and \(\Gamma^{+}_{5}\) along \(\Gamma\)-Z, producing the topological band inversion at Z (Fig.~\ref{fig2}(d)).

\begin{figure*} 
\begin{centering}
\includegraphics[width=2\columnwidth]{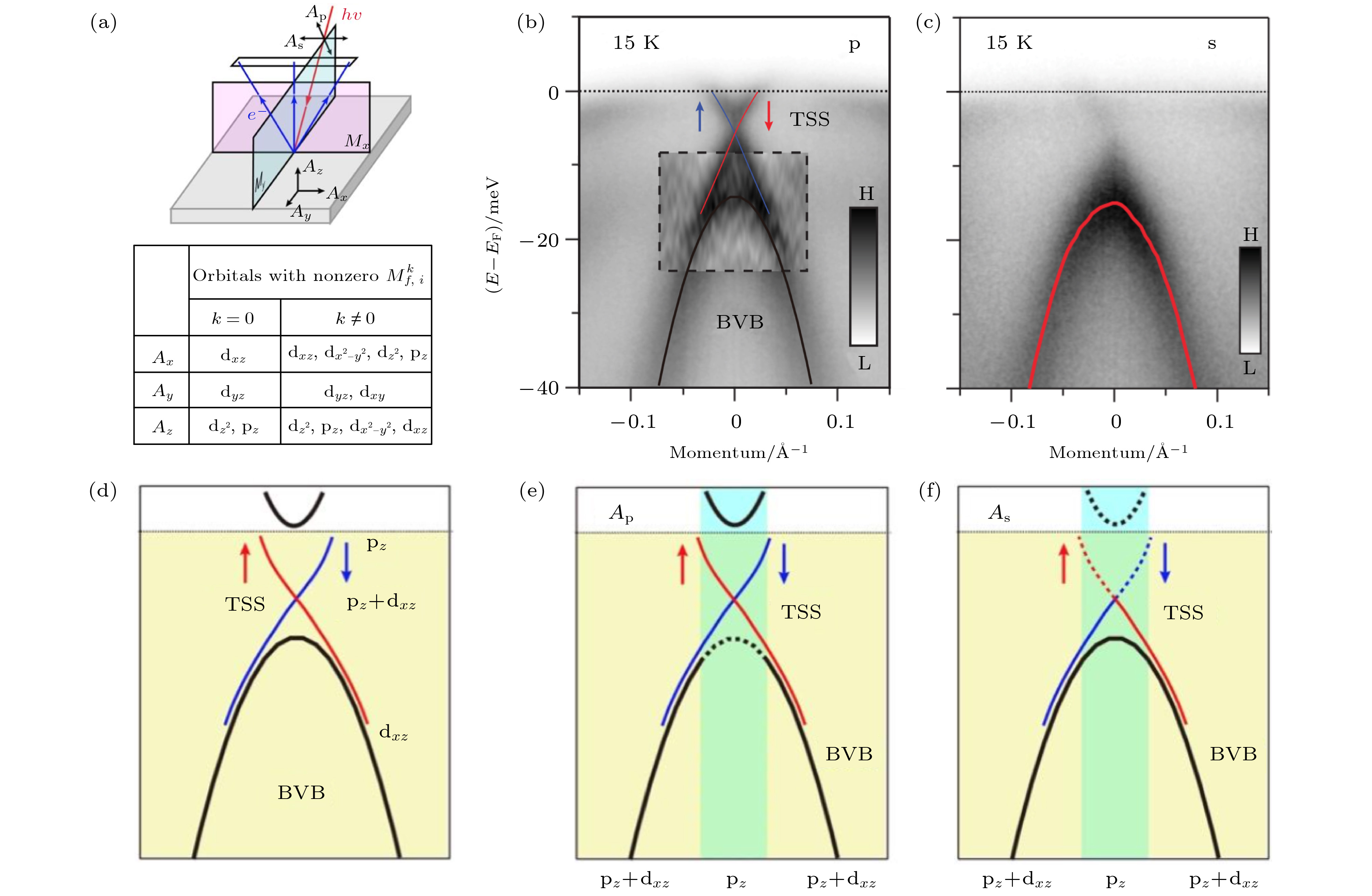}
\par\end{centering}
\caption{\textbf{Experimental observation of the linear-dispersed Dirac surface states in FeTe\(_{0.55}\)Se\(_{0.45}\)}. (a) matrix element effect which defines the selection rule of ARPES intensity, depends on the relationship between photon polarization and electron orbitals; (b) Dirac surface states observed under p-polarization; (c) \(d_{xz}\) bulk bands observed under s-polarization; (d) orbital characters around  \(\Gamma\) in FeTe\(_{0.55}\)Se\(_{0.45}\); (e), (f) orbital character determined by the matrix element analysis under p- and s-polarization, respectively. The orbital characters marked at the bottom represent the active orbitals under certain polarization and momentum. The dashed parts in the band structure represent the intensity suppressed by the selection rule~\citep{100-zhang2018observation}.}
\label{fig3}
\end{figure*}

With SOC, the \(\Gamma^{-}_{2}\) band evolves to be \(\Gamma^{-}_{6}\), which is composed of the \(p_{z}\) orbital; the \(\Gamma^{+}_{4}\) band evolves to be \(\Gamma^{+}_{7}\), which is composed of the \(d_{xy}\) orbital; and the \(\Gamma^{+}_{5}\) band evolves to be \(\Gamma^{+}_{6}\) and \(\Gamma^{+}_{7}\), which are composed of the \(d_{xz}\) orbital and the \(d_{yz}\) orbital respectively. DFT calculations of FeTe\(_{0.5}\)Se\(_{0.5}\) single crystals show~\citep{87-wang2015topological} that the \(d_{xy}\) band (\(\Gamma^{+}_{7}\)) is above the \(d_{yz}\) band (\(\Gamma^{+}_{7}\)), both of them cross the Fermi level around \(\Gamma\). It leads to three intersections between the odd-parity \(p_{z}\) band and the even-parity bands along \(\Gamma\)-Z. However, ultra-high resolution ARPES experiment observed only one bulk band (\(d_{yz}\)) crossing the Fermi level at \(\Gamma\) (Fig.~\ref{fig2}(e))~\citep{101-zhang2019multiple}. This discrepancy invokes the investigation of the realistic details of the band inversion along \(\Gamma\)-Z. Owing to orbital-dependent mass renormalization, the band width of the \(d_{xy}\) orbital is very small~\citep{81-yin2011kinetic}. It pushes the \(d_{xy}\) band sinking below the Fermi level, leads to the absence of the \(d_{xy}\) Fermi surface. In addition, the \(d_{xy}\) band intersects with the \(d_{yz}\) band below the Fermi level, and a hybridization gap is clearly observed by ARPES around the intersection (Fig.~\ref{fig2}(e))~\citep{101-zhang2019multiple}, which is an expected result owing to the same irreducible representation of the two bands (\(\Gamma^{+}_{7}\)). Later on, theorists reproduced the realistic band structure (with SOC) of FeTe\(_{0.55}\)Se\(_{0.45}\) single crystals by using the \(k\cdot p\) model (Fig.~\ref{fig2}(f))~\citep{101-zhang2019multiple}. Different from the DFT calculation, there are only two topological intersections along \(\Gamma\)-Z, one is a gapped intersection (\(p_{z}\) and \(d_{xz}\)), in which a strong topological insulator state occurs, the other is a protected intersection (\(p_{z}\) and \(d_{yz}\)), which is a topological Dirac semimetal state protected by the rotational symmetry. The strong topological insulator state is protected by the size of the hybridization gap. Tellurium substitution enhances the hybridization gap due to a larger SOC. It is beneficial for emergence of Dirac surface states.

\subsection{Discovery of topological insulator state}\label{sec:2.2}

Several pieces of evidence of Dirac surface states were observed by ARPES, confirming the topological insulator state in FeTe\(_{0.55}\)Se\(_{0.45}\) single crystals~\citep{100-zhang2018observation}. First, The linear-dispersed Dirac surface state was clearly observed in ARPES spectra. Second, the orbital characters were identified by the matrix element analysis in a polarization dependent ARPES~\citep{76-richard2015arpes,110-wang2012orbital}. Third, the signature of spin-momentum locking was directly observed through spin-resolved experiments~\citep{25-hasan2010colloquium,26-qi2011topological,27-hsieh2009tunable}. Fourth, an ultra-low temperature ARPES measurement found an isotropic superconducting gap opening on the Dirac surface state below \(T_\mathrm{c}\) (14.5 K). Thus the requirements of Fu-Kane model~\citep{24-fu2008superconducting} are fulfilled in a single material platform of FeTe\(_{0.55}\)Se\(_{0.45}\), rather than a heterostructure as studied previously (Fig.~\ref{fig1}(a)). Furthermore, the \(E_\mathrm{F}\) of the Dirac surface state was found to be very small in FeTe\(_{0.55}\)Se\(_{0.45}\), indicates a potential for a clear observation of vortex MZM.

ARPES is capable to distinguish both the momentum and the energy of electrons simultaneously, so it is widely used in band structure measurements on topological materials~\citep{111-lv2019angle}. It is well known that, the spectral intensity of ARPES~\citep{112-hufner2013photoelectron,113-damascelli2004probing} is
\begin{equation}\label{eq:4}
I(\boldsymbol{k}, \omega)=I_{0}\left|M_{f, i}^{k}\right|^{2} \boldsymbol{A}(\boldsymbol{k}, \omega) f(\omega)
\end{equation}
where \(f(\omega)\) is the Fermi-Dirac function;
\begin{equation}\label{eq:5}
\begin{aligned}
& \boldsymbol{A}(\boldsymbol{k}, \omega)=-\frac{1}{\pi} \operatorname{Im} G^{-}(\boldsymbol{k}, \omega) \\
=& \sum_{m}\left|\left\langle\psi_{m}^{N-1}\left|c_{k}\right| \psi_{i}^{N}\right\rangle\right|^{2} \delta\left(\omega-E_{m}^{N-1}-E_{i}^{N}\right)
\end{aligned}
\end{equation}
is the spectral function of the material;
\begin{equation}
M_{f, i}^{k}=\left\langle\psi_{f}^{k}|\boldsymbol{A} \cdot \boldsymbol{p}| \psi_{i}^{k}\right\rangle
\end{equation}
is the ARPES matrix element, and determines the orbital selection rule of ARPES experiments. The ARPES signal exists only if \(M_{f, i}^{k}\) is nonzero. Since the emission photoelectron is a plane wave, \(\psi_{f}^{k}\) is an even function relative to the signal acquisition plane (namely the slit plane of the ARPES analyzer, see the pink plane in Fig.~\ref{fig3}(a)). In order to get nonzero ARPES signal, the parity of photon polarization (\(\boldsymbol{A}\cdot\boldsymbol{p}\)) and the orbital \(\psi_{i}^{k}\) should be same. In the case of normal emission, the wavefunction \(\psi_{f}^{k}\) is also an even function with respect to the plane vertical to the signal acquisition plane (the blue plane in Fig.~\ref{fig3}(a)). The parity of \(\boldsymbol{A}\cdot\boldsymbol{p}\) and \(\psi_{i}^{k}\) relative to the vertical plane also affects the selection rule. We summarize the orbital selection rule under different experimental conditions in the table of Fig.~\ref{fig3}(a)~\citep{100-zhang2018observation}.

The topological insulator state in Fe(Te,Se) single crystals is originated from the topological band inversion of \(p_{z}\) and \(d_{xz}\) bands. Those two orbitals compose the main orbital characters of the Dirac surface states due to the SOC. The expected orbital characters around  \(\Gamma\) are shown in Fig.~\ref{fig3}(d). The experimental geometry of the p-polarization contains both \(\boldsymbol{A}_{y}\) and \(\boldsymbol{A}_{z}\), which is beneficial to the observation of \(p_{z}\) orbital components, but suppresses the signal of the \(d_{xz}\) orbital at  \(\Gamma\) (Fig.~\ref{fig3}(e)). Experimental results are shown in Fig.~\ref{fig3}(b). A cone-like band structure can be clearly observed, but the top of the hole band is rather ambiguous which can only be distinguished in the second derivative. On the other hand, the experimental geometry of s-polarization only contains the \(\boldsymbol{A}_{x}\) polarized photons, which strongly suppresses the signal of the \(p_{z}\) orbital near  \(\Gamma\), so that the topological surface states are almost invisible. But the s-polarization has a good selectivity to the \(d_{xz}\) orbital, and indeed the hole-like \(d_{xz}\) band was well resolved in the experiment (Fig.~\ref{fig3}(c)). These results are fully consistent with the matrix-element analysis which resolves the orbital characters of each band. In summary, the hole-like bulk band below the SOC gap is mainly composed of the \(d_{xz}\) orbital, the upper half branch of the Dirac surface states is mainly composed of the \(p_{z}\) orbital, the lower half branch is composed of the \(p_{z}\) and \(d_{xz}\) orbitals, and the electron-like bulk band above the SOC gap can be inferred as the \(p_{z}\) orbital (Fig.~\ref{fig3}(d)).

\begin{figure*} 
\begin{centering}
\includegraphics[width=2\columnwidth]{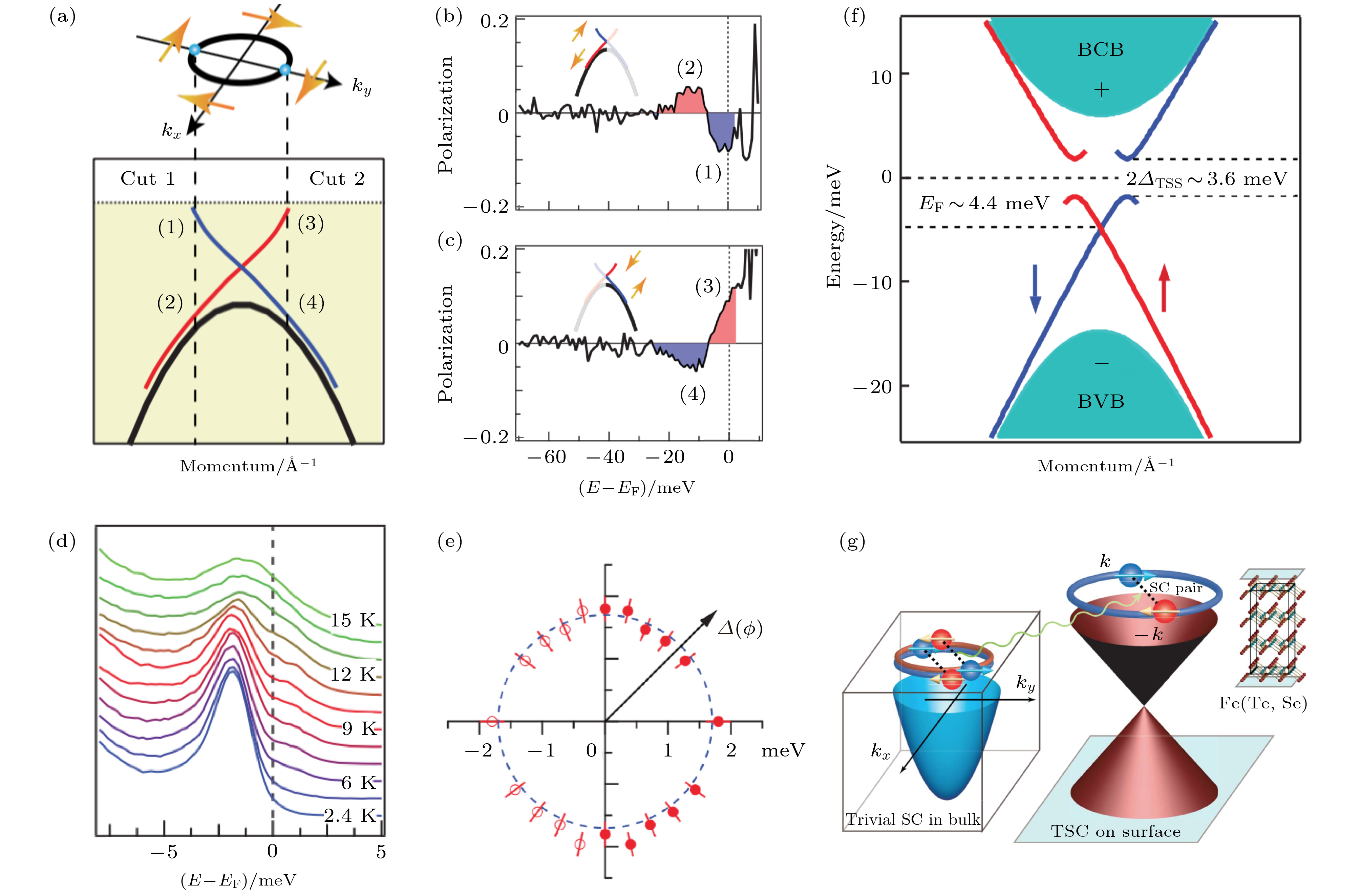}
\par\end{centering}
\caption{\textbf{Spin-momentum locking and isotropic superconducting gap on the Dirac surface state}. (a) spin-momentum locking feature in FeTe\(_{0.55}\)Se\(_{0.45}\) single crystals~\citep{100-zhang2018observation}; (b), (c) spin-resolved ARPES data measured along Cut 1 and Cut 2 in panel (a), respectively~\citep{100-zhang2018observation}; (d) temperature-dependent energy distribution curves measured at \(k_\mathrm{F}\) of the Dirac surface state indicates a superconducting gap of the Dirac surface state opening below 14.5 K, which is the bulk \(T_\mathrm{c}\)~\citep{100-zhang2018observation}; (e) isotropic superconducting gap on the Dirac surface state~\citep{100-zhang2018observation}; (f) summary of the main observations of the Dirac surface state on FeTe\(_{0.55}\)Se\(_{0.45}\) single crystals, \(i.e.\), topological band inversion, linear dispersion, spin-momentum locking, large superconducting gap, small Fermi energy~\citep{102-wang2018evidence}; (g) Dirac surface state of FeTe\(_{0.55}\)Se\(_{0.45}\) single crystals acquires an effective spinless pairing due to the proximity effect from the \(s\pm\) wave bulk superconductivity, which satisfies all of the requirements of Fu-Kane model~\citep{100-zhang2018observation}.}
\label{fig4}
\end{figure*}

Spin-momentum locking is a fingerprint of the Dirac surface state. The Dirac electron obtains a \(\pi\)-flux spin Berry phase on the Fermi surface (Fig.~\ref{fig4}(a)), which is decisive evidence of topological insulator state. As shown in Fig.~\ref{fig4}(a), a pair of spin-polarized energy distribution curves (Cut 1 and Cut 2) are measured at the selected Fermi points of the Dirac surface states (Fig.~\ref{fig4}(b) and Fig.~\ref{fig4}(c)). The spin signals are reversed respect to the Dirac point, indicating that the spin polarization of the upper branch and lower branch of the Dirac surface states are different. In addition, the spins of Cut 1 and Cut 2 are opposite at the same energy. These results identify the spin-momentum locking of the observed linear-dispersed bands, providing strong evidence of the Dirac surface state.
 
FeTe\(_{0.55}\)Se\(_{0.45}\) single crystals is regarded as a \(s\pm\)-wave superconductor~\citep{77-hanaguri2010unconventional,114-chen2019direct} (the momentum resolved \(s\pm\)-wave superconducting order parameters is shown in Fig.~\ref{fig1}(c)). Since the Dirac surface state is mainly concentrated in the center of the Brillouin zone (\(k_{\mathrm{F}} \approx 0.02 \mathrm{\AA}^{-1}\)), the superconductivity of the Dirac surface state, which is proximitized by the bulk states at different momenta, can only have a conventional \(s\)-wave. An early ARPES experiment observed two superconducting gaps (\(\varDelta_{1}\) = 1.7 meV, \(\varDelta_{2}\) = 2.5 meV) near \(\Gamma\), and one superconducting gap (\(\varDelta_{3}\) = 4.2 meV) near M~\citep{115-miao2012isotropic}. By incorporating the next-next-nearest-neighbor antiferromagnetic exchange coupling (\(J_{3}\)) in a strong coupling model~\citep{66-seo2008pairing}, a theoretical gap equation, that is \(\varDelta=\left|J_{2} \cos\left(k_{x}\right) \cos\left(k_{y}\right)-J_{3}\left[\cos \left(2 k_{x}\right)+\cos \left(2 k_{y}\right)\right] / 2\right|\), was proposed to understand the observed three gaps in the experiment. It was found that \(\varDelta_{2}\) and \(\varDelta_{3}\) are scaled very well with the gap function. The extracted \(J_{2}\)/\(J_{3}\) values are close to the experimental results of neutron scattering~\citep{116-lipscombe2011spin}. However, the minimum gap \(\varDelta_{1}\) has the smallest \(k_\mathrm{F}\) among the three gaps, and greatly deviates from the expectations of the \(J_{1}\)-\(J_{2}\)-\(J_{3}\) model~\citep{115-miao2012isotropic}. These results suggest that \(\varDelta_{2}\) and \(\varDelta_{3}\) follow \(s\pm\)-wave pairing, while \(\varDelta_{1}\) may be with a different origin. 

Due to the improvement of ARPES resolution in the recent years, we achieved clearer observation of the superconducting gaps on FeTe\(_{0.55}\)Se\(_{0.45}\)~\citep{100-zhang2018observation}. New experiments show that the Dirac surface state opens a superconducting gap of 1.8 meV (\(\varDelta_{\mathrm{surface}}\) = 1.8 meV) at low temperature, with an isotropic distribution along the Fermi surface (Fig.~\ref{fig4}(e)). By measuring the temperature dependent energy distribution curves at the \(k_\mathrm{F}\) of the Dirac surface state, the gap closing temperature is resolved at about 14.5 K. \(\varDelta_{\mathrm{surface}}\), which is at a similar value as \(\varDelta_{1}\) in the earlier experiment, differs from the behavior of \(s\pm\)-wave, but conforms to the picture of the induced gap by \(k\)-proximity effect. It provides a reasonable understanding for the ill-scaling of \(\varDelta_{1}\) to \(J_{1}\)-\(J_{2}\)-\(J_{3}\) model. Combined with different experimental measurements~\citep{77-hanaguri2010unconventional,82-rinott2017tuning,99-yin2015observation,100-zhang2018observation,101-zhang2019multiple,102-wang2018evidence,115-miao2012isotropic,116-lipscombe2011spin,117-homes2010electronic,118-escudero2015energy,119-wu2020superconductivity}, we summarize the superconducting gaps of FeTe\(_{0.55}\)Se\(_{0.45}\) single crystals as follows: near \(\Gamma\), \(\varDelta_{\mathrm{surface}}\) = 1.8 meV and \(\varDelta^{d_{yz}}_{\mathrm{Bulk}}\) = 2.5 meV; near M, \(\varDelta^{M}_{\mathrm{Bulk}}\) = 4 meV. It is worth noting that the surface-sensitive experimental methods (\(e.g.\) ARPES and STM) are easier to observe the superconducting gap of the surface states~\citep{119-wu2020superconductivity}. 

The experimental parameters of the topological insulator state of FeTe\(_{0.55}\)Se\(_{0.45}\) are summarized in Fig.~\ref{fig4}(f). Besides the basic features, two details are worth to be especially mentioned. First, the SOC gap of the topological insulator state is about 20 meV which is much smaller than that of other strong topological insulators (\(e.g.\) about 300 meV in Bi\(_{2}\)Se\(_{3}\))~\citep{25-hasan2010colloquium}. 
Secondly, the Dirac point of the topological surface states appears near the Fermi level, which eliminates the requirement of tuning the chemical potential. The Fermi energy of the Dirac surface state is only about 4.4 meV. 
The ratio of \(\varDelta/E_\mathrm{F}\) (about 0.5) is very large in FeTe\(_{0.55}\)Se\(_{0.45}\) single crystals, which is more than two orders of magnitude larger than that in Bi\(_{2}\)Se\(_{3}\) (about 10\(^{-3}\) to 10\(^{-2}\)). 
This ideal parameter brings a hope for the realization of MZM with a large topological gap. In this sense, Fe(Te,Se) is a gift from nature~\citep{120-Lee2019spontan}.

\begin{figure*} 
\begin{centering}
\includegraphics[width=2\columnwidth]{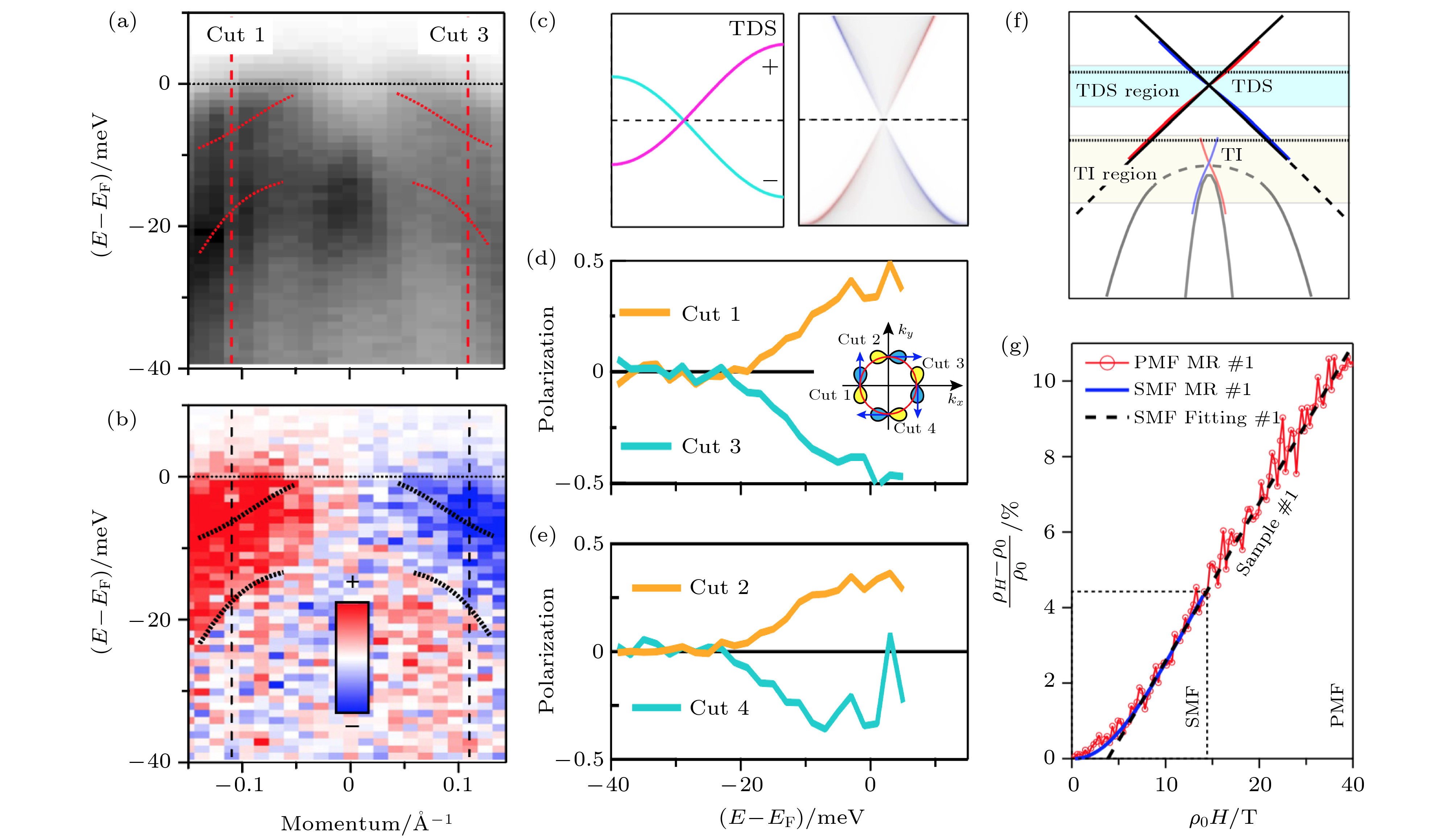}
\par\end{centering}
\caption{\textbf{Evidence of the Dirac semimetal phase in FeTe\(_{0.55}\)Se\(_{0.45}\) single crystals}. (a), (b) spin-integrated and spin-resolved ARPES spectrum around \(\Gamma\), respectively; (c) projected band structure on the (001) surface of a \(C_{4}\) symmetry protected Dirac semimetal. The spin-polarized surface states are mixed with the bulk bands; (d), (e) spin polarization of the \(d_{yz}\) bulk band measured on four representative \(k_\mathrm{F}\) points around the Fermi surface (as indicates in the inset). It is clear that the \(d_{yz}\) bulk band has the helical spin texture; (f) summary of the topological band structure along the in-plane momentum. There is a strong topological insulator phase around the Fermi level and a topological Dirac semimetal phase above it; (g) transport evidence of the Dirac semimetal phase in FeTe\(_{0.55}\)Se\(_{0.45}\) single crystals. The linear transverse magnetoresistance indicates the incorporation of bulk Dirac electrons. The transport experiments were performed at 16 K. The PMF and SMF represent pulse and static magnetic fields, respectively~\citep{101-zhang2019multiple}.}
\label{fig5}
\end{figure*} 

\subsection{Discovery of Dirac semimetal state}\label{sec:2.3}
Topological Dirac semimetal is protected by the rotational symmetry of the principal axis~\citep{121-yang2014classification}. As shown in Fig.~\ref{fig2}(f), the Dirac semimetal state is composed by the \(p_{z}\) and \(d_{yz}\) orbitals in FeTe\(_{0.55}\)Se\(_{0.45}\) single crystals. Since the Dirac point is above the Fermi level, the band structure of the bulk Dirac cone cannot be directly measured by ARPES. However, the Dirac semimetal state can be identified in ARPES experiments through the anomalous spin-polarization on a bulk band, owing to the overlap between the Fermi arc and the bulk Dirac bands on the (001) surface (Fig.~\ref{fig5}(c))~\citep{122-xu2015unconventional,123-jozwiak2016spin,124-neupane2015surface}. As mentioned above, due to the large effective mass, the \(d_{xy}\) band anticrosses with the \(d_{yz}\) band below the Fermi level. The upper branch of the \(d_{yz}\) band extends upwards participating in the formation of the Dirac semimetal state, while the lower branch connects to the \(d_{xy}\) orbital below the Fermi level (Fig.~\ref{fig5}(a)). A spin-resolved ARPES experiment observed spin polarization in the upper branch of the \(d_{yz}\) band with the opposite spin on the left and right sides (Fig.~\ref{fig5}(b)). The helical spin texture of the Fermi surface was further confirmed by spin polarization measurements of the \(d_{yz}\) band along both the \(k_{x}\) and \(k_{y}\) directions (Fig.~\ref{fig5}(d) and Fig.~\ref{fig5}(e)). In addition, there is no spin polarization in the lower branch of the \(d_{yz}\) band, which is clearly shown in the spectra below -20 meV in Fig.~\ref{fig5}(d) and Fig.~\ref{fig5}(e). It partially excludes other non-topological origins of the spin polarization. These observations provide evidence of the projected Fermi arcs on the (001) surface, support the existence of the Dirac semimetal states of Fe(Te,Se) above the Fermi level.

Owing to the absence of the Fermi surface of the \(d_{xy}\) band, the Dirac band may dominate the transport behavior on Fe(Te,Se). As shown in Fig.~\ref{fig5}(g), the magnetoresistance shows semi-classical (quadratic) behavior below 6 T, but becomes linear between 6 and 40 T. Although the anomalous linear magnetoresistance can be explained by several conventional mechanisms, such as the averaging effect in polycrystalline materials with one-dimensional Fermi surface~\citep{125-abrikosov1998quantum} and heavily disordered systems~\citep{126-parish2003non}, they are not consistent with the conditions of FeTe\(_{0.55}\)Se\(_{0.45}\) single crystals. 

In the following we show that the magnetoresistance behavior can be well explained by the Dirac semimetal state. Linear magnetoresistance also exists under the quantum limit of Landau levels. The electron degeneracy on the Landau level increases with increasing magnetic field (the degeneracy density is \(g = eB/h\)). When the magnetic field is larger than the threshold of quantum limit, all the carriers occupy the lowest Landau level~\citep{125-abrikosov1998quantum,127-abrikosov2000quantum}. A strong magnetic field is not the only requirement for realization of the quantum limit, the energy gap between the lowest Landau level and the first excited state (\(\Delta E_\mathrm{LL}\)) should be also larger than the Fermi energy (\(E_\mathrm{F}\)) and thermal fluctuation (\(k_\mathrm{B}T\))~\citep{128-sun2014multiband}. For the conventional parabolic bands, the spectrum of Landau level is the type of harmonic oscillators, namely
\(E_{n}=\hbar eB/m^{*}(n+1/2)-\mu\), where \(m^{*}\) is the effective mass and \(\mu\) is the chemical potential. The quantum limit condition is, \(\Delta E_\mathrm{LL}=\hbar eB/m^{*}>k_\mathrm{B}T\). In this case, the critical magnetic field of Fe(Te,Se) is estimated to be about 60 T (the parameters used in the estimation are \(T\) = 16 K, \(m^{*}\) = \(5m_{e}\))~\citep{81-yin2011kinetic,100-zhang2018observation,101-zhang2019multiple}, which is at odd with the experimental conditions (\(T\) = 16 K, \(B_{\mathrm{max}}\) = 40 T). For the Dirac bands, the spectrum of Landau level is \(E_{n}=E_{\mathrm{D}}+\operatorname{sgn}(n) \nu_{\mathrm{F}} \sqrt{2 \hbar e B|n|}\), where \(E_\mathrm{D}\) is the energy of the Dirac point, and \(\nu_\mathrm{F}\) is the Fermi velocity. Since \(\Delta E_{\mathrm{LL}}=\nu_{\mathrm{F}} \sqrt{2 \hbar e B}\) the critical magnetic field is estimated to be about 2.8 T (the parameters are \(T\) = 16 K and \(k_\mathrm{F}\) = 0.1 \(\mathrm{\AA}^{-1}\)), that agrees with the experimental observations very well. Note that in an earlier magnetoresistance measurement the linear magnetoresistance occurs at 2 T~\citep{128-sun2014multiband}, which is even closer to our estimation. 

\begin{figure*} 
\begin{centering}
\includegraphics[width=2\columnwidth]{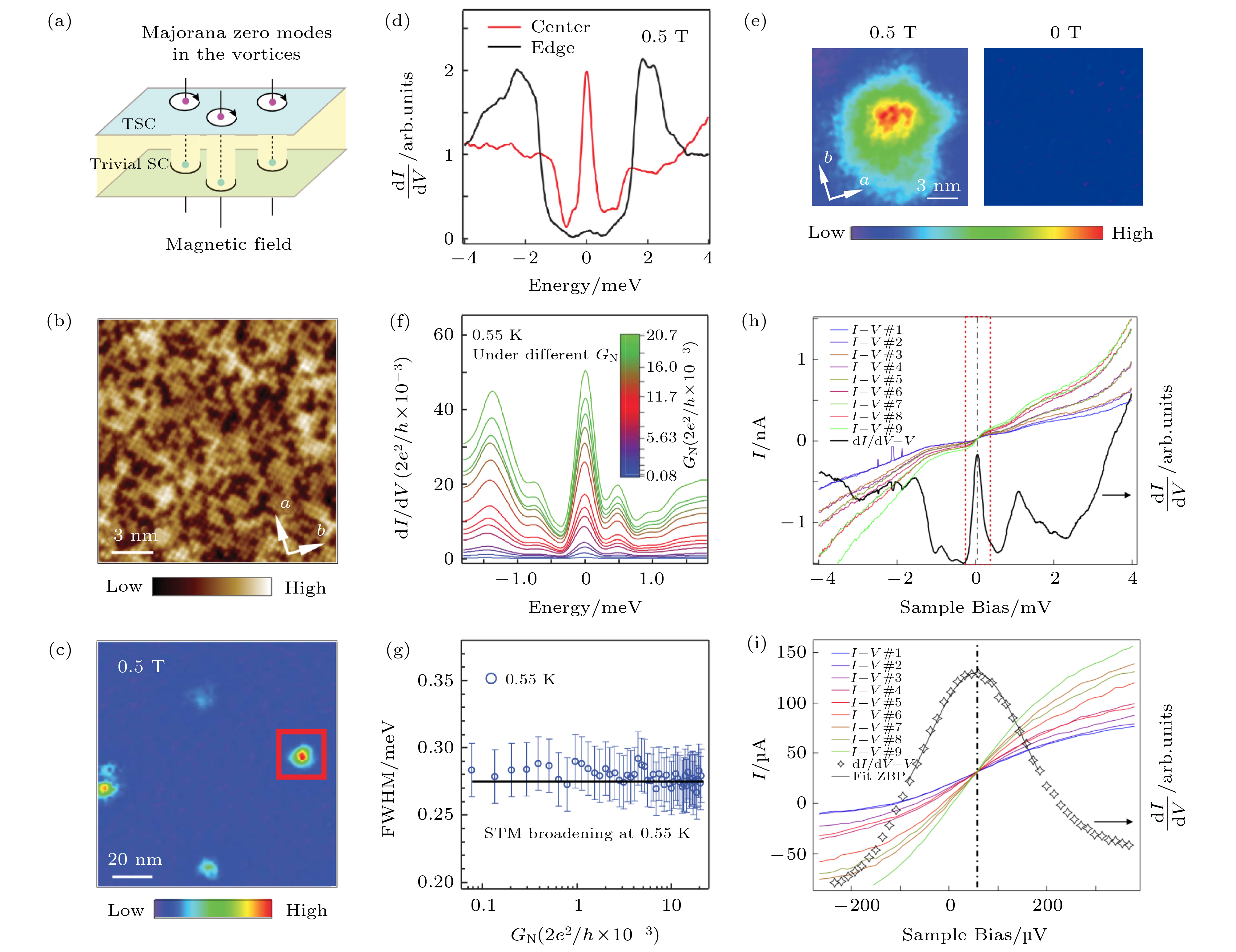}
\par\end{centering}
\caption{\textbf{Discovery of vortex Majorana zero mode in FeTe\(_{0.55}\)Se\(_{0.45}\) single crystals}. (a) theoretical prediction of vortex MZMs in FeTe\(_{0.55}\)Se\(_{0.45}\) single crystals~\citep{100-zhang2018observation}; (b) STM topography of FeTe\(_{0.55}\)Se\(_{0.45}\) single crystals~\citep{102-wang2018evidence}; (c) zero-bias conductance map which shows vortex lattice~\citep{102-wang2018evidence}; (d) a sharp zero-bias conductance peak measured at the center of a vortex. In order to make sure the observation of a zero-energy vortex bound state, three careful checks are listed as follows. Firstly, to make sure that the signal measured is indeed from vortex bound states~\citep{102-wang2018evidence}, ZBC maps after and before applying a magnetic field was measured in (e). It shows the local environment of the vortex is clean and free of impurities~\citep{102-wang2018evidence}; (f) ZBCP is stable under different tunneling barriers. Secondly, to make sure that the observed ZBCP is truly a single peak~\citep{102-wang2018evidence}; (g) FWHM of ZBCP measured under different tunneling barriers; Thirdly, (h), (i) the observed ZBCP is truly a zero mode~\citep{102-wang2018evidence}. (h) simultaneous measurements of \(I\)(\(V\)) curve and \(\mathrm{d} I / \mathrm{d} V\) curve on the center of a vortex core~\citep{102-wang2018evidence}, (i) enlarged display of red box area in (h)~\citep{102-wang2018evidence}.}
\label{fig6}
\end{figure*}

Later on, the Dirac surface states of Fe(Te,Se) were verified by other experimental groups~\citep{129-rameau2019interplay,130-lohani2020band}. The topological band structure of Fe(Te,Se) single crystals is summarized in Fig.~\ref{fig5}(f), multiple topological states provide good opportunities for creating MZM in iron-based superconductors. It is worth noting that only the states around the Fermi level influence quasiparticle excitations, emergence of MZM does care about the circumstance of the Fermi surface. Iron-based superconductors have high mobility of normal carriers, which indicates that the chemical potential is difficult to be controlled by conventional methods such as the solid gating effect~\citep{131-zhu2017tuning}. In Fe(Te,Se) single crystals, the topological insulator state appears at the Fermi level.
As a result, in the following sections we will focus on the topological insulator state and introduce the vortex MZM emerges from it (Fig.~\ref{fig6}(a)).

\section{Vortex Majorana zero modes}\label{sec:3}
Since both the bulk and the surface bands have large value of \(\varDelta^{2}/E_\mathrm{F}\), the topological gap between the vortex MZM and the lowest excitations is expected to be relatively large in FeTe\(_{0.55}\)Se\(_{0.45}\) single crystals, which is beneficial for the observation of a pure MZM. STM technique has excellent energy and spatial resolution, as well as multi-dimensional adjustability including temperature, magnetic field, and tunneling conditions~\citep{132-binnig1987scanning,133-chen1993introduction}. In the case of single-particle tunneling in vacuum, the tunneling current between the tip and the sample is
\begin{equation}\label{eq:7}
\begin{aligned}
& I_{\mathrm{t}}(V, T, x, y, S) \\
\approx & \frac{4 \pi e}{\hbar} \int_{-\infty}^{+\infty} \mathrm{d} E \cdot \tau(S, V, E) \cdot \rho_{\mathrm{t}}(E-e V) \\
& \times \rho_{\mathrm{s}}(E, x, y) \cdot(f(E-e V, T)-f(E, T))
\end{aligned}
\end{equation}
where \(V\) is the bias voltage; \(T\) is the temperature; \(x\), \(y\) is the in-plane position of the sample; \(S\) is the distance between the tip and the sample; \(\tau\) is the tunneling matrix element. When the tip wavefunction is \(s\)-wave, the matrix element can be written as \(\tau(S, V, E)=\exp \left(-2 \sqrt{m_{\mathrm{e}}\left(\phi_{\mathrm{t}}+\phi_{\mathrm{s}}-2 E+e V\right)} S / {\hbar}\right)\), where \(\phi_\mathrm{t}\) and \(\phi_\mathrm{s}\) are the work functions of the tip and the sample respectively; \(\rho_\mathrm{t}\) is the tip DOS; \(\rho_\mathrm{s}\) is the spatially resolved local DOS of the samples; \(f(E, T)\) is the Fermi-Dirac function. When the bias voltage is small, \(\tau\) is approximately irrelevant to the bias voltage \(V\). When the tip position is fixed and the tip DOS is a constant, the local DOS of the sample can be directly measured by the differential conductance. It can be easily proved under the conditions mentioned above, that \({\mathrm{d}I}/{\mathrm{d}V}(V, x, y) \propto \rho_{\mathrm{s}}(V, x, y)\). STM is an ideal technique to validate the existence of vortex MZMs, which appear as a zero-energy peak on the density of states (DOS)~\citep{24-fu2008superconducting,100-zhang2018observation}, by discovering the ZBCPs on the \(\mathrm{d}I/\mathrm{d}V\) spectrum.

FeTe\(_{0.55}\)Se\(_{0.45}\) single crystals was cleaved \(in\)-\(situ\) under ultra-high vacuum~\citep{102-wang2018evidence}. The fresh surface was studied by STM under ultra-low temperature. Since the Se-Fe-Se ``sandwich'' is weakly bonded through vdW interaction, it is cleaved in between the vdW gap, resulting in a Se/Te atomic bare surface. An atomic-resolved topography is shown in Fig.~\ref{fig6}(b), in which Te atoms with a large atomic radius are shown as the bright spots, and Se atoms with a relatively small atomic radius as the dark spots. That indicates a good spatial resolution realized in the experiments, beneficial for the measurements on the wavefunction behavior of the vortex bound states. Under a magnetic field of 0.5 T perpendicular to the sample surface, a vortex lattice can be observed on the zero-bias conductance map (Fig.~\ref{fig6}(c)), which is an important feature of the type-II superconductors~\citep{10-abrikosov2004nobel}. The vortex size measured in STM experiment reflects the value of coherence length. It is different from other scanning probe microscopies such as scanning magnetic microscopes and scanning Hall microscopes where the vortex size is controlled by penetration depth (\(\lambda\)). At the vortex center, a very sharp ZBCP appears on the \(\mathrm{d} I / \mathrm{d} V\) spectrum, isolated in the middle of the superconducting gap. It is most likely a single zero-energy vortex bound state. At the position far away from the vortex center, the \(\mathrm{d} I / \mathrm{d} V\) spectrum recovers the shape of superconducting gap (Fig.~\ref{fig6}(d)). The observation of zero-energy vortex bound state is an important signature of the existence of MZM~\citep{24-fu2008superconducting}. STM technique with higher spatial accuracy improves the reliability of the ZBCPs in Fe(Te,Se), which is an advantage comparing to transport methods~\citep{33-lutchyn2018majorana,134-antipov2018effects}.

\subsection{Experimental validations}\label{sec:3.1}
In order to verify the observation of a true MZM, the signal of ZBCP should be experimentally determined as a real single zero-energy vortex bound state. The experimental tests were performed as follows.

First, the observed subgap state should be a real vortex bound state, rather than a quasiparticle peak caused by impurities, inhomogeneities or disorders. Before measuring the vortex, the material is characterized under the zero field. By performing the measurements of topography, zero-field \(\mathrm{d} I / \mathrm{d} V\) spectrum, and zero-field zero-bias conductance map, the good areas can be identified by the feature of ordered surface, the good spectral feature of the superconducting gap and a nearly vanishing conductance at the zero-bias, respectively. These areas were selected for the further studies under magnetic fields. For example, the vortex shown in the left panel of Fig.~\ref{fig6}(e) supports a ZBCP under 2 T (Fig.~\ref{fig6}(d)). As a check, a zero-field zero-bias conductance map was measured at the same spatial locations (right panel of Fig.~\ref{fig6}(e)). The zero-bias DOS in the region almost vanishes without applying magnetic field, indicating no superconducting quasiparticle caused by other effects. Thus the DOS observed in Fig.~\ref{fig6}(d) is totally caused by the vortex. For a further verification, a variable-tunnel-coupled \(\mathrm{d} I / \mathrm{d} V\) measurement was performed at the vortex (Fig.~\ref{fig6}(f)). Under the feedback regulation, the tip height of STM is adjusted accordingly as per the barrier conductance. The tunneling coupling can be manifested by the barrier conductance, which is defined as \(G_{\mathrm{N}} \equiv I_{\mathrm{t}}/V_{\mathrm{s}}\). A large conductance corresponds to a small tip-sample distance and strong tunneling coupling. It was observed that the ZBCPs remain robust over two orders of magnitude in the barrier conductance (in unit of \(2e^{2}/h\)), which is consistent with the behavior of vortex bound states. On the contrary, for the quasiparticle states caused by impurities, inhomogeneities, or disorders, the energy of the peak changes with the tunneling coupling. For example, the impurity bound states share some similar conditions as the Andreev bound states in a quantum dot of hybrid nanowires. The tunneling coupling can influence the chemical potential, and can change the peak energy too~\citep{135-lee2014spin}.

Second, the observed subgap states should be a true single quantum state, rather than a mixture of multiple near-zero energy peaks. By fitting the spectra in Fig.~\ref{fig6}(f), the full width at half maximum (FWHM) of the ZBCPs was extracted under different tunnel barriers. The measured FWHMs are less than 0.29 meV and basically remain unchanged with the change of the tunnel barrier (Fig.~\ref{fig6}(g)). Compared to the total energy broadening of the system which is about 0.28 meV~\citep{102-wang2018evidence}, it shows that the measured width of ZBCP is mainly determined by the system resolution, strongly supports the observed ZBCP as a true single quantum state. The measurements displayed in Fig.~\ref{fig6}(f) are within the weak tunneling regime. A scaling function (see Section~\ref{sec:5} for details) is capable to fit the conductance values under different tunnel barriers~\citep{102-wang2018evidence}. It can be estimated that the intrinsic Majorana width, which is determined by the tunneling coupling (\(\varGamma_{\mathrm{t}}\)), is approximately \(2 \varGamma_{\mathrm{t}} \approx \frac{170 G_{\mathrm{N}}}{e^{2} / h}(\mu \mathrm{eV})\), much smaller than the thermal broadening (\(3.5k_\mathrm{B}T\)) and the resolution of the STM. Therefore the FWHM of ZBCP is unchanged as a function of tunnel barrier (Fig.~\ref{fig6}(g)), manifesting the width evolution of a single quantum state under the weak tunneling regime. In addition, it is worth noting that under the strong tunneling regime the tunneling coupling increases faster than a linear function of \(G_{\mathrm{N}}\), and the specific relation alters as a different thermal and dissipative broadening (see Section~\ref{sec:5.1} for details).

Third, the observed subgap states should be a true zero mode. The problem of zero-bias offset exists in STM. Generally, the zero voltage picked from the voltmeter is not the physical zero bias. In order to ensure that the measured ZBCP is a true zero mode, one needs to calibrate the zero drift carefully. The current and the voltage have a linear relationship at small bias voltages, thus the intersection of the \(I(V)\) curves measured under different tunnel barriers is the true zero point of current and voltage. As shown in Fig.~\ref{fig6}(h), the \(I(V)\) curves at different tunnel barriers are measured at the vortex center. A \(\mathrm{d} I / \mathrm{d} V\) curve outputting from the lock-in amplifier is displayed simultaneously. The area marked by a red box is enlarged and displayed in Fig.~\ref{fig6}(i). Obviously, the peak position of the \(\mathrm{d} I / \mathrm{d} V\) curve is exactly consistent with the intersection of the \(I(V)\) curves with the error bar less than the sampling interval of the \(\mathrm{d} I / \mathrm{d} V\) curve (14 \(\mu\)eV).

\subsection{Exam alternative explanations}\label{sec:3.2}

The experimental validations demonstrate the observation of single isolated zero-energy vortex bound state in a material with the superconducting topological surface state, which fulfills the requirements of vortex MZM in Fu-Kane model. However, the ZBCP is a common feature appearing in the experiments of hybrid superconducting structures, that obscures the Majorana explanations. Although we performed experimental validations as mentioned above, for the sake of rigorousness, here we list and discuss alternative mechanisms that can lead to a ZBCP. These mechanisms caused difficulties in identifying a ZBCP as a MZM in other systems. 

1) coherent Andreev reflection~\citep{136-van1992excess,137-marmorkos1993three,138-kim2018zero}. The most important mechanism in this category is the reflectionless tunneling. On a disordered superconductor-normal metal interface, scattering centers can cause the mirror-reflected electrons to shoot back to the sample again. The holes generated by the Andreev reflection can return along the incident path of the electrons. The phase conjugation of hole and electron induces an excess current at the zero bias. This effect can be destroyed by applying an external magnetic field which breaks the phase conjugation.

2) incoherent Andreev reflection~\citep{139-nguyen1992anomalous,140-xiong1993subgap}. A ZBCP can be induced by a cumulative effect. The possibility of incoherent Andreev reflection is very small in the weak tunneling regime, which is at odds with the observation of a strong ZBCP in vortex.

3) Kondo effect~\citep{141-kouwenhoven2001revival,142-lee2012zero,143-churchill2013superconductor,144-shen1968zero}. The zero-energy DOS can be induced by the spin-flip resonance when the conduction electrons couple to a localized impurity with a degenerate quantum ground state. The ZBCP of Kondo resonance splits under a magnetic field~\citep{144-shen1968zero}.

4) Josephson current~\citep{145-ternes2006subgap,146-levy2013experimental,147-naaman2001fluctuation,148-naaman2004subharmonic}. Cooper-pair tunneling induces a sharp ZBCP and a negative conductance at slightly lager bias voltages on the \(\mathrm{d} I / \mathrm{d} V\) spectrum. It can be well explained by the Josephson model with thermal fluctuations~\citep{147-naaman2001fluctuation,148-naaman2004subharmonic}. STM experiments use a metallic tip which cannot induce Josephson currents.

5) disorder-induced zero mode~\citep{149-hikami1980spin,150-pikulin2012zero,151-bagrets2012class,152-liu2012zero,153-pan2020generic}. The most important mechanism in this category is the class-\(D\) weak antilocalization (WAL). When the size of the system (\(L\)) is longer than the mean free path (\(l\)), but shorter than the phase coherence length (\(l_{\phi}\)), it is in the regime of quantum diffusive transport. In conventional materials, quantum diffusive transport can lead to weak localization (WL), which is proposed as the precursor of Anderson’s localization in some literatures~\citep{154-anderson1958absence}. In topological materials, the additional geometric phase (\(\pi\)) reserves the interference condition of quantum diffusive transport, resulting in WAL and ZBCP~\citep{155-checkelsky2009quantum}. Normally, a magnetic field destroys the electronic coherence, and eliminates the conventional WL and WAL. However, in the class-\(D\) superconductors (with time-reversal symmetry broken), the electronic coherence can be realized in an additional pathway, that is the electron and hole path of the Andreev reflection, it can induce WAL and ZBCPs even under a large magnetic field~\citep{150-pikulin2012zero,151-bagrets2012class}. In principle, this mechanism can appear in Fe(Te,Se) single crystals, but is avoidable by selecting an good area without disorder, which is not difficult for the STM technique with high spatial resolution.

6) charge potential fluctuation induced zero-energy Andreev bound states~\citep{156-kells2012near,157-liu2017andreev,158-moore2018two,159-moore2018quantized}. This phenomenon commonly exists in quantum dots and the ends of hybrid nanowires~\citep{34-prada2019andreev,160-deng2016majorana}. In principle, it can also exist in a Fe(Te,Se) single crystals. The risk of this mechanism can be effectively avoided with the help of the zero-field test, a good area can be selected for measurements. If the ZBCP is caused by a smooth potential, it should already appear in the zero-field.

7) Yu-Shiba-Rusinov state of a magnetic impurity~\citep{161-heinrich2018single,162-hatter2015magnetic}. In experiments, a good area without magnetic impurity can be chosen for vortex measurement to avoid this risk (Section~\ref{sec:3.1}).

8) zero-energy surface Andreev bound state of unconventional superconductors~\citep{163-kashiwaya2000tunnelling,164-kashiwaya1996theory,165-lofwander2001andreev,166-hu1994midgap,167-tanaka2002theory,168-tanaka2018surface,169-kobayashi2015fragile,170-tamura2017theory,171-hsieh2012majorana,172-wei1998directional,173-kashiwaya2011edge,174-sasaki2011topological}. The \(\pi\) phase interference between particle and hole trajectories could induce a zero-energy Andreev bound state. It occurs when the tunneling direction meets horizontal line node of the superconductor. For instance, a sharp ZBCP shows up in the tunneling spectrum of a \(d\)-wave superconductor measured along \(\langle 110\rangle\). The gap structure of Fe(Te,Se) does not support this mechanism.

9) packed near-zero-energy vortex bound states. When the level spacing of the vortex bound states is very small, multiple near-zero-energy bound states are crowding near the zero energy. Under the condition of limited resolution, the multiple vortex bound states are convoluted to be a fake ZBCP~\citep{13-caroli1964bound,14-hess1989scanning,15-gygi1990electronic,175-hess1990vortex}. This phenomenon is commonly observed in conventional superconductors (such as NbSe\(_{2}\)). In Section~\ref{sec:6}, we introduce the observation of the discrete levels of vortex bound states in Fe(Te,Se), which safely excludes this concern.

\section{Majorana mechanism of emergence}\label{sec:4}
The simplification of Fu-Kane model avoids secondary factors that interfere the elaboration of Majorana physics, and demonstrates emergence of zero-dimensional vortex MZM from a two-dimensional superconducting Dirac surface states~\citep{24-fu2008superconducting}. However, single crystals such as Fe(Te,Se) are not as perfect as the theory. The three-dimensionality of the real materials has influence on MZM. Under a vertical magnetic field, there are not only zero-dimensional vortices on the surface, but also one-dimensional vortex lines throughout the bulk~\citep{12-suderow2014imaging}. The zero-dimensional vortex on the surface is actually the ends of the one-dimensional vortex lines. Moreover, FeTe\(_{0.55}\)Se\(_{0.45}\) single crystal is not a perfect topological insulator. The coexistence of Dirac surface states and bulk bands (Fig.~\ref{fig2}(e)) is an inherent requirement of the superconducting self-proximity effect. There are multiple Fermi surfaces formed by the Dirac surface states and other bulk bands~\citep{100-zhang2018observation}. Furthermore, FeTe\(_{0.55}\)Se\(_{0.45}\) single crystals suffers multiple imperfections including disorders, inhomogeneities, and defects, which induce quantum dissipation of the quasiparticle bound states. Thus a one-by-one experimental investigation on those factors can help our understanding of emergence of the vortex MZM, by clarifying the promoting and destructive issues of formation of MZM. 

In this section, with the help of wavefunction behavior and quasiparticle poisoning of vortex MZM~\citep{102-wang2018evidence}, we introduce the mechanism of emergence of vortex MZM in realistic circumstance of Fe(Te,Se). It shows that MZM is produced by the Dirac surface states, and damaged by the material-imperfection-induced quasiparticle poisoning. We also compare the two-dimensional Fu-Kane model~\citep{24-fu2008superconducting} with a three-dimensional vortex line model in the last part of this section~\citep{176-hosur2011majorana,177-chiu2011vortex,178-hung2013vortex}. The behavior of vortex MZM can be well described by Fu-Kane model at sufficient low temperatures.
  
\begin{figure*} 
\begin{centering}
\includegraphics[width=2\columnwidth]{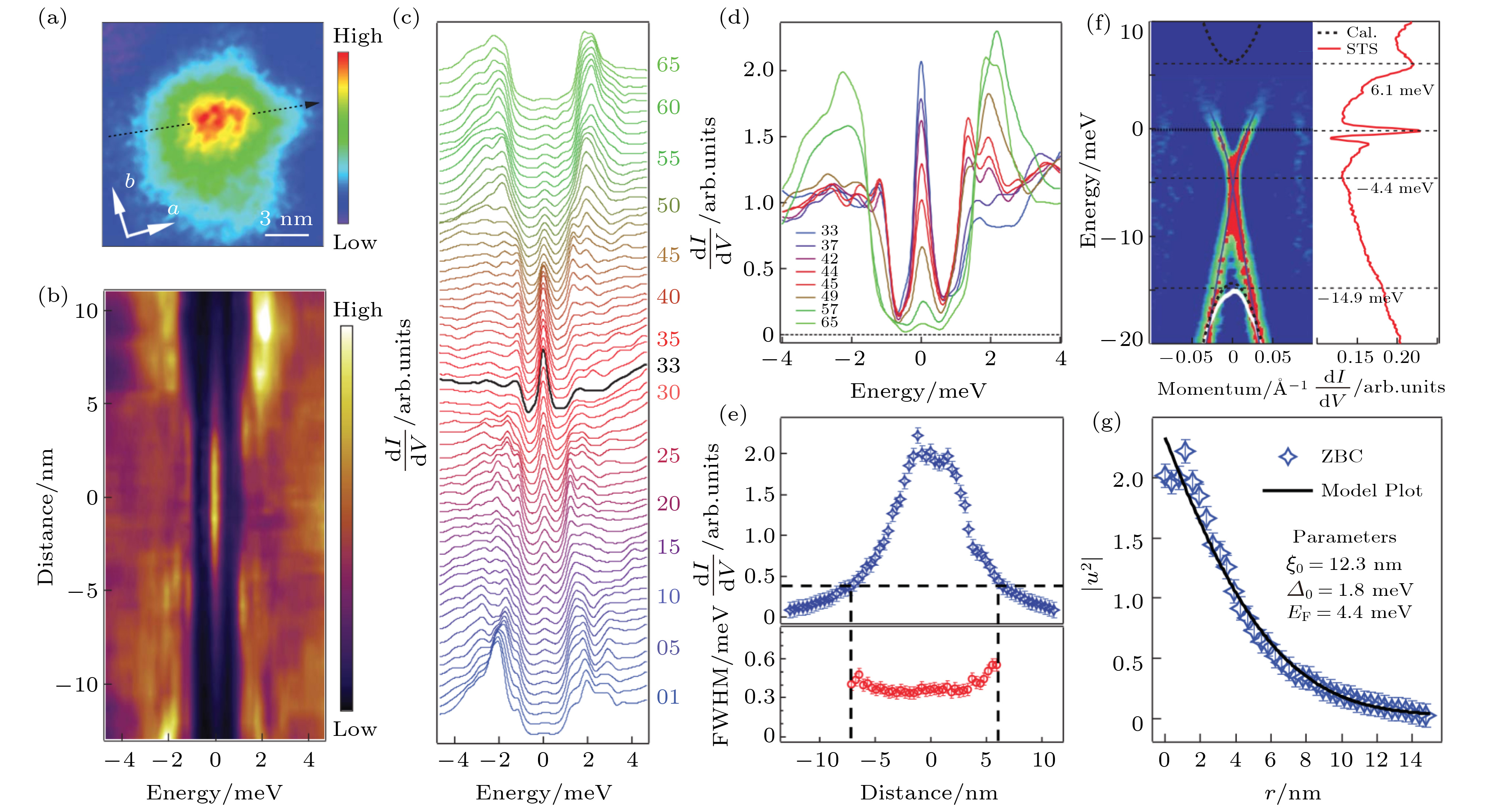}
\par\end{centering}
\caption{\textbf{Wavefunction of vortex Majorana zero mode}. (a) zero-bias conductance map of a topological vortex; (b) \(\mathrm{d} I / \mathrm{d} V\)\((r, V)\) line-cut intensity plot along the black dashed line indicated in (a); (c) waterfall-like plot of (b) with 65 spectra; (d) overlapping display of eight \(\mathrm{d} I / \mathrm{d} V\) spectra selected from (c); (e) spatial dependence of the height (top) and FWHM (bottom) of the ZBCP; (f) comparison between ARPES and STS results, \(\varDelta_{0}\) = 1.8 meV, \(E_\mathrm{F}\) = 4.4 meV, \(\xi\) = \(\nu_\mathrm{F}/\varDelta_{0}\) = 123 \(\mathrm{\AA}\); (g) comparison between the measured ZBCP peak intensity and a theoretical calculation of MZM spatial profile with the parameters extracted from (f)~\citep{102-wang2018evidence}.}
\label{fig7}
\end{figure*} 

\subsection{Wavefunction of vortex MZM}\label{sec:4.1}
A MZM wavefunction occupies a certain spatial volume which is determined by the coherence length of the superconducting Dirac fermions. ZBCPs can be observed within the certain spatial range around the vortex center. The conductance of ZBCP can be calculated by modular square of MZM wavefunction~\citep{24-fu2008superconducting,179-wang2017topological}, that is
\begin{equation}\label{eq:8}
|u|^{2}=\left|u_{1}(r)\right|^{2}+\left|u_{2}(r)\right|^{2}
\end{equation}
where
\begin{equation}\label{eq:9}
u_{1}(r)=\mathrm{J}_{0}\left(\frac{E_{\mathrm{F}} r}{\nu_{\mathrm{F}}}\right) \exp \left[-\int_{0}^{r} \frac{\varDelta\left(r^{\prime}\right)}{\nu_{\mathrm{F}}} \mathrm{d} r^{\prime}\right](i+1)
\end{equation}
\begin{equation}\label{eq:10}
u_{2}(r)=\mathrm{J}_{1}\left(\frac{E_{\mathrm{F}} r}{\nu_{\mathrm{F}}}\right) \exp \left[-\int_{0}^{r} \frac{\varDelta\left(r^{\prime}\right)}{\nu_{\mathrm{F}}} \mathrm{d} r^{\prime}\right](i-1)
\end{equation}
where \(r\) is the distance from the vortex center; \(\mathrm{J}_{i}(x)\) is a Bessel function; \(\varDelta(r)\) is taken as a step function~\citep{102-wang2018evidence}. Thus the modular square of MZM wavefunction is,
\begin{equation}\label{eq:11}
|u|^{2}=C \sum_{i=0}^{1}\left[\mathrm{~J}_{i}\left(\frac{E_{\mathrm{F}} r}{\varDelta_{0} \xi}\right) \mathrm{e}^{-\frac{(r-80)^{2}}{\xi}}\right]^{2}
\end{equation}
Note that the truncation threshold of \(\varDelta(r)\) does not affect the results. \(C\) is a normalized constant; \(\varDelta_{0}\), \(E_\mathrm{F}\) and \(\xi\) represent the superconducting gap, the Fermi energy and the coherence length of the Dirac surface states, respectively.

The \(\mathrm{d} I / \mathrm{d} V\) spectra were carefully studied along the line crossing a vortex (Fig.~\ref{fig7}(a)). It was found that a subgap ZBCP remains at the zero energy as measuring at various spatial positions, and the intensity of ZBCP decreases when moving away from the vortex center (Figs.~\ref{fig7}(b)-(d)). These observations are fully consistent with the theoretical expectation of MZM. The spatial distribution of the zero-bias conductance of ZBCP was extracted in Fig.~\ref{fig7}(e). In order to determine the parameters of the Dirac surface states, we compared a wide-range \(\mathrm{d} I / \mathrm{d} V\) spectrum with the ARPES spectrum (Fig.~\ref{fig7}(f)). The features at 6.1 meV and -14.9 meV of the \(\mathrm{d} I / \mathrm{d} V\) spectrum are in good agreement with the bottom of the conduction band and the top of the valence band shown in the ARPES measurement respectively. A linear depression, which is the DOS feature of a Dirac point~\citep{180-principi2012tunneling}, appears at -4.4 meV of the \(\mathrm{d} I / \mathrm{d} V\) spectrum, which is coincident with the position of the Dirac point observed in ARPES. Based on these results, the experimental parameters of the Dirac surface states can be determined as follows, \(\varDelta_{0}\) = 1.8 meV, \(E_\mathrm{F}\) = 4.4 meV, \(\xi\) = \(\nu_\mathrm{F}/\varDelta_{0}\) = 123 \(\mathrm{\AA}\). By substituting these values into the MZM wavefunction (Eq.(\ref{eq:11})), the Fu-Kane model fully reproduces the spatial distribution of ZBCP observed in our STM experiment (Fig.~\ref{fig7}(g)). It attributes the ZBCP to an underlying Dirac surface state, thus supporting the appearance of Majorana zero mode in the vortex.

Although the vortex line penetrates throughout the three-dimensional bulk, the two-dimensional approximation of Fu-Kane model is still valid for the analysis of behaviors of the vortex MZM under sufficient low-temperatures. In this situation, the three-dimensionality is not important for the Majorana modes. A MZM can be regarded as a topological quasiparticle appearing on the point defect (vortex) of a two-dimensional topological superconductor. The influences of the one-dimensional vortex line on emergence of vortex MZM will be discussed later (Section~\ref{sec:4.3} and Section~\ref{sec:7}). The appearance of one-dimensional dispersive vortex bound states can affect the evolution of vortex MZM on the surface, manifesting the three-dimensionality of the material.
\begin{figure*} 
\begin{centering}
\includegraphics[width=2\columnwidth]{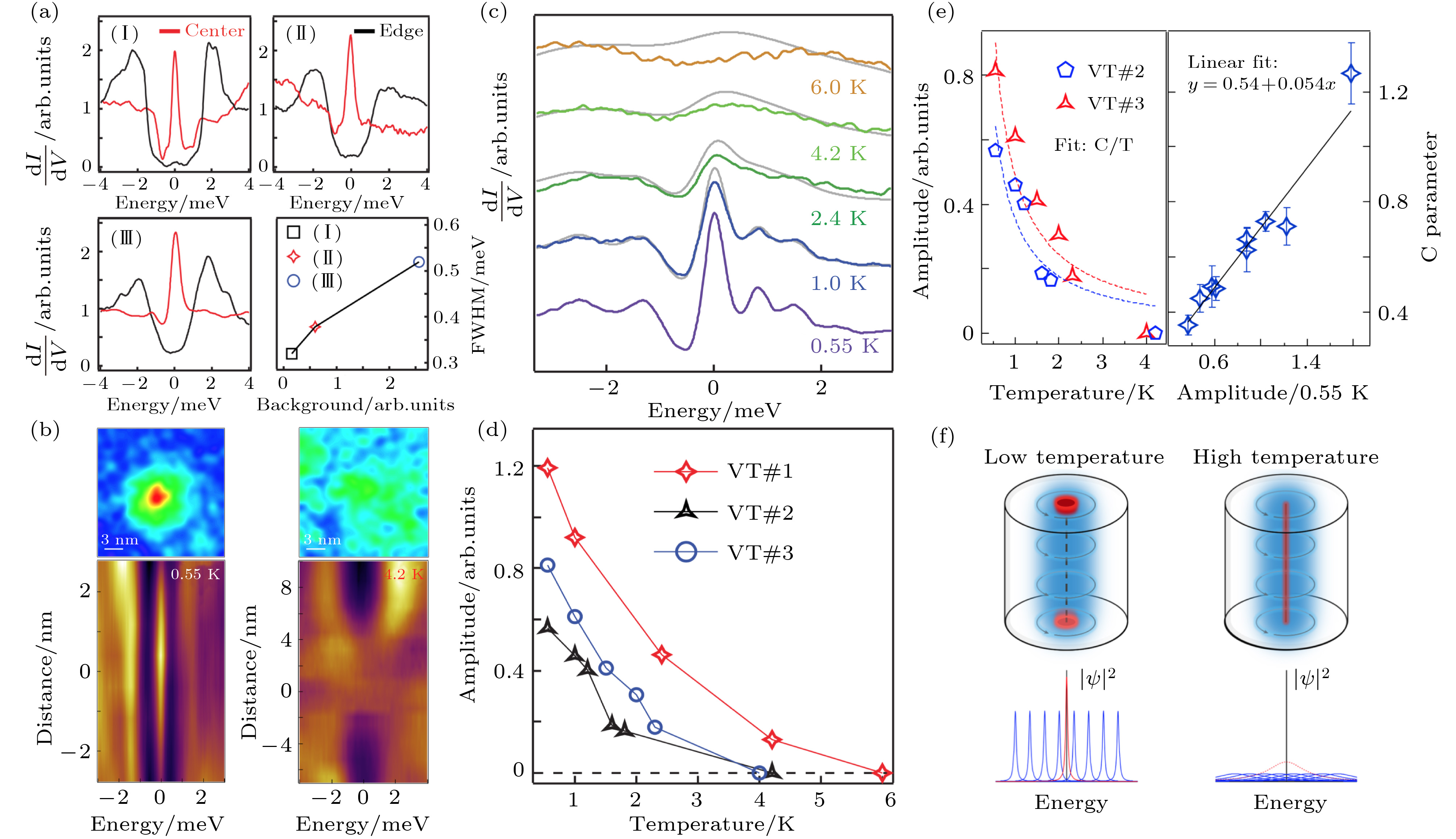}
\par\end{centering}
\caption{\textbf{Quasiparticle poisoning of vortex Majorana zero modes}. (a) three vortex Majorana zero modes measured on different locations, the FWHM of ZBCP at the center of the vortex core is larger when the SC gap around the vortex core is softer; (b) zero-bias conductance map of vortex and line-cut intensity plot of Majorana zero modes measured at 0.55 K (left) and 4.2 K (right), respectively; (c) temperature evolution of ZBCPs in a vortex core. The gray curves are the numerically broadened 0.55 K-data at each temperature; (d) amplitude of the ZBCPs of three vortex MZMs at different temperatures. The amplitude is defined as the peak-valley difference of the ZBCP; (e) left panel: \(C/T\) fitting of the amplitude of Majorana ZBCPs at different temperatures. Right panel: summary on several temperature evolution measurements; (f) schematic explanation of the temperature effect on Majorana ZBCPs. The red line is the vortex MZM and the blue line is the bound state of body vortex~\citep{102-wang2018evidence}.}
\label{fig8}
\end{figure*} 
\subsection{Evidence of quasiparticle poisoning}\label{sec:4.2}
In perfect \(s\)-wave superconductors, conduction electrons condense to Cooper pairs completely at zero kelvin, leads to a hard gap without any subgap quasiparticles~\citep{181-de2018superconductivity}. When temperature is raised, thermally-excited quasiparticles gradually occupy the subgap states. At a finite temperature, the equilibrium quasiparticle density is
\begin{equation}\label{eq:12}
n_{\mathrm{qp}}^{\mathrm{eq}}=\rho_\mathrm{N}(0) \sqrt{2 \pi k_{\mathrm{B}} T \varDelta} \exp \left[-\varDelta /\left(k_{\mathrm{B}} T\right)\right]
\end{equation}
where \(\rho_\mathrm{N}(0)\) is the normal state DOS at the Fermi level. The quasiparticle lifetime can be estimated as follows,
\begin{equation}\label{eq:13}
\tau_{\mathrm{qp}}=\frac{\tau_{0}}{\sqrt{\pi}}\left(\frac{k_{\mathrm{B}} T_{\mathrm{c}}}{2 \varDelta}\right)^{5 / 2} \sqrt{\frac{T_{\mathrm{c}}}{T}} \exp \left(\frac{\varDelta}{k_{\mathrm{B}} T}\right) \propto \frac{1}{n_{\mathrm{qp}}}
\end{equation}

However, multiple experiments have found that the quasiparticle density far exceeds the equilibrium value (Eq.(\ref{eq:12})), implying a reduction of quasiparticle lifetime~\citep{182-joyez1994observation,183-aumentado2004nonequilibrium}. The non-equilibrium occupation of superconducting quasiparticles applies severe disturbance on experiments, but its origin has yet to be well understood. As a result, it was called ``quasiparticle poisoning"~\citep{182-joyez1994observation,183-aumentado2004nonequilibrium,184-cheng2012topological,185-rainis2012majorana,186-colbert2014proposal,187-sarma2016infer,188-van2015one,189-higginbotham2015parity,190-albrecht2017transport}. The degree of quasiparticle poisoning can be characterized by the ``hardness'' of the superconducting gap. A hard gap has steep gap edges and low subgap DOS, indicating negligible amounts of unpaired quasiparticles.

Quasiparticle poisoning is a dissipation process that the coherent quantum states couple to a fermionic bath. It reduces the lifetime and weakens the signals of the quantum states in the DOS spectrum~\citep{184-cheng2012topological,185-rainis2012majorana,186-colbert2014proposal,187-sarma2016infer}. The conservation of fermion parity is one of the prerequisites of the topological quantum computation bases on Majorana qubits. Quasiparticle poisoning changes fermion parity randomly, causes information loss of Majorana qubits. It constrains the speed of gate operations of MZMs which should be safely within the lifetime determined by the poisoning rate~\citep{184-cheng2012topological,185-rainis2012majorana,186-colbert2014proposal,187-sarma2016infer,188-van2015one,189-higginbotham2015parity,190-albrecht2017transport}. Moreover, from experimental view, the quasiparticle poisoning reduces the zero-bias conductance and increase the energy broadening of MZM, causing the disappearance of MZM signal when the experimental resolution is limited. Quasiparticle poisoning is one of the most destructive factors to MZM and topological quantum computation. In iron-based superconductors, the evidence of quasiparticle poisoning was carefully studied. In addition to the basic quasiparticle poisoning, experiments show that the bulk bands may provide additional quasiparticle poisoning at high temperatures, as introduced below.  

First, the spatial variation of basic quasiparticle poisoning is studied at the STM base temperature (0.55 K). As shown in Fig.~\ref{fig8}(a), the \(\mathrm{d} I / \mathrm{d} V\) spectra of the vortex center (red curve) and vortex edge (black curve) are measured on three vortices. A sharp ZBCP appears at the vortex center, and a gap feature appears at the edge. The basic quasiparticle poisoning can be quantified by the subgap DOS extracted from the spectra of vortex edge, and specifically, the quasiparticle background was defined as the integral of \(\mathrm{d} I / \mathrm{d} V\) over the range from -1 to 1 meV. By comparing the FWHM of ZBCP at the vortex center and the quasiparticle background at the vortex edge, we found that stronger poisoning correlates with larger energy broadening of MZM, indicates a shorter quasiparticle lifetime (the lower right of Fig.~\ref{fig8}(a)). It makes the ZBCP broadening of some vortex MZMs much larger than the energy resolution of the system. At the positions with weak poisoning, the broadening of MZM is close to the resolution limit of the system. Obviously, the degree of quasiparticle poisoning of vortex MZM is spatially nonuniform, which is related to the intrinsic inhomogeneity of FeTe\(_{0.55}\)Se\(_{0.45}\) single crystals (Section~\ref{sec:6.3} and Section~\ref{sec:7} for details).

Next, we introduce the additional quasiparticle poisoning effect which is investigated by temperature dependent measurements on several selected vortices. In order to minimize the influence of vortex creep~\citep{191-klein2014vortex,192-eley2017universal}, the temperature was ramping up very slowly which ensures that the selected vortices stay at the same positions during the temperature-variable experiments. The left and right panels of Fig.~\ref{fig8}(b) show the vortex map (upper) and the \(\mathrm{d} I / \mathrm{d} V\)\((r, V)\) across the vortex (lower) at 0.55 K and 4.20 K respectively. Surprisingly, the spectral feature of vortex MZM disappears completely at 4.2 K, which is much lower than the gap-closing temperature of the Dirac surface states. (as shown in Section~\ref{sec:2.2}, the gap-closing temperature of the Dirac surface states is about 14.5 K). The early destruction of vortex MZM cannot be explained by a simplest model (\(e.g.\) Fu-Kane model). As depicted in Eq.(\ref{eq:11}), temperature modulates the intensity of vortex MZM through the superconducting gap, which basically remains unchanged below \(T_\mathrm{c}/2\)~\citep{181-de2018superconductivity}. Thus the \(\mathrm{d} I / \mathrm{d} V\) spectra of vortex MZM are expected to follow the thermal convolution below 8 K for FeTe\(_{0.55}\)Se\(_{0.45}\) single crystals. As shown in Fig.~\ref{fig8}(c), the temperature evolution of the ZBCPs was measured at the center of a same vortex. The width of the ZBCPs is jointly determined by the tunneling coupling, instrumental resolution, thermal broadening and the poisoning. The first two issues do not change with temperature, providing that the experimental setups remain the same. If the quasiparticle poisoning is also unchanged, the \(\mathrm{d} I / \mathrm{d} V\) spectrum below \(T_\mathrm{c}/2\) can be numerically reproduced by convolving the base-temperature \(\mathrm{d} I / \mathrm{d} V\) spectrum with the derivative of the Fermi-Dirac function. The numerical results after convolution are shown as the gray curves in Fig.~\ref{fig8}(c). It shows that the suppression of the ZBCP exceeds the degree determined by the thermal effect at high temperatures. This observation was reproduced independently on nine vortices, indicates that additional destructive factors may emerge at high temperatures, responsible to the early disappearance of vortex MZM (Fig.~\ref{fig8}(b)). 

In order to make a quantitative investigation, we define ZBCP amplitude as the conductance difference between peak and valley. A nonzero ZBCP amplitude is required for an observable MZM in experiment. Temperature evolutions of ZBCP amplitude are extracted on three vortices. As shown in Fig.~\ref{fig8}(d), the extracted ZBCP amplitude of those vortices goes zero at about 3 K. This value is compatible to the minigap (\(\delta\)) of the one-dimensional vortex~\citep{99-yin2015observation}, that \(\delta=\left.\left(\Delta^{2} / E_{\mathrm{F}}\right)_{\mathrm{Bulk}} \approx\left(k_{\mathrm{B}} T\right)\right|_{T}=3 \mathrm{~K}\), indicates that the excessive suppression of vortex MZM is correlated to the bulk bands. The quasiparticle density increases exponentially with the temperature, and these thermally-excited quasiparticles can disturb the vortex MZM on the surface. The bulk quasiparticles are easier to be thermally excited in FeTe\(_{0.55}\)Se\(_{0.45}\) single crystals, owing to a smaller bulk \(\varDelta^{2}/E_\mathrm{F}\) comparing to the Dirac surface state. It supplies a new fermionic bath for additional quasiparticle poisoning, causing the excessive suppression of the vortex MZM with the rising temperature (the right of Fig.~\ref{fig8}(f)). When \(k_\mathrm{B}T \approx \delta\), the one-dimensional vortex line is somewhat conducting, and the thermally-excited bulk quasiparticles strongly mix with the vortex MZM, and pull the MZMs more inward into the bulk, or even annihilate them by meeting their partners on each end of the vortex line. In addition, at sufficient low temperatures (\(k_\mathrm{B}T  \ll \delta\)), the one-dimensional vortex line become a good insulator for superconducting quasiparticles, and the vortex MZM can be approximated as emerging on a point defect of an isolated two-dimensional topological superconductor. The three-dimensionality of the material can be neglected in this situation.

Temperature-dependent measurements have been done on nine vortices. By fitting the ZBCP amplitudes with \(C/T\) (see a theory in Section~\ref{sec:5}), \(C\) parameters can be extracted on each vortex, which reflects the survival temperature of MZMs. Two examples of this fitting are shown in the left panel of Fig.~\ref{fig8}(e). The extracted \(C\) parameters were summarized in the right panel of Fig.~\ref{fig8}(e) as a function of the amplitude at the base temperature (0.55 K). The data converge very well, demonstrating a positive correlation between the \(C\) parameters and the ZBCP amplitude at the base temperature. Owing to the inhomogeneity, basic quasiparticle poisoning is nonuniform (Fig.~\ref{fig8}(a)). It produces different degree of poisoning at different positions, which is manifested by different ZBCP amplitudes at the base temperature. Reducing quasiparticle poisoning is helpful to retain MZMs at higher temperatures.

\begin{figure*} 
\begin{centering}
\includegraphics[width=2\columnwidth]{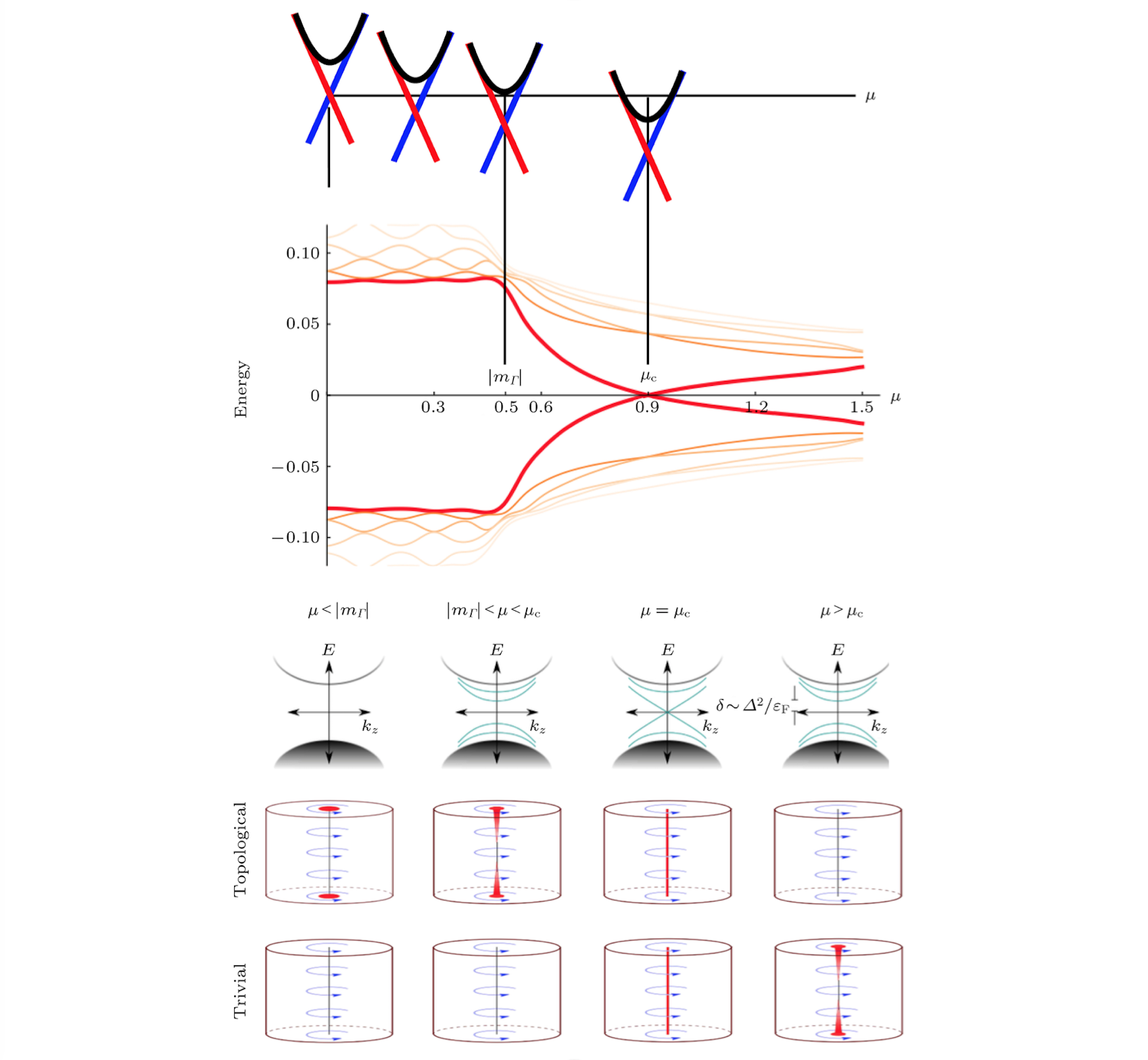}
\par\end{centering}
\caption{\textbf{Topological vortex phase transition in the three-dimensional vortex line model}. The first line: Evolution of the band structure of a topological material by tuning the chemical potential. The second line: Evolution of the low-energy bound state at \(k_{z}\) = 0 with different chemical potentials. The third line: The \(k_{z}\) dispersion of the low-energy bound states. The fourth and fifth lines: evolutions of vortex Majorana zero modes with different chemical potentials for a topological insulator and a normal insulator, respectively. This figure is mainly adapted from Ref.~\citep{176-hosur2011majorana}.}
\label{fig9}
\end{figure*} 

\subsection{Picture vortex MZM: two- versus three-dimensional model}\label{sec:4.3}
Although the low-temperature behavior of MZM can be well described by the two-dimensional Fu-Kane model, the three-dimensionality of the real materials, especially the coexistence of Dirac surface states and bulk bands, does impact emergence of the vortex MZM on the sample surface. A more realistic three-dimensional model is highly required, not only for extending the Fu-Kane model, but also for the precise understanding of vortex MZMs in real materials.

After the proposal of Fu-Kane model, people soon noticed that the realistic properties of a material may complicate the fate of MZM. In most topological insulators, their bulk is a good metal rather than an insulator~\citep{25-hasan2010colloquium}, which raises a question ``Does the vortex MZM still exist in this situation?"~\citep{176-hosur2011majorana,193-wray2010observation}. To answer the question, a three-dimensional model was built by taking the electron-doped Bi\(_{2}\)Se\(_{2}\) as an example~\citep{176-hosur2011majorana,177-chiu2011vortex,178-hung2013vortex}. This model studied the topological phase transition of the one-dimensional vortex lines along the c-axis. Here we summarize the evolution of dispersive bound states along the vortex line in the three-dimensional model (Fig.~\ref{fig9}); 1) when the chemical potential is below the bulk band gap \(\left(\mu<\left|m_{\Gamma}\right|\right)\), the vortex line is completely insulating, and there is no subgap bound state; 2) when the chemical potential rises and reaches the edge of bulk bands \(\left(\mu=\left|m_{\Gamma}\right|\right)\), some dispersive vortex bound states appear along the vortex line, and there is a very small quasiparticle gap between different energy levels \((\delta \approx\left(\Delta^{2} / E_{\mathrm{F}}\right)_{\mathrm{Bulk}})\). In this case, the vortex line can be insulating only at an extremely low temperatures (lower than \(\delta\)); 3) as the chemical potential continues to increase, the quasiparticle gap continues to decrease, and the Berry phase of bulk band varies accordingly; when \(\mu = \mu_{c}\), the Berry phase of bulk band reaches \(\pi\), the dispersive bound states become gapless, and causes a topological vortex phase transition; 4) when \(\mu > \mu_{c}\), quasiparticle gap reopens, identifying a new topological phase.

The critical points, where the topological vortex phase transition occurs, can be directly determined by calculating the Berry phase of the bulk bands or the evolution of the bound states in a vortex line~\citep{176-hosur2011majorana}. However, the identification of the topological regime, whether before or after the critical points, relies on the presence or absence of the Dirac surface state. In order to make this point clearly, we next discuss the vortex bound states for the cases of superconducting topological insulators and normal insulators, and analyze their behaviors in the views of two- and three-dimensional models. 1) when the normal state is a topological insulator. In the Fu-Kane model, a vortex MZM exists at any chemical potential, changing the chemical potential only changes the quasiparticle gap of MZMs, the vortex MZM never disappears. In the three-dimensional model, the vortex MZM can only exist in a limited energy window near the Dirac point. The vortex line is a one-dimensional topological superconductor belonging to the class-\(D\)~\citep{194-altland1997nonstandard,195-schnyder2008classification,196-teo2010topological,197-chiu2016classification}. When the chemical potential is located near the Dirac point, a vortex MZM appears as a boundary state of the topological superconductor. With increasing the chemical potential, a one-dimensional bound state appears in the vortex line as soon as a bulk band is involved. As the c-axis localization length of MZM is inversely proportional to \(\delta\)~\citep{176-hosur2011majorana}, the appearance of the dispersive bound state drags the wavefunction of vortex MZM inward into the bulk. At the critical chemical potential where the topological phase transition occurs, the vortex MZM disappears. 2) when the normal state is a normal insulator. In this case, the two-dimensional model is vacuum, there is no MZM at any chemical potential. In the three-dimensional model, no MZMs exist near the zero chemical potential, but with the change of the chemical potential, a topological vortex phase transition occurs at the critical point where the Berry phase of bulk band crossing \(\pi\). After the phase transition, a vortex MZM appears on the sample surface~\citep{176-hosur2011majorana}.

The concept of topological vortex phase transition plays a vital role on the understanding of the three-dimensionality in a real material. Recently, the three-dimensional model has been extended to more specific circumstances, including 1) the vortex phase transitions in Fe(Te,Se) single crystals~\citep{198-xu2016topological}; 2) the vortex phase transitions when the topological bands are coupled with trivial bands. In this case, the topological regime deforms, but still includes the Dirac points~\citep{199-chiu2012stabilization}; 3) the topological vortex phase transitions induced by Zeeman coupling~\citep{200-ghazaryan2020effect}; 4) the topological vortex phase transitions of the weak topological insulator~\citep{201-qin2019topological} and the Dirac semimetal~\citep{202-konig2019crystalline,203-qin2019quasi}.

In this section, we explain in details about the mechanism of emergence of vortex MZM in a three-dimensional material Fe(Te,Se). We show that the Dirac surface state is the direct contributor to emergence of vortex MZM. At sufficient low temperatures, the behavior of vortex MZM can be well described by the Fu-Kane model, and the three-dimensionality of Fe(Te,Se) can be ignored on the context of vortex MZM. On the other hand, the quasiparticle poisoning effect damages the MZM, particularly, the bulk bands introduce additional quasiparticle poisoning at high temperatures, which accelerates the disappearance of the vortex MZM. The bulk bands also affect emergence of the vortex MZM through the topological vortex phase transition. In a material with inhomogeneity, the quantum critical point may be reached on some areas, leading to the coexistence of two topological phases in a same piece of sample. After a thorough introduction of the mechanism of emergence, the physics properties of vortex MZM will be discussed in the following sections.

\section{Majorana nature of vortex zero mode}\label{sec:5}
The self-conjugation of the creation and the annihilation operators (\(\gamma^{\dagger}\) = \(\gamma\)) is the essential nature of MZM. A MZM can be expressed in the form of Bogoliubov quasiparticle by the operators of complex fermion (\(c^{\dagger}\) = \(c\)),
\begin{equation}\label{eq:14}
\gamma=u c^{\dagger}+v c
\end{equation}

The wavefunctions of electron component and hole component are required to be equal (\(u^{*} = v\)), which is a direct representation of the Majorana nature of MZM~\citep{1_nayak2008non,2_wilczek2009majorana,3_alicea2012new,4_beenakker2013search,5_elliott2015colloquium,6_kitaev1997quantum,7_kitaev2003fault,8_kitaev2006anyons}. In this section, we introduce both the theoretical and experimental investigations of the intrinsic quantum conductance of Majorana modes (\(2e^{2}/h\)). As proved in Law-Lee-Ng theory~\citep{204-law2009majorana}, the quantized Majorana conductance is a consequence of resonant Andreev reflection which is induced by the particle-hole equivalence of Majorana modes. The quantized Majorana conductance is immune to variations of the tunneling barrier under ideal conditions. In Reference~\citep{105-zhu2020nearly}, stable conductance plateaus were observed on the vortex zero mode of the FeTe\(_{0.55}\)Se\(_{0.45}\) single crystals by means of variable-tunnel-coupled STS. It was found that the plateau behavior is unique to the MZM and is absent at other states, including the finite-energy CdGM states, the zero-field zero-bias conductance, and the continuum state outside the superconducting gap. Remarkably, a nearly \(2e^{2}/h\)-quantized conductance plateau was observed in one vortex. These results indicate the appearance of Majorana-induced resonant Andreev reflection (MIRAR), imply the Majorana nature of vortex MZM.

\begin{figure*} 
\begin{centering}
\includegraphics[width=2\columnwidth]{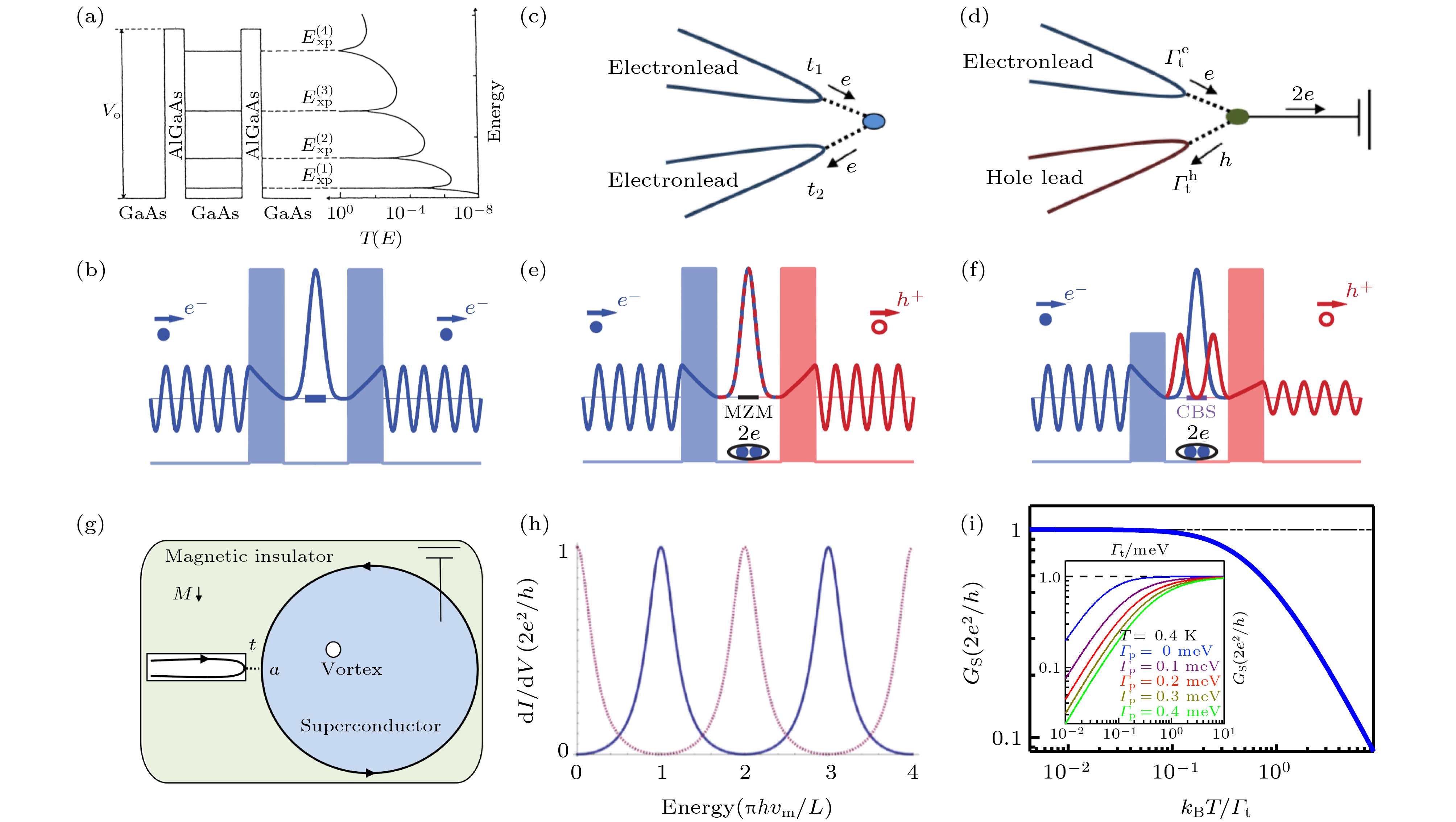}
\par\end{centering}
\caption{\textbf{Resonant Andreev reflection induced quantized Majorana conductance}. (a) conventional electron resonant tunneling in a semiconductor heterostructure~\citep{207-chang2012resonant}; (b) wavefunction of conventional resonant tunneling~\citep{105-zhu2020nearly}; (c) two-tip cross-tunneling can be regarded as a replacement of the semiconductor heterostructure for realizing semiconductor heterostructure under the condition of equal hopping amplitude around the two tips (\(t_{1} = t_{2}\))~\citep{204-law2009majorana}; (d) Majorana-induced resonant Andreev reflection (MIRAR) can be regarded as a superconducting version of the conventional resonant tunneling in the particle-hole Hilbert space. Here a single lead plays both roles of electron and hole lead~\citep{204-law2009majorana}. Due to the particle-hole equivalence property, Majorana modes couple to the incident electron and the reflected hole with equal tunneling coupling strength, which satisfies the resonant condition (\(\varGamma^{\mathrm{e}}_{\mathrm{t}}\) = \(\varGamma^{\mathrm{h}}_{\mathrm{t}}\); \(\varGamma_{\mathrm{t}}\) = \(2\pi \rho_{0}|t|^{2}\), \(\rho_{0}\) being the related density of states); (e), (f) wavefunction of Andreev reflection mediated by MZM and a conventional Andreev bound state, respectively~\citep{105-zhu2020nearly}; (g) material setup used in Law-Lee-Ng model~\citep{204-law2009majorana}; (h) theoretically calculated resonant quantum conductance of Majorana modes~\citep{204-law2009majorana}; (i) theoretically calculated Majorana conductance at finite temperature and poisoning rate.}
\label{fig10}
\end{figure*} 

\subsection{Theory of Majorana conductance}\label{sec:5.1}
MIRAR is a superconducting version of classical resonant tunneling~\citep{205-tsu1973tunneling,206-chang1974resonant,207-chang2012resonant}. A double-barrier quantum well can be constructed by band engineering in a semiconductor heterostructure (Fig.~\ref{fig10}(a)). Quasibound states appear in the quantum well with quantization in energy. Under the condition of symmetric tunnel barrier, the transmission coefficient of electrons tunneling across the quantum well is,
\begin{equation}\label{eq:15}
T(E)=\left(\frac{\hbar}{2 \tau_{\mathrm{qs}}}\right)^{2} /\left[\left(E-E_{n}\right)^{2}+\left(\frac{\hbar}{2 \tau_{\mathrm{qs}}}\right)^{2}\right]
\end{equation}
where \(\tau_\mathrm{qs}\), \(E_{n}\) are the lifetime and energy of the quasibound states, respectively. Perfect transmission occurs when the incident electron is aligned with the energy level of the quasibound state. This phenomenon is similar with the resonant transmission in an optical Fabry-Perot cavity. The barriers of the double quantum well like two mirrors of the electron wave. When the electron energy is aligned with \(E_{n}\), the reflected electron waves are eliminated by constructive interference, leads to a resonant transmission coefficient of 1. Classical resonant tunneling can also occur between two identical tips (Fig.~\ref{fig10}(c)), when the tunneling amplitudes between the two tips are equal (\(t_{1}\) = \(t_{2}\)). A schematic of classical resonant tunneling is shown in Fig.~\ref{fig10}(b). It is worth noting that the relationship between the tunneling coupling strength (\(\varGamma_{\mathrm{t}}\)) and the tunneling amplitude (\(t\)) is \(\varGamma_{\mathrm{t}}\) = \(2\pi \rho_{0}|t|^{2}\), where \(\rho_{0}\) refers to the DOS related to the coupling.

Resonant Andreev reflection can be easily derived from classical resonant tunneling by replacing the outgoing electron with a hole (the electron barrier of the outgoing part is also replaced with a hole barrier) (Fig.~\ref{fig10}(d))~\citep{204-law2009majorana}. A single tip can act as both the lead of incident electron and reflective hole, which guarantees \(t_{\mathrm{e}}\) = \(t_{\mathrm{h}}\) in Andreev reflection. In order to realize the resonant condition (\(\varGamma^{\mathrm{e}}_{\mathrm{t}}\) = \(\varGamma^{\mathrm{h}}_{\mathrm{t}}\)), the wavefunctions of the electron and hole components should be equal (\(\rho_{\mathrm{e}}\) = \(\rho_{\mathrm{h}}\)). This requirement is naturally satisfied by Majorana symmetry (Fig.~\ref{fig10}(e)). MZM-mediated Andreev reflection is a resonant process. Under the ideal conditions, Majorana conductance is irrelevant to the strength of tunneling coupling. On the contrary, conventional Andreev bound states (such as the subgap CdGM state) can not guarantee the particle-hole equivalence, thus the conditions of resonant Andreev reflection are not satisfied (Fig.~\ref{fig10}(f)).

Law-Lee-Ng theory calculate the MIRAR induced quantum conductance by taking chiral Majorana modes as an example~\citep{56-fu2009probing,57-akhmerov2009electrically}. As shown in Fig.~\ref{fig10}(g), a superconductor island is put on top of a topological insulator, while the area outside is covered by a magnetic insulator. Effective spinless \(p_{x}+ip_{y}\) superconductivity appears under the island, outside the island a Zeeman gap opens at the Dirac points (\(E_{Z} = g\mu_{\mathrm{B}}M\)), where \(g\) is the Lande-\(g\) factor, \(\mu_{\mathrm{B}}\) is the Bohr magnetic moment, and \(M\) is the effective magnetization. When \(E_{\mathrm{z}}>2 \sqrt{\varDelta^{2}+\mu^{2}}\), chiral Majorana modes appear on the boundary of the superconducting island. The quantized energy level of the chiral Majorana modes is
\begin{equation}\label{eq:16}
E_{m}=m \frac{2 \pi}{L} \hbar \nu_{\mathrm{m}}+\frac{\pi}{L} \hbar \nu_{\mathrm{m}}+\frac{n \pi}{L} \hbar \nu_{\mathrm{m}}.
\end{equation}
The corresponding quantized momentum is
\begin{equation}\label{eq:17}
k_{m}=(2 m+1+n) \pi / L
\end{equation}
where \(m\) is an integer; \(L\) is the perimeter of the superconducting island; \(\nu_{\mathrm{m}}\) is the Fermi velocity of the chiral Majorana mode; \(n\) is the number of vortices appearing in the superconducting island. In Eq.(\ref{eq:16}) and Eq.(\ref{eq:17}), the first term is derived from periodic boundary conditions; the second term is the contribution of the spin Berry phase of the Dirac surface state; the third term is derived from the vortex. The chiral Majorana mode is tunnel-coupled with a single lead at point \(a\) (Fig.~\ref{fig10}(g)) with the coupling amplitude of \(t\). The total Hamiltonian of the system is \(H_{\mathrm{LLN}} = H_{\mathrm{L}} + H_{\mathrm{M}} + H_{\mathrm{T}}\). The lead term is
\begin{equation}\label{eq:18}
H_{\mathrm{L}}=-\mathrm{i} \nu_{\mathrm{f}} \sum_{\sigma=\uparrow \downarrow} \int_{-\infty}^{+\infty} \psi_{\sigma}^{\dagger}(x) \partial_{x} \psi_{\sigma}(x) \mathrm{d} x
\end{equation}
where \(\nu_\mathrm{F}\) is the Fermi velocity of the electrons and \(\psi_{\sigma}(x)\) is the fermion field. The Majorana term is 
\begin{equation}\label{eq:19}
H_{\mathrm{M}}=-\mathrm{i} \nu_{\mathrm{m}} \int_{0}^{L} \eta^{\dagger}(l) \partial_{l} \eta(l) \mathrm{d} l.
\end{equation}
The coupling term is
\begin{equation}\label{eq:20}
H_{\mathrm{T}}=-\mathrm{i} \frac{1}{\sqrt{2}} \operatorname{t\eta}(a)\left[\left(f \psi_{\uparrow}(0)+g \psi_{\downarrow}(0)\right)+\text {h.c.}\right]
\end{equation}
where \(\eta(a)\) is the Majorana field at \(a\), \(f\) and \(g\) are complex numbers with a modulus of 1. 

By a transformation on fermion field as follows,
\begin{equation}\label{eq:21}
\left(\begin{array}{l}
\psi_{1} \\
\psi_{2}
\end{array}\right)=\frac{1}{\sqrt{2}}\left(\begin{array}{cc}
f & g \\
f & -g
\end{array}\right)\left(\begin{array}{l}
\psi_{\uparrow} \\
\psi_{\downarrow}
\end{array}\right)
\end{equation}
the Hamiltonian of the system can be simplified as \(H^{'}_{\mathrm{LLN}} = H_{\mathrm{L}}^{'} + H_{\mathrm{M}} + H_{\mathrm{T}}^{'}\), where
\begin{equation}\label{eq:22}
H_{\mathrm{L}}^{\prime} =-\mathrm{i} \hbar \nu_{\mathrm{f}} \int_{-\infty}^{+\infty} \psi_{1}^{\dagger}(x) \partial_{x} \psi_{1}(x) \mathrm{d} x
\end{equation}

\begin{equation}\label{eq:23}
H_{\mathrm{T}}^{\prime} =-\mathrm{i} t \eta(a)\left(\psi_{1}(0)+g \psi_{1}^{\dagger}(0)\right)
\end{equation}

It is apparently that the Majorana modes only couple to one spin (\(\psi_{1}\)) in the process of Andreev reflection, while the other spin (\(\psi_{2}\)) is completely decoupled~\citep{204-law2009majorana,208-he2014selective,209-haim2015signatures,210-kawakami2015evolution,211-hu2016theory}. The Andreev process mediated by Majorana mode is equal-spin. The incident electrons and the reflected holes are denoted as (\(\psi_{1,k}(-)\)) and (\(\psi_{1,-k}^{\dagger}(+)\)), respectively. The electron-hole scattering can be described in the form of \(\boldsymbol{S}\)-matrix,
\begin{equation}\label{eq:24}
\left(\begin{array}{c}
\psi_{1, k}(+) \\
\psi_{1,-k}^{\dagger}(+)
\end{array}\right)=\boldsymbol{S}\left(\begin{array}{c}
\psi_{1, k}(-) \\
\psi_{1,-k}^{\dagger}(-)
\end{array}\right)
\end{equation}
\(\boldsymbol{S}\)-matrix can be calculated by \(H^{'}_{\mathrm{LLN}}\), that
\begin{equation}\label{eq:25}
\begin{array}{c}
\boldsymbol{S}=\left(\begin{array}{cc}
s^{\mathrm{ee}} & s^{\mathrm{eh}} \\
s^{\mathrm{he}} & s^{\mathrm{hh}}
\end{array}\right) \\
=\frac{1}{Z}\left(\begin{array}{ccc}
i \sin \left[\frac{\theta(k, n)}{2}\right] & -\tilde{t}^{2} \cos \left[\frac{\theta(k, n)}{2}\right] \\
-\tilde{t}^{2} \cos \left[\frac{\theta(k, n)}{2}\right] & i \sin \left[\frac{\theta(k, n)}{2}\right]
\end{array}\right)
\end{array}
\end{equation}
where \(\tilde{t}=\frac{t}{2 \sqrt{\hbar \nu_{\mathrm{f}} \hbar \nu_{\mathrm{m}}}} ;~\theta(k, n)=k L+\pi+n \pi\) is the phase of the Majorana mode, \(Z=i \sin \left[\frac{\theta(k, n)}{2}\right]+\tilde{t}^{2} \cos \left[\frac{\theta(k, n)}{2}\right] \). Resonant Andreev tunneling (\(|s^{he}|^{2} = 1\)) occurs when \(\theta(k, n)/2 = m\pi\) (\(\pi\) is an integer). In this condition, \(k\) is just the quantized Majorana momentum (\(k = (2m - 1 - n)\pi/L\)). In other words, resonant Andreev reflection occurs when the incident electrons are aligned with the quantized energy level of the chiral Majorana mode. On resonance, the probability of the particle-hole reflection is \( 100 \%\) and irrelevant to \(t\). 

The tunneling current on resonance can be calculated by \(\boldsymbol{S}\)-matrix, 
\begin{equation}\label{eq:26}
I=\frac{2 e}{h} \int_{0}^{e V}\left\{\varGamma_{\mathrm{t}}^{2} /\left[\left(E-E_{m}\right)^{2}+\varGamma_{\mathrm{t}}^{2}\right]\right\} \mathrm{d} E
\end{equation}
where \(\varGamma_{\mathrm{t}}=2 t^{2} \hbar \nu_{\mathrm{m}} / L\) is the tunneling coupling strength. The differential conductance is
\begin{equation}\label{eq:27}
\frac{\mathrm{d} I}{\mathrm{~d} V}(V)=\frac{2 e^{2}}{h} \frac{\varGamma_{\mathrm{t}}^{2}}{\varGamma_{\mathrm{t}}^{2}+\left(e V-E_{m}\right)^{2}}
\end{equation}
The \(2e^{2}/h\)-quantized resonant conductance of Majorana modes is derived by the Law-Lee-Ng theory (Fig.~\ref{fig10}(h)). It is worth noting that the quantum conductance is a universal property of Majorana quasiparticles, rooted on the Majorana nature of particles-holes equivalence, and irrelevant to the details used in the model, including the forms of Majorana mode (edge mode~\citep{204-law2009majorana} or zero mode~\citep{212-flensberg2010tunneling,213-wimmer2011quantum,214-fidkowski2012universal}), the theoretical technique (\(\boldsymbol{S}\)-matrix~\citep{204-law2009majorana,214-fidkowski2012universal} or Green function~\citep{212-flensberg2010tunneling}), and the coupling strength (single particle tunneling~\citep{204-law2009majorana,212-flensberg2010tunneling} or quantum ballistic transport~\citep{213-wimmer2011quantum}). Therefore, it is a direct manifestation of essential properties of Majorana symmetry.

Law-Lee-Ng theory investigates the Majorana conductance in a dissipationless system at absolute zero temperature. With those ideal conditions, the zero-bias conductance of MZM should be at the quantum value (\(2e^{2}/h\)). A quantized plateau should appear in a variable-tunnel-coupled measurement of Majorana modes. However, the conductance of Majorana modes is much smaller than \(2e^{2}/h\) in many experiments~\citep{33-lutchyn2018majorana}, indicates the influence of imperfect conditions on the conductance behavior of Majorana modes~\citep{212-flensberg2010tunneling,213-wimmer2011quantum,214-fidkowski2012universal,215-sengupta2001midgap,216-nichele2017scaling,217-setiawan2017electron,218-pientka2012enhanced}. In the following, we consider quasiparticle poisoning effect and thermal effect, to investigate their influence on the behavior of the zero-bias conductance. 

First, we analysis the Majorana conductance in a dissipationless system at a finite temperature. The temperature evolution of Majorana conductance is,
\begin{equation}\label{eq:28}
G_{\mathrm{s}}=\frac{2 e^{2}}{h} \int_{-\infty}^{+\infty} \frac{\varGamma_{\mathrm{t}}^{2}}{\varGamma_{\mathrm{t}}^{2}+E^{2}} \frac{1}{4 k_{\mathrm{B}} T \cosh ^{2}\left[E /\left(2 k_{\mathrm{B}} T\right)\right]} \mathrm{d} E
\end{equation}
The zero-bias conductance of MZM satisfies a \(k_\mathrm{B}T/\varGamma_{\mathrm{t}}\) scaling behavior (Fig.~\ref{fig10}(i))~\citep{215-sengupta2001midgap,216-nichele2017scaling,217-setiawan2017electron}, that is \(G_{\mathrm{s}}=\frac{2 e^{2}}{h} f\left(k_{\mathrm{B}} T / \varGamma_{\mathrm{t}}\right)\), where \(f(x)\) is the scaling function. In a dissipationless system the quantized conductance can be observed only when the tunneling coupling is much larger than the thermal broadening.

Next, we analysis the Majorana conductance in a dissipative system at a finite temperature. At the zero kelvin, the conductance of MZM with quasiparticle poisoning is

\begin{equation}\label{eq:29}
\frac{\mathrm{d} I}{\mathrm{~d} V}(V)=\frac{2 e^{2}}{h} \frac{\left(\varGamma_{\mathrm{t}}+\varGamma_{\mathrm{p}}\right) \varGamma_{\mathrm{t}}}{\left(\varGamma_{\mathrm{t}}+\varGamma_{\mathrm{p}}\right)^{2}+(e V)^{2}}
\end{equation}
where \(\varGamma_{\mathrm{p}} = 2\pi \rho_{\mathrm{p}}|t_{\mathrm{p}}|^{2}\) is the coupling strength between the fermionic bath and the MZM~\citep{186-colbert2014proposal}. The quasiparticle poisoning effect weakens the zero-bias conductance of MZM by a fraction of \(\varGamma_{\mathrm{t}}/(\varGamma_{\mathrm{t}} + \varGamma_{\mathrm{p}})\). Similarly, the MZM conductance at a finite temperature can be obtained by convolving the derivative of Fermi-Dirac function \(\left(\mathrm{d} f_{\mathrm{FD}} / \mathrm{d} E\right)
\) with the zero-temperature conductance. In a dissipative system, the zero-bias conductance no longer follows the \(k_\mathrm{B}T/\varGamma_{\mathrm{t}}\) scaling behavior. We calculate the evolution of zero-bias conductance under different dissipation strength (inset of Fig.~\ref{fig10}(i)). Dissipation causes the decrease in \(G_{\mathrm{s}}\), but \(G_{\mathrm{s}}\) still approaches to the quantum conductance when \(\Gamma_{\mathrm{t}} \gg \Gamma_{\mathrm{p}}\).

In summary, the width of MZM is determined by the temperature broadening \((3.5 k_\mathrm{B}T)\), the dissipation and the tunneling coupling broadening \((2(\varGamma_{\mathrm{p}} + \varGamma_{\mathrm{t}}))\). A quantized Majorana conductance plateau can be experimentally observed only when \(\varGamma_{\mathrm{t}}\) is much larger than \(k_\mathrm{B}T\) and \(\varGamma_{\mathrm{p}}\).

\subsection{Conductance plateau of vortex zero modes}\label{sec:5.2}

The variable-tunnel-coupled experiment takes the advantages of the feedback regulation of STM, which controls the tip height by the setpoint of tunneling current (\(I_{\mathrm{t}}\)) and bias voltage \((V_{\mathrm{s}})\). It can be used to change the tunneling coupling continuously in an experiment. The conductance of the tunnel barrier (\(G_{\mathrm{N}} \equiv I_{\mathrm{t}}/V_{\mathrm{s}}\)) is regarded as a measure of tunneling coupling (\(\varGamma_{\mathrm{t}}\)) between the tip and the sample (\(G_{\mathrm{N}}\) is positively correlated with \(\varGamma_{\mathrm{t}}\)). Quantized Majorana conductance can be hopefully explored by achieving a large value of \(G_{\mathrm{N}}\), in which \(\varGamma_{\mathrm{t}}\) is much larger than \(k_\mathrm{B}T\) and \(\varGamma_{\mathrm{p}}\).

\begin{figure*} 
\begin{centering}
\includegraphics[width=2\columnwidth]{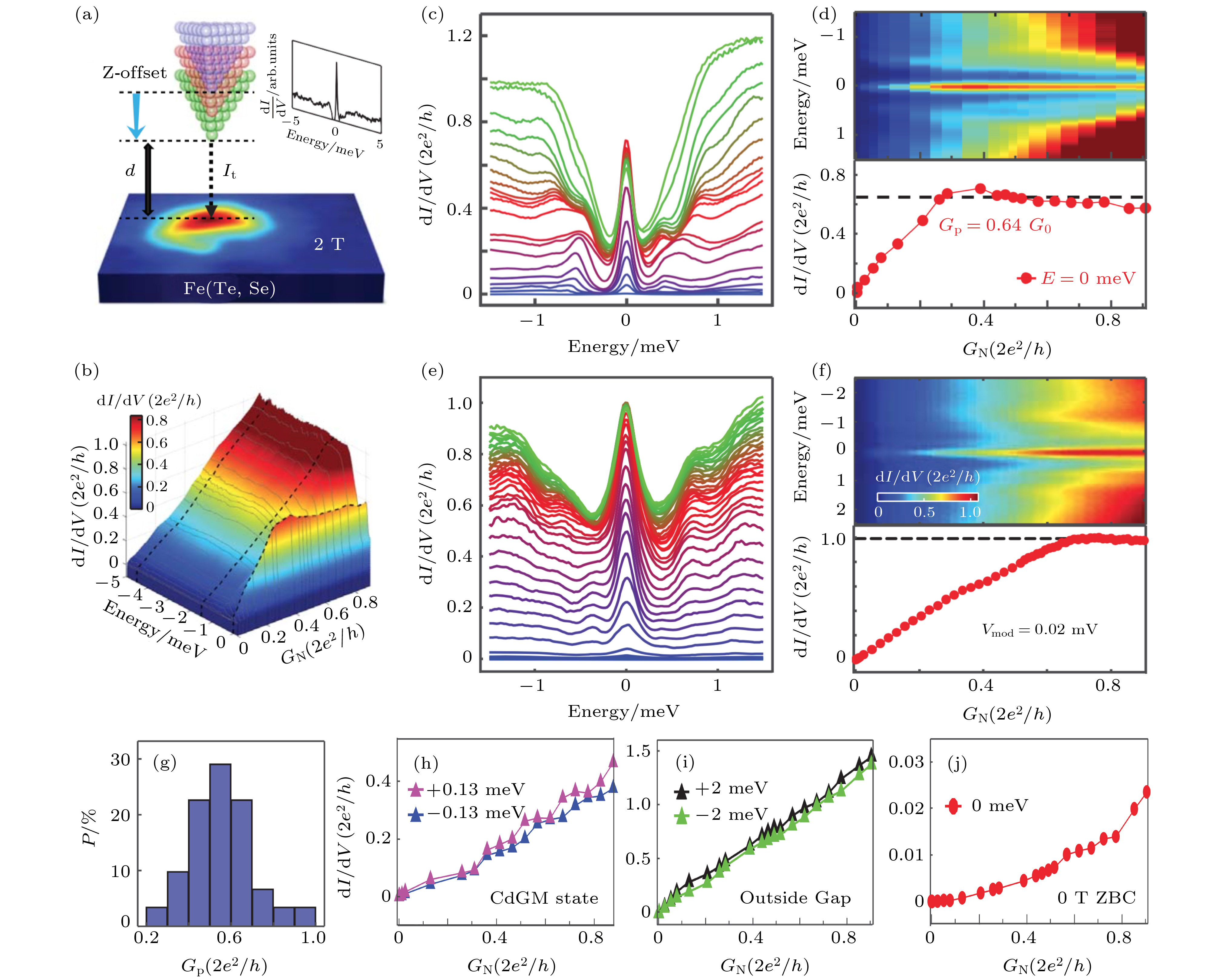}
\par\end{centering}
\caption{\textbf{Variable-tunnel-coupling STM method and observation of conductance plateau of vortex Majorana zero modes}. (a) the tunneling coupling strength can be changed by the tip-sample separation distance under the effect of STM regulation loop; (b) three-dimensional plot of tunneling coupling dependent measurement, \(\mathrm{d} I / \mathrm{d} V\)\((E, G_{\mathrm{N}})\), which shows a zero-bias conductance plateau; (c), (d) general phenomena observed on Majorana conductance of FeTe\(_{0.55}\)Se\(_{0.45}\), \(i.e.\), non-quantized plateau; (e), (f) rare case of nearly quantized plateau in Majorana conductance of FeTe\(_{0.55}\)Se\(_{0.45}\); (g) histogram of the plateau conductance (\(G_{\mathrm{P}}\)) from 31 sets of data; (h)–(j) conductance evolution under different tunneling couplings, showing no plateau feature measured on the finite-energy CdGM states, the continuum outside the superconducting gap, and the zero-filed superconducting state, respectively~\citep{105-zhu2020nearly}.}
\label{fig11}
\end{figure*}

By positioning the STM tip directly above the vortex MZM in a FeTe\(_{0.55}\)Se\(_{0.45}\) single crystal (Fig.~\ref{fig11}(a)), the evolution of the zero-bias conductance of MZM was measured continuously with increasing the tunneling coupling strength~\citep{105-zhu2020nearly}. As shown in Fig.~\ref{fig11}(b), the zero-bias conductance of MZM tends to be saturated when the tunneling coupling is strong enough \((G_{\mathrm{N}} \approx 0.3G_{0}, G_{0} = 2e^{2}/h)\). As the tunneling coupling is further increased, the zero-bias conductance of MZM shows a plateau behavior, while the conductance of the electron continuum outside the superconducting gap keeps increasing. It implies that tunneling coupling between the tip and the vortex MZM may be unconventional. In order to exclude other possible trivial mechanisms of the zero-bias conductance plateau, the conductance behavior of the CdGM bound states at finite energies (Fig.~\ref{fig11}(h)), the continuum outside the superconducting gap under the zero field (Fig.~\ref{fig11}(i)), and the zero-field zero-bias conductance (Fig.~\ref{fig11}(j)) were measured with the change of tunneling coupling strength. Those measurement show that no conductance plateau appears on those cases. The conductance measurement under the zero field eliminates the possibility of quantum ballistic transport~\citep{219-van1988quantized,220-kammhuber2016conductance,221-beenakker1992quantum,222-kjaergaard2016quantized,223-zhang2017ballistic,224-gul2018ballistic}. In addition, the wavefunction of CdGM states is electron–hole inequivalence (Fig.~\ref{fig10}(f)), which is at odds with the requirements of the resonant Andreev reflection. A repeatable experiment show that the conductance plateau behavior is unique to the vortex MZM.

In Ref.~\citep{105-zhu2020nearly}, variable-tunnel-coupled STS measurements were performed on 60 topological vortices, among which 29 vortices became unstable in the process of approaching the tip, possibly owing to vortex creep. On the other 31 vortices, the zero-bias conductance of the MZM shows a plateau behavior after the tunneling coupling exceeding a threshold. The plateau conductance (\(G_{\mathrm{P}}\)) of the 31 vortex MZMs are summarized in Fig.~\ref{fig11}(g). It is obviously that in most cases the zero-bias conductance reaches a plateau at a non-universal value (\(0.2G_{0}\) to \(1.0G_{0}\)), the central value of the \(G_{\mathrm{P}}\) distribution is about 0.6\(G_{0}\). A typical case of non-quantized conductance plateau is shown in Fig.~\ref{fig11}(c) and Fig.~\ref{fig11}(d). However, among the 31 measurements, in one vortex, the zero-bias conductance saturates on a nearly quantized value when the barrier conductance \(G_{\mathrm{N}}\) = 0.7\(G_{0}\) (Fig.~\ref{fig11}(e) and Fig.~\ref{fig11}(f)). It is consistent with the theoretical expectation of MIRAR, and may present the Majorana nature of MZM.

The non-universal plateau conductance is a common behavior of vortex MZMs in FeTe\(_{0.55}\)Se\(_{0.45}\) single crystals. Further experiments~\citep{105-zhu2020nearly} found that, artificially increasing the instrumental broadening (increasing the lock-in excitation voltage) can reduce the plateau conductance of vortex MZM. Furthermore, by comparing the behavior of vortex MZM at different positions, it was found that the basic quasiparticle poisoning suppresses the plateau conductance. We note that the non-universal plateau conductance cannot be described by the theory explained in Section~\ref{sec:5.1}. Although the conductance plateau behavior is unique to the vortex MZM, which strongly implies the appearance of MIRAR, there is no direct experimental evidence requiring that the non-universal conductance plateau must be induced by Majorana modes. Further theoretical and experimental studies are needed to reach a fully understanding of the non-universal plateau conductance.

The discovery of zero-bias conductance plateau of vortex MZMs, especially the nearly quantized case (Fig.~\ref{fig11}(e) and Fig.~\ref{fig11}(f)), indicates a direct measurement of the Majorana nature on vortex MZM, which supports a Majorana origin of the observed ZBCP in the vortex of FeTe\(_{0.55}\)Se\(_{0.45}\) single crystals.

\section{Topological nature of vortex zero mode}\label{sec:6}

MZM is characterized by non-trivial topological invariants~\citep{194-altland1997nonstandard,195-schnyder2008classification,196-teo2010topological,197-chiu2016classification}. For intrinsic topological superconductors (\(e.g.\) \(p_{x}+ip_{y}\) superconductor~\citep{16-read2000paired} and \(p\)-wave Kitaev chains~\citep{225-kitaev2001unpaired}), the topological properties of their quasiparticle spectra can be demonstrated by the mapping from \(\boldsymbol{S}^{2}\) (\(k\)-space) to \(\boldsymbol{S}^{2}\) (spinor space). For connate topological superconductors (\(e.g.\) Fe(Te,Se) single crystals), the vortex MZM can be regarded as the end mode of an one-dimensional topological superconductor (the vortex line) in the view of three-dimensional model, or as the bound state on the topological defects (the vortex) in the view of two-dimensional model (Section~\ref{sec:4.3}). However, emergence of MZM in a vortex requires neither the topologically non-trivial superconductivity, nor the topologically non-trivial band structure~\citep{226-tsui2019classification}, but needs to integrate all the ingredients, \(i.e.\) band topology, superconductivity and vorticity, to reach a non-trivial global topological invariant. For a system with \(N\) Fermi surfaces, the topological invariant of a vortex MZM can be presented as~\citep{227-chan2017generic,228-qi2010topological}, 
\begin{equation}
\mathbb{Z}_{2}=\sum_{i}^{N} m_{i} w_{i}~ (\bmod~2)
\end{equation} 
where \(w_{i}\) is the winding of the quasiparticle spectrum on the \(i\)-th Fermi surface, \(m_{i}\) is the vorticity of the \(i\)-th Fermi surface. When \(\mathbb{Z}_{2}\) = 1, a vortex MZM can emerge. Note that \(\mathbb{Z}_{2}\) is an emergent topological invariant, which indicates the opportunity for emergence of a vortex MZM in trivial materials~\citep{226-tsui2019classification,227-chan2017generic,229-yan2017majorana}. For instance, the fractional vortex of multi-component superconductors (\(e.g.\) spin triplet superconductors, pair density wave, and nematic superconductors) can satisfy the requirements of emergence of vortex MZM~\citep{227-chan2017generic,230-agterberg2008dislocations}.

Since a vortex has an intrinsic winding number, the nontrivial \(\mathbb{Z}_{2}\) in the Fu-Kane hamiltonian is mainly attributed to the appearance of Dirac surface states~\citep{24-fu2008superconducting}. It inspired an experimental method for identifying the topological nature of vortex MZM by investigating the influence of Dirac surface states on the behaviors of vortex bound states. Dirac surface states possess an intrinsic spin angular momentum, which leads to a half-integer level shift on the vortex bound states level sequence. It not only produces a zero-energy Majorana mode, but concomitantly changes the DOS spatial pattern sequence of vortex bound states. These characteristics directly reflect the topological nature of vortex MZM. In this section, we discuss the behaviors of vortex bound states with an underlying superconducting Dirac surface state, and introduce the experimental observations of the topological nature of vortex MZM.

\subsection{Vortex bound state with topology}\label{sec:6.1}

\begin{figure*} 
\begin{centering}
\includegraphics[width=2\columnwidth]{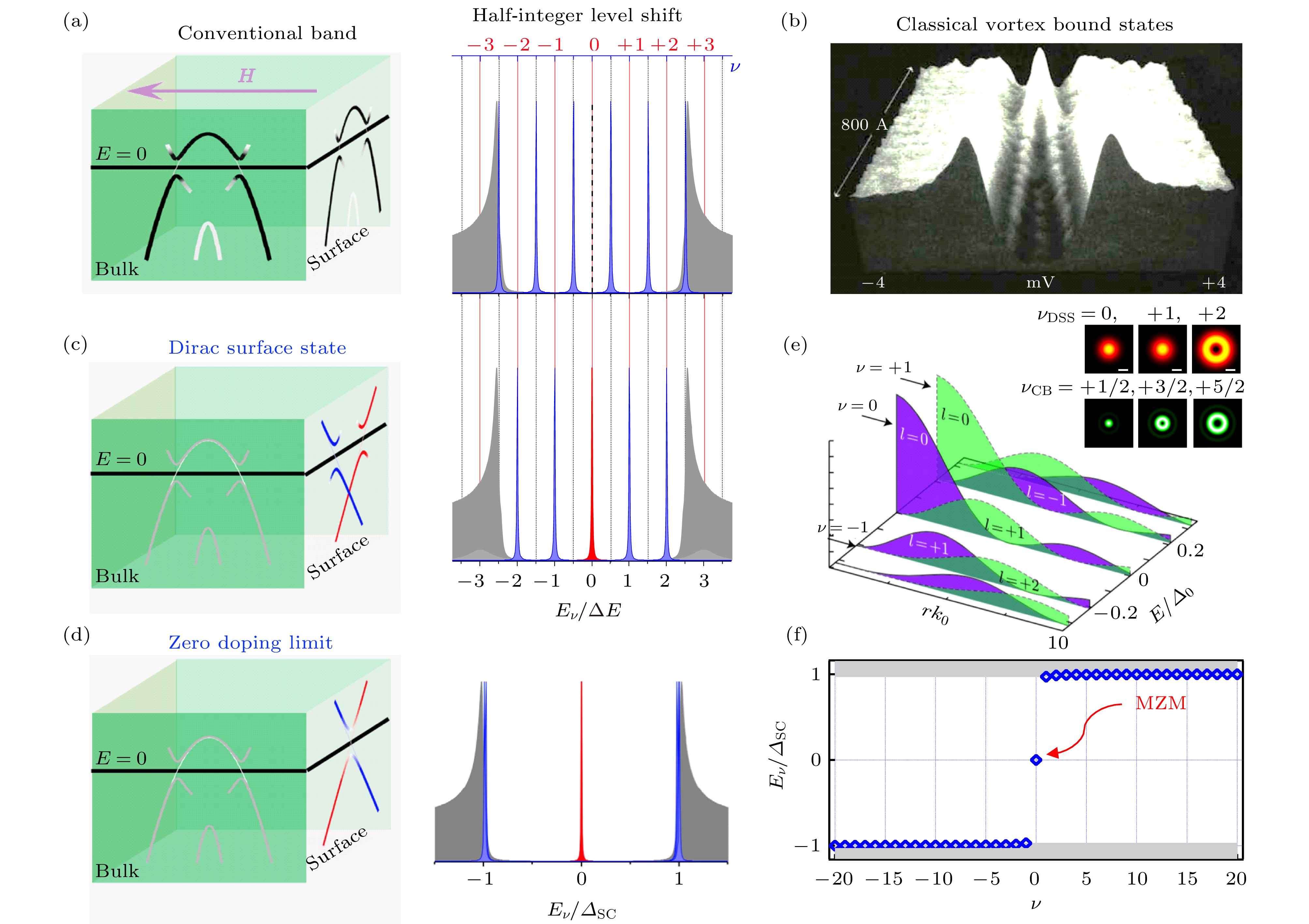}
\par\end{centering}
\caption{\textbf{Surface Dirac electron induced the half-integer level shift of the bound states}. (a) half-odd-integer quantized level sequences of the bound states in a conventional \(s\)-wave superconductor. There are only parabolic bulk bands involved~\citep{104-kong2019half}; (b) the quantum limit is difficult to reach in conventional \(s\)-wave superconductors so that a large ZBCP observed in the center of vortex core is generally due to multiple overlapping of densely packed non-zero peaks~\citep{231-khurana1990stm}; (c) integer quantized level sequences of the bound state in Fu-Kane model. The intrinsic spin Berry phase carried by the Dirac surface states induces the half-integer level shift~\citep{104-kong2019half}; (d) the zero-doping limit is defined as the situation that the chemical potential approaches the energy of the Dirac point. In this case, a vortex MZM is the only allowed subgap bound state~\citep{104-kong2019half}; (e) theoretically calculated angular momentum resolved wavefunction of the BdG eigenstate, the blue and green curves are spin-down and spin-up components, respectively~\citep{210-kawakami2015evolution}. Inset: calculated spin-integrated 2D local density of states of three lowest levels of the bound states in the cases of (c) and (a), respectively~\citep{104-kong2019half}; (f) theoretically calculated eigenvalue of BdG Hamiltonian near the zero chemical potential limit.}
\label{fig12}
\end{figure*}

In conventional \(s\)-wave superconductors, the total angular momenta (\(\nu\)) of vortex bound states are half-integers (\(\nu\) = \(\pm\)1/2, \(\pm\)3/2, \(\pm\)5/2···) owing to the underlying parabolic band structure. When the level spacing is not too large, the vortex bound states follow the half-integer level sequence (Fig.~\ref{fig12}(a)), \(i.e.\), \(E_{\nu}  \approx \nu\varDelta^{2}/E_\mathrm{F}\), with no zero-energy modes. It is a common behavior of conventional \(s\)-wave superconductors. However, a pronounced ZBCP was observed at the vortex center of a conventional superconductors NbSe\(_{2}\) in an early STM experiment. When moving away from the vortex center, the ZBCP splits into two symmetrical peaks, displaying a spatially-dispersive distribution (Fig.~\ref{fig12}(b))~\citep{14-hess1989scanning,175-hess1990vortex,231-khurana1990stm}. These experimental observations are apparently at odds with the theoretical prediction which requires non-zero and discrete solutions in energies. However, in nodal superconductors (such as \(d\)-wave superconductors), the subgap quasiparticles mix with the continuum outside the gap, leading to a spatially-dispersive distribution of the vortex DOS~\citep{232-kopnin1997flux,233-franz1998self,234-berthod2017observation,235-berthod2016vortex}. Although it seemly matches the experimental results, it is apparently inconsistent with the fully-gapped superconductivity of NbSe\(_{2}\).

The anomalous behavior of vortex bound states is caused by the realistic condition that the experimental temperature fails to reach the quantum limit of vortex bound states~\citep{15-gygi1990electronic}. When the thermal broadening is smaller than the level spacing of the vortex bound states~\citep{236-hayashi1998low}, the quantum limit condition is reached. The threshold temperature of quantum limit (\(T_\mathrm{QL}\)) is estimated as, \(T_\mathrm{QL} = T_\mathrm{c}\varDelta/E_\mathrm{F}\). When the experimental temperature exceeds \(T_\mathrm{QL}\), thermal broadening dominates the behavior of vortex bound states, so their energy levels overlap with each other, leading to the spatially-dispersive distribution. Discrete vortex bound states can be only observed in the quantum limit. The realization of quantum limit is a prerequisite for investigating the influence of Dirac surface states on the behaviors of vortex bound states. Conventional superconductors usually have a large Fermi energy (\(E_{F} \approx \) 2 to 10 eV) and a small superconducting gap (\(\varDelta \approx\) 1 meV). It leads to an extremely small \(T_{\mathrm{QL}}\), which is difficult to be reached in experiments. The realization of quantum limit prefers a material with a high critical temperature and a small Fermi energy. As discussed in Section~\ref{sec:1}, several compounds of iron-based superconductors naturally meet these requirements (Fig.~\ref{fig1}(a))~\citep{237-shan2011observation,238-hanaguri2012scanning,239-chen2018discrete,240-berthod2018signatures,241-hanaguri2019quantum,242-chen2020observation}.

A superconducting vortex is a magnetic field induced topological defect in a type-II superconductor. The amplitude of the superconducting gap gradually decreases to zero along the radial direction towards the vortex center. The phase of superconducting gap winds around the vortex which is accordant with the supercurrent circulating at the boundary (London penetration depth). Generally, a superconducting vortex can be written as
\begin{equation}\label{eq:30}
\varDelta(\boldsymbol{r})=\varDelta_{0} h(\boldsymbol{r}) \mathrm{e}^{\mathrm{i} m \phi(x, y)}
\end{equation}
where \(\varDelta_{0}\) is the gap amplitude at the region far away from the vortex; \(h(r)\) is the radial distribution function of the gap amplitude, it can be written as \(h(r) = \mathrm{tanh}(r/\xi)\) or \(h(r) = r/\sqrt{r^{2}+\xi^{2}}\); \(m\), the vorticity, which is an integer and represents how many turns that the superconducting phase changes as it goes around a vortex per circle. \(\phi(x, y)\) is the spatial distribution of the superconducting phase. For example, a single vortex can be expressed as  \(\phi(x, y) = \mathrm{arctan}(y/x)\), and the vortex-antivortex pair can be expressed as \(\phi(x, y) = \mathrm{arctan}[2ay/(x^{2}+y^{2}- a^{2})]\), in which the vortex and the antivortex locate at \((a, 0)\) and \((-a, 0)\), respectively~\citep{243-chang2014majorana}. Note that the definition of the vortex and the antivortex depends on the positive and negative value of the vortex winding numbers, respectively. The vortex winding number is \( Z_{v}=\oint_{c} \arg [\Delta(\boldsymbol{r})] \mathrm{d} s\), where \(c\) is a closed loop around the vortex center~\citep{244-lee2016structure}.

The vortex bound states induced by a superconducting Dirac surface state can be derived by incorporating the surface Dirac cone in the BdG equation with a single vortex~\citep{24-fu2008superconducting,207-chang2012resonant,208-he2014selective},
\begin{equation}\label{eq:31}
H_{\mathrm{BdG}}(\boldsymbol{r})=\left(\begin{array}{cc}
H_{\mathrm{TI}}^{\text {Surface }}(\boldsymbol{r}) & \varDelta(\boldsymbol{r}) \\
\varDelta(\boldsymbol{r})^{\dagger} & {\left[-H_{\mathrm{TI}}^{\mathrm{Surface}}(\boldsymbol{r})\right]^{*}}
\end{array}\right)
\end{equation}

Due to the rotational symmetry of the vortex, the vertical component (\(K_{z}\)) of the total angular momentum is a good quantum number, namely \([K_{z}, H_{\mathrm{BdG}}] = 0\), the energy eigenvalues of vortex bound states can be expressed by the eigenvalues (\(\nu\)) of \(K_{z}\). When \(\varDelta/E_\mathrm{F}\) is not too large, the level sequence of vortex bound states can be approximately expressed as
\begin{equation}\label{eq:32}
E_{\nu} \cong-m \cdot \operatorname{sgn}(\mu) \cdot \nu \frac{\varDelta^{2}}{E_{\mathrm{F}}}
\end{equation}
where \(m\) is the vorticity; sgn(\(\mu\)) is the sign of the chemical potential of the Dirac surface states; \(\nu\) is the vertical component of the total angular momentum. It has been proved that with the incorporation of Dirac electrons, the angular momentum is,
\begin{equation}\label{eq:33}
\nu=l_{z}+\frac{1}{2}\left(S_{z}-m\right)
\end{equation}
\(l_{z}\) is the orbital angular momentum which is an integer. \(S_{z}\) is the spin angular momentum, which is +1 (-1) for the spin-up (down) component~\citep{18-volovik1999fermion,210-kawakami2015evolution,211-hu2016theory}. Different from the case of conventional \(s\)-wave superconductors, the angular momentum (\(\nu\)) of the vortex bound states shown in Eq.(\ref{eq:33}) can be any integers \((\nu = 0, \pm 1, \pm 2, \pm 3, ···)\). Accordingly, the energy levels of vortex bound states inherit the integer quantization, in which \(\nu = 0\) level is the vortex MZM (Fig.~\ref{fig12}(c)). This exotic behavior is caused by the additional \(S_{z}\) introduced by the Dirac surface states~\citep{104-kong2019half}. As a contrast, the angular momentum in a conventional \(s\)-wave superconductor is
\begin{equation}\label{eq:34}
\nu=l_{z}-m / 2 
\end{equation}
This Dirac-surface-state-induced half-integer level shift of vortex bound states is a direct expression of the topological nature of the vortex MZM.

The half-integer level shift of the vortex bound states is accompanied by a change of the DOS pattern sequence on each level. As we introduced in Section~\ref{sec:4.1}, the theoretical wavefunction of a vortex MZM is proportional to Bessel function \(\mathrm{J}_{i}(x)\), of which the order is determined by the orbital angular momentum of vortex bound states, namely \(i = l_{z}\). According to the properties of Bessel function, the maximum of the \(l_{z} = 0\) component appears at the vortex center. With increasing \(l_{z}\), the DOS maximum gradually moves away from the vortex center, leading to a low-intensity valley. Combining with the analysis shown in Eq.(\ref{eq:32})-Eq.(\ref{eq:34}), we obtain the behaviors of angular momentum \((\nu, l_{z}, S_{z})\) and wavefunction on each level. In Fig.~\ref{fig12}(e), we show a theoretical simulation of the DOS distribution of vortex bound states under the conditions of \(m = -1\) and \(\mu > 0\), where the purple (green) curves represent the spin-down (up) component~\citep{210-kawakami2015evolution}. By summing up all the components on each energy level, we obtain the DOS pattern sequence of vortex bound states, which is observable by the constant-energy conductance map of a STM measurement. It can be found that the lowest energy level (\(\nu \) = 0) and one of the two second lowest levels (\(\nu \) = +1 or -1) are of ``solid circle pattern'' while the rest are of ``hollow ring patter'' (the top row of Fig.~\ref{fig12}(e)). On the contrary, in the case of conventional \(s\)-wave superconductors, only one of the two lowest energy levels (\(\nu\)= +1/2 or -1/2) has ``solid circle pattern'' (the bottom row of Fig.~\ref{fig12}(e)). This difference in the pattern sequence is an important feature of the half-integer level shift induced by the Dirac surface state~\citep{104-kong2019half}.

By implementing the wavefunction and angular momentum analysis, we derived some conclusions of the influence of Dirac surface states on the vortex bound states, as listed below: 1) reversing the direction of magnetic field changes the sign of vorticity, consequently changing the positive and negative correspondence between \(\nu\) and \(E_{\nu}\), but it does not change the sign of the energy of the second solid-circle-pattern level~\citep{106-liu2019new}; 2) the sign of the chemical potential determines the sign of the energy of the second solid-circle-pattern level. Specifically, the second solid-circle-pattern level appears in the unoccupied side (positive energy) when \(\mu > 0\), and vice versa; 3) at the vortex center, the spin of the zero-energy level is always parallel to the magnetic field, while the spin of the second solid-circle-pattern level is always anti-parallel to the magnetic field. It is the spin-resolved property of the vortex MZM~\citep{62-sun2016majorana,208-he2014selective,209-haim2015signatures,210-kawakami2015evolution,211-hu2016theory}.

In the last part of this section, we focus on the behavior of vortex bound states under the zero-doping limit. As mentioned previously, the integer quantization is robust in term of the angular momentum, but not robust in term of energy. When \(\varDelta/E_\mathrm{F}\) is very large, the energy of the second lowest vortex bound states will be very close to the superconducting gap edge, and this would induce quantum confinement on the higher levels that locate between the gap edge and the second lowest level. Thus the energies of the vortex bound states are no longer integer quantized.

Theoretically, when the chemical potential approaches to the Dirac point (\(E_{F} \rightarrow 0)\), which is referred as the zero-doping limit, the MZM is the only allowed subgap bound state (Fig.~\ref{fig12}(d))~\citep{246-jackiw1981zero,247-ghaemi2012near}. The energy eigenvalue of the BdG equation has analytical solutions under some special circumstances. For example, when \(h(r) = r/\sqrt{r^{2}+\xi^{2}}\), the energies of vortex bound states in a single vortex \((|m| = 1)\) can be derived as,
\begin{equation}\label{eq:35}
E_{\nu}=-m \frac{\varDelta_{0} \nu / k_{\mathrm{F}}}{\sqrt{\left(\nu / k_{\mathrm{F}}\right)^{2}+(\xi)^{2}}} \frac{K_{0}\left(2 \sqrt{\left[\nu /\left(k_{\mathrm{F}} \xi\right)\right]^{2}+1}\right)}{K_{1}\left(2 \sqrt{\left[\nu /\left(k_{\mathrm{F}} \xi\right)\right]^{2}+1}\right)}
\end{equation}
where \(K_{i}\) is the McDonald function~\citep{245-khaymovich2009vortex}. By using Eq.(\ref{eq:35}), we calculate the level spectrum of vortex bound states when \(m = -1\) and \(\mu \rightarrow 0^{+}\) as shown in Fig.~\ref{fig12}(f), finding that only an isolated vortex MZM exists within the superconducting gap, while other non-zero vortex bound states are pushed to the gap edge. The topological gap of vortex MZM is the largest under the zero-doping limit.

It is worth noting that when \(\mu\) is exact zero, an additional pseudo-chiral symmetry appears in the Fu-Kane hamiltonian, it changes the topological classification of vortices from class-\(D\) to class-\(B\)\(D\)\(I\), and the topological invariant from \(\mathbb{Z}_{2}\) to \(\mathbb{Z}\)~\citep{196-teo2010topological}. Majorana hybridization of the vortex MZM is forbidden in class-\(B\)\(D\)\(I\)~\citep{248-cheng2010tunneling,249-cheng2009splitting,250-biswas2013majorana}, resulting in failure of Majorana topological qubits. However, the pseudo-chiral symmetry guarantees the degeneracy of multiple vortex MZMs. A Majorana flat band appears in a lattice formed by the vortex MZMs, and the four-body Majorana interactions play an important role in the Majorana flat bands~\citep{251-chiu2015strongly,252-liu2015electronic}. It provides a rare opportunity to generate novel quantum phenomena~\citep{253-rahmani2019interacting}, including SYK model and Majorana fractional quantum Hall effect. The MZM interaction is beyond our scope here. The readers interested in this topic may refer to Ref.~\citep{253-rahmani2019interacting}.

\begin{figure*} 
\begin{centering}
\includegraphics[width=2\columnwidth]{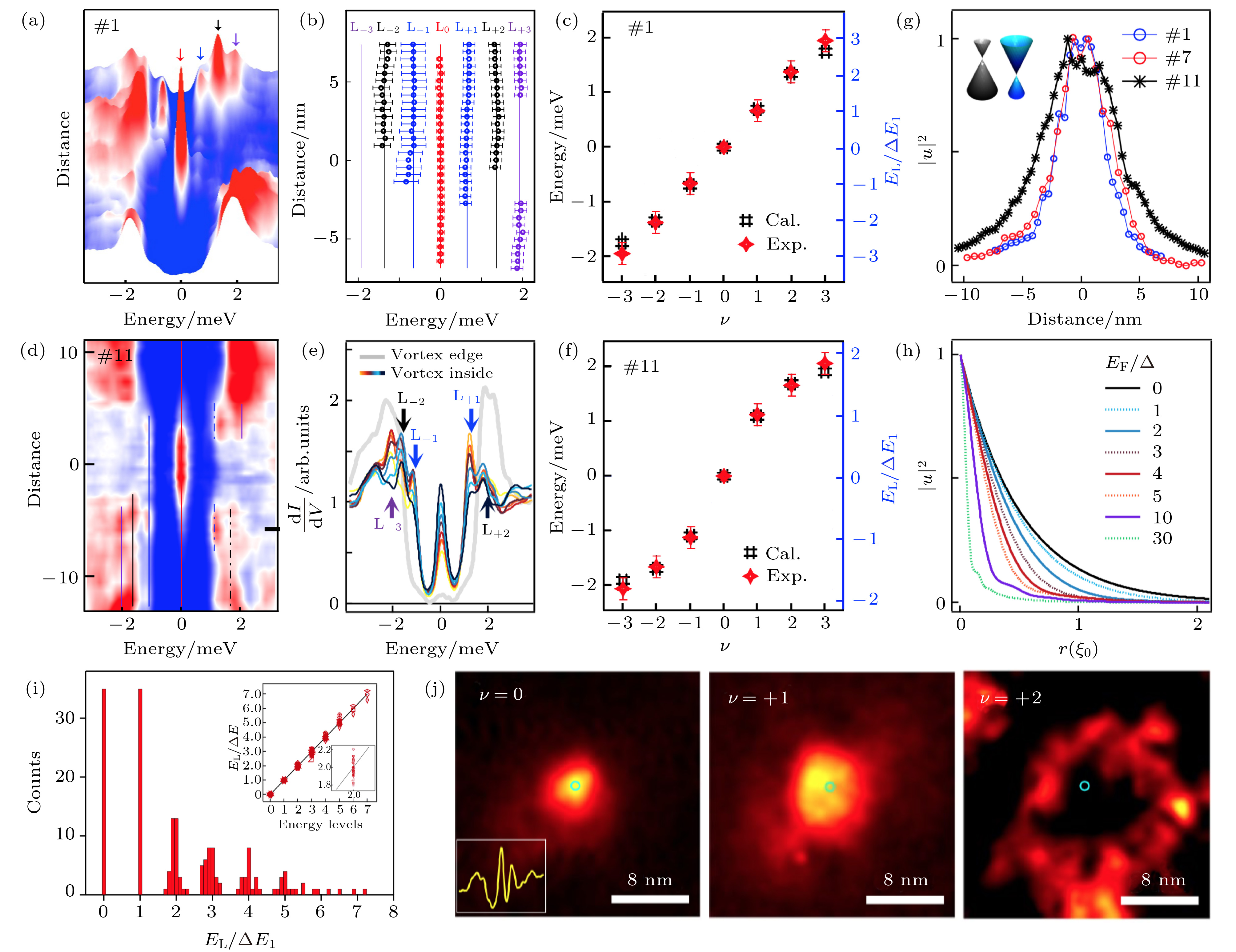}
\par\end{centering}
\caption{\textbf{Observation of integer quantized bound states}. (a) \(\mathrm{d} I / \mathrm{d} V\)\((r, V)\) line-cut intensity plot measured on a topological vortex1. Integer quantized bound states are clearly observed; (b) peak positions extracted from (a); (c) comparison between experimentally observed and theoretically calculated level energy in topological vortex1; (d) same as (a), but measured on vortex11, which is close to the zero chemical potential limit; (e) overlapping display of \(\mathrm{d} I / \mathrm{d} V\) spectra selected from (d); (f) same as (c), but shows the case of vortex11; (g) comparison of observed MZM line profile in topological vortices under integer quantization (open circles) and near the zero chemical potential limit (dark stars); (h) calculated MZM wavefunction under different chemical potential by Fu-Kane model; (i) histogram of averaged level energies normalized by the first level spacing, \(i.e.\), the ratio \(E_{\mathrm{L}}/\Delta E\). The statistical analysis is performed among all the 35 topological vortices which show integer quantized CdGMs levels; (j) experimentally observed spatial pattern of the lowest three levels of bound state in a topological vortex~\citep{104-kong2019half}.}
\label{fig13}
\end{figure*}

\subsection{Observation of integer-quantized topological vortex}\label{sec:6.2}
The realization of quantized vortex bound states in Fe(Te,Se)~\citep{239-chen2018discrete} enables the measurements of Dirac-surface-state-induced integer quantization, which reflects the topological nature of vortex MZM. In Section~\ref{sec:4} and Section~\ref{sec:5}, we only focus on the properties of the vortex MZM located at the zero energy, but behaviors of other vortex bound states that accompany the presence of the MZM were missed previously~\citep{102-wang2018evidence,105-zhu2020nearly}. With a deep understanding of the topological nature of vortex MZM, the characteristics of the global level sequence of vortex bound states have been further studied. In this section, we introduce the relevant experimental results.

On the basis of previous work, the vortices of Fe(Te,Se) were further studied in experiments. By paying more attention to the global behavior of vortex bound states rather than focusing only on the zero modes, we found that both the vortex MZM and other finite-energy vortex bound states exist in the topological vortices. At extremely low temperatures, they all show discrete features, indicating realization of the quantum limit in FeTe\(_{0.55}\)Se\(_{0.45}\) single crystals (Fig.~\ref{fig13}(a)). The energy/spatial positions of the vortex bound states (vortex1) were extracted in Fig.~\ref{fig13}(b) which clearly shows the level quantization. In Fig.~\ref{fig13}(c), the energy of each levels (\(E_{\mathrm{L}}\)) was normalized by the level spacing (\(\Delta E_{1}\)). An integer-quantized level sequence can be clearly observed. The energy eigenvalues of the vortex bound states can be numerically calculated as a function of angular momentum, which was performed on vortex1, and the results are fully consistent with experiment (Fig.~\ref{fig13}(c)). Furthermore, a statistical analysis was done among all the topological vortices measured. Figure~\ref{fig13}(i) is a histogram that shows the distribution of normalized level energies of 35 vortices, and integer quantization was further supported by the statistical analysis.

The integer-quantized level sequence is attributed to an additional half-integer angular momentum introduced by the Dirac surface states (Section~\ref{sec:6.1}). In addition to the integer quantization,
the Dirac surface states also produce double ``solid circle pattern'' levels in the DOS spectra of the vortex bound states. In order to ensure that the integer level sequence observed in Fig.~\ref{fig13}(a)-Fig.~\ref{fig13}(c) does come from the underlying Dirac surface state, rather than a coincidence, the constant energy conductance maps were measured experimentally at the energy of the 0-, +1- and +2-levels of the vortex bound states (Fig.~\ref{fig13}(j)). It shows that the first two levels have solid-circle DOS pattern, and the +2-level is hollow-ring-like. Considering the positive chemical potential of the Dirac surface states in FeTe\(_{0.55}\)Se\(_{0.45}\), the second ``solid-circle'' level should appear on the unoccupied side (positive bias)~\citep{207-chang2012resonant}. The experimental results are fully consistent with the theoretical expectations (Fig.~\ref{fig12}(e)). The observation of the double ``solid circle'' spatial pattern indicates that the observed integer-quantized bound states in FeTe\(_{0.55}\)Se\(_{0.45}\) single crystals are emerged from the superconducting Dirac surface states. We note that the clear observation of the hollow-ring-like pattern at \(\nu = +2\) level is attributed to the tiny \(k_\mathrm{F}\) (about \(0.02 \mathrm{\AA}\)) of the Dirac surface state (Fig.~\ref{fig3}). The ring radium of the levels with \(l_{z} \neq 0\) is proportional to \(1/k_\mathrm{F}\) (Fig.~\ref{fig12}(e)). A smaller \(k_\mathrm{F}\) value results in a larger spatial oscillation period, which is easier to be observed on the constant energy conductance map of STM. However, \(k_\mathrm{F}\) of the bulk states of Fe(Te,Se) is about \(0.1 \mathrm{\AA}\), corresponding to a small spatial oscillation period in the order of 1 nm. It is difficult to survive under spatial inhomogeneity.

In some vortices, a prominent MZM was isolated in the middle of the superconducting gap, there seemly no other bound states except the MZM, obviously violates the integer quantized level sequence. In Fig.~\ref{fig13}(d)-Fig.~\ref{fig13}(f), we reanalyze a topological vortex belong to this case (vortex11 displayed in Fig.~\ref{fig7}). It can be found that the prominent MZM is not the only subgap state, there are three discrete vortex bound states near the gap edge. It is obvious that the vortex bound states do not conform to the integer quantization (Fig.~\ref{fig13}(f)). A numerical simulation, following the same model used in Fig.~\ref{fig13}(c) but with a smaller chemical potential, fully reproduced experimental results, indicates vortex11 is near the zero doping limit. Furthermore, the zero-bias conductance of MZM was extracted on the three vortices (Fig.~\ref{fig13}(g)). It shows that the spatial extension of the MZM near the zero-doping limit is wider. We calculated the analytical Majorana wavefunction (\(|u|^{2}(r)\)) under different \(E_\mathrm{F}\) value (Fig.~\ref{fig13}(h)), and found that smaller \(E_\mathrm{F}\) values correspond to larger spatial distribution of vortex MZM~\citep{24-fu2008superconducting,210-kawakami2015evolution}. It is fully consistent with the experiments, indicates that the break down of integer quantization in vortex11 is a result of near zero-doping limit.

\begin{figure*} 
\begin{centering}
\includegraphics[width=2\columnwidth]{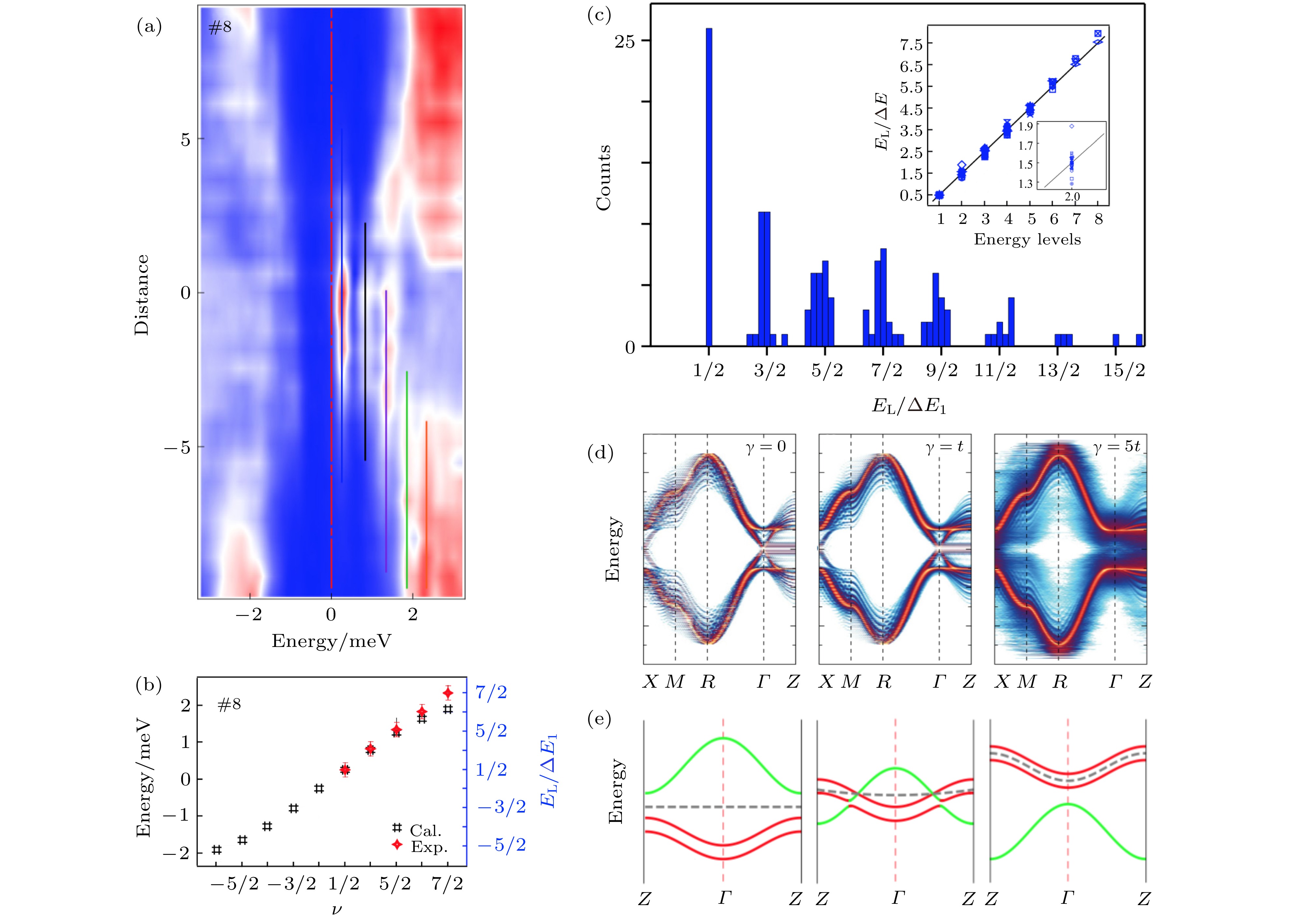}
\par\end{centering}
\caption{\textbf{Inhomogeneity of a material enables coexistence of ordinary and topological vortices}. (a) \(\mathrm{d} I / \mathrm{d} V\)\((r, V)\) line-cut intensity plot measured on ordinary vortex8. Half-odd-integer quantized bound states are clearly observed~\citep{104-kong2019half}; (b) comparison between experimentally observed and theoretically calculated level energy in ordinary vortex8~\citep{104-kong2019half}; (c) histogram of averaged level energies normalized by the first level spacing, \(i.e.\), the ratio \(E_{\mathrm{L}}/\Delta E\). The statistical analysis is performed among all the 26 ordinary vortices which show the half-odd-integer quantized CdGM levels~\citep{104-kong2019half}; (d) surface disorder transforms a strong topological insulator to a normal insulator. The scattering potentials gradually increase from left to right~\citep{262-schubert2012fate}; (e) concentration of the dopants could drive a strong topological insulator to be a normal insulator or a weak topological insulator in Fe(Te,Se). The bands in green (red) represent \(p_{z}\) (\(d_{xz}/d_{yz}\)) orbital with odd (even) parity~\citep{201-qin2019topological}.}
\label{fig14}
\end{figure*}

\subsection{Inhomogeneity enables ordinary vortex}\label{sec:6.3}

Even though topology dictates the existence of two types of discrete bound state spectra, \(i.e.\), integer quantization in a topological vortex and half-integer quantization in an ordinary vortex, in ordinary circumstances a given material belongs to just one of the classes. This restricts a single sample to one type of spectrum and thus forbids a direct comparison. However, the intrinsic inhomogeneity of FeTe\(_{0.55}\)Se\(_{0.45}\) single crystals~\citep{254-beidenkopf2011spatial,255-kotta2020spectromicroscopic,256-martin2008observation,257-rhodes2019disorder,258-he2011nanoscale,259-lin2013direct,260-singh2013spatial,261-massee2015imaging}, while playing a destructive role on emergence of vortex MZM (Section~\ref{sec:4}), enriches the sample properties, so that a direct comparison between topological and ordinary vortices can be realized in a same region, provides a rare opportunity to identify the half-integer level shift.

Intrinsic inhomogeneity is identified in FeTe\(_{0.55}\)Se\(_{0.45}\) single crystals on multiple aspects, including chemical composition, chemical potential, disorder/scattering potential, superconducting gap, and etc. Some influences of them have been demonstrated in the previous sections, that 1) spatial fluctuations of the superconducting gap and the chemical potential can induce the near-zero-doping limit behavior in some topological vortices (Section~\ref{sec:6.2}), 2) stronger spatial fluctuations of the chemical potential can drive a topological phase transition of the vortex line (Section~\ref{sec:4.3}), 3) inhomogeneous disorder/scattering potential may contribute to nonuniform quasiparticle poisoning, induces different zero-bias conductance of MZM on different vortices (Section~\ref{sec:4.2}). In those cases, inhomogeneity directly affects the behavior of low-energy superconducting quasiparticles, and we call them ``weak inhomogeneity'' in this review. In contrast, the ``strong inhomogeneity'' not only affects low-lying quasiparticles directly, but also indirectly modify the underlying band structure. 

\begin{figure*} 
\begin{centering}
\includegraphics[width=2\columnwidth]{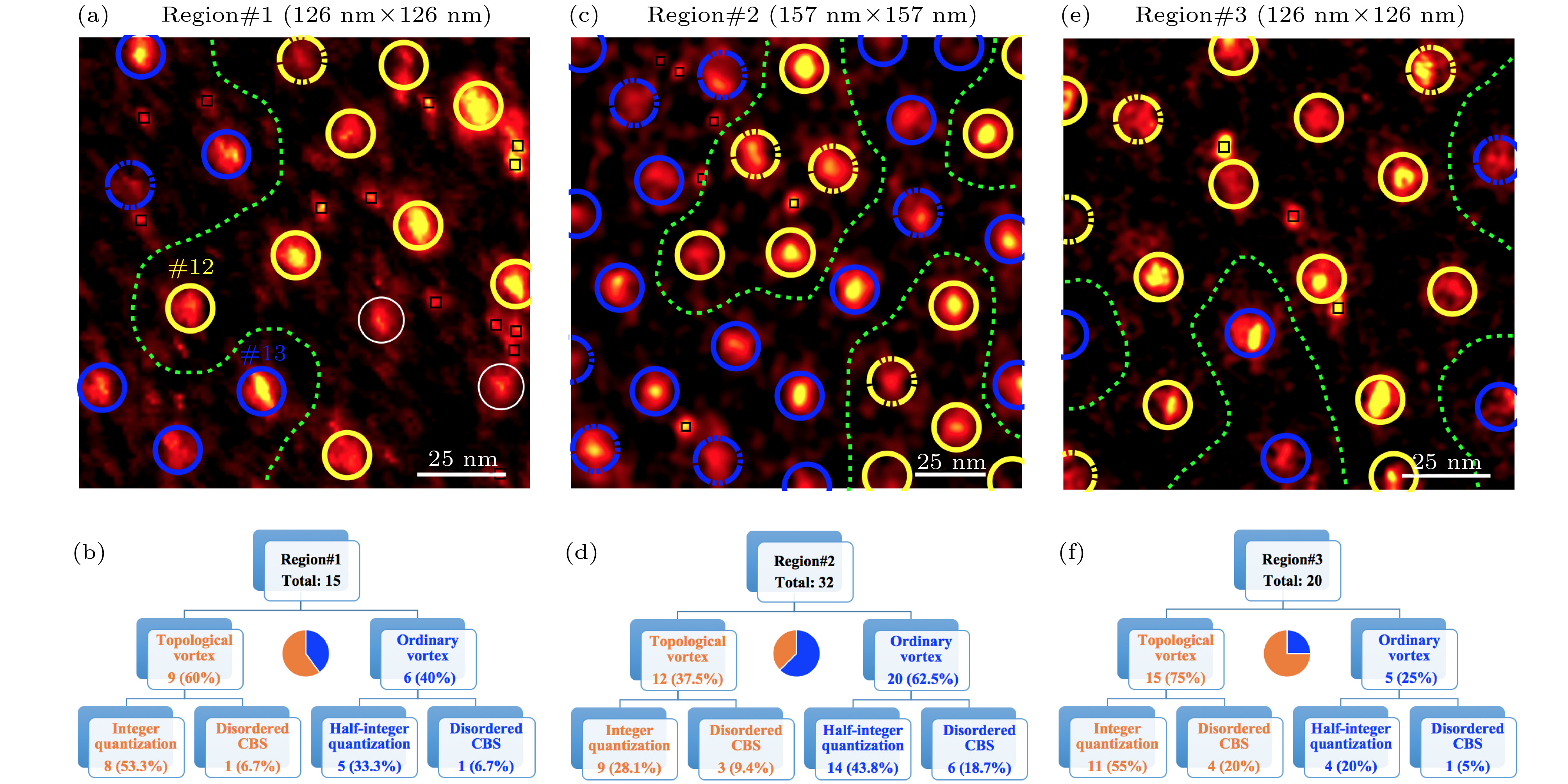}
\par\end{centering}
\caption{\textbf{Spatial distribution of two classes of vortices}. (a), (c), (e) zero-bias conductance maps of three well-separated regions. The yellow solid circles mark the vortices with ZBCPs and integer quantized CdGM levels, the yellow dashed circles mark the vortices with ZBCPs but their CdGM level sequences cannot be fitted to integer quantization, and the blue solid circles mark the vortices without ZBCPs and half-integer quantized CdGM levels, and blue dashed circles mark the vortices without ZBCPs or half-integer quantized CBS levels. The green dashed lines encircle the same class of vortices. Topological vortices and ordinary vortices usually group together, which indicates topological region and trivial region coexist on a sample surface due to spatial inhomogeneity; (b), (d), (f) summary of the ratio of different types of vortices in the three regions, respectively. The data in the three regions are measured at 40 mK and 2.0 T~\citep{104-kong2019half}.}
\label{fig15}
\end{figure*}

\begin{figure*} 
\begin{centering}
\includegraphics[width=2\columnwidth]{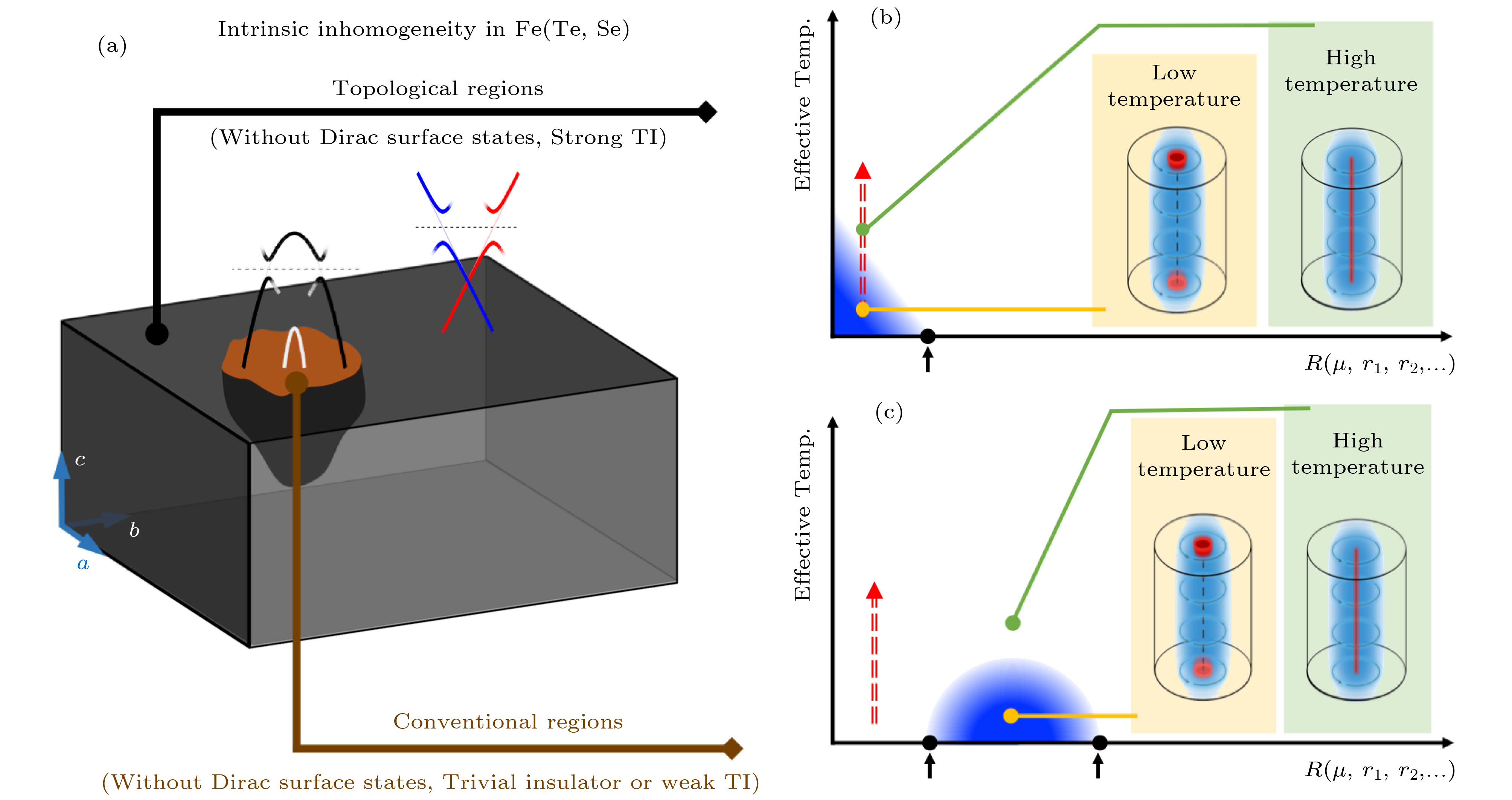}
\par\end{centering}
\caption{\textbf{Mechanism of presence or absence of MZMs in Fe(Te,Se)}. (a) Fe(Te,Se) single crystals are intrinsically inhomogeneous. The disappearance of Dirac surface states is possible in some regions of the (001) surface (brown color). In the conventional regions, the corresponding bulk states can be normal insulators or weak topological insulators. Consequently, the Dirac surface state moves deeper into the bulk and goes around the conventional region, as indicated by the gray boundary inside the crystal. In other topological regions (gray color), where the Dirac surface states remain intact, the corresponding bulk states are still in the strong topological insulating phase; (b) schematic phase diagram of vortex MZMs appearing in topological regions (topological vortices). The gradient blue areas in (b) and (c) indicate the phase sector that MZMs can be detected by STM/S experiments. In the dark blue sector, the Majorana wavefunction is more localized on the sample surface, while in brighter positions, the Majorana wavefunction strongly hybridizes with the bulk quasiparticles and moves deeper beneath the surface, leading to a weak ZBCP signal measured by STM/S. The vertical axis demonstrates the evolution of MZMs as a function of effective temperature which can be represented by the extrinsic broadening of observed ZBCPs. The horizontal axis demonstrates the MZMs evolution as a function of quantum parameters, \(e.g.\), the chemical potential (\(\mu\)) measured from the Dirac point. The black dots with an arrow indicate the quantum critical points in which a vortex phase transition happens. Across the critical point, the vortex line turns to be topologically trivial and MZMs disappear in the topological region. The red dashed line indicates the achievable region in experiments; (c) a schematic phase diagram of vortex MZMs appearing in conventional regions (ordinary vortices). There are no MZMs in our measurements in those vortices. The observable MZMs can only exist above the critical points when the vortex phase transition turns the trivial vortex line into a 1D topological superconductor in the conventional region~\citep{104-kong2019half}.}
\label{fig16}
\end{figure*}

In a FeTe\(_{0.55}\)Se\(_{0.45}\) single crystal, strong inhomogeneity can destroy strong topological insulator states in some portions. A single piece of sample can be regarded as a topological crystal embedded by many non-topological nanocrystals. The Dirac surface state is not fully seated on the bare surface of the sample, but wiggles along the new boundaries between topological and non-topological crystals (Fig.~\ref{fig16}(a)). It leads to the disappearance of the Dirac surface state on some non-topological regions where the half-integer-quantized ordinary vortices can appear. Experimental studies found that topological or ordinary vortices usually appear in groups, It supports the picture of disappearance of the Dirac surface states on some surface regions of FeTe\(_{0.55}\)Se\(_{0.45}\) (Fig.~\ref{fig15}). Two possible mechanisms of ``strong inhomogeneity'' which can locally destroy the strong topological insulator states are described below.

1) disorder induced non-magnetic scattering drives the transition to a normal insulator state. Generally, the strong topological insulator states are protected by the time-reversal symmetry, and it cannot be destroyed by any non-magnetic impurities~\citep{25-hasan2010colloquium}. However, the topological protection is only valid under the condition of weak impurity scattering. When the non-magnetic scattering potential is comparable to the inverted SOC gap of the bulk bands, the strong topological insulator state can be destroyed. A theoretical calculation confirms the disappearance of the Dirac surface state under the condition of strong scattering potential (Fig.~\ref{fig14}(d)). The SOC gap of the bulk bands in FeTe\(_{0.55}\)Se\(_{0.45}\) is estimated to be about 20 meV (Fig.~\ref{fig4}(f)), which is much smaller than that of the classical topological insulator Bi\(_{2}\)Se\(_{3}\) (about 300 meV). Although the topological protection of the Dirac surface state is robust and universal within the topological phase, the topological phase itself is not robust owing to the small SOC gap of FeTe\(_{0.55}\)Se\(_{0.45}\). The strong inhomogeneity of the material breaks the topological band structure at some regions, in which ordinary vortices appear~\citep{262-schubert2012fate,263-sacksteder2015modification,264-pan2020physical}.

2) The composition fluctuation of the anionic atomic drives the transition to normal insulator or weak topological insulator state~\citep{265-fu2007topological,266-noguchi2019weak}~ (Fig.~\ref{fig14}(e)). As shown in Section~\ref{sec:2.1}, there is no topological bands in Se-rich Fe(Te,Se) single crystals. With proper doping concentration of Te atoms, a topological band inversion occurs at Z, the material enters a strong topological insulator state. However, it is worth noting that in an overdopped sample, the topological band inversion also occurs at \(\Gamma\), so the material becomes a weak topological insulator, and the Dirac surface state only exists on the side surface~\citep{265-fu2007topological,266-noguchi2019weak}. Therefore, the overdoping of both Se and Te atoms may eliminate the Dirac surface states on the (001) surface, resulting in the appearance of ordinary vortex in some regions~\citep{201-qin2019topological,267-shi2017fete1,268-peng2019observation}.

Ordinary vortex was observed on the surface of FeTe\(_{0.55}\)Se\(_{0.45}\) single crystals (Fig.~\ref{fig14}(a)–Fig.~\ref{fig14}(c)). After implementing a standard analysis (used in Section~\ref{sec:6.2}) on 26 ordinary vortices, we identified a good half-integer level sequence of the vortex bound states. This result can also be fully reproduced by numerical simulation. Owing to the strong inhomogeneity of FeTe\(_{0.55}\)Se\(_{0.45}\), these ordinary vortices emerge at some regions on the (001) surface without Dirac electrons, coexisting with topological vortices in a same piece of sample. A direct comparison between the two classes of vortices under different topologies displays the half-integer level shift of vortex bound state (Fig.~\ref{fig12}(a) \(\&\) (c)), and reveals the topological nature of vortex MZM.

\begin{figure*} 
\begin{centering}
\includegraphics[width=2\columnwidth]{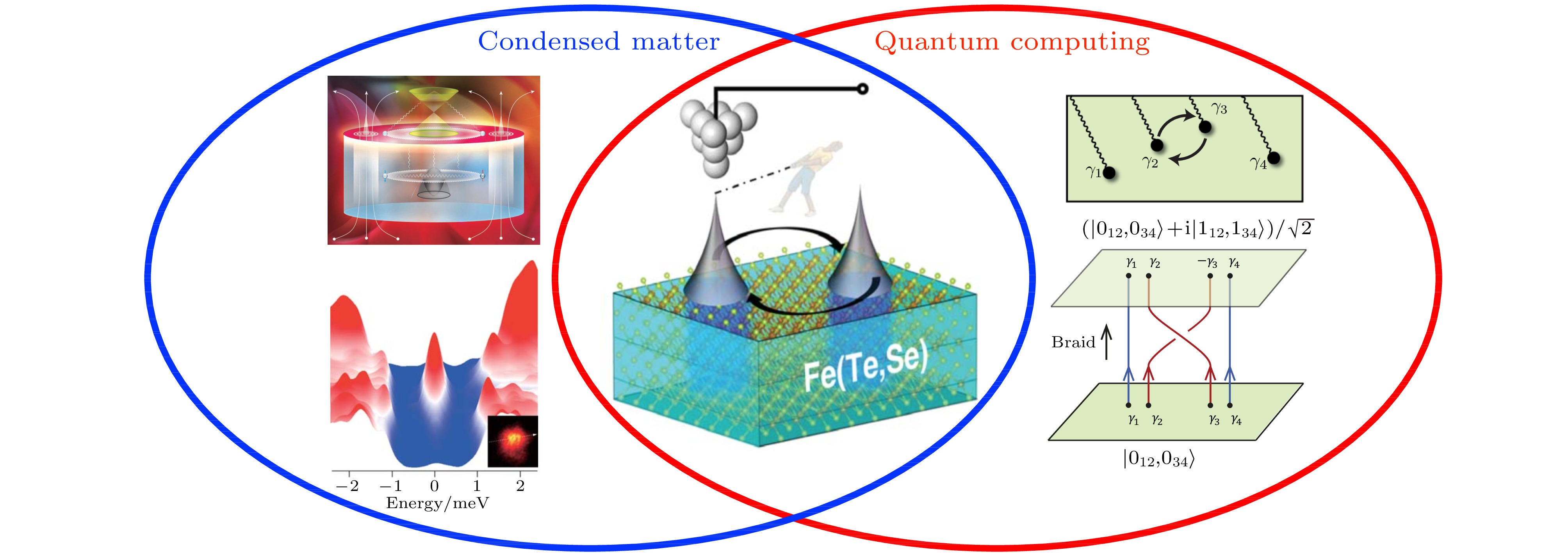}
\par\end{centering}
\caption{\textbf{Braiding vortex MZMs and topological quantum computation}. Left-top panel: surface effective spinless \(p_{x}+ip_{y}\) pairing induced by \(k\)-proximity effect from the bulk bands in Fe(Te,Se)~\citep{84-hao2019topological}. Left-bottom panel: pristine vortex MZM observed in Fe(Te,Se)~\citep{103-kong2019majorana}. Middle panel: It is possible to use a STM tip to manipulate the vortex MZMs on the surface of Fe(Te,Se)~\citep{102-wang2018evidence}. Right panel: Topological qubit built by braiding four vortex MZMs~\citep{3_alicea2012new}.}
\label{fig17}
\end{figure*}

\section{From quantum physics to quantum engineering}\label{sec:7}
In a real material, the three-dimensionality and intrinsic inhomogeneity induce variable behaviors and complex distributions of vortices, including two distinct classes of vortices and disappearance of MZM in some vortices. As every coin has two sides, the imperfection of samples plays different roles depending on the topics concerned. On the perspective of physics study, it is a gift of nature, which enables a direct comparison of the vortex bound states under two different topologies. But from the view of practical applications, it becomes a nightmare for the exploration of topological quantum computation. In the previous sections, we describe in details the basic properties and experimental observations of the vortex MZM in iron-based superconductors. In combination with the simplified theoretical models, that only taking the perfect cases into account, the analysis always captures the key nature of exotic phenomena (Section~\ref{sec:4}). This is the methods of physics research. However, real materials are not as simple as the theoretical models, the imperfections in materials and environments introduce other factors, which interfere with the target physical phenomena. Those problems must be solved in developing practical applications such as topological quantum computation. This is the task of quantum engineering. Therefore, in order to realize the ultimate goal of topological quantum computation, it is of profound significance to experimentally study the morphology of vortex MZM on the sample surface and theoretically explore the physics mechanisms that affect presence or absence of vortex MZM.

Here we introduce the distribution characteristics of vortex MZMs on FeTe\(_{0.55}\)Se\(_{0.45}\)~\citep{104-kong2019half}. Early experiments were performed at a higher temperature (about 450 mK), under such a condition, the percentage of the vortices that host MZM is less than 20\(\%\), which is difficult to conduct statistical studies~\citep{102-wang2018evidence}. The investigation of quasiparticle poisoning (Section~\ref{sec:4.2}) shows that the signal of vortex MZMs is stronger at lower temperatures. It stimulated new experiments conducted under more extreme conditions. As shown in Fig.~\ref{fig15}, the vortex morphology was studied on three regions which were randomly selected and far away from each other. All the vortices appearing on the three regions were carefully studied by spatial dependent \(\mathrm{d} I / \mathrm{d} V\) measurements of the vortex bound states (in total 76 vortices were measured). By performing standard data analysis as used in Fig.~\ref{fig13} and Fig.~\ref{fig14}, their types were identified on each vortex. As marked in Figs.~\ref{fig15}(a), (c), (e), the yellow circles represent the topological vortices with vortex MZM, the blue circles represent the ordinary vortices without vortex MZM. The statistical results of the three regions are listed in Figs.~\ref{fig15}(b), (d), (f). We derived three conclusions from these data: 1) the occurrence probability of vortex MZM has large spatial fluctuation. It shows that the probability varies from 37\(\%\) to 75\(\%\) depending on the regions. Since the three regions are randomly selected, the large-scale inhomogeneity should play an important role on the occurrence probability of vortex MZM, thus emphasizes the importance of sample quality. 2) most of the vortices (about 76\(\%\), among all three regions) show observable quantization sequences of vortex bound states, either integer quantization in topological vortices or half-odd-integer quantization in ordinary vortices. 3) ordinary or topological vortices appear in groups. In Fig.~\ref{fig15}, the vortices within the same class are encircled by green dotted lines. It supports the picture that the Dirac surface states disappear on some surface areas, while remaining intact in the others.

The disappearance of MZM in some vortices arose controversies on the field of iron-based Majorana platform. In order to understand these phenomena, here we systematically summarize the possible microscopic mechanisms which can influence the presence or absence of vortex MZM~\citep{104-kong2019half}.

First of all, strong inhomogeneity eliminates the Dirac surface states on some areas of the (001) surface. Topological regions (with Dirac surface states) and conventional regions (without Dirac surface states) can coexist on sample surface, in which topological vortices (with vortex MZM) and ordinary vortices (without vortex MZM) appear, respectively.

Second, the following effects affect presence or absence of vortex MZM at a given spatial position.

1) vortex topological phase transition changes the topological invariant~\citep{176-hosur2011majorana} (see a detailed discussion of the influence of topological phase transition of vortex lines in Section~\ref{sec:4.3}). For the vortices in a topological region, the occurrence of topological phase transition destroys the vortex MZM, while in a conventional region, the phase transition creates the vortex MZM.

2) low-lying quasiparticles suppress the vortex MZM. First, the vortex MZM is protected by the bulk mini gap (\(\delta\)) of the one-dimensional vortex lines. The vertical decay length of the Majorana wavefunction is proportional to 1/\(\delta\). On approaching to the critical point of quantum phase transition, the Majorana wavefunction goes deeply into the bulk of the material, thus become invisible in STM measurementa. On the quantum critical point, the one-dimensional vortex line becomes gapless. The two MZMs on the top and bottom surfaces annihilate with each other. Second, the quasiparticle poisoning is not uniform owing to the intrinsic inhomogeneity of the material. It results in different intensities of vortex MZM measured at the same experimental temperature. Third, additional quasiparticle poisoning exists at higher temperatures (Section~\ref{sec:4.2}), which induces the rapid smearing of vortex MZMs with raising temperatures.

These mechanisms are summarized into a phase diagram of vortex MZM when the magnetic field is week (Fig.~\ref{fig16}), in which the red dot line is an indication of the phase region which can be covered by current experiments.

Finally, it is worth pointing out that under high magnetic fields the in-plane Majorana hybridization plays a vital role on controlling presence or absence of a vortex MZM. It was observed that the occurrence probability of the vortex MZM decreases with the increasing magnetic field~\citep{269-machida2019zero,270-chen2019observation}, this phenomenon was later explained by a theoretical simulation of Majorana hybridization in a disordered vortex lattice~\citep{271-chiu2020scalable}.

\section{Conclusion and outlook}\label{sec:8}
In this article, we have present a systematic and comprehensive review on vortex MZMs in Fe(Te,Se) single crystals, ranging from the origin idea to research progresses, from classical theories to new experiments, from band structure to quasiparticles, and from fundamental physics to realistic details. We aim to bridge the gap between the well-established Majorana theories and the emerging  ``iron home'' of Majoranas, to help the readers to thoroughly understand and reasonably evaluate the emergent vortex MZMs in the iron-based Majorana platform.

Since Fe(Te,Se) was first discovered as a carrier of Majorana modes, the topological properties of iron-based superconductors soon become a hot topic in condensed matter physics. Over the past few years, a large number of theories and experiments have emerged. Here we try to summarize them as follows, 1) independent verifications of appearance of vortex MZM in Fe(Te,Se) single crystals~\citep{269-machida2019zero,270-chen2019observation,271-chiu2020scalable,272-wang2020evidence}. 2) new developments in theories of the vortex topological phase transition in iron-based superconductors~\citep{200-ghazaryan2020effect,201-qin2019topological,202-konig2019crystalline,203-qin2019quasi}. 3) discovery of more topological compounds of iron-based superconductors. Topological band structure has been found to be universal in iron-based superconductors~\citep{101-zhang2019multiple}. Vortex MZMs were observed in (Li,Fe)OHFeSe~\citep{273-liu2018robust,274-chen2019quantized} and CaKFe\(_{4}\)As\(_{4}\)~\citep{106-liu2019new}; 4) exploration of new compounds of iron-based superconductors which may support vortex MZM at higher temperatures. Experimental studies identified a topological band structure in a high-\(T_\mathrm{c}\) superconductor Fe(Te,Se) monolayer~\citep{267-shi2017fete1,268-peng2019observation}, indicating that the monolayer of iron-based superconductor may be a high-temperature Majorana platform above the liquid helium temperature~\citep{71-lee2018routes}. 5) developments of defect states, both in theory and experiment. Evidence of the Majorana mode has been reported on point-like impurities~\citep{99-yin2015observation,275-liu2020zero,276-jiang2019quantum,277-zhang2020field,278-fan2020reversible}, step edges~\citep{279-gray2019evidence}, domain walls~\citep{272-wang2020evidence}, and atomic line defects~\citep{280-chen2020atomic,281-wu2020topological,282-zhang2020atomic}. 6) evidence of time-reversal symmetry breaking~\citep{283-zaki2019time,284-hu2020pairing,285-wu2020topological}. 7) possibilities of intrinsic topological superconductivity~\citep{286-kawakami2019topological,287-luo2020topological}. 8) studies of heterostructures that combine a Fe(Te,Se) single crystal with a topological material~\citep{288-chen2018superconductivity,289-zhao2018superconducting,290-bao2018visualizing,291-chen2019zero,292-rachmilowitz2019proximity}. 9) improvement of sample quality~\citep{293-dong2019bulk}. 10) design of Majorana Kramer pairs which preserve the time-reversal symmetry. Majorana Kramer pairs are predicted to appear on the boundary of some iron-based superconductors, providing the \(s\pm\) superconducting pairing applied on the topological surface state~\citep{294-zhang2019helical,295-wu2020boundary}. 11) design of high-order MZMs~\citep{296-zhang2019higher,297-wu2019high,298-chen2019lattice,299-wu2020pursuit}. 12) exploration of Majorana braiding~\citep{300-liu2019protocol,301-november2019scheme,302-posske2020vortex}. 13) Majorana research on other connate topological superconductors~\citep{303-yuan2019evidence,304-xia2020bulk,305-zhang2020hass}. The booming developments of this field not only open up a new horizon for Majorana physics but also strongly support the research on high-temperature superconductivity.

The vortex of iron-based superconductors is emerging as one of the most reliable platform for Majorana zero modes. On the basis of deep understanding of their properties, hybridizing, braiding, fusing MZMs and reading out the quantum information of Majorana qubits become important research directions in the future. The realization of these goals requires joint efforts on the aspects of theory, material, and technique. A practical braiding strategy should be theoretically designed under the real situation of vortex MZM~\citep{306-beenakker2019search}; The sample quality should be optimized to improve the surviving temperature of MZM, and achieve homogeneous electronic environments in the bulk and on the surface; A controllable technique for vortex manipulation should be explored, which is capable for fast braiding within the quasiparticle lifetime. Those advances could remove the barriers for construction of topological qubits and merge the two large fields (Fig.~\ref{fig17}), that are the condensed matter physics and the quantum computation, in a small piece of crystals of iron-based superconductors.

\begin{acknowledgments}
This project is supported by the National Natural Science Foundation of China (Grant Nos. 11888101, 11674371), the Strategic Priority Research Program of Chinese Academy of Sciences, China (Grant Nos. XDB28000000, XDB07000000), and the Beijing Municipal Science \& Technology Commission, China (Grant No. Z191100007219012). This preprint is selected in CNKI Journal Translation Project \url{https://jtp.cnki.net/bilingual/Navi/Detail?pykm=WLXB&year=2020&issue=11}, of which the original version was published in Chinese (Acta Physica Sinica \(\boldsymbol{69}\), 110301 (2020)).

\end{acknowledgments}

\bibliography{apssamp}

\begin{thebibliography}{306}%
\makeatletter
\providecommand \@ifxundefined [1]{%
 \@ifx{#1\undefined}
}%
\providecommand \@ifnum [1]{%
 \ifnum #1\expandafter \@firstoftwo
 \else \expandafter \@secondoftwo
 \fi
}%
\providecommand \@ifx [1]{%
 \ifx #1\expandafter \@firstoftwo
 \else \expandafter \@secondoftwo
 \fi
}%
\providecommand \natexlab [1]{#1}%
\providecommand \enquote  [1]{``#1''}%
\providecommand \bibnamefont  [1]{#1}%
\providecommand \bibfnamefont [1]{#1}%
\providecommand \citenamefont [1]{#1}%
\providecommand \href@noop [0]{\@secondoftwo}%
\providecommand \href [0]{\begingroup \@sanitize@url \@href}%
\providecommand \@href[1]{\@@startlink{#1}\@@href}%
\providecommand \@@href[1]{\endgroup#1\@@endlink}%
\providecommand \@sanitize@url [0]{\catcode `\\12\catcode `\$12\catcode
  `\&12\catcode `\#12\catcode `\^12\catcode `\_12\catcode `\%12\relax}%
\providecommand \@@startlink[1]{}%
\providecommand \@@endlink[0]{}%
\providecommand \url  [0]{\begingroup\@sanitize@url \@url }%
\providecommand \@url [1]{\endgroup\@href {#1}{\urlprefix }}%
\providecommand \urlprefix  [0]{URL }%
\providecommand \Eprint [0]{\href }%
\providecommand \doibase [0]{https://doi.org/}%
\providecommand \selectlanguage [0]{\@gobble}%
\providecommand \bibinfo  [0]{\@secondoftwo}%
\providecommand \bibfield  [0]{\@secondoftwo}%
\providecommand \translation [1]{[#1]}%
\providecommand \BibitemOpen [0]{}%
\providecommand \bibitemStop [0]{}%
\providecommand \bibitemNoStop [0]{.\EOS\space}%
\providecommand \EOS [0]{\spacefactor3000\relax}%
\providecommand \BibitemShut  [1]{\csname bibitem#1\endcsname}%
\let\auto@bib@innerbib\@empty
\bibitem [{\citenamefont {Nayak}\ \emph {et~al.}(2008)\citenamefont {Nayak},
  \citenamefont {Simon}, \citenamefont {Stern}, \citenamefont {Freedman},\ and\
  \citenamefont {Sarma}}]{1_nayak2008non}%
  \BibitemOpen
  \bibfield  {author} {\bibinfo {author} {\bibfnamefont {C.}~\bibnamefont
  {Nayak}}, \bibinfo {author} {\bibfnamefont {S.~H.}\ \bibnamefont {Simon}},
  \bibinfo {author} {\bibfnamefont {A.}~\bibnamefont {Stern}}, \bibinfo
  {author} {\bibfnamefont {M.}~\bibnamefont {Freedman}},\ and\ \bibinfo
  {author} {\bibfnamefont {S.~D.}\ \bibnamefont {Sarma}},\ }\bibfield  {title}
  {\bibinfo {title} {Non-{A}belian anyons and topological quantum
  computation},\ }\href@noop {} {\bibfield  {journal} {\bibinfo  {journal}
  {Rev. Mod. Phys.}\ }\textbf {\bibinfo {volume} {80}},\ \bibinfo {pages}
  {1083} (\bibinfo {year} {2008})}\BibitemShut {NoStop}%
\bibitem [{\citenamefont {Wilczek}(2009)}]{2_wilczek2009majorana}%
  \BibitemOpen
  \bibfield  {author} {\bibinfo {author} {\bibfnamefont {F.}~\bibnamefont
  {Wilczek}},\ }\bibfield  {title} {\bibinfo {title} {Majorana returns},\
  }\href@noop {} {\bibfield  {journal} {\bibinfo  {journal} {Nat. Phys.}\
  }\textbf {\bibinfo {volume} {5}},\ \bibinfo {pages} {614} (\bibinfo {year}
  {2009})}\BibitemShut {NoStop}%
\bibitem [{\citenamefont {Alicea}(2012)}]{3_alicea2012new}%
  \BibitemOpen
  \bibfield  {author} {\bibinfo {author} {\bibfnamefont {J.}~\bibnamefont
  {Alicea}},\ }\bibfield  {title} {\bibinfo {title} {New directions in the
  pursuit of {M}ajorana fermions in solid state systems},\ }\href@noop {}
  {\bibfield  {journal} {\bibinfo  {journal} {Rep. Prog. Phys.}\ }\textbf
  {\bibinfo {volume} {75}},\ \bibinfo {pages} {076501} (\bibinfo {year}
  {2012})}\BibitemShut {NoStop}%
\bibitem [{\citenamefont {Beenakker}(2013)}]{4_beenakker2013search}%
  \BibitemOpen
  \bibfield  {author} {\bibinfo {author} {\bibfnamefont {C.}~\bibnamefont
  {Beenakker}},\ }\bibfield  {title} {\bibinfo {title} {Search for {M}ajorana
  fermions in superconductors},\ }\href@noop {} {\bibfield  {journal} {\bibinfo
   {journal} {Annu. Rev. Condens. Matter Phys.}\ }\textbf {\bibinfo {volume}
  {4}},\ \bibinfo {pages} {113} (\bibinfo {year} {2013})}\BibitemShut {NoStop}%
\bibitem [{\citenamefont {Elliott}\ and\ \citenamefont
  {Franz}(2015)}]{5_elliott2015colloquium}%
  \BibitemOpen
  \bibfield  {author} {\bibinfo {author} {\bibfnamefont {S.~R.}\ \bibnamefont
  {Elliott}}\ and\ \bibinfo {author} {\bibfnamefont {M.}~\bibnamefont
  {Franz}},\ }\bibfield  {title} {\bibinfo {title} {Colloquium: {M}ajorana
  fermions in nuclear, particle, and solid-state physics},\ }\href@noop {}
  {\bibfield  {journal} {\bibinfo  {journal} {Rev. Mod. Phys.}\ }\textbf
  {\bibinfo {volume} {87}},\ \bibinfo {pages} {137} (\bibinfo {year}
  {2015})}\BibitemShut {NoStop}%
\bibitem [{\citenamefont {Kitaev}(1997)}]{6_kitaev1997quantum}%
  \BibitemOpen
  \bibfield  {author} {\bibinfo {author} {\bibfnamefont {A.~Y.}\ \bibnamefont
  {Kitaev}},\ }\bibfield  {title} {\bibinfo {title} {Quantum computations:
  algorithms and error correction},\ }\href@noop {} {\bibfield  {journal}
  {\bibinfo  {journal} {Russ. Math. Surv.}\ }\textbf {\bibinfo {volume} {52}},\
  \bibinfo {pages} {1191} (\bibinfo {year} {1997})}\BibitemShut {NoStop}%
\bibitem [{\citenamefont {Kitaev}(2003)}]{7_kitaev2003fault}%
  \BibitemOpen
  \bibfield  {author} {\bibinfo {author} {\bibfnamefont {A.~Y.}\ \bibnamefont
  {Kitaev}},\ }\bibfield  {title} {\bibinfo {title} {Fault-tolerant quantum
  computation by anyons},\ }\href@noop {} {\bibfield  {journal} {\bibinfo
  {journal} {Ann. Phys.}\ }\textbf {\bibinfo {volume} {303}},\ \bibinfo {pages}
  {2} (\bibinfo {year} {2003})}\BibitemShut {NoStop}%
\bibitem [{\citenamefont {Kitaev}(2006)}]{8_kitaev2006anyons}%
  \BibitemOpen
  \bibfield  {author} {\bibinfo {author} {\bibfnamefont {A.}~\bibnamefont
  {Kitaev}},\ }\bibfield  {title} {\bibinfo {title} {Anyons in an exactly
  solved model and beyond},\ }\href@noop {} {\bibfield  {journal} {\bibinfo
  {journal} {Ann. Phys.}\ }\textbf {\bibinfo {volume} {321}},\ \bibinfo {pages}
  {2} (\bibinfo {year} {2006})}\BibitemShut {NoStop}%
\bibitem [{\citenamefont {Aasen}\ \emph {et~al.}(2016)\citenamefont {Aasen},
  \citenamefont {Hell}, \citenamefont {Mishmash}, \citenamefont {Higginbotham},
  \citenamefont {Danon}, \citenamefont {Leijnse}, \citenamefont {Jespersen},
  \citenamefont {Folk}, \citenamefont {Marcus}, \citenamefont {Flensberg} \emph
  {et~al.}}]{9_aasen2016milestones}%
  \BibitemOpen
  \bibfield  {author} {\bibinfo {author} {\bibfnamefont {D.}~\bibnamefont
  {Aasen}}, \bibinfo {author} {\bibfnamefont {M.}~\bibnamefont {Hell}},
  \bibinfo {author} {\bibfnamefont {R.~V.}\ \bibnamefont {Mishmash}}, \bibinfo
  {author} {\bibfnamefont {A.}~\bibnamefont {Higginbotham}}, \bibinfo {author}
  {\bibfnamefont {J.}~\bibnamefont {Danon}}, \bibinfo {author} {\bibfnamefont
  {M.}~\bibnamefont {Leijnse}}, \bibinfo {author} {\bibfnamefont {T.~S.}\
  \bibnamefont {Jespersen}}, \bibinfo {author} {\bibfnamefont {J.~A.}\
  \bibnamefont {Folk}}, \bibinfo {author} {\bibfnamefont {C.~M.}\ \bibnamefont
  {Marcus}}, \bibinfo {author} {\bibfnamefont {K.}~\bibnamefont {Flensberg}},
  \emph {et~al.},\ }\bibfield  {title} {\bibinfo {title} {Milestones toward
  {M}ajorana-based quantum computing},\ }\href@noop {} {\bibfield  {journal}
  {\bibinfo  {journal} {Phys. Rev. X}\ }\textbf {\bibinfo {volume} {6}},\
  \bibinfo {pages} {031016} (\bibinfo {year} {2016})}\BibitemShut {NoStop}%
\bibitem [{\citenamefont {Abrikosov}(2004)}]{10-abrikosov2004nobel}%
  \BibitemOpen
  \bibfield  {author} {\bibinfo {author} {\bibfnamefont {A.}~\bibnamefont
  {Abrikosov}},\ }\bibfield  {title} {\bibinfo {title} {Nobel lecture:
  Type-{II} superconductors and the vortex lattice},\ }\href@noop {} {\bibfield
   {journal} {\bibinfo  {journal} {Rev. Mod. Phys.}\ }\textbf {\bibinfo
  {volume} {76}},\ \bibinfo {pages} {975} (\bibinfo {year} {2004})}\BibitemShut
  {NoStop}%
\bibitem [{\citenamefont {Blatter}\ \emph {et~al.}(1994)\citenamefont
  {Blatter}, \citenamefont {Feigel'man}, \citenamefont {Geshkenbein},
  \citenamefont {Larkin},\ and\ \citenamefont
  {Vinokur}}]{11-blatter1994vortices}%
  \BibitemOpen
  \bibfield  {author} {\bibinfo {author} {\bibfnamefont {G.}~\bibnamefont
  {Blatter}}, \bibinfo {author} {\bibfnamefont {M.~V.}\ \bibnamefont
  {Feigel'man}}, \bibinfo {author} {\bibfnamefont {V.~B.}\ \bibnamefont
  {Geshkenbein}}, \bibinfo {author} {\bibfnamefont {A.~I.}\ \bibnamefont
  {Larkin}},\ and\ \bibinfo {author} {\bibfnamefont {V.~M.}\ \bibnamefont
  {Vinokur}},\ }\bibfield  {title} {\bibinfo {title} {Vortices in
  high-temperature superconductors},\ }\href@noop {} {\bibfield  {journal}
  {\bibinfo  {journal} {Rev. Mod. Phys.}\ }\textbf {\bibinfo {volume} {66}},\
  \bibinfo {pages} {1125} (\bibinfo {year} {1994})}\BibitemShut {NoStop}%
\bibitem [{\citenamefont {Suderow}\ \emph {et~al.}(2014)\citenamefont
  {Suderow}, \citenamefont {Guillam{\'o}n}, \citenamefont {Rodrigo},\ and\
  \citenamefont {Vieira}}]{12-suderow2014imaging}%
  \BibitemOpen
  \bibfield  {author} {\bibinfo {author} {\bibfnamefont {H.}~\bibnamefont
  {Suderow}}, \bibinfo {author} {\bibfnamefont {I.}~\bibnamefont
  {Guillam{\'o}n}}, \bibinfo {author} {\bibfnamefont {J.~G.}\ \bibnamefont
  {Rodrigo}},\ and\ \bibinfo {author} {\bibfnamefont {S.}~\bibnamefont
  {Vieira}},\ }\bibfield  {title} {\bibinfo {title} {Imaging superconducting
  vortex cores and lattices with a scanning tunneling microscope},\ }\href@noop
  {} {\bibfield  {journal} {\bibinfo  {journal} {Supercond. Sci. Tech.}\
  }\textbf {\bibinfo {volume} {27}},\ \bibinfo {pages} {063001} (\bibinfo
  {year} {2014})}\BibitemShut {NoStop}%
\bibitem [{\citenamefont {Caroli}\ \emph {et~al.}(1964)\citenamefont {Caroli},
  \citenamefont {De~Gennes},\ and\ \citenamefont
  {Matricon}}]{13-caroli1964bound}%
  \BibitemOpen
  \bibfield  {author} {\bibinfo {author} {\bibfnamefont {C.}~\bibnamefont
  {Caroli}}, \bibinfo {author} {\bibfnamefont {P.}~\bibnamefont {De~Gennes}},\
  and\ \bibinfo {author} {\bibfnamefont {J.}~\bibnamefont {Matricon}},\
  }\bibfield  {title} {\bibinfo {title} {Bound fermion states on a vortex line
  in a type-{II} superconductor},\ }\href@noop {} {\bibfield  {journal}
  {\bibinfo  {journal} {Phys. Lett.}\ }\textbf {\bibinfo {volume} {9}},\
  \bibinfo {pages} {307} (\bibinfo {year} {1964})}\BibitemShut {NoStop}%
\bibitem [{\citenamefont {Hess}\ \emph {et~al.}(1989)\citenamefont {Hess},
  \citenamefont {Robinson}, \citenamefont {Dynes}, \citenamefont {Valles~Jr},\
  and\ \citenamefont {Waszczak}}]{14-hess1989scanning}%
  \BibitemOpen
  \bibfield  {author} {\bibinfo {author} {\bibfnamefont {H.}~\bibnamefont
  {Hess}}, \bibinfo {author} {\bibfnamefont {R.}~\bibnamefont {Robinson}},
  \bibinfo {author} {\bibfnamefont {R.}~\bibnamefont {Dynes}}, \bibinfo
  {author} {\bibfnamefont {J.}~\bibnamefont {Valles~Jr}},\ and\ \bibinfo
  {author} {\bibfnamefont {J.}~\bibnamefont {Waszczak}},\ }\bibfield  {title}
  {\bibinfo {title} {Scanning-tunneling-microscope observation of the abrikosov
  flux lattice and the density of states near and inside a fluxoid},\
  }\href@noop {} {\bibfield  {journal} {\bibinfo  {journal} {Phys. Rev. Lett.}\
  }\textbf {\bibinfo {volume} {62}},\ \bibinfo {pages} {214} (\bibinfo {year}
  {1989})}\BibitemShut {NoStop}%
\bibitem [{\citenamefont {Gygi}\ and\ \citenamefont
  {Schluter}(1990)}]{15-gygi1990electronic}%
  \BibitemOpen
  \bibfield  {author} {\bibinfo {author} {\bibfnamefont {F.}~\bibnamefont
  {Gygi}}\ and\ \bibinfo {author} {\bibfnamefont {M.}~\bibnamefont
  {Schluter}},\ }\bibfield  {title} {\bibinfo {title} {Electronic tunneling
  into an isolated vortex in a clean type-{II} superconductor},\ }\href@noop {}
  {\bibfield  {journal} {\bibinfo  {journal} {Phys. Rev. B}\ }\textbf {\bibinfo
  {volume} {41}},\ \bibinfo {pages} {822} (\bibinfo {year} {1990})}\BibitemShut
  {NoStop}%
\bibitem [{\citenamefont {Read}\ and\ \citenamefont
  {Green}(2000)}]{16-read2000paired}%
  \BibitemOpen
  \bibfield  {author} {\bibinfo {author} {\bibfnamefont {N.}~\bibnamefont
  {Read}}\ and\ \bibinfo {author} {\bibfnamefont {D.}~\bibnamefont {Green}},\
  }\bibfield  {title} {\bibinfo {title} {Paired states of fermions in two
  dimensions with breaking of parity and time-reversal symmetries and the
  fractional quantum {H}all effect},\ }\href@noop {} {\bibfield  {journal}
  {\bibinfo  {journal} {Phys. Rev. B}\ }\textbf {\bibinfo {volume} {61}},\
  \bibinfo {pages} {10267} (\bibinfo {year} {2000})}\BibitemShut {NoStop}%
\bibitem [{\citenamefont {Ivanov}(2001)}]{17-ivanov2001non}%
  \BibitemOpen
  \bibfield  {author} {\bibinfo {author} {\bibfnamefont {D.~A.}\ \bibnamefont
  {Ivanov}},\ }\bibfield  {title} {\bibinfo {title} {Non-{A}belian statistics
  of half-quantum vortices in p-wave superconductors},\ }\href@noop {}
  {\bibfield  {journal} {\bibinfo  {journal} {Phys. Rev. Lett.}\ }\textbf
  {\bibinfo {volume} {86}},\ \bibinfo {pages} {268} (\bibinfo {year}
  {2001})}\BibitemShut {NoStop}%
\bibitem [{\citenamefont {Volovik}(1999)}]{18-volovik1999fermion}%
  \BibitemOpen
  \bibfield  {author} {\bibinfo {author} {\bibfnamefont {G.}~\bibnamefont
  {Volovik}},\ }\bibfield  {title} {\bibinfo {title} {Fermion zero modes on
  vortices in chiral superconductors},\ }\href@noop {} {\bibfield  {journal}
  {\bibinfo  {journal} {JETP Lett.}\ }\textbf {\bibinfo {volume} {70}},\
  \bibinfo {pages} {609} (\bibinfo {year} {1999})}\BibitemShut {NoStop}%
\bibitem [{\citenamefont {Senthil}\ and\ \citenamefont
  {Fisher}(2000)}]{19-senthil2000quasiparticle}%
  \BibitemOpen
  \bibfield  {author} {\bibinfo {author} {\bibfnamefont {T.}~\bibnamefont
  {Senthil}}\ and\ \bibinfo {author} {\bibfnamefont {M.~P.}\ \bibnamefont
  {Fisher}},\ }\bibfield  {title} {\bibinfo {title} {Quasiparticle localization
  in superconductors with spin-orbit scattering},\ }\href@noop {} {\bibfield
  {journal} {\bibinfo  {journal} {Phys. Rev. B}\ }\textbf {\bibinfo {volume}
  {61}},\ \bibinfo {pages} {9690} (\bibinfo {year} {2000})}\BibitemShut
  {NoStop}%
\bibitem [{\citenamefont {Stone}\ and\ \citenamefont
  {Roy}(2004)}]{20-stone2004edge}%
  \BibitemOpen
  \bibfield  {author} {\bibinfo {author} {\bibfnamefont {M.}~\bibnamefont
  {Stone}}\ and\ \bibinfo {author} {\bibfnamefont {R.}~\bibnamefont {Roy}},\
  }\bibfield  {title} {\bibinfo {title} {Edge modes, edge currents, and gauge
  invariance in p$_{x}$+ i p$_{y}$ superfluids and superconductors},\
  }\href@noop {} {\bibfield  {journal} {\bibinfo  {journal} {Phys. Rev. B}\
  }\textbf {\bibinfo {volume} {69}},\ \bibinfo {pages} {184511} (\bibinfo
  {year} {2004})}\BibitemShut {NoStop}%
\bibitem [{\citenamefont {Willett}(2013)}]{21-willett2013quantum}%
  \BibitemOpen
  \bibfield  {author} {\bibinfo {author} {\bibfnamefont {R.}~\bibnamefont
  {Willett}},\ }\bibfield  {title} {\bibinfo {title} {The quantum {H}all effect
  at 5/2 filling factor},\ }\href@noop {} {\bibfield  {journal} {\bibinfo
  {journal} {Rep. Prog. Phys.}\ }\textbf {\bibinfo {volume} {76}},\ \bibinfo
  {pages} {076501} (\bibinfo {year} {2013})}\BibitemShut {NoStop}%
\bibitem [{\citenamefont {Moore}\ and\ \citenamefont
  {Read}(1991)}]{22-moore1991nonabelions}%
  \BibitemOpen
  \bibfield  {author} {\bibinfo {author} {\bibfnamefont {G.}~\bibnamefont
  {Moore}}\ and\ \bibinfo {author} {\bibfnamefont {N.}~\bibnamefont {Read}},\
  }\bibfield  {title} {\bibinfo {title} {Nonabelions in the fractional quantum
  {H}all effect},\ }\href@noop {} {\bibfield  {journal} {\bibinfo  {journal}
  {Nucl. Phys. B}\ }\textbf {\bibinfo {volume} {360}},\ \bibinfo {pages} {362}
  (\bibinfo {year} {1991})}\BibitemShut {NoStop}%
\bibitem [{\citenamefont {Mackenzie}\ \emph {et~al.}(2017)\citenamefont
  {Mackenzie}, \citenamefont {Scaffidi}, \citenamefont {Hicks},\ and\
  \citenamefont {Maeno}}]{23-mackenzie2017even}%
  \BibitemOpen
  \bibfield  {author} {\bibinfo {author} {\bibfnamefont {A.~P.}\ \bibnamefont
  {Mackenzie}}, \bibinfo {author} {\bibfnamefont {T.}~\bibnamefont {Scaffidi}},
  \bibinfo {author} {\bibfnamefont {C.~W.}\ \bibnamefont {Hicks}},\ and\
  \bibinfo {author} {\bibfnamefont {Y.}~\bibnamefont {Maeno}},\ }\bibfield
  {title} {\bibinfo {title} {Even odder after twenty-three years: the
  superconducting order parameter puzzle of {S}r$_{2}${R}u{O}$_{4}$},\
  }\href@noop {} {\bibfield  {journal} {\bibinfo  {journal} {npj Quantum
  Mater.}\ }\textbf {\bibinfo {volume} {2}},\ \bibinfo {pages} {40} (\bibinfo
  {year} {2017})}\BibitemShut {NoStop}%
\bibitem [{\citenamefont {Hasan}\ and\ \citenamefont
  {Kane}(2010)}]{25-hasan2010colloquium}%
  \BibitemOpen
  \bibfield  {author} {\bibinfo {author} {\bibfnamefont {M.~Z.}\ \bibnamefont
  {Hasan}}\ and\ \bibinfo {author} {\bibfnamefont {C.~L.}\ \bibnamefont
  {Kane}},\ }\bibfield  {title} {\bibinfo {title} {Colloquium: topological
  insulators},\ }\href@noop {} {\bibfield  {journal} {\bibinfo  {journal} {Rev.
  Mod. Phys.}\ }\textbf {\bibinfo {volume} {82}},\ \bibinfo {pages} {3045}
  (\bibinfo {year} {2010})}\BibitemShut {NoStop}%
\bibitem [{\citenamefont {Qi}\ and\ \citenamefont
  {Zhang}(2011)}]{26-qi2011topological}%
  \BibitemOpen
  \bibfield  {author} {\bibinfo {author} {\bibfnamefont {X.-L.}\ \bibnamefont
  {Qi}}\ and\ \bibinfo {author} {\bibfnamefont {S.-C.}\ \bibnamefont {Zhang}},\
  }\bibfield  {title} {\bibinfo {title} {Topological insulators and
  superconductors},\ }\href@noop {} {\bibfield  {journal} {\bibinfo  {journal}
  {Rev. Mod. Phys.}\ }\textbf {\bibinfo {volume} {83}},\ \bibinfo {pages}
  {1057} (\bibinfo {year} {2011})}\BibitemShut {NoStop}%
\bibitem [{\citenamefont {Hsieh}\ \emph {et~al.}(2009)\citenamefont {Hsieh},
  \citenamefont {Xia}, \citenamefont {Qian}, \citenamefont {Wray},
  \citenamefont {Dil}, \citenamefont {Meier}, \citenamefont {Osterwalder},
  \citenamefont {Patthey}, \citenamefont {Checkelsky}, \citenamefont {Ong}
  \emph {et~al.}}]{27-hsieh2009tunable}%
  \BibitemOpen
  \bibfield  {author} {\bibinfo {author} {\bibfnamefont {D.}~\bibnamefont
  {Hsieh}}, \bibinfo {author} {\bibfnamefont {Y.}~\bibnamefont {Xia}}, \bibinfo
  {author} {\bibfnamefont {D.}~\bibnamefont {Qian}}, \bibinfo {author}
  {\bibfnamefont {L.}~\bibnamefont {Wray}}, \bibinfo {author} {\bibfnamefont
  {J.}~\bibnamefont {Dil}}, \bibinfo {author} {\bibfnamefont {F.}~\bibnamefont
  {Meier}}, \bibinfo {author} {\bibfnamefont {J.}~\bibnamefont {Osterwalder}},
  \bibinfo {author} {\bibfnamefont {L.}~\bibnamefont {Patthey}}, \bibinfo
  {author} {\bibfnamefont {J.}~\bibnamefont {Checkelsky}}, \bibinfo {author}
  {\bibfnamefont {N.~P.}\ \bibnamefont {Ong}}, \emph {et~al.},\ }\bibfield
  {title} {\bibinfo {title} {A tunable topological insulator in the spin
  helical {D}irac transport regime},\ }\href@noop {} {\bibfield  {journal}
  {\bibinfo  {journal} {Nature}\ }\textbf {\bibinfo {volume} {460}},\ \bibinfo
  {pages} {1101} (\bibinfo {year} {2009})}\BibitemShut {NoStop}%
\bibitem [{\citenamefont {Ding}\ \emph {et~al.}(2008)\citenamefont {Ding},
  \citenamefont {Richard}, \citenamefont {Nakayama}, \citenamefont {Sugawara},
  \citenamefont {Arakane}, \citenamefont {Sekiba}, \citenamefont {Takayama},
  \citenamefont {Souma}, \citenamefont {Sato}, \citenamefont {Takahashi} \emph
  {et~al.}}]{74-ding2008observation}%
  \BibitemOpen
  \bibfield  {author} {\bibinfo {author} {\bibfnamefont {H.}~\bibnamefont
  {Ding}}, \bibinfo {author} {\bibfnamefont {P.}~\bibnamefont {Richard}},
  \bibinfo {author} {\bibfnamefont {K.}~\bibnamefont {Nakayama}}, \bibinfo
  {author} {\bibfnamefont {K.}~\bibnamefont {Sugawara}}, \bibinfo {author}
  {\bibfnamefont {T.}~\bibnamefont {Arakane}}, \bibinfo {author} {\bibfnamefont
  {Y.}~\bibnamefont {Sekiba}}, \bibinfo {author} {\bibfnamefont
  {A.}~\bibnamefont {Takayama}}, \bibinfo {author} {\bibfnamefont
  {S.}~\bibnamefont {Souma}}, \bibinfo {author} {\bibfnamefont
  {T.}~\bibnamefont {Sato}}, \bibinfo {author} {\bibfnamefont {T.}~\bibnamefont
  {Takahashi}}, \emph {et~al.},\ }\bibfield  {title} {\bibinfo {title}
  {Observation of fermi-surface--dependent nodeless superconducting gaps in
  {B}a$_{0.6}${K}$_{0.4}${F}e$_{2}${A}s$_{2}$},\ }\href@noop {} {\bibfield
  {journal} {\bibinfo  {journal} {EPL (Europhysics Letters)}\ }\textbf
  {\bibinfo {volume} {83}},\ \bibinfo {pages} {47001} (\bibinfo {year}
  {2008})}\BibitemShut {NoStop}%
\bibitem [{\citenamefont {Richard}\ \emph {et~al.}(2015)\citenamefont
  {Richard}, \citenamefont {Qian},\ and\ \citenamefont
  {Ding}}]{76-richard2015arpes}%
  \BibitemOpen
  \bibfield  {author} {\bibinfo {author} {\bibfnamefont {P.}~\bibnamefont
  {Richard}}, \bibinfo {author} {\bibfnamefont {T.}~\bibnamefont {Qian}},\ and\
  \bibinfo {author} {\bibfnamefont {H.}~\bibnamefont {Ding}},\ }\bibfield
  {title} {\bibinfo {title} {{ARPES} measurements of the superconducting gap of
  {F}e-based superconductors and their implications to the pairing mechanism},\
  }\href@noop {} {\bibfield  {journal} {\bibinfo  {journal} {J Phys.: Condens.
  Mat.}\ }\textbf {\bibinfo {volume} {27}},\ \bibinfo {pages} {293203}
  (\bibinfo {year} {2015})}\BibitemShut {NoStop}%
\bibitem [{\citenamefont {Yin}\ \emph {et~al.}(2011)\citenamefont {Yin},
  \citenamefont {Haule},\ and\ \citenamefont {Kotliar}}]{81-yin2011kinetic}%
  \BibitemOpen
  \bibfield  {author} {\bibinfo {author} {\bibfnamefont {Z.}~\bibnamefont
  {Yin}}, \bibinfo {author} {\bibfnamefont {K.}~\bibnamefont {Haule}},\ and\
  \bibinfo {author} {\bibfnamefont {G.}~\bibnamefont {Kotliar}},\ }\bibfield
  {title} {\bibinfo {title} {Kinetic frustration and the nature of the magnetic
  and paramagnetic states in iron pnictides and iron chalcogenides},\
  }\href@noop {} {\bibfield  {journal} {\bibinfo  {journal} {Nat. Mater.}\
  }\textbf {\bibinfo {volume} {10}},\ \bibinfo {pages} {932} (\bibinfo {year}
  {2011})}\BibitemShut {NoStop}%
\bibitem [{\citenamefont {Rinott}\ \emph {et~al.}(2017)\citenamefont {Rinott},
  \citenamefont {Chashka}, \citenamefont {Ribak}, \citenamefont {Rienks},
  \citenamefont {Taleb-Ibrahimi}, \citenamefont {Le~Fevre}, \citenamefont
  {Bertran}, \citenamefont {Randeria},\ and\ \citenamefont
  {Kanigel}}]{82-rinott2017tuning}%
  \BibitemOpen
  \bibfield  {author} {\bibinfo {author} {\bibfnamefont {S.}~\bibnamefont
  {Rinott}}, \bibinfo {author} {\bibfnamefont {K.}~\bibnamefont {Chashka}},
  \bibinfo {author} {\bibfnamefont {A.}~\bibnamefont {Ribak}}, \bibinfo
  {author} {\bibfnamefont {E.~D.}\ \bibnamefont {Rienks}}, \bibinfo {author}
  {\bibfnamefont {A.}~\bibnamefont {Taleb-Ibrahimi}}, \bibinfo {author}
  {\bibfnamefont {P.}~\bibnamefont {Le~Fevre}}, \bibinfo {author}
  {\bibfnamefont {F.}~\bibnamefont {Bertran}}, \bibinfo {author} {\bibfnamefont
  {M.}~\bibnamefont {Randeria}},\ and\ \bibinfo {author} {\bibfnamefont
  {A.}~\bibnamefont {Kanigel}},\ }\bibfield  {title} {\bibinfo {title} {Tuning
  across the {BCS}-{BEC} crossover in the multiband superconductor
  {F}e$_{1+y}${S}e$_{x}${T}e$_{1-x}$: An angle-resolved photoemission study},\
  }\href@noop {} {\bibfield  {journal} {\bibinfo  {journal} {Sci. Adv.}\
  }\textbf {\bibinfo {volume} {3}},\ \bibinfo {pages} {e1602372} (\bibinfo
  {year} {2017})}\BibitemShut {NoStop}%
\bibitem [{\citenamefont {Lutchyn}\ \emph {et~al.}(2010)\citenamefont
  {Lutchyn}, \citenamefont {Sau},\ and\ \citenamefont
  {Sarma}}]{28-lutchyn2010majorana}%
  \BibitemOpen
  \bibfield  {author} {\bibinfo {author} {\bibfnamefont {R.~M.}\ \bibnamefont
  {Lutchyn}}, \bibinfo {author} {\bibfnamefont {J.~D.}\ \bibnamefont {Sau}},\
  and\ \bibinfo {author} {\bibfnamefont {S.~D.}\ \bibnamefont {Sarma}},\
  }\bibfield  {title} {\bibinfo {title} {Majorana fermions and a topological
  phase transition in semiconductor-superconductor heterostructures},\
  }\href@noop {} {\bibfield  {journal} {\bibinfo  {journal} {Phys. Rev. Lett.}\
  }\textbf {\bibinfo {volume} {105}},\ \bibinfo {pages} {077001} (\bibinfo
  {year} {2010})}\BibitemShut {NoStop}%
\bibitem [{\citenamefont {Oreg}\ \emph {et~al.}(2010)\citenamefont {Oreg},
  \citenamefont {Refael},\ and\ \citenamefont
  {Von~Oppen}}]{29-oreg2010helical}%
  \BibitemOpen
  \bibfield  {author} {\bibinfo {author} {\bibfnamefont {Y.}~\bibnamefont
  {Oreg}}, \bibinfo {author} {\bibfnamefont {G.}~\bibnamefont {Refael}},\ and\
  \bibinfo {author} {\bibfnamefont {F.}~\bibnamefont {Von~Oppen}},\ }\bibfield
  {title} {\bibinfo {title} {Helical liquids and {M}ajorana bound states in
  quantum wires},\ }\href@noop {} {\bibfield  {journal} {\bibinfo  {journal}
  {Phys. Rev. Lett.}\ }\textbf {\bibinfo {volume} {105}},\ \bibinfo {pages}
  {177002} (\bibinfo {year} {2010})}\BibitemShut {NoStop}%
\bibitem [{\citenamefont {Potter}\ and\ \citenamefont
  {Lee}(2010)}]{30-potter2010multichannel}%
  \BibitemOpen
  \bibfield  {author} {\bibinfo {author} {\bibfnamefont {A.~C.}\ \bibnamefont
  {Potter}}\ and\ \bibinfo {author} {\bibfnamefont {P.~A.}\ \bibnamefont
  {Lee}},\ }\bibfield  {title} {\bibinfo {title} {Multichannel generalization
  of {K}itaev’s {M}ajorana end states and a practical route to realize them
  in thin films},\ }\href@noop {} {\bibfield  {journal} {\bibinfo  {journal}
  {Phys. Rev. Lett.}\ }\textbf {\bibinfo {volume} {105}},\ \bibinfo {pages}
  {227003} (\bibinfo {year} {2010})}\BibitemShut {NoStop}%
\bibitem [{\citenamefont {Mourik}\ \emph {et~al.}(2012)\citenamefont {Mourik},
  \citenamefont {Zuo}, \citenamefont {Frolov}, \citenamefont {Plissard},
  \citenamefont {Bakkers},\ and\ \citenamefont
  {Kouwenhoven}}]{31-mourik2012signatures}%
  \BibitemOpen
  \bibfield  {author} {\bibinfo {author} {\bibfnamefont {V.}~\bibnamefont
  {Mourik}}, \bibinfo {author} {\bibfnamefont {K.}~\bibnamefont {Zuo}},
  \bibinfo {author} {\bibfnamefont {S.~M.}\ \bibnamefont {Frolov}}, \bibinfo
  {author} {\bibfnamefont {S.}~\bibnamefont {Plissard}}, \bibinfo {author}
  {\bibfnamefont {E.~P.}\ \bibnamefont {Bakkers}},\ and\ \bibinfo {author}
  {\bibfnamefont {L.~P.}\ \bibnamefont {Kouwenhoven}},\ }\bibfield  {title}
  {\bibinfo {title} {Signatures of {M}ajorana fermions in hybrid
  superconductor-semiconductor nanowire devices},\ }\href@noop {} {\bibfield
  {journal} {\bibinfo  {journal} {Science}\ }\textbf {\bibinfo {volume}
  {336}},\ \bibinfo {pages} {1003} (\bibinfo {year} {2012})}\BibitemShut
  {NoStop}%
\bibitem [{\citenamefont {Gazibegovic}\ \emph {et~al.}(2017)\citenamefont
  {Gazibegovic}, \citenamefont {Car}, \citenamefont {Zhang}, \citenamefont
  {Balk}, \citenamefont {Logan}, \citenamefont {de~Moor}, \citenamefont
  {Cassidy}, \citenamefont {Schmits}, \citenamefont {Xu}, \citenamefont {Wang}
  \emph {et~al.}}]{32-gazibegovic2017epitaxy}%
  \BibitemOpen
  \bibfield  {author} {\bibinfo {author} {\bibfnamefont {S.}~\bibnamefont
  {Gazibegovic}}, \bibinfo {author} {\bibfnamefont {D.}~\bibnamefont {Car}},
  \bibinfo {author} {\bibfnamefont {H.}~\bibnamefont {Zhang}}, \bibinfo
  {author} {\bibfnamefont {S.~C.}\ \bibnamefont {Balk}}, \bibinfo {author}
  {\bibfnamefont {J.~A.}\ \bibnamefont {Logan}}, \bibinfo {author}
  {\bibfnamefont {M.~W.}\ \bibnamefont {de~Moor}}, \bibinfo {author}
  {\bibfnamefont {M.~C.}\ \bibnamefont {Cassidy}}, \bibinfo {author}
  {\bibfnamefont {R.}~\bibnamefont {Schmits}}, \bibinfo {author} {\bibfnamefont
  {D.}~\bibnamefont {Xu}}, \bibinfo {author} {\bibfnamefont {G.}~\bibnamefont
  {Wang}}, \emph {et~al.},\ }\bibfield  {title} {\bibinfo {title} {Epitaxy of
  advanced nanowire quantum devices},\ }\href@noop {} {\bibfield  {journal}
  {\bibinfo  {journal} {Nature}\ }\textbf {\bibinfo {volume} {548}},\ \bibinfo
  {pages} {434} (\bibinfo {year} {2017})}\BibitemShut {NoStop}%
\bibitem [{\citenamefont {Lutchyn}\ \emph {et~al.}(2018)\citenamefont
  {Lutchyn}, \citenamefont {Bakkers}, \citenamefont {Kouwenhoven},
  \citenamefont {Krogstrup}, \citenamefont {Marcus},\ and\ \citenamefont
  {Oreg}}]{33-lutchyn2018majorana}%
  \BibitemOpen
  \bibfield  {author} {\bibinfo {author} {\bibfnamefont {R.~M.}\ \bibnamefont
  {Lutchyn}}, \bibinfo {author} {\bibfnamefont {E.~P.}\ \bibnamefont
  {Bakkers}}, \bibinfo {author} {\bibfnamefont {L.~P.}\ \bibnamefont
  {Kouwenhoven}}, \bibinfo {author} {\bibfnamefont {P.}~\bibnamefont
  {Krogstrup}}, \bibinfo {author} {\bibfnamefont {C.~M.}\ \bibnamefont
  {Marcus}},\ and\ \bibinfo {author} {\bibfnamefont {Y.}~\bibnamefont {Oreg}},\
  }\bibfield  {title} {\bibinfo {title} {Majorana zero modes in
  superconductor--semiconductor heterostructures},\ }\href@noop {} {\bibfield
  {journal} {\bibinfo  {journal} {Nat. Rev. Mater.}\ }\textbf {\bibinfo
  {volume} {3}},\ \bibinfo {pages} {52} (\bibinfo {year} {2018})}\BibitemShut
  {NoStop}%
\bibitem [{\citenamefont {Prada}\ \emph {et~al.}(2019)\citenamefont {Prada},
  \citenamefont {San-Jose}, \citenamefont {de~Moor}, \citenamefont {Geresdi},
  \citenamefont {Lee}, \citenamefont {Klinovaja}, \citenamefont {Loss},
  \citenamefont {Nyg{\aa}rd}, \citenamefont {Aguado},\ and\ \citenamefont
  {Kouwenhoven}}]{34-prada2019andreev}%
  \BibitemOpen
  \bibfield  {author} {\bibinfo {author} {\bibfnamefont {E.}~\bibnamefont
  {Prada}}, \bibinfo {author} {\bibfnamefont {P.}~\bibnamefont {San-Jose}},
  \bibinfo {author} {\bibfnamefont {M.~W.}\ \bibnamefont {de~Moor}}, \bibinfo
  {author} {\bibfnamefont {A.}~\bibnamefont {Geresdi}}, \bibinfo {author}
  {\bibfnamefont {E.~J.}\ \bibnamefont {Lee}}, \bibinfo {author} {\bibfnamefont
  {J.}~\bibnamefont {Klinovaja}}, \bibinfo {author} {\bibfnamefont
  {D.}~\bibnamefont {Loss}}, \bibinfo {author} {\bibfnamefont {J.}~\bibnamefont
  {Nyg{\aa}rd}}, \bibinfo {author} {\bibfnamefont {R.}~\bibnamefont {Aguado}},\
  and\ \bibinfo {author} {\bibfnamefont {L.~P.}\ \bibnamefont {Kouwenhoven}},\
  }\bibfield  {title} {\bibinfo {title} {From {A}ndreev to {M}ajorana bound
  states in hybrid superconductor-semiconductor nanowires},\ }\href@noop {}
  {\bibfield  {journal} {\bibinfo  {journal} {arXiv:1911.04512}\ } (\bibinfo
  {year} {2019})}\BibitemShut {NoStop}%
\bibitem [{\citenamefont {Potter}\ and\ \citenamefont
  {Lee}(2012)}]{35-potter2012topological}%
  \BibitemOpen
  \bibfield  {author} {\bibinfo {author} {\bibfnamefont {A.~C.}\ \bibnamefont
  {Potter}}\ and\ \bibinfo {author} {\bibfnamefont {P.~A.}\ \bibnamefont
  {Lee}},\ }\bibfield  {title} {\bibinfo {title} {Topological superconductivity
  and {M}ajorana fermions in metallic surface states},\ }\href@noop {}
  {\bibfield  {journal} {\bibinfo  {journal} {Phys. Rev. B}\ }\textbf {\bibinfo
  {volume} {85}},\ \bibinfo {pages} {094516} (\bibinfo {year}
  {2012})}\BibitemShut {NoStop}%
\bibitem [{\citenamefont {Manna}\ \emph {et~al.}(2020)\citenamefont {Manna},
  \citenamefont {Wei}, \citenamefont {Xie}, \citenamefont {Law}, \citenamefont
  {Lee},\ and\ \citenamefont {Moodera}}]{36-manna2020signature}%
  \BibitemOpen
  \bibfield  {author} {\bibinfo {author} {\bibfnamefont {S.}~\bibnamefont
  {Manna}}, \bibinfo {author} {\bibfnamefont {P.}~\bibnamefont {Wei}}, \bibinfo
  {author} {\bibfnamefont {Y.}~\bibnamefont {Xie}}, \bibinfo {author}
  {\bibfnamefont {K.~T.}\ \bibnamefont {Law}}, \bibinfo {author} {\bibfnamefont
  {P.~A.}\ \bibnamefont {Lee}},\ and\ \bibinfo {author} {\bibfnamefont {J.~S.}\
  \bibnamefont {Moodera}},\ }\bibfield  {title} {\bibinfo {title} {Signature of
  a pair of {M}ajorana zero modes in superconducting gold surface states},\
  }\href@noop {} {\bibfield  {journal} {\bibinfo  {journal} {Proc. Natl. Acad.
  Sci. U.S.A.}\ }\textbf {\bibinfo {volume} {117}},\ \bibinfo {pages} {8775}
  (\bibinfo {year} {2020})}\BibitemShut {NoStop}%
\bibitem [{\citenamefont {Xie}\ \emph {et~al.}(2020)\citenamefont {Xie},
  \citenamefont {Law},\ and\ \citenamefont {Lee}}]{37-xie2020topological}%
  \BibitemOpen
  \bibfield  {author} {\bibinfo {author} {\bibfnamefont {Y.-M.}\ \bibnamefont
  {Xie}}, \bibinfo {author} {\bibfnamefont {K.}~\bibnamefont {Law}},\ and\
  \bibinfo {author} {\bibfnamefont {P.~A.}\ \bibnamefont {Lee}},\ }\bibfield
  {title} {\bibinfo {title} {Topological superconductivity in
  {E}u{S}/{A}u/superconductor heterostructures},\ }\href@noop {} {\bibfield
  {journal} {\bibinfo  {journal} {arXiv:2003.07052}\ } (\bibinfo {year}
  {2020})}\BibitemShut {NoStop}%
\bibitem [{\citenamefont {Braunecker}\ \emph {et~al.}(2010)\citenamefont
  {Braunecker}, \citenamefont {Japaridze}, \citenamefont {Klinovaja},\ and\
  \citenamefont {Loss}}]{38-braunecker2010spin}%
  \BibitemOpen
  \bibfield  {author} {\bibinfo {author} {\bibfnamefont {B.}~\bibnamefont
  {Braunecker}}, \bibinfo {author} {\bibfnamefont {G.~I.}\ \bibnamefont
  {Japaridze}}, \bibinfo {author} {\bibfnamefont {J.}~\bibnamefont
  {Klinovaja}},\ and\ \bibinfo {author} {\bibfnamefont {D.}~\bibnamefont
  {Loss}},\ }\bibfield  {title} {\bibinfo {title} {Spin-selective {P}eierls
  transition in interacting one-dimensional conductors with spin-orbit
  interaction},\ }\href@noop {} {\bibfield  {journal} {\bibinfo  {journal}
  {Phys. Rev. B}\ }\textbf {\bibinfo {volume} {82}},\ \bibinfo {pages} {045127}
  (\bibinfo {year} {2010})}\BibitemShut {NoStop}%
\bibitem [{\citenamefont {Nadj-Perge}\ \emph {et~al.}(2013)\citenamefont
  {Nadj-Perge}, \citenamefont {Drozdov}, \citenamefont {Bernevig},\ and\
  \citenamefont {Yazdani}}]{39-nadj2013proposal}%
  \BibitemOpen
  \bibfield  {author} {\bibinfo {author} {\bibfnamefont {S.}~\bibnamefont
  {Nadj-Perge}}, \bibinfo {author} {\bibfnamefont {I.}~\bibnamefont {Drozdov}},
  \bibinfo {author} {\bibfnamefont {B.~A.}\ \bibnamefont {Bernevig}},\ and\
  \bibinfo {author} {\bibfnamefont {A.}~\bibnamefont {Yazdani}},\ }\bibfield
  {title} {\bibinfo {title} {Proposal for realizing {M}ajorana fermions in
  chains of magnetic atoms on a superconductor},\ }\href@noop {} {\bibfield
  {journal} {\bibinfo  {journal} {Phys. Rev. B}\ }\textbf {\bibinfo {volume}
  {88}},\ \bibinfo {pages} {020407} (\bibinfo {year} {2013})}\BibitemShut
  {NoStop}%
\bibitem [{\citenamefont {Nadj-Perge}\ \emph {et~al.}(2014)\citenamefont
  {Nadj-Perge}, \citenamefont {Drozdov}, \citenamefont {Li}, \citenamefont
  {Chen}, \citenamefont {Jeon}, \citenamefont {Seo}, \citenamefont {MacDonald},
  \citenamefont {Bernevig},\ and\ \citenamefont
  {Yazdani}}]{40-nadj2014observation}%
  \BibitemOpen
  \bibfield  {author} {\bibinfo {author} {\bibfnamefont {S.}~\bibnamefont
  {Nadj-Perge}}, \bibinfo {author} {\bibfnamefont {I.~K.}\ \bibnamefont
  {Drozdov}}, \bibinfo {author} {\bibfnamefont {J.}~\bibnamefont {Li}},
  \bibinfo {author} {\bibfnamefont {H.}~\bibnamefont {Chen}}, \bibinfo {author}
  {\bibfnamefont {S.}~\bibnamefont {Jeon}}, \bibinfo {author} {\bibfnamefont
  {J.}~\bibnamefont {Seo}}, \bibinfo {author} {\bibfnamefont {A.~H.}\
  \bibnamefont {MacDonald}}, \bibinfo {author} {\bibfnamefont {B.~A.}\
  \bibnamefont {Bernevig}},\ and\ \bibinfo {author} {\bibfnamefont
  {A.}~\bibnamefont {Yazdani}},\ }\bibfield  {title} {\bibinfo {title}
  {Observation of {M}ajorana fermions in ferromagnetic atomic chains on a
  superconductor},\ }\href@noop {} {\bibfield  {journal} {\bibinfo  {journal}
  {Science}\ }\textbf {\bibinfo {volume} {346}},\ \bibinfo {pages} {602}
  (\bibinfo {year} {2014})}\BibitemShut {NoStop}%
\bibitem [{\citenamefont {Jeon}\ \emph {et~al.}(2017)\citenamefont {Jeon},
  \citenamefont {Xie}, \citenamefont {Li}, \citenamefont {Wang}, \citenamefont
  {Bernevig},\ and\ \citenamefont {Yazdani}}]{41-jeon2017distinguishing}%
  \BibitemOpen
  \bibfield  {author} {\bibinfo {author} {\bibfnamefont {S.}~\bibnamefont
  {Jeon}}, \bibinfo {author} {\bibfnamefont {Y.}~\bibnamefont {Xie}}, \bibinfo
  {author} {\bibfnamefont {J.}~\bibnamefont {Li}}, \bibinfo {author}
  {\bibfnamefont {Z.}~\bibnamefont {Wang}}, \bibinfo {author} {\bibfnamefont
  {B.~A.}\ \bibnamefont {Bernevig}},\ and\ \bibinfo {author} {\bibfnamefont
  {A.}~\bibnamefont {Yazdani}},\ }\bibfield  {title} {\bibinfo {title}
  {Distinguishing a {M}ajorana zero mode using spin-resolved measurements},\
  }\href@noop {} {\bibfield  {journal} {\bibinfo  {journal} {Science}\ }\textbf
  {\bibinfo {volume} {358}},\ \bibinfo {pages} {772} (\bibinfo {year}
  {2017})}\BibitemShut {NoStop}%
\bibitem [{\citenamefont {Desjardins}\ \emph {et~al.}(2019)\citenamefont
  {Desjardins}, \citenamefont {Contamin}, \citenamefont {Delbecq},
  \citenamefont {Dartiailh}, \citenamefont {Bruhat}, \citenamefont {Cubaynes},
  \citenamefont {Viennot}, \citenamefont {Mallet}, \citenamefont {Rohart},
  \citenamefont {Thiaville} \emph {et~al.}}]{42-desjardins2019synthetic}%
  \BibitemOpen
  \bibfield  {author} {\bibinfo {author} {\bibfnamefont {M.}~\bibnamefont
  {Desjardins}}, \bibinfo {author} {\bibfnamefont {L.}~\bibnamefont
  {Contamin}}, \bibinfo {author} {\bibfnamefont {M.}~\bibnamefont {Delbecq}},
  \bibinfo {author} {\bibfnamefont {M.}~\bibnamefont {Dartiailh}}, \bibinfo
  {author} {\bibfnamefont {L.}~\bibnamefont {Bruhat}}, \bibinfo {author}
  {\bibfnamefont {T.}~\bibnamefont {Cubaynes}}, \bibinfo {author}
  {\bibfnamefont {J.}~\bibnamefont {Viennot}}, \bibinfo {author} {\bibfnamefont
  {F.}~\bibnamefont {Mallet}}, \bibinfo {author} {\bibfnamefont
  {S.}~\bibnamefont {Rohart}}, \bibinfo {author} {\bibfnamefont
  {A.}~\bibnamefont {Thiaville}}, \emph {et~al.},\ }\bibfield  {title}
  {\bibinfo {title} {Synthetic spin--orbit interaction for {M}ajorana
  devices},\ }\href@noop {} {\bibfield  {journal} {\bibinfo  {journal} {Nat.
  Mater.}\ }\textbf {\bibinfo {volume} {18}},\ \bibinfo {pages} {1060}
  (\bibinfo {year} {2019})}\BibitemShut {NoStop}%
\bibitem [{\citenamefont {R{\"o}ntynen}\ and\ \citenamefont
  {Ojanen}(2015)}]{43-rontynen2015topological}%
  \BibitemOpen
  \bibfield  {author} {\bibinfo {author} {\bibfnamefont {J.}~\bibnamefont
  {R{\"o}ntynen}}\ and\ \bibinfo {author} {\bibfnamefont {T.}~\bibnamefont
  {Ojanen}},\ }\bibfield  {title} {\bibinfo {title} {Topological
  superconductivity and high chern numbers in 2{D} ferromagnetic {S}hiba
  lattices},\ }\href@noop {} {\bibfield  {journal} {\bibinfo  {journal} {Phys.
  Rev. Lett.}\ }\textbf {\bibinfo {volume} {114}},\ \bibinfo {pages} {236803}
  (\bibinfo {year} {2015})}\BibitemShut {NoStop}%
\bibitem [{\citenamefont {Li}\ \emph {et~al.}(2016)\citenamefont {Li},
  \citenamefont {Neupert}, \citenamefont {Wang}, \citenamefont {MacDonald},
  \citenamefont {Yazdani},\ and\ \citenamefont {Bernevig}}]{44-li2016two}%
  \BibitemOpen
  \bibfield  {author} {\bibinfo {author} {\bibfnamefont {J.}~\bibnamefont
  {Li}}, \bibinfo {author} {\bibfnamefont {T.}~\bibnamefont {Neupert}},
  \bibinfo {author} {\bibfnamefont {Z.}~\bibnamefont {Wang}}, \bibinfo {author}
  {\bibfnamefont {A.}~\bibnamefont {MacDonald}}, \bibinfo {author}
  {\bibfnamefont {A.}~\bibnamefont {Yazdani}},\ and\ \bibinfo {author}
  {\bibfnamefont {B.~A.}\ \bibnamefont {Bernevig}},\ }\bibfield  {title}
  {\bibinfo {title} {Two-dimensional chiral topological superconductivity in
  {S}hiba lattices},\ }\href@noop {} {\bibfield  {journal} {\bibinfo  {journal}
  {Nat. Commun.}\ }\textbf {\bibinfo {volume} {7}},\ \bibinfo {pages} {12297}
  (\bibinfo {year} {2016})}\BibitemShut {NoStop}%
\bibitem [{\citenamefont {Rachel}\ \emph {et~al.}(2017)\citenamefont {Rachel},
  \citenamefont {Mascot}, \citenamefont {Cocklin}, \citenamefont {Vojta},\ and\
  \citenamefont {Morr}}]{45-rachel2017quantized}%
  \BibitemOpen
  \bibfield  {author} {\bibinfo {author} {\bibfnamefont {S.}~\bibnamefont
  {Rachel}}, \bibinfo {author} {\bibfnamefont {E.}~\bibnamefont {Mascot}},
  \bibinfo {author} {\bibfnamefont {S.}~\bibnamefont {Cocklin}}, \bibinfo
  {author} {\bibfnamefont {M.}~\bibnamefont {Vojta}},\ and\ \bibinfo {author}
  {\bibfnamefont {D.~K.}\ \bibnamefont {Morr}},\ }\bibfield  {title} {\bibinfo
  {title} {Quantized charge transport in chiral {M}ajorana edge modes},\
  }\href@noop {} {\bibfield  {journal} {\bibinfo  {journal} {Phys. Rev. B}\
  }\textbf {\bibinfo {volume} {96}},\ \bibinfo {pages} {205131} (\bibinfo
  {year} {2017})}\BibitemShut {NoStop}%
\bibitem [{\citenamefont {M{\'e}nard}\ \emph {et~al.}(2017)\citenamefont
  {M{\'e}nard}, \citenamefont {Guissart}, \citenamefont {Brun}, \citenamefont
  {Leriche}, \citenamefont {Trif}, \citenamefont {Debontridder}, \citenamefont
  {Demaille}, \citenamefont {Roditchev}, \citenamefont {Simon},\ and\
  \citenamefont {Cren}}]{46-menard2017two}%
  \BibitemOpen
  \bibfield  {author} {\bibinfo {author} {\bibfnamefont {G.~C.}\ \bibnamefont
  {M{\'e}nard}}, \bibinfo {author} {\bibfnamefont {S.}~\bibnamefont
  {Guissart}}, \bibinfo {author} {\bibfnamefont {C.}~\bibnamefont {Brun}},
  \bibinfo {author} {\bibfnamefont {R.~T.}\ \bibnamefont {Leriche}}, \bibinfo
  {author} {\bibfnamefont {M.}~\bibnamefont {Trif}}, \bibinfo {author}
  {\bibfnamefont {F.}~\bibnamefont {Debontridder}}, \bibinfo {author}
  {\bibfnamefont {D.}~\bibnamefont {Demaille}}, \bibinfo {author}
  {\bibfnamefont {D.}~\bibnamefont {Roditchev}}, \bibinfo {author}
  {\bibfnamefont {P.}~\bibnamefont {Simon}},\ and\ \bibinfo {author}
  {\bibfnamefont {T.}~\bibnamefont {Cren}},\ }\bibfield  {title} {\bibinfo
  {title} {Two-dimensional topological superconductivity in {P}b/{C}o/{S}i
  (111)},\ }\href@noop {} {\bibfield  {journal} {\bibinfo  {journal} {Nat.
  Commun.}\ }\textbf {\bibinfo {volume} {8}},\ \bibinfo {pages} {2040}
  (\bibinfo {year} {2017})}\BibitemShut {NoStop}%
\bibitem [{\citenamefont {M{\'e}nard}\ \emph {et~al.}(2019)\citenamefont
  {M{\'e}nard}, \citenamefont {Brun}, \citenamefont {Leriche}, \citenamefont
  {Trif}, \citenamefont {Debontridder}, \citenamefont {Demaille}, \citenamefont
  {Roditchev}, \citenamefont {Simon},\ and\ \citenamefont
  {Cren}}]{47-menard2019yu}%
  \BibitemOpen
  \bibfield  {author} {\bibinfo {author} {\bibfnamefont {G.~C.}\ \bibnamefont
  {M{\'e}nard}}, \bibinfo {author} {\bibfnamefont {C.}~\bibnamefont {Brun}},
  \bibinfo {author} {\bibfnamefont {R.}~\bibnamefont {Leriche}}, \bibinfo
  {author} {\bibfnamefont {M.}~\bibnamefont {Trif}}, \bibinfo {author}
  {\bibfnamefont {F.}~\bibnamefont {Debontridder}}, \bibinfo {author}
  {\bibfnamefont {D.}~\bibnamefont {Demaille}}, \bibinfo {author}
  {\bibfnamefont {D.}~\bibnamefont {Roditchev}}, \bibinfo {author}
  {\bibfnamefont {P.}~\bibnamefont {Simon}},\ and\ \bibinfo {author}
  {\bibfnamefont {T.}~\bibnamefont {Cren}},\ }\bibfield  {title} {\bibinfo
  {title} {{Y}u-{S}hiba-{R}usinov bound states versus topological edge states
  in {P}b/{S}i (111)},\ }\href@noop {} {\bibfield  {journal} {\bibinfo
  {journal} {Eur. Phys. J : Spec. Top.}\ }\textbf {\bibinfo {volume} {227}},\
  \bibinfo {pages} {2303} (\bibinfo {year} {2019})}\BibitemShut {NoStop}%
\bibitem [{\citenamefont {Palacio-Morales}\ \emph {et~al.}(2019)\citenamefont
  {Palacio-Morales}, \citenamefont {Mascot}, \citenamefont {Cocklin},
  \citenamefont {Kim}, \citenamefont {Rachel}, \citenamefont {Morr},\ and\
  \citenamefont {Wiesendanger}}]{48-palacio2019atomic}%
  \BibitemOpen
  \bibfield  {author} {\bibinfo {author} {\bibfnamefont {A.}~\bibnamefont
  {Palacio-Morales}}, \bibinfo {author} {\bibfnamefont {E.}~\bibnamefont
  {Mascot}}, \bibinfo {author} {\bibfnamefont {S.}~\bibnamefont {Cocklin}},
  \bibinfo {author} {\bibfnamefont {H.}~\bibnamefont {Kim}}, \bibinfo {author}
  {\bibfnamefont {S.}~\bibnamefont {Rachel}}, \bibinfo {author} {\bibfnamefont
  {D.~K.}\ \bibnamefont {Morr}},\ and\ \bibinfo {author} {\bibfnamefont
  {R.}~\bibnamefont {Wiesendanger}},\ }\bibfield  {title} {\bibinfo {title}
  {Atomic-scale interface engineering of {M}ajorana edge modes in a 2{D}
  magnet-superconductor hybrid system},\ }\href@noop {} {\bibfield  {journal}
  {\bibinfo  {journal} {Sci. Adv.}\ }\textbf {\bibinfo {volume} {5}},\ \bibinfo
  {pages} {eaav6600} (\bibinfo {year} {2019})}\BibitemShut {NoStop}%
\bibitem [{\citenamefont {Qi}\ \emph {et~al.}(2010{\natexlab{a}})\citenamefont
  {Qi}, \citenamefont {Hughes},\ and\ \citenamefont {Zhang}}]{49-qi2010chiral}%
  \BibitemOpen
  \bibfield  {author} {\bibinfo {author} {\bibfnamefont {X.-L.}\ \bibnamefont
  {Qi}}, \bibinfo {author} {\bibfnamefont {T.~L.}\ \bibnamefont {Hughes}},\
  and\ \bibinfo {author} {\bibfnamefont {S.-C.}\ \bibnamefont {Zhang}},\
  }\bibfield  {title} {\bibinfo {title} {Chiral topological superconductor from
  the quantum {H}all state},\ }\href@noop {} {\bibfield  {journal} {\bibinfo
  {journal} {Phys. Rev. B}\ }\textbf {\bibinfo {volume} {82}},\ \bibinfo
  {pages} {184516} (\bibinfo {year} {2010}{\natexlab{a}})}\BibitemShut
  {NoStop}%
\bibitem [{\citenamefont {Chen}\ \emph
  {et~al.}(2018{\natexlab{a}})\citenamefont {Chen}, \citenamefont {He},
  \citenamefont {Xu},\ and\ \citenamefont {Law}}]{50-chen2018emergent}%
  \BibitemOpen
  \bibfield  {author} {\bibinfo {author} {\bibfnamefont {C.-Z.}\ \bibnamefont
  {Chen}}, \bibinfo {author} {\bibfnamefont {J.~J.}\ \bibnamefont {He}},
  \bibinfo {author} {\bibfnamefont {D.-H.}\ \bibnamefont {Xu}},\ and\ \bibinfo
  {author} {\bibfnamefont {K.}~\bibnamefont {Law}},\ }\bibfield  {title}
  {\bibinfo {title} {Emergent {J}osephson current of n=1 chiral topological
  superconductor in quantum anomalous {H}all insulator/superconductor
  heterostructures},\ }\href@noop {} {\bibfield  {journal} {\bibinfo  {journal}
  {Phys. Rev. B}\ }\textbf {\bibinfo {volume} {98}},\ \bibinfo {pages} {165439}
  (\bibinfo {year} {2018}{\natexlab{a}})}\BibitemShut {NoStop}%
\bibitem [{\citenamefont {Lian}\ \emph {et~al.}(2018)\citenamefont {Lian},
  \citenamefont {Sun}, \citenamefont {Vaezi}, \citenamefont {Qi},\ and\
  \citenamefont {Zhang}}]{51-lian2018topological}%
  \BibitemOpen
  \bibfield  {author} {\bibinfo {author} {\bibfnamefont {B.}~\bibnamefont
  {Lian}}, \bibinfo {author} {\bibfnamefont {X.-Q.}\ \bibnamefont {Sun}},
  \bibinfo {author} {\bibfnamefont {A.}~\bibnamefont {Vaezi}}, \bibinfo
  {author} {\bibfnamefont {X.-L.}\ \bibnamefont {Qi}},\ and\ \bibinfo {author}
  {\bibfnamefont {S.-C.}\ \bibnamefont {Zhang}},\ }\bibfield  {title} {\bibinfo
  {title} {Topological quantum computation based on chiral {M}ajorana
  fermions},\ }\href@noop {} {\bibfield  {journal} {\bibinfo  {journal} {Proc.
  Natl. Acad. Sci. U.S.A.}\ }\textbf {\bibinfo {volume} {115}},\ \bibinfo
  {pages} {10938} (\bibinfo {year} {2018})}\BibitemShut {NoStop}%
\bibitem [{\citenamefont {He}\ \emph {et~al.}(2017)\citenamefont {He},
  \citenamefont {Pan}, \citenamefont {Stern}, \citenamefont {Burks},
  \citenamefont {Che}, \citenamefont {Yin}, \citenamefont {Wang}, \citenamefont
  {Lian}, \citenamefont {Zhou}, \citenamefont {Choi} \emph
  {et~al.}}]{52-he2017chiral}%
  \BibitemOpen
  \bibfield  {author} {\bibinfo {author} {\bibfnamefont {Q.~L.}\ \bibnamefont
  {He}}, \bibinfo {author} {\bibfnamefont {L.}~\bibnamefont {Pan}}, \bibinfo
  {author} {\bibfnamefont {A.~L.}\ \bibnamefont {Stern}}, \bibinfo {author}
  {\bibfnamefont {E.~C.}\ \bibnamefont {Burks}}, \bibinfo {author}
  {\bibfnamefont {X.}~\bibnamefont {Che}}, \bibinfo {author} {\bibfnamefont
  {G.}~\bibnamefont {Yin}}, \bibinfo {author} {\bibfnamefont {J.}~\bibnamefont
  {Wang}}, \bibinfo {author} {\bibfnamefont {B.}~\bibnamefont {Lian}}, \bibinfo
  {author} {\bibfnamefont {Q.}~\bibnamefont {Zhou}}, \bibinfo {author}
  {\bibfnamefont {E.~S.}\ \bibnamefont {Choi}}, \emph {et~al.},\ }\bibfield
  {title} {\bibinfo {title} {Chiral {M}ajorana fermion modes in a quantum
  anomalous {H}all insulator--superconductor structure},\ }\href@noop {}
  {\bibfield  {journal} {\bibinfo  {journal} {Science}\ }\textbf {\bibinfo
  {volume} {357}},\ \bibinfo {pages} {294} (\bibinfo {year}
  {2017})}\BibitemShut {NoStop}%
\bibitem [{\citenamefont {Shen}\ \emph {et~al.}(2020)\citenamefont {Shen},
  \citenamefont {Lyu}, \citenamefont {Gao}, \citenamefont {Xie}, \citenamefont
  {Chen}, \citenamefont {Cho}, \citenamefont {Atanov}, \citenamefont {Chen},
  \citenamefont {Liu}, \citenamefont {Hu} \emph
  {et~al.}}]{53-shen2020spectroscopic}%
  \BibitemOpen
  \bibfield  {author} {\bibinfo {author} {\bibfnamefont {J.}~\bibnamefont
  {Shen}}, \bibinfo {author} {\bibfnamefont {J.}~\bibnamefont {Lyu}}, \bibinfo
  {author} {\bibfnamefont {J.~Z.}\ \bibnamefont {Gao}}, \bibinfo {author}
  {\bibfnamefont {Y.-M.}\ \bibnamefont {Xie}}, \bibinfo {author} {\bibfnamefont
  {C.-Z.}\ \bibnamefont {Chen}}, \bibinfo {author} {\bibfnamefont {C.-W.}\
  \bibnamefont {Cho}}, \bibinfo {author} {\bibfnamefont {O.}~\bibnamefont
  {Atanov}}, \bibinfo {author} {\bibfnamefont {Z.}~\bibnamefont {Chen}},
  \bibinfo {author} {\bibfnamefont {K.}~\bibnamefont {Liu}}, \bibinfo {author}
  {\bibfnamefont {Y.~J.}\ \bibnamefont {Hu}}, \emph {et~al.},\ }\bibfield
  {title} {\bibinfo {title} {Spectroscopic fingerprint of chiral {M}ajorana
  modes at the edge of a quantum anomalous {H}all insulator/superconductor
  heterostructure},\ }\href@noop {} {\bibfield  {journal} {\bibinfo  {journal}
  {Proc. Natl. Acad. Sci. U.S.A.}\ }\textbf {\bibinfo {volume} {117}},\
  \bibinfo {pages} {238} (\bibinfo {year} {2020})}\BibitemShut {NoStop}%
\bibitem [{\citenamefont {Kayyalha}\ \emph {et~al.}(2020)\citenamefont
  {Kayyalha}, \citenamefont {Xiao}, \citenamefont {Zhang}, \citenamefont
  {Shin}, \citenamefont {Jiang}, \citenamefont {Wang}, \citenamefont {Zhao},
  \citenamefont {Xiao}, \citenamefont {Zhang}, \citenamefont {Fijalkowski}
  \emph {et~al.}}]{54-kayyalha2020absence}%
  \BibitemOpen
  \bibfield  {author} {\bibinfo {author} {\bibfnamefont {M.}~\bibnamefont
  {Kayyalha}}, \bibinfo {author} {\bibfnamefont {D.}~\bibnamefont {Xiao}},
  \bibinfo {author} {\bibfnamefont {R.}~\bibnamefont {Zhang}}, \bibinfo
  {author} {\bibfnamefont {J.}~\bibnamefont {Shin}}, \bibinfo {author}
  {\bibfnamefont {J.}~\bibnamefont {Jiang}}, \bibinfo {author} {\bibfnamefont
  {F.}~\bibnamefont {Wang}}, \bibinfo {author} {\bibfnamefont {Y.-F.}\
  \bibnamefont {Zhao}}, \bibinfo {author} {\bibfnamefont {R.}~\bibnamefont
  {Xiao}}, \bibinfo {author} {\bibfnamefont {L.}~\bibnamefont {Zhang}},
  \bibinfo {author} {\bibfnamefont {K.~M.}\ \bibnamefont {Fijalkowski}}, \emph
  {et~al.},\ }\bibfield  {title} {\bibinfo {title} {Absence of evidence for
  chiral {M}ajorana modes in quantum anomalous {H}all-superconductor devices},\
  }\href@noop {} {\bibfield  {journal} {\bibinfo  {journal} {Science}\ }\textbf
  {\bibinfo {volume} {367}},\ \bibinfo {pages} {64} (\bibinfo {year}
  {2020})}\BibitemShut {NoStop}%
\bibitem [{\citenamefont {Fu}\ and\ \citenamefont
  {Kane}(2008)}]{24-fu2008superconducting}%
  \BibitemOpen
  \bibfield  {author} {\bibinfo {author} {\bibfnamefont {L.}~\bibnamefont
  {Fu}}\ and\ \bibinfo {author} {\bibfnamefont {C.~L.}\ \bibnamefont {Kane}},\
  }\bibfield  {title} {\bibinfo {title} {Superconducting proximity effect and
  {M}ajorana fermions at the surface of a topological insulator},\ }\href@noop
  {} {\bibfield  {journal} {\bibinfo  {journal} {Phys. Rev. Lett.}\ }\textbf
  {\bibinfo {volume} {100}},\ \bibinfo {pages} {096407} (\bibinfo {year}
  {2008})}\BibitemShut {NoStop}%
\bibitem [{\citenamefont {J{\"a}ck}\ \emph {et~al.}(2019)\citenamefont
  {J{\"a}ck}, \citenamefont {Xie}, \citenamefont {Li}, \citenamefont {Jeon},
  \citenamefont {Bernevig},\ and\ \citenamefont
  {Yazdani}}]{55-jack2019observation}%
  \BibitemOpen
  \bibfield  {author} {\bibinfo {author} {\bibfnamefont {B.}~\bibnamefont
  {J{\"a}ck}}, \bibinfo {author} {\bibfnamefont {Y.}~\bibnamefont {Xie}},
  \bibinfo {author} {\bibfnamefont {J.}~\bibnamefont {Li}}, \bibinfo {author}
  {\bibfnamefont {S.}~\bibnamefont {Jeon}}, \bibinfo {author} {\bibfnamefont
  {B.~A.}\ \bibnamefont {Bernevig}},\ and\ \bibinfo {author} {\bibfnamefont
  {A.}~\bibnamefont {Yazdani}},\ }\bibfield  {title} {\bibinfo {title}
  {Observation of a {M}ajorana zero mode in a topologically protected edge
  channel},\ }\href@noop {} {\bibfield  {journal} {\bibinfo  {journal}
  {Science}\ }\textbf {\bibinfo {volume} {364}},\ \bibinfo {pages} {1255}
  (\bibinfo {year} {2019})}\BibitemShut {NoStop}%
\bibitem [{\citenamefont {Fu}\ and\ \citenamefont
  {Kane}(2009)}]{56-fu2009probing}%
  \BibitemOpen
  \bibfield  {author} {\bibinfo {author} {\bibfnamefont {L.}~\bibnamefont
  {Fu}}\ and\ \bibinfo {author} {\bibfnamefont {C.~L.}\ \bibnamefont {Kane}},\
  }\bibfield  {title} {\bibinfo {title} {Probing neutral {M}ajorana fermion
  edge modes with charge transport},\ }\href@noop {} {\bibfield  {journal}
  {\bibinfo  {journal} {Phys. Rev. Lett.}\ }\textbf {\bibinfo {volume} {102}},\
  \bibinfo {pages} {216403} (\bibinfo {year} {2009})}\BibitemShut {NoStop}%
\bibitem [{\citenamefont {Akhmerov}\ \emph {et~al.}(2009)\citenamefont
  {Akhmerov}, \citenamefont {Nilsson},\ and\ \citenamefont
  {Beenakker}}]{57-akhmerov2009electrically}%
  \BibitemOpen
  \bibfield  {author} {\bibinfo {author} {\bibfnamefont {A.}~\bibnamefont
  {Akhmerov}}, \bibinfo {author} {\bibfnamefont {J.}~\bibnamefont {Nilsson}},\
  and\ \bibinfo {author} {\bibfnamefont {C.}~\bibnamefont {Beenakker}},\
  }\bibfield  {title} {\bibinfo {title} {Electrically detected interferometry
  of {M}ajorana fermions in a topological insulator},\ }\href@noop {}
  {\bibfield  {journal} {\bibinfo  {journal} {Phys. Rev. Lett.}\ }\textbf
  {\bibinfo {volume} {102}},\ \bibinfo {pages} {216404} (\bibinfo {year}
  {2009})}\BibitemShut {NoStop}%
\bibitem [{\citenamefont {Williams}\ \emph {et~al.}(2012)\citenamefont
  {Williams}, \citenamefont {Bestwick}, \citenamefont {Gallagher},
  \citenamefont {Hong}, \citenamefont {Cui}, \citenamefont {Bleich},
  \citenamefont {Analytis}, \citenamefont {Fisher},\ and\ \citenamefont
  {Goldhaber-Gordon}}]{58-williams2012unconventional}%
  \BibitemOpen
  \bibfield  {author} {\bibinfo {author} {\bibfnamefont {J.}~\bibnamefont
  {Williams}}, \bibinfo {author} {\bibfnamefont {A.}~\bibnamefont {Bestwick}},
  \bibinfo {author} {\bibfnamefont {P.}~\bibnamefont {Gallagher}}, \bibinfo
  {author} {\bibfnamefont {S.~S.}\ \bibnamefont {Hong}}, \bibinfo {author}
  {\bibfnamefont {Y.}~\bibnamefont {Cui}}, \bibinfo {author} {\bibfnamefont
  {A.~S.}\ \bibnamefont {Bleich}}, \bibinfo {author} {\bibfnamefont
  {J.}~\bibnamefont {Analytis}}, \bibinfo {author} {\bibfnamefont
  {I.}~\bibnamefont {Fisher}},\ and\ \bibinfo {author} {\bibfnamefont
  {D.}~\bibnamefont {Goldhaber-Gordon}},\ }\bibfield  {title} {\bibinfo {title}
  {Unconventional {J}osephson effect in hybrid superconductor-topological
  insulator devices},\ }\href@noop {} {\bibfield  {journal} {\bibinfo
  {journal} {Phys. Rev. Lett.}\ }\textbf {\bibinfo {volume} {109}},\ \bibinfo
  {pages} {056803} (\bibinfo {year} {2012})}\BibitemShut {NoStop}%
\bibitem [{\citenamefont {Wang}\ \emph
  {et~al.}(2012{\natexlab{a}})\citenamefont {Wang}, \citenamefont {Liu},
  \citenamefont {Xu}, \citenamefont {Yang}, \citenamefont {Miao}, \citenamefont
  {Yao}, \citenamefont {Gao}, \citenamefont {Shen}, \citenamefont {Ma},
  \citenamefont {Chen} \emph {et~al.}}]{59-wang2012coexistence}%
  \BibitemOpen
  \bibfield  {author} {\bibinfo {author} {\bibfnamefont {M.-X.}\ \bibnamefont
  {Wang}}, \bibinfo {author} {\bibfnamefont {C.}~\bibnamefont {Liu}}, \bibinfo
  {author} {\bibfnamefont {J.-P.}\ \bibnamefont {Xu}}, \bibinfo {author}
  {\bibfnamefont {F.}~\bibnamefont {Yang}}, \bibinfo {author} {\bibfnamefont
  {L.}~\bibnamefont {Miao}}, \bibinfo {author} {\bibfnamefont {M.-Y.}\
  \bibnamefont {Yao}}, \bibinfo {author} {\bibfnamefont {C.}~\bibnamefont
  {Gao}}, \bibinfo {author} {\bibfnamefont {C.}~\bibnamefont {Shen}}, \bibinfo
  {author} {\bibfnamefont {X.}~\bibnamefont {Ma}}, \bibinfo {author}
  {\bibfnamefont {X.}~\bibnamefont {Chen}}, \emph {et~al.},\ }\bibfield
  {title} {\bibinfo {title} {The coexistence of superconductivity and
  topological order in the {B}i$_{2}${S}e$_{3}$ thin films},\ }\href@noop {}
  {\bibfield  {journal} {\bibinfo  {journal} {Science}\ }\textbf {\bibinfo
  {volume} {336}},\ \bibinfo {pages} {52} (\bibinfo {year}
  {2012}{\natexlab{a}})}\BibitemShut {NoStop}%
\bibitem [{\citenamefont {Xu}\ \emph {et~al.}(2014)\citenamefont {Xu},
  \citenamefont {Liu}, \citenamefont {Wang}, \citenamefont {Ge}, \citenamefont
  {Liu}, \citenamefont {Yang}, \citenamefont {Chen}, \citenamefont {Liu},
  \citenamefont {Xu}, \citenamefont {Gao} \emph
  {et~al.}}]{60-xu2014artificial}%
  \BibitemOpen
  \bibfield  {author} {\bibinfo {author} {\bibfnamefont {J.-P.}\ \bibnamefont
  {Xu}}, \bibinfo {author} {\bibfnamefont {C.}~\bibnamefont {Liu}}, \bibinfo
  {author} {\bibfnamefont {M.-X.}\ \bibnamefont {Wang}}, \bibinfo {author}
  {\bibfnamefont {J.}~\bibnamefont {Ge}}, \bibinfo {author} {\bibfnamefont
  {Z.-L.}\ \bibnamefont {Liu}}, \bibinfo {author} {\bibfnamefont
  {X.}~\bibnamefont {Yang}}, \bibinfo {author} {\bibfnamefont {Y.}~\bibnamefont
  {Chen}}, \bibinfo {author} {\bibfnamefont {Y.}~\bibnamefont {Liu}}, \bibinfo
  {author} {\bibfnamefont {Z.-A.}\ \bibnamefont {Xu}}, \bibinfo {author}
  {\bibfnamefont {C.-L.}\ \bibnamefont {Gao}}, \emph {et~al.},\ }\bibfield
  {title} {\bibinfo {title} {Artificial topological superconductor by the
  proximity effect},\ }\href@noop {} {\bibfield  {journal} {\bibinfo  {journal}
  {Phys. Rev. Lett.}\ }\textbf {\bibinfo {volume} {112}},\ \bibinfo {pages}
  {217001} (\bibinfo {year} {2014})}\BibitemShut {NoStop}%
\bibitem [{\citenamefont {Xu}\ \emph {et~al.}(2015{\natexlab{a}})\citenamefont
  {Xu}, \citenamefont {Wang}, \citenamefont {Liu}, \citenamefont {Ge},
  \citenamefont {Yang}, \citenamefont {Liu}, \citenamefont {Xu}, \citenamefont
  {Guan}, \citenamefont {Gao}, \citenamefont {Qian} \emph
  {et~al.}}]{61-xu2015experimental}%
  \BibitemOpen
  \bibfield  {author} {\bibinfo {author} {\bibfnamefont {J.-P.}\ \bibnamefont
  {Xu}}, \bibinfo {author} {\bibfnamefont {M.-X.}\ \bibnamefont {Wang}},
  \bibinfo {author} {\bibfnamefont {Z.~L.}\ \bibnamefont {Liu}}, \bibinfo
  {author} {\bibfnamefont {J.-F.}\ \bibnamefont {Ge}}, \bibinfo {author}
  {\bibfnamefont {X.}~\bibnamefont {Yang}}, \bibinfo {author} {\bibfnamefont
  {C.}~\bibnamefont {Liu}}, \bibinfo {author} {\bibfnamefont {Z.~A.}\
  \bibnamefont {Xu}}, \bibinfo {author} {\bibfnamefont {D.}~\bibnamefont
  {Guan}}, \bibinfo {author} {\bibfnamefont {C.~L.}\ \bibnamefont {Gao}},
  \bibinfo {author} {\bibfnamefont {D.}~\bibnamefont {Qian}}, \emph {et~al.},\
  }\bibfield  {title} {\bibinfo {title} {Experimental detection of a {M}ajorana
  mode in the core of a magnetic vortex inside a topological
  insulator-superconductor {B}i$_{2}${T}e$_{3}$/{N}b{S}e$_{2}$
  heterostructure},\ }\href@noop {} {\bibfield  {journal} {\bibinfo  {journal}
  {Phys. Rev. Lett.}\ }\textbf {\bibinfo {volume} {114}},\ \bibinfo {pages}
  {017001} (\bibinfo {year} {2015}{\natexlab{a}})}\BibitemShut {NoStop}%
\bibitem [{\citenamefont {Sun}\ \emph {et~al.}(2016)\citenamefont {Sun},
  \citenamefont {Zhang}, \citenamefont {Hu}, \citenamefont {Li}, \citenamefont
  {Wang}, \citenamefont {Ma}, \citenamefont {Xu}, \citenamefont {Gao},
  \citenamefont {Guan}, \citenamefont {Li} \emph
  {et~al.}}]{62-sun2016majorana}%
  \BibitemOpen
  \bibfield  {author} {\bibinfo {author} {\bibfnamefont {H.-H.}\ \bibnamefont
  {Sun}}, \bibinfo {author} {\bibfnamefont {K.-W.}\ \bibnamefont {Zhang}},
  \bibinfo {author} {\bibfnamefont {L.-H.}\ \bibnamefont {Hu}}, \bibinfo
  {author} {\bibfnamefont {C.}~\bibnamefont {Li}}, \bibinfo {author}
  {\bibfnamefont {G.-Y.}\ \bibnamefont {Wang}}, \bibinfo {author}
  {\bibfnamefont {H.-Y.}\ \bibnamefont {Ma}}, \bibinfo {author} {\bibfnamefont
  {Z.-A.}\ \bibnamefont {Xu}}, \bibinfo {author} {\bibfnamefont {C.-L.}\
  \bibnamefont {Gao}}, \bibinfo {author} {\bibfnamefont {D.-D.}\ \bibnamefont
  {Guan}}, \bibinfo {author} {\bibfnamefont {Y.-Y.}\ \bibnamefont {Li}}, \emph
  {et~al.},\ }\bibfield  {title} {\bibinfo {title} {Majorana zero mode detected
  with spin selective {A}ndreev reflection in the vortex of a topological
  superconductor},\ }\href@noop {} {\bibfield  {journal} {\bibinfo  {journal}
  {Phys. Rev. Lett.}\ }\textbf {\bibinfo {volume} {116}},\ \bibinfo {pages}
  {257003} (\bibinfo {year} {2016})}\BibitemShut {NoStop}%
\bibitem [{\citenamefont {Takei}\ \emph {et~al.}(2013)\citenamefont {Takei},
  \citenamefont {Fregoso}, \citenamefont {Hui}, \citenamefont {Lobos},\ and\
  \citenamefont {Sarma}}]{63-takei2013soft}%
  \BibitemOpen
  \bibfield  {author} {\bibinfo {author} {\bibfnamefont {S.}~\bibnamefont
  {Takei}}, \bibinfo {author} {\bibfnamefont {B.~M.}\ \bibnamefont {Fregoso}},
  \bibinfo {author} {\bibfnamefont {H.-Y.}\ \bibnamefont {Hui}}, \bibinfo
  {author} {\bibfnamefont {A.~M.}\ \bibnamefont {Lobos}},\ and\ \bibinfo
  {author} {\bibfnamefont {S.~D.}\ \bibnamefont {Sarma}},\ }\bibfield  {title}
  {\bibinfo {title} {Soft superconducting gap in semiconductor {M}ajorana
  nanowires},\ }\href@noop {} {\bibfield  {journal} {\bibinfo  {journal} {Phys.
  Rev. Lett.}\ }\textbf {\bibinfo {volume} {110}},\ \bibinfo {pages} {186803}
  (\bibinfo {year} {2013})}\BibitemShut {NoStop}%
\bibitem [{\citenamefont {Mazin}\ \emph {et~al.}(2008)\citenamefont {Mazin},
  \citenamefont {Singh}, \citenamefont {Johannes},\ and\ \citenamefont
  {Du}}]{64-mazin2008unconventional}%
  \BibitemOpen
  \bibfield  {author} {\bibinfo {author} {\bibfnamefont {I.}~\bibnamefont
  {Mazin}}, \bibinfo {author} {\bibfnamefont {D.~J.}\ \bibnamefont {Singh}},
  \bibinfo {author} {\bibfnamefont {M.}~\bibnamefont {Johannes}},\ and\
  \bibinfo {author} {\bibfnamefont {M.-H.}\ \bibnamefont {Du}},\ }\bibfield
  {title} {\bibinfo {title} {Unconventional superconductivity with a sign
  reversal in the order parameter of {L}a{F}e{A}s{O}$_{1- x}${F}$_{x}$},\
  }\href@noop {} {\bibfield  {journal} {\bibinfo  {journal} {Phys. Rev. Lett.}\
  }\textbf {\bibinfo {volume} {101}},\ \bibinfo {pages} {057003} (\bibinfo
  {year} {2008})}\BibitemShut {NoStop}%
\bibitem [{\citenamefont {Kuroki}\ \emph {et~al.}(2008)\citenamefont {Kuroki},
  \citenamefont {Onari}, \citenamefont {Arita}, \citenamefont {Usui},
  \citenamefont {Tanaka}, \citenamefont {Kontani},\ and\ \citenamefont
  {Aoki}}]{65-kuroki2008unconventional}%
  \BibitemOpen
  \bibfield  {author} {\bibinfo {author} {\bibfnamefont {K.}~\bibnamefont
  {Kuroki}}, \bibinfo {author} {\bibfnamefont {S.}~\bibnamefont {Onari}},
  \bibinfo {author} {\bibfnamefont {R.}~\bibnamefont {Arita}}, \bibinfo
  {author} {\bibfnamefont {H.}~\bibnamefont {Usui}}, \bibinfo {author}
  {\bibfnamefont {Y.}~\bibnamefont {Tanaka}}, \bibinfo {author} {\bibfnamefont
  {H.}~\bibnamefont {Kontani}},\ and\ \bibinfo {author} {\bibfnamefont
  {H.}~\bibnamefont {Aoki}},\ }\bibfield  {title} {\bibinfo {title}
  {Unconventional pairing originating from the disconnected {F}ermi surfaces of
  superconducting {L}a{F}e{A}s{O}$_{1- x}${F}$_{x}$},\ }\href@noop {}
  {\bibfield  {journal} {\bibinfo  {journal} {Phys. Rev. Lett.}\ }\textbf
  {\bibinfo {volume} {101}},\ \bibinfo {pages} {087004} (\bibinfo {year}
  {2008})}\BibitemShut {NoStop}%
\bibitem [{\citenamefont {Seo}\ \emph {et~al.}(2008)\citenamefont {Seo},
  \citenamefont {Bernevig},\ and\ \citenamefont {Hu}}]{66-seo2008pairing}%
  \BibitemOpen
  \bibfield  {author} {\bibinfo {author} {\bibfnamefont {K.}~\bibnamefont
  {Seo}}, \bibinfo {author} {\bibfnamefont {B.~A.}\ \bibnamefont {Bernevig}},\
  and\ \bibinfo {author} {\bibfnamefont {J.}~\bibnamefont {Hu}},\ }\bibfield
  {title} {\bibinfo {title} {Pairing symmetry in a two-orbital exchange
  coupling model of oxypnictides},\ }\href@noop {} {\bibfield  {journal}
  {\bibinfo  {journal} {Phys. Rev. Lett.}\ }\textbf {\bibinfo {volume} {101}},\
  \bibinfo {pages} {206404} (\bibinfo {year} {2008})}\BibitemShut {NoStop}%
\bibitem [{\citenamefont {Chen}\ \emph {et~al.}(2009)\citenamefont {Chen},
  \citenamefont {Yang}, \citenamefont {Zhou},\ and\ \citenamefont
  {Zhang}}]{67-chen2009strong}%
  \BibitemOpen
  \bibfield  {author} {\bibinfo {author} {\bibfnamefont {W.-Q.}\ \bibnamefont
  {Chen}}, \bibinfo {author} {\bibfnamefont {K.-Y.}\ \bibnamefont {Yang}},
  \bibinfo {author} {\bibfnamefont {Y.}~\bibnamefont {Zhou}},\ and\ \bibinfo
  {author} {\bibfnamefont {F.-C.}\ \bibnamefont {Zhang}},\ }\bibfield  {title}
  {\bibinfo {title} {Strong coupling theory for superconducting iron
  pnictides},\ }\href@noop {} {\bibfield  {journal} {\bibinfo  {journal} {Phys.
  Rev. Lett.}\ }\textbf {\bibinfo {volume} {102}},\ \bibinfo {pages} {047006}
  (\bibinfo {year} {2009})}\BibitemShut {NoStop}%
\bibitem [{\citenamefont {Maier}\ \emph {et~al.}(2011)\citenamefont {Maier},
  \citenamefont {Graser}, \citenamefont {Hirschfeld},\ and\ \citenamefont
  {Scalapino}}]{68-maier2011d}%
  \BibitemOpen
  \bibfield  {author} {\bibinfo {author} {\bibfnamefont {T.}~\bibnamefont
  {Maier}}, \bibinfo {author} {\bibfnamefont {S.}~\bibnamefont {Graser}},
  \bibinfo {author} {\bibfnamefont {P.}~\bibnamefont {Hirschfeld}},\ and\
  \bibinfo {author} {\bibfnamefont {D.}~\bibnamefont {Scalapino}},\ }\bibfield
  {title} {\bibinfo {title} {d-wave pairing from spin fluctuations in the
  {K}$_{x} ${F}e$_{2-y}${S}e$_{2}$ superconductors},\ }\href@noop {} {\bibfield
   {journal} {\bibinfo  {journal} {Phys. Rev. B}\ }\textbf {\bibinfo {volume}
  {83}},\ \bibinfo {pages} {100515} (\bibinfo {year} {2011})}\BibitemShut
  {NoStop}%
\bibitem [{\citenamefont {Khodas}\ and\ \citenamefont
  {Chubukov}(2012)}]{69-khodas2012interpocket}%
  \BibitemOpen
  \bibfield  {author} {\bibinfo {author} {\bibfnamefont {M.}~\bibnamefont
  {Khodas}}\ and\ \bibinfo {author} {\bibfnamefont {A.}~\bibnamefont
  {Chubukov}},\ }\bibfield  {title} {\bibinfo {title} {Interpocket pairing and
  gap symmetry in {F}e-based superconductors with only electron pockets},\
  }\href@noop {} {\bibfield  {journal} {\bibinfo  {journal} {Phys. Rev. Lett.}\
  }\textbf {\bibinfo {volume} {108}},\ \bibinfo {pages} {247003} (\bibinfo
  {year} {2012})}\BibitemShut {NoStop}%
\bibitem [{\citenamefont {Agterberg}\ \emph {et~al.}(2017)\citenamefont
  {Agterberg}, \citenamefont {Shishidou}, \citenamefont {O’Halloran},
  \citenamefont {Brydon},\ and\ \citenamefont
  {Weinert}}]{70-agterberg2017resilient}%
  \BibitemOpen
  \bibfield  {author} {\bibinfo {author} {\bibfnamefont {D.}~\bibnamefont
  {Agterberg}}, \bibinfo {author} {\bibfnamefont {T.}~\bibnamefont
  {Shishidou}}, \bibinfo {author} {\bibfnamefont {J.}~\bibnamefont
  {O’Halloran}}, \bibinfo {author} {\bibfnamefont {P.}~\bibnamefont
  {Brydon}},\ and\ \bibinfo {author} {\bibfnamefont {M.}~\bibnamefont
  {Weinert}},\ }\bibfield  {title} {\bibinfo {title} {Resilient nodeless d-wave
  superconductivity in monolayer {F}e{S}e},\ }\href@noop {} {\bibfield
  {journal} {\bibinfo  {journal} {Phys. Rev. Lett.}\ }\textbf {\bibinfo
  {volume} {119}},\ \bibinfo {pages} {267001} (\bibinfo {year}
  {2017})}\BibitemShut {NoStop}%
\bibitem [{\citenamefont {Lee}(2018)}]{71-lee2018routes}%
  \BibitemOpen
  \bibfield  {author} {\bibinfo {author} {\bibfnamefont {D.-H.}\ \bibnamefont
  {Lee}},\ }\bibfield  {title} {\bibinfo {title} {Routes to high-temperature
  superconductivity: A lesson from {F}e{S}e/{S}r{T}i{O}$_{3}$},\ }\href@noop {}
  {\bibfield  {journal} {\bibinfo  {journal} {Annu. Rev. Condens. Matter
  Phys.}\ }\textbf {\bibinfo {volume} {9}},\ \bibinfo {pages} {261} (\bibinfo
  {year} {2018})}\BibitemShut {NoStop}%
\bibitem [{\citenamefont {Paglione}\ and\ \citenamefont
  {Greene}(2010)}]{72-paglione2010high}%
  \BibitemOpen
  \bibfield  {author} {\bibinfo {author} {\bibfnamefont {J.}~\bibnamefont
  {Paglione}}\ and\ \bibinfo {author} {\bibfnamefont {R.~L.}\ \bibnamefont
  {Greene}},\ }\bibfield  {title} {\bibinfo {title} {High-temperature
  superconductivity in iron-based materials},\ }\href@noop {} {\bibfield
  {journal} {\bibinfo  {journal} {Nat. Phys.}\ }\textbf {\bibinfo {volume}
  {6}},\ \bibinfo {pages} {645} (\bibinfo {year} {2010})}\BibitemShut {NoStop}%
\bibitem [{\citenamefont {Chen}\ \emph {et~al.}(2014)\citenamefont {Chen},
  \citenamefont {Dai}, \citenamefont {Feng}, \citenamefont {Xiang},\ and\
  \citenamefont {Zhang}}]{73-chen2014iron}%
  \BibitemOpen
  \bibfield  {author} {\bibinfo {author} {\bibfnamefont {X.}~\bibnamefont
  {Chen}}, \bibinfo {author} {\bibfnamefont {P.}~\bibnamefont {Dai}}, \bibinfo
  {author} {\bibfnamefont {D.}~\bibnamefont {Feng}}, \bibinfo {author}
  {\bibfnamefont {T.}~\bibnamefont {Xiang}},\ and\ \bibinfo {author}
  {\bibfnamefont {F.-C.}\ \bibnamefont {Zhang}},\ }\bibfield  {title} {\bibinfo
  {title} {Iron-based high transition temperature superconductors},\
  }\href@noop {} {\bibfield  {journal} {\bibinfo  {journal} {Natl. Sci. Rev.}\
  }\textbf {\bibinfo {volume} {1}},\ \bibinfo {pages} {371} (\bibinfo {year}
  {2014})}\BibitemShut {NoStop}%
\bibitem [{\citenamefont {Hirschfeld}\ \emph {et~al.}(2011)\citenamefont
  {Hirschfeld}, \citenamefont {Korshunov},\ and\ \citenamefont
  {Mazin}}]{75-hirschfeld2011gap}%
  \BibitemOpen
  \bibfield  {author} {\bibinfo {author} {\bibfnamefont {P.}~\bibnamefont
  {Hirschfeld}}, \bibinfo {author} {\bibfnamefont {M.}~\bibnamefont
  {Korshunov}},\ and\ \bibinfo {author} {\bibfnamefont {I.}~\bibnamefont
  {Mazin}},\ }\bibfield  {title} {\bibinfo {title} {Gap symmetry and structure
  of {F}e-based superconductors},\ }\href@noop {} {\bibfield  {journal}
  {\bibinfo  {journal} {Rep. Prog. Phys.}\ }\textbf {\bibinfo {volume} {74}},\
  \bibinfo {pages} {124508} (\bibinfo {year} {2011})}\BibitemShut {NoStop}%
\bibitem [{\citenamefont {Hanaguri}\ \emph {et~al.}(2010)\citenamefont
  {Hanaguri}, \citenamefont {Niitaka}, \citenamefont {Kuroki},\ and\
  \citenamefont {Takagi}}]{77-hanaguri2010unconventional}%
  \BibitemOpen
  \bibfield  {author} {\bibinfo {author} {\bibfnamefont {T.}~\bibnamefont
  {Hanaguri}}, \bibinfo {author} {\bibfnamefont {S.}~\bibnamefont {Niitaka}},
  \bibinfo {author} {\bibfnamefont {K.}~\bibnamefont {Kuroki}},\ and\ \bibinfo
  {author} {\bibfnamefont {H.}~\bibnamefont {Takagi}},\ }\bibfield  {title}
  {\bibinfo {title} {Unconventional s-wave superconductivity in
  {F}e({S}e,{T}e)},\ }\href@noop {} {\bibfield  {journal} {\bibinfo  {journal}
  {Science}\ }\textbf {\bibinfo {volume} {328}},\ \bibinfo {pages} {474}
  (\bibinfo {year} {2010})}\BibitemShut {NoStop}%
\bibitem [{\citenamefont {Liu}\ \emph {et~al.}(2019{\natexlab{a}})\citenamefont
  {Liu}, \citenamefont {Wang}, \citenamefont {Gao}, \citenamefont {Liu},
  \citenamefont {Liu}, \citenamefont {Wang},\ and\ \citenamefont
  {Wang}}]{78-liu2019spectroscopic}%
  \BibitemOpen
  \bibfield  {author} {\bibinfo {author} {\bibfnamefont {C.}~\bibnamefont
  {Liu}}, \bibinfo {author} {\bibfnamefont {Z.}~\bibnamefont {Wang}}, \bibinfo
  {author} {\bibfnamefont {Y.}~\bibnamefont {Gao}}, \bibinfo {author}
  {\bibfnamefont {X.}~\bibnamefont {Liu}}, \bibinfo {author} {\bibfnamefont
  {Y.}~\bibnamefont {Liu}}, \bibinfo {author} {\bibfnamefont {Q.-H.}\
  \bibnamefont {Wang}},\ and\ \bibinfo {author} {\bibfnamefont
  {J.}~\bibnamefont {Wang}},\ }\bibfield  {title} {\bibinfo {title}
  {Spectroscopic imaging of quasiparticle bound states induced by strong
  nonmagnetic scatterings in one-unit-cell {F}e{S}e/{S}r{T}i{O}$_{3}$},\
  }\href@noop {} {\bibfield  {journal} {\bibinfo  {journal} {Phys. Rev. Lett.}\
  }\textbf {\bibinfo {volume} {123}},\ \bibinfo {pages} {036801} (\bibinfo
  {year} {2019}{\natexlab{a}})}\BibitemShut {NoStop}%
\bibitem [{\citenamefont {Liu}\ \emph {et~al.}(2019{\natexlab{b}})\citenamefont
  {Liu}, \citenamefont {Wang}, \citenamefont {Ye}, \citenamefont {Chen},
  \citenamefont {Liu}, \citenamefont {Wang}, \citenamefont {Wang},\ and\
  \citenamefont {Wang}}]{79-liu2019detection}%
  \BibitemOpen
  \bibfield  {author} {\bibinfo {author} {\bibfnamefont {C.}~\bibnamefont
  {Liu}}, \bibinfo {author} {\bibfnamefont {Z.}~\bibnamefont {Wang}}, \bibinfo
  {author} {\bibfnamefont {S.}~\bibnamefont {Ye}}, \bibinfo {author}
  {\bibfnamefont {C.}~\bibnamefont {Chen}}, \bibinfo {author} {\bibfnamefont
  {Y.}~\bibnamefont {Liu}}, \bibinfo {author} {\bibfnamefont {Q.}~\bibnamefont
  {Wang}}, \bibinfo {author} {\bibfnamefont {Q.-H.}\ \bibnamefont {Wang}},\
  and\ \bibinfo {author} {\bibfnamefont {J.}~\bibnamefont {Wang}},\ }\bibfield
  {title} {\bibinfo {title} {Detection of bosonic mode as a signature of
  magnetic excitation in one-unit-cell {F}e{S}e on {S}r{T}i{O}$_{3}$},\
  }\href@noop {} {\bibfield  {journal} {\bibinfo  {journal} {Nano Lett.}\
  }\textbf {\bibinfo {volume} {19}},\ \bibinfo {pages} {3464} (\bibinfo {year}
  {2019}{\natexlab{b}})}\BibitemShut {NoStop}%
\bibitem [{\citenamefont {Chen}\ \emph
  {et~al.}(2020{\natexlab{a}})\citenamefont {Chen}, \citenamefont {Liu},
  \citenamefont {Liu},\ and\ \citenamefont {Wang}}]{80-chen2020bosonic}%
  \BibitemOpen
  \bibfield  {author} {\bibinfo {author} {\bibfnamefont {C.}~\bibnamefont
  {Chen}}, \bibinfo {author} {\bibfnamefont {C.}~\bibnamefont {Liu}}, \bibinfo
  {author} {\bibfnamefont {Y.}~\bibnamefont {Liu}},\ and\ \bibinfo {author}
  {\bibfnamefont {J.}~\bibnamefont {Wang}},\ }\bibfield  {title} {\bibinfo
  {title} {Bosonic mode and impurity-scattering in monolayer {F}e({T}e,{S}e)
  high-temperature superconductors},\ }\href@noop {} {\bibfield  {journal}
  {\bibinfo  {journal} {Nano Lett.}\ }\textbf {\bibinfo {volume} {20}},\
  \bibinfo {pages} {2056} (\bibinfo {year} {2020}{\natexlab{a}})}\BibitemShut
  {NoStop}%
\bibitem [{\citenamefont {Hao}\ and\ \citenamefont {Hu}(2018)}]{83-hnwlxb}%
  \BibitemOpen
  \bibfield  {author} {\bibinfo {author} {\bibfnamefont {N.}~\bibnamefont
  {Hao}}\ and\ \bibinfo {author} {\bibfnamefont {J.}~\bibnamefont {Hu}},\
  }\bibfield  {title} {\bibinfo {title} {Research progress of topological
  quantum states in iron-based superconductor},\ }\href@noop {} {\bibfield
  {journal} {\bibinfo  {journal} {Acta Phys. Sin.}\ }\textbf {\bibinfo {volume}
  {67}} (\bibinfo {year} {2018})}\BibitemShut {NoStop}%
\bibitem [{\citenamefont {Hao}\ and\ \citenamefont
  {Hu}(2019)}]{84-hao2019topological}%
  \BibitemOpen
  \bibfield  {author} {\bibinfo {author} {\bibfnamefont {N.}~\bibnamefont
  {Hao}}\ and\ \bibinfo {author} {\bibfnamefont {J.}~\bibnamefont {Hu}},\
  }\bibfield  {title} {\bibinfo {title} {Topological quantum states of matter
  in iron-based superconductors: from concept to material realization},\
  }\href@noop {} {\bibfield  {journal} {\bibinfo  {journal} {Natl. Sci. Rev.}\
  }\textbf {\bibinfo {volume} {6}},\ \bibinfo {pages} {213} (\bibinfo {year}
  {2019})}\BibitemShut {NoStop}%
\bibitem [{\citenamefont {Hao}\ and\ \citenamefont
  {Hu}(2014)}]{85-hao2014topological}%
  \BibitemOpen
  \bibfield  {author} {\bibinfo {author} {\bibfnamefont {N.}~\bibnamefont
  {Hao}}\ and\ \bibinfo {author} {\bibfnamefont {J.}~\bibnamefont {Hu}},\
  }\bibfield  {title} {\bibinfo {title} {Topological phases in the single-layer
  {F}e{S}e},\ }\href@noop {} {\bibfield  {journal} {\bibinfo  {journal} {Phys.
  Rev. X}\ }\textbf {\bibinfo {volume} {4}},\ \bibinfo {pages} {031053}
  (\bibinfo {year} {2014})}\BibitemShut {NoStop}%
\bibitem [{\citenamefont {Wu}\ \emph {et~al.}(2016{\natexlab{a}})\citenamefont
  {Wu}, \citenamefont {Qin}, \citenamefont {Liang}, \citenamefont {Fan},\ and\
  \citenamefont {Hu}}]{86-wu2016topological}%
  \BibitemOpen
  \bibfield  {author} {\bibinfo {author} {\bibfnamefont {X.}~\bibnamefont
  {Wu}}, \bibinfo {author} {\bibfnamefont {S.}~\bibnamefont {Qin}}, \bibinfo
  {author} {\bibfnamefont {Y.}~\bibnamefont {Liang}}, \bibinfo {author}
  {\bibfnamefont {H.}~\bibnamefont {Fan}},\ and\ \bibinfo {author}
  {\bibfnamefont {J.}~\bibnamefont {Hu}},\ }\bibfield  {title} {\bibinfo
  {title} {Topological characters in {F}e({T}e$_{1-x}$,{S}e$_{x}$) thin
  films},\ }\href@noop {} {\bibfield  {journal} {\bibinfo  {journal} {Phys.
  Rev. B}\ }\textbf {\bibinfo {volume} {93}},\ \bibinfo {pages} {115129}
  (\bibinfo {year} {2016}{\natexlab{a}})}\BibitemShut {NoStop}%
\bibitem [{\citenamefont {Wang}\ \emph {et~al.}(2015)\citenamefont {Wang},
  \citenamefont {Zhang}, \citenamefont {Xu}, \citenamefont {Zeng},
  \citenamefont {Miao}, \citenamefont {Xu}, \citenamefont {Qian}, \citenamefont
  {Weng}, \citenamefont {Richard}, \citenamefont {Fedorov} \emph
  {et~al.}}]{87-wang2015topological}%
  \BibitemOpen
  \bibfield  {author} {\bibinfo {author} {\bibfnamefont {Z.}~\bibnamefont
  {Wang}}, \bibinfo {author} {\bibfnamefont {P.}~\bibnamefont {Zhang}},
  \bibinfo {author} {\bibfnamefont {G.}~\bibnamefont {Xu}}, \bibinfo {author}
  {\bibfnamefont {L.}~\bibnamefont {Zeng}}, \bibinfo {author} {\bibfnamefont
  {H.}~\bibnamefont {Miao}}, \bibinfo {author} {\bibfnamefont {X.}~\bibnamefont
  {Xu}}, \bibinfo {author} {\bibfnamefont {T.}~\bibnamefont {Qian}}, \bibinfo
  {author} {\bibfnamefont {H.}~\bibnamefont {Weng}}, \bibinfo {author}
  {\bibfnamefont {P.}~\bibnamefont {Richard}}, \bibinfo {author} {\bibfnamefont
  {A.~V.}\ \bibnamefont {Fedorov}}, \emph {et~al.},\ }\bibfield  {title}
  {\bibinfo {title} {Topological nature of the {F}e{S}e$_{0.5}${T}e$_{0.5}$
  superconductor},\ }\href@noop {} {\bibfield  {journal} {\bibinfo  {journal}
  {Phys. Rev. B}\ }\textbf {\bibinfo {volume} {92}},\ \bibinfo {pages} {115119}
  (\bibinfo {year} {2015})}\BibitemShut {NoStop}%
\bibitem [{\citenamefont {Wu}\ \emph {et~al.}(2015)\citenamefont {Wu},
  \citenamefont {Qin}, \citenamefont {Liang}, \citenamefont {Le}, \citenamefont
  {Fan},\ and\ \citenamefont {Hu}}]{88-wu2015cafeas}%
  \BibitemOpen
  \bibfield  {author} {\bibinfo {author} {\bibfnamefont {X.}~\bibnamefont
  {Wu}}, \bibinfo {author} {\bibfnamefont {S.}~\bibnamefont {Qin}}, \bibinfo
  {author} {\bibfnamefont {Y.}~\bibnamefont {Liang}}, \bibinfo {author}
  {\bibfnamefont {C.}~\bibnamefont {Le}}, \bibinfo {author} {\bibfnamefont
  {H.}~\bibnamefont {Fan}},\ and\ \bibinfo {author} {\bibfnamefont
  {J.}~\bibnamefont {Hu}},\ }\bibfield  {title} {\bibinfo {title}
  {{C}a{F}e{A}s$_{2}$: A staggered intercalation of quantum spin {H}all and
  high-temperature superconductivity},\ }\href@noop {} {\bibfield  {journal}
  {\bibinfo  {journal} {Phys. Rev. B}\ }\textbf {\bibinfo {volume} {91}},\
  \bibinfo {pages} {081111} (\bibinfo {year} {2015})}\BibitemShut {NoStop}%
\bibitem [{\citenamefont {Ran}\ \emph {et~al.}(2009)\citenamefont {Ran},
  \citenamefont {Wang}, \citenamefont {Zhai}, \citenamefont {Vishwanath},\ and\
  \citenamefont {Lee}}]{89-ran2009nodal}%
  \BibitemOpen
  \bibfield  {author} {\bibinfo {author} {\bibfnamefont {Y.}~\bibnamefont
  {Ran}}, \bibinfo {author} {\bibfnamefont {F.}~\bibnamefont {Wang}}, \bibinfo
  {author} {\bibfnamefont {H.}~\bibnamefont {Zhai}}, \bibinfo {author}
  {\bibfnamefont {A.}~\bibnamefont {Vishwanath}},\ and\ \bibinfo {author}
  {\bibfnamefont {D.-H.}\ \bibnamefont {Lee}},\ }\bibfield  {title} {\bibinfo
  {title} {Nodal spin density wave and band topology of the {F}e{A}s-based
  materials},\ }\href@noop {} {\bibfield  {journal} {\bibinfo  {journal} {Phys.
  Rev. B}\ }\textbf {\bibinfo {volume} {79}},\ \bibinfo {pages} {014505}
  (\bibinfo {year} {2009})}\BibitemShut {NoStop}%
\bibitem [{\citenamefont {Morinari}\ \emph {et~al.}(2010)\citenamefont
  {Morinari}, \citenamefont {Kaneshita},\ and\ \citenamefont
  {Tohyama}}]{90-morinari2010topological}%
  \BibitemOpen
  \bibfield  {author} {\bibinfo {author} {\bibfnamefont {T.}~\bibnamefont
  {Morinari}}, \bibinfo {author} {\bibfnamefont {E.}~\bibnamefont
  {Kaneshita}},\ and\ \bibinfo {author} {\bibfnamefont {T.}~\bibnamefont
  {Tohyama}},\ }\bibfield  {title} {\bibinfo {title} {Topological and transport
  properties of {D}irac fermions in an antiferromagnetic metallic phase of
  iron-based superconductors},\ }\href@noop {} {\bibfield  {journal} {\bibinfo
  {journal} {Phys. Rev. Lett.}\ }\textbf {\bibinfo {volume} {105}},\ \bibinfo
  {pages} {037203} (\bibinfo {year} {2010})}\BibitemShut {NoStop}%
\bibitem [{\citenamefont {Richard}\ \emph {et~al.}(2010)\citenamefont
  {Richard}, \citenamefont {Nakayama}, \citenamefont {Sato}, \citenamefont
  {Neupane}, \citenamefont {Xu}, \citenamefont {Bowen}, \citenamefont {Chen},
  \citenamefont {Luo}, \citenamefont {Wang}, \citenamefont {Dai} \emph
  {et~al.}}]{91-richard2010observation}%
  \BibitemOpen
  \bibfield  {author} {\bibinfo {author} {\bibfnamefont {P.}~\bibnamefont
  {Richard}}, \bibinfo {author} {\bibfnamefont {K.}~\bibnamefont {Nakayama}},
  \bibinfo {author} {\bibfnamefont {T.}~\bibnamefont {Sato}}, \bibinfo {author}
  {\bibfnamefont {M.}~\bibnamefont {Neupane}}, \bibinfo {author} {\bibfnamefont
  {Y.-M.}\ \bibnamefont {Xu}}, \bibinfo {author} {\bibfnamefont
  {J.}~\bibnamefont {Bowen}}, \bibinfo {author} {\bibfnamefont
  {G.}~\bibnamefont {Chen}}, \bibinfo {author} {\bibfnamefont {J.}~\bibnamefont
  {Luo}}, \bibinfo {author} {\bibfnamefont {N.}~\bibnamefont {Wang}}, \bibinfo
  {author} {\bibfnamefont {X.}~\bibnamefont {Dai}}, \emph {et~al.},\ }\bibfield
   {title} {\bibinfo {title} {Observation of {D}irac cone electronic dispersion
  in {B}a{F}e$_{2}${A}s$_{2}$},\ }\href@noop {} {\bibfield  {journal} {\bibinfo
   {journal} {Phys. Rev. Lett.}\ }\textbf {\bibinfo {volume} {104}},\ \bibinfo
  {pages} {137001} (\bibinfo {year} {2010})}\BibitemShut {NoStop}%
\bibitem [{\citenamefont {Huynh}\ \emph {et~al.}(2011)\citenamefont {Huynh},
  \citenamefont {Tanabe},\ and\ \citenamefont {Tanigaki}}]{92-huynh2011both}%
  \BibitemOpen
  \bibfield  {author} {\bibinfo {author} {\bibfnamefont {K.~K.}\ \bibnamefont
  {Huynh}}, \bibinfo {author} {\bibfnamefont {Y.}~\bibnamefont {Tanabe}},\ and\
  \bibinfo {author} {\bibfnamefont {K.}~\bibnamefont {Tanigaki}},\ }\bibfield
  {title} {\bibinfo {title} {Both electron and hole {D}irac cone states in
  {B}a{F}e$_{2}${A}s$_{2}$ confirmed by magnetoresistance},\ }\href@noop {}
  {\bibfield  {journal} {\bibinfo  {journal} {Phys. Rev. Lett.}\ }\textbf
  {\bibinfo {volume} {106}},\ \bibinfo {pages} {217004} (\bibinfo {year}
  {2011})}\BibitemShut {NoStop}%
\bibitem [{\citenamefont {Hao}\ \emph {et~al.}(2017)\citenamefont {Hao},
  \citenamefont {Zheng}, \citenamefont {Zhang},\ and\ \citenamefont
  {Shen}}]{93-hao2017topological}%
  \BibitemOpen
  \bibfield  {author} {\bibinfo {author} {\bibfnamefont {N.}~\bibnamefont
  {Hao}}, \bibinfo {author} {\bibfnamefont {F.}~\bibnamefont {Zheng}}, \bibinfo
  {author} {\bibfnamefont {P.}~\bibnamefont {Zhang}},\ and\ \bibinfo {author}
  {\bibfnamefont {S.-Q.}\ \bibnamefont {Shen}},\ }\bibfield  {title} {\bibinfo
  {title} {Topological crystalline antiferromagnetic state in tetragonal fes},\
  }\href@noop {} {\bibfield  {journal} {\bibinfo  {journal} {Phys. Rev. B}\
  }\textbf {\bibinfo {volume} {96}},\ \bibinfo {pages} {165102} (\bibinfo
  {year} {2017})}\BibitemShut {NoStop}%
\bibitem [{\citenamefont {Wu}\ \emph {et~al.}(2016{\natexlab{b}})\citenamefont
  {Wu}, \citenamefont {Liang}, \citenamefont {Fan},\ and\ \citenamefont
  {Hu}}]{94-wu2016nematic}%
  \BibitemOpen
  \bibfield  {author} {\bibinfo {author} {\bibfnamefont {X.}~\bibnamefont
  {Wu}}, \bibinfo {author} {\bibfnamefont {Y.}~\bibnamefont {Liang}}, \bibinfo
  {author} {\bibfnamefont {H.}~\bibnamefont {Fan}},\ and\ \bibinfo {author}
  {\bibfnamefont {J.}~\bibnamefont {Hu}},\ }\bibfield  {title} {\bibinfo
  {title} {Nematic orders and nematicity-driven topological phase transition in
  {F}e{S}e},\ }\href@noop {} {\bibfield  {journal} {\bibinfo  {journal}
  {arXiv:1603.02055}\ } (\bibinfo {year} {2016}{\natexlab{b}})}\BibitemShut
  {NoStop}%
\bibitem [{\citenamefont {Tan}\ \emph {et~al.}(2016)\citenamefont {Tan},
  \citenamefont {Fang}, \citenamefont {Xie}, \citenamefont {Feng},
  \citenamefont {Wen}, \citenamefont {Song}, \citenamefont {Chen},
  \citenamefont {Zhang}, \citenamefont {Zhang}, \citenamefont {Luo} \emph
  {et~al.}}]{95-tan2016observation}%
  \BibitemOpen
  \bibfield  {author} {\bibinfo {author} {\bibfnamefont {S.}~\bibnamefont
  {Tan}}, \bibinfo {author} {\bibfnamefont {Y.}~\bibnamefont {Fang}}, \bibinfo
  {author} {\bibfnamefont {D.}~\bibnamefont {Xie}}, \bibinfo {author}
  {\bibfnamefont {W.}~\bibnamefont {Feng}}, \bibinfo {author} {\bibfnamefont
  {C.}~\bibnamefont {Wen}}, \bibinfo {author} {\bibfnamefont {Q.}~\bibnamefont
  {Song}}, \bibinfo {author} {\bibfnamefont {Q.}~\bibnamefont {Chen}}, \bibinfo
  {author} {\bibfnamefont {W.}~\bibnamefont {Zhang}}, \bibinfo {author}
  {\bibfnamefont {Y.}~\bibnamefont {Zhang}}, \bibinfo {author} {\bibfnamefont
  {L.}~\bibnamefont {Luo}}, \emph {et~al.},\ }\bibfield  {title} {\bibinfo
  {title} {Observation of {D}irac cone band dispersions in {F}e{S}e thin films
  by photoemission spectroscopy},\ }\href@noop {} {\bibfield  {journal}
  {\bibinfo  {journal} {Phys. Rev. B}\ }\textbf {\bibinfo {volume} {93}},\
  \bibinfo {pages} {104513} (\bibinfo {year} {2016})}\BibitemShut {NoStop}%
\bibitem [{\citenamefont {Watson}\ \emph {et~al.}(2016)\citenamefont {Watson},
  \citenamefont {Kim}, \citenamefont {Rhodes}, \citenamefont {Eschrig},
  \citenamefont {Hoesch}, \citenamefont {Haghighirad},\ and\ \citenamefont
  {Coldea}}]{96-watson2016evidence}%
  \BibitemOpen
  \bibfield  {author} {\bibinfo {author} {\bibfnamefont {M.}~\bibnamefont
  {Watson}}, \bibinfo {author} {\bibfnamefont {T.}~\bibnamefont {Kim}},
  \bibinfo {author} {\bibfnamefont {L.}~\bibnamefont {Rhodes}}, \bibinfo
  {author} {\bibfnamefont {M.}~\bibnamefont {Eschrig}}, \bibinfo {author}
  {\bibfnamefont {M.}~\bibnamefont {Hoesch}}, \bibinfo {author} {\bibfnamefont
  {A.}~\bibnamefont {Haghighirad}},\ and\ \bibinfo {author} {\bibfnamefont
  {A.}~\bibnamefont {Coldea}},\ }\bibfield  {title} {\bibinfo {title} {Evidence
  for unidirectional nematic bond ordering in {F}e{S}e},\ }\href@noop {}
  {\bibfield  {journal} {\bibinfo  {journal} {Phys. Rev. B}\ }\textbf {\bibinfo
  {volume} {94}},\ \bibinfo {pages} {201107} (\bibinfo {year}
  {2016})}\BibitemShut {NoStop}%
\bibitem [{\citenamefont {Phan}\ \emph {et~al.}(2017)\citenamefont {Phan},
  \citenamefont {Nakayama}, \citenamefont {Sugawara}, \citenamefont {Sato},
  \citenamefont {Urata}, \citenamefont {Tanabe}, \citenamefont {Tanigaki},
  \citenamefont {Nabeshima}, \citenamefont {Imai}, \citenamefont {Maeda} \emph
  {et~al.}}]{97-phan2017effects}%
  \BibitemOpen
  \bibfield  {author} {\bibinfo {author} {\bibfnamefont {G.}~\bibnamefont
  {Phan}}, \bibinfo {author} {\bibfnamefont {K.}~\bibnamefont {Nakayama}},
  \bibinfo {author} {\bibfnamefont {K.}~\bibnamefont {Sugawara}}, \bibinfo
  {author} {\bibfnamefont {T.}~\bibnamefont {Sato}}, \bibinfo {author}
  {\bibfnamefont {T.}~\bibnamefont {Urata}}, \bibinfo {author} {\bibfnamefont
  {Y.}~\bibnamefont {Tanabe}}, \bibinfo {author} {\bibfnamefont
  {K.}~\bibnamefont {Tanigaki}}, \bibinfo {author} {\bibfnamefont
  {F.}~\bibnamefont {Nabeshima}}, \bibinfo {author} {\bibfnamefont
  {Y.}~\bibnamefont {Imai}}, \bibinfo {author} {\bibfnamefont {A.}~\bibnamefont
  {Maeda}}, \emph {et~al.},\ }\bibfield  {title} {\bibinfo {title} {Effects of
  strain on the electronic structure, superconductivity, and nematicity in
  {F}e{S}e studied by angle-resolved photoemission spectroscopy},\ }\href@noop
  {} {\bibfield  {journal} {\bibinfo  {journal} {Phys. Rev. B}\ }\textbf
  {\bibinfo {volume} {95}},\ \bibinfo {pages} {224507} (\bibinfo {year}
  {2017})}\BibitemShut {NoStop}%
\bibitem [{\citenamefont {Zhang}\ \emph {et~al.}(2014)\citenamefont {Zhang},
  \citenamefont {Richard}, \citenamefont {Xu}, \citenamefont {Xu},
  \citenamefont {Ma}, \citenamefont {Qian}, \citenamefont {Fedorov},
  \citenamefont {Denlinger}, \citenamefont {Gu},\ and\ \citenamefont
  {Ding}}]{98-zhang2014observation}%
  \BibitemOpen
  \bibfield  {author} {\bibinfo {author} {\bibfnamefont {P.}~\bibnamefont
  {Zhang}}, \bibinfo {author} {\bibfnamefont {P.}~\bibnamefont {Richard}},
  \bibinfo {author} {\bibfnamefont {N.}~\bibnamefont {Xu}}, \bibinfo {author}
  {\bibfnamefont {Y.-M.}\ \bibnamefont {Xu}}, \bibinfo {author} {\bibfnamefont
  {J.}~\bibnamefont {Ma}}, \bibinfo {author} {\bibfnamefont {T.}~\bibnamefont
  {Qian}}, \bibinfo {author} {\bibfnamefont {A.}~\bibnamefont {Fedorov}},
  \bibinfo {author} {\bibfnamefont {J.}~\bibnamefont {Denlinger}}, \bibinfo
  {author} {\bibfnamefont {G.}~\bibnamefont {Gu}},\ and\ \bibinfo {author}
  {\bibfnamefont {H.}~\bibnamefont {Ding}},\ }\bibfield  {title} {\bibinfo
  {title} {Observation of an electron band above the {F}ermi level in
  {F}e{T}e$_{0.55}${S}e$_{0.45}$ from in-situ surface doping},\ }\href@noop {}
  {\bibfield  {journal} {\bibinfo  {journal} {Appl. Phys. Lett.}\ }\textbf
  {\bibinfo {volume} {105}},\ \bibinfo {pages} {172601} (\bibinfo {year}
  {2014})}\BibitemShut {NoStop}%
\bibitem [{\citenamefont {Yin}\ \emph {et~al.}(2015)\citenamefont {Yin},
  \citenamefont {Wu}, \citenamefont {Wang}, \citenamefont {Ye}, \citenamefont
  {Gong}, \citenamefont {Hou}, \citenamefont {Shan}, \citenamefont {Li},
  \citenamefont {Liang}, \citenamefont {Wu} \emph
  {et~al.}}]{99-yin2015observation}%
  \BibitemOpen
  \bibfield  {author} {\bibinfo {author} {\bibfnamefont {J.}~\bibnamefont
  {Yin}}, \bibinfo {author} {\bibfnamefont {Z.}~\bibnamefont {Wu}}, \bibinfo
  {author} {\bibfnamefont {J.}~\bibnamefont {Wang}}, \bibinfo {author}
  {\bibfnamefont {Z.}~\bibnamefont {Ye}}, \bibinfo {author} {\bibfnamefont
  {J.}~\bibnamefont {Gong}}, \bibinfo {author} {\bibfnamefont {X.}~\bibnamefont
  {Hou}}, \bibinfo {author} {\bibfnamefont {L.}~\bibnamefont {Shan}}, \bibinfo
  {author} {\bibfnamefont {A.}~\bibnamefont {Li}}, \bibinfo {author}
  {\bibfnamefont {X.}~\bibnamefont {Liang}}, \bibinfo {author} {\bibfnamefont
  {X.}~\bibnamefont {Wu}}, \emph {et~al.},\ }\bibfield  {title} {\bibinfo
  {title} {Observation of a robust zero-energy bound state in iron-based
  superconductor {F}e({T}e,{S}e)},\ }\href@noop {} {\bibfield  {journal}
  {\bibinfo  {journal} {Nat. Phys.}\ }\textbf {\bibinfo {volume} {11}},\
  \bibinfo {pages} {543} (\bibinfo {year} {2015})}\BibitemShut {NoStop}%
\bibitem [{\citenamefont {Zhang}\ \emph {et~al.}(2018)\citenamefont {Zhang},
  \citenamefont {Yaji}, \citenamefont {Hashimoto}, \citenamefont {Ota},
  \citenamefont {Kondo}, \citenamefont {Okazaki}, \citenamefont {Wang},
  \citenamefont {Wen}, \citenamefont {Gu}, \citenamefont {Ding} \emph
  {et~al.}}]{100-zhang2018observation}%
  \BibitemOpen
  \bibfield  {author} {\bibinfo {author} {\bibfnamefont {P.}~\bibnamefont
  {Zhang}}, \bibinfo {author} {\bibfnamefont {K.}~\bibnamefont {Yaji}},
  \bibinfo {author} {\bibfnamefont {T.}~\bibnamefont {Hashimoto}}, \bibinfo
  {author} {\bibfnamefont {Y.}~\bibnamefont {Ota}}, \bibinfo {author}
  {\bibfnamefont {T.}~\bibnamefont {Kondo}}, \bibinfo {author} {\bibfnamefont
  {K.}~\bibnamefont {Okazaki}}, \bibinfo {author} {\bibfnamefont
  {Z.}~\bibnamefont {Wang}}, \bibinfo {author} {\bibfnamefont {J.}~\bibnamefont
  {Wen}}, \bibinfo {author} {\bibfnamefont {G.}~\bibnamefont {Gu}}, \bibinfo
  {author} {\bibfnamefont {H.}~\bibnamefont {Ding}}, \emph {et~al.},\
  }\bibfield  {title} {\bibinfo {title} {Observation of topological
  superconductivity on the surface of an iron-based superconductor},\
  }\href@noop {} {\bibfield  {journal} {\bibinfo  {journal} {Science}\ }\textbf
  {\bibinfo {volume} {360}},\ \bibinfo {pages} {182} (\bibinfo {year}
  {2018})}\BibitemShut {NoStop}%
\bibitem [{\citenamefont {Zhang}\ \emph
  {et~al.}(2019{\natexlab{a}})\citenamefont {Zhang}, \citenamefont {Wang},
  \citenamefont {Wu}, \citenamefont {Yaji}, \citenamefont {Ishida},
  \citenamefont {Kohama}, \citenamefont {Dai}, \citenamefont {Sun},
  \citenamefont {Bareille}, \citenamefont {Kuroda} \emph
  {et~al.}}]{101-zhang2019multiple}%
  \BibitemOpen
  \bibfield  {author} {\bibinfo {author} {\bibfnamefont {P.}~\bibnamefont
  {Zhang}}, \bibinfo {author} {\bibfnamefont {Z.}~\bibnamefont {Wang}},
  \bibinfo {author} {\bibfnamefont {X.}~\bibnamefont {Wu}}, \bibinfo {author}
  {\bibfnamefont {K.}~\bibnamefont {Yaji}}, \bibinfo {author} {\bibfnamefont
  {Y.}~\bibnamefont {Ishida}}, \bibinfo {author} {\bibfnamefont
  {Y.}~\bibnamefont {Kohama}}, \bibinfo {author} {\bibfnamefont
  {G.}~\bibnamefont {Dai}}, \bibinfo {author} {\bibfnamefont {Y.}~\bibnamefont
  {Sun}}, \bibinfo {author} {\bibfnamefont {C.}~\bibnamefont {Bareille}},
  \bibinfo {author} {\bibfnamefont {K.}~\bibnamefont {Kuroda}}, \emph
  {et~al.},\ }\bibfield  {title} {\bibinfo {title} {Multiple topological states
  in iron-based superconductors},\ }\href@noop {} {\bibfield  {journal}
  {\bibinfo  {journal} {Nat. Phys.}\ }\textbf {\bibinfo {volume} {15}},\
  \bibinfo {pages} {41} (\bibinfo {year} {2019}{\natexlab{a}})}\BibitemShut
  {NoStop}%
\bibitem [{\citenamefont {Wang}\ \emph {et~al.}(2018)\citenamefont {Wang},
  \citenamefont {Kong}, \citenamefont {Fan}, \citenamefont {Chen},
  \citenamefont {Zhu}, \citenamefont {Liu}, \citenamefont {Cao}, \citenamefont
  {Sun}, \citenamefont {Du}, \citenamefont {Schneeloch} \emph
  {et~al.}}]{102-wang2018evidence}%
  \BibitemOpen
  \bibfield  {author} {\bibinfo {author} {\bibfnamefont {D.}~\bibnamefont
  {Wang}}, \bibinfo {author} {\bibfnamefont {L.}~\bibnamefont {Kong}}, \bibinfo
  {author} {\bibfnamefont {P.}~\bibnamefont {Fan}}, \bibinfo {author}
  {\bibfnamefont {H.}~\bibnamefont {Chen}}, \bibinfo {author} {\bibfnamefont
  {S.}~\bibnamefont {Zhu}}, \bibinfo {author} {\bibfnamefont {W.}~\bibnamefont
  {Liu}}, \bibinfo {author} {\bibfnamefont {L.}~\bibnamefont {Cao}}, \bibinfo
  {author} {\bibfnamefont {Y.}~\bibnamefont {Sun}}, \bibinfo {author}
  {\bibfnamefont {S.}~\bibnamefont {Du}}, \bibinfo {author} {\bibfnamefont
  {J.}~\bibnamefont {Schneeloch}}, \emph {et~al.},\ }\bibfield  {title}
  {\bibinfo {title} {Evidence for {M}ajorana bound states in an iron-based
  superconductor},\ }\href@noop {} {\bibfield  {journal} {\bibinfo  {journal}
  {Science}\ }\textbf {\bibinfo {volume} {362}},\ \bibinfo {pages} {333}
  (\bibinfo {year} {2018})}\BibitemShut {NoStop}%
\bibitem [{\citenamefont {Kong}\ and\ \citenamefont
  {Ding}(2019)}]{103-kong2019majorana}%
  \BibitemOpen
  \bibfield  {author} {\bibinfo {author} {\bibfnamefont {L.}~\bibnamefont
  {Kong}}\ and\ \bibinfo {author} {\bibfnamefont {H.}~\bibnamefont {Ding}},\
  }\bibfield  {title} {\bibinfo {title} {Majorana gets an iron twist},\
  }\href@noop {} {\bibfield  {journal} {\bibinfo  {journal} {Natl. Sci. Rev.}\
  }\textbf {\bibinfo {volume} {6}},\ \bibinfo {pages} {196} (\bibinfo {year}
  {2019})}\BibitemShut {NoStop}%
\bibitem [{\citenamefont {Kong}\ \emph {et~al.}(2019)\citenamefont {Kong},
  \citenamefont {Zhu}, \citenamefont {Papaj}, \citenamefont {Chen},
  \citenamefont {Cao}, \citenamefont {Isobe}, \citenamefont {Xing},
  \citenamefont {Liu}, \citenamefont {Wang}, \citenamefont {Fan} \emph
  {et~al.}}]{104-kong2019half}%
  \BibitemOpen
  \bibfield  {author} {\bibinfo {author} {\bibfnamefont {L.}~\bibnamefont
  {Kong}}, \bibinfo {author} {\bibfnamefont {S.}~\bibnamefont {Zhu}}, \bibinfo
  {author} {\bibfnamefont {M.}~\bibnamefont {Papaj}}, \bibinfo {author}
  {\bibfnamefont {H.}~\bibnamefont {Chen}}, \bibinfo {author} {\bibfnamefont
  {L.}~\bibnamefont {Cao}}, \bibinfo {author} {\bibfnamefont {H.}~\bibnamefont
  {Isobe}}, \bibinfo {author} {\bibfnamefont {Y.}~\bibnamefont {Xing}},
  \bibinfo {author} {\bibfnamefont {W.}~\bibnamefont {Liu}}, \bibinfo {author}
  {\bibfnamefont {D.}~\bibnamefont {Wang}}, \bibinfo {author} {\bibfnamefont
  {P.}~\bibnamefont {Fan}}, \emph {et~al.},\ }\bibfield  {title} {\bibinfo
  {title} {Half-integer level shift of vortex bound states in an iron-based
  superconductor},\ }\href@noop {} {\bibfield  {journal} {\bibinfo  {journal}
  {Nat. Phys.}\ }\textbf {\bibinfo {volume} {15}},\ \bibinfo {pages} {1181}
  (\bibinfo {year} {2019})}\BibitemShut {NoStop}%
\bibitem [{\citenamefont {Zhu}\ \emph {et~al.}(2020)\citenamefont {Zhu},
  \citenamefont {Kong}, \citenamefont {Cao}, \citenamefont {Chen},
  \citenamefont {Papaj}, \citenamefont {Du}, \citenamefont {Xing},
  \citenamefont {Liu}, \citenamefont {Wang}, \citenamefont {Shen} \emph
  {et~al.}}]{105-zhu2020nearly}%
  \BibitemOpen
  \bibfield  {author} {\bibinfo {author} {\bibfnamefont {S.}~\bibnamefont
  {Zhu}}, \bibinfo {author} {\bibfnamefont {L.}~\bibnamefont {Kong}}, \bibinfo
  {author} {\bibfnamefont {L.}~\bibnamefont {Cao}}, \bibinfo {author}
  {\bibfnamefont {H.}~\bibnamefont {Chen}}, \bibinfo {author} {\bibfnamefont
  {M.}~\bibnamefont {Papaj}}, \bibinfo {author} {\bibfnamefont
  {S.}~\bibnamefont {Du}}, \bibinfo {author} {\bibfnamefont {Y.}~\bibnamefont
  {Xing}}, \bibinfo {author} {\bibfnamefont {W.}~\bibnamefont {Liu}}, \bibinfo
  {author} {\bibfnamefont {D.}~\bibnamefont {Wang}}, \bibinfo {author}
  {\bibfnamefont {C.}~\bibnamefont {Shen}}, \emph {et~al.},\ }\bibfield
  {title} {\bibinfo {title} {Nearly quantized conductance plateau of vortex
  zero mode in an iron-based superconductor},\ }\href@noop {} {\bibfield
  {journal} {\bibinfo  {journal} {Science}\ }\textbf {\bibinfo {volume}
  {367}},\ \bibinfo {pages} {189} (\bibinfo {year} {2020})}\BibitemShut
  {NoStop}%
\bibitem [{\citenamefont {Liu}\ \emph {et~al.}(2019{\natexlab{c}})\citenamefont
  {Liu}, \citenamefont {Cao}, \citenamefont {Zhu}, \citenamefont {Kong},
  \citenamefont {Wang}, \citenamefont {Papaj}, \citenamefont {Zhang},
  \citenamefont {Liu}, \citenamefont {Chen}, \citenamefont {Li} \emph
  {et~al.}}]{106-liu2019new}%
  \BibitemOpen
  \bibfield  {author} {\bibinfo {author} {\bibfnamefont {W.}~\bibnamefont
  {Liu}}, \bibinfo {author} {\bibfnamefont {L.}~\bibnamefont {Cao}}, \bibinfo
  {author} {\bibfnamefont {S.}~\bibnamefont {Zhu}}, \bibinfo {author}
  {\bibfnamefont {L.}~\bibnamefont {Kong}}, \bibinfo {author} {\bibfnamefont
  {G.}~\bibnamefont {Wang}}, \bibinfo {author} {\bibfnamefont {M.}~\bibnamefont
  {Papaj}}, \bibinfo {author} {\bibfnamefont {P.}~\bibnamefont {Zhang}},
  \bibinfo {author} {\bibfnamefont {Y.}~\bibnamefont {Liu}}, \bibinfo {author}
  {\bibfnamefont {H.}~\bibnamefont {Chen}}, \bibinfo {author} {\bibfnamefont
  {G.}~\bibnamefont {Li}}, \emph {et~al.},\ }\bibfield  {title} {\bibinfo
  {title} {A new {M}ajorana platform in an {F}e-{A}s bilayer superconductor},\
  }\href@noop {} {\bibfield  {journal} {\bibinfo  {journal} {arXiv:1907.00904}\
  } (\bibinfo {year} {2019}{\natexlab{c}})}\BibitemShut {NoStop}%
\bibitem [{\citenamefont {Bahramy}\ \emph {et~al.}(2018)\citenamefont
  {Bahramy}, \citenamefont {Clark}, \citenamefont {Yang}, \citenamefont {Feng},
  \citenamefont {Bawden}, \citenamefont {Riley}, \citenamefont {Markovi{\'c}},
  \citenamefont {Mazzola}, \citenamefont {Sunko}, \citenamefont {Biswas} \emph
  {et~al.}}]{107-bahramy2018ubiquitous}%
  \BibitemOpen
  \bibfield  {author} {\bibinfo {author} {\bibfnamefont {M.}~\bibnamefont
  {Bahramy}}, \bibinfo {author} {\bibfnamefont {O.}~\bibnamefont {Clark}},
  \bibinfo {author} {\bibfnamefont {B.-J.}\ \bibnamefont {Yang}}, \bibinfo
  {author} {\bibfnamefont {J.}~\bibnamefont {Feng}}, \bibinfo {author}
  {\bibfnamefont {L.}~\bibnamefont {Bawden}}, \bibinfo {author} {\bibfnamefont
  {J.}~\bibnamefont {Riley}}, \bibinfo {author} {\bibfnamefont
  {I.}~\bibnamefont {Markovi{\'c}}}, \bibinfo {author} {\bibfnamefont
  {F.}~\bibnamefont {Mazzola}}, \bibinfo {author} {\bibfnamefont
  {V.}~\bibnamefont {Sunko}}, \bibinfo {author} {\bibfnamefont
  {D.}~\bibnamefont {Biswas}}, \emph {et~al.},\ }\bibfield  {title} {\bibinfo
  {title} {Ubiquitous formation of bulk {D}irac cones and topological surface
  states from a single orbital manifold in transition-metal dichalcogenides},\
  }\href@noop {} {\bibfield  {journal} {\bibinfo  {journal} {Nat. Mater.}\
  }\textbf {\bibinfo {volume} {17}},\ \bibinfo {pages} {21} (\bibinfo {year}
  {2018})}\BibitemShut {NoStop}%
\bibitem [{\citenamefont {Clark}\ \emph {et~al.}(2018)\citenamefont {Clark},
  \citenamefont {Neat}, \citenamefont {Okawa}, \citenamefont {Bawden},
  \citenamefont {Markovi{\'c}}, \citenamefont {Mazzola}, \citenamefont {Feng},
  \citenamefont {Sunko}, \citenamefont {Riley}, \citenamefont {Meevasana} \emph
  {et~al.}}]{108-clark2018fermiology}%
  \BibitemOpen
  \bibfield  {author} {\bibinfo {author} {\bibfnamefont {O.}~\bibnamefont
  {Clark}}, \bibinfo {author} {\bibfnamefont {M.}~\bibnamefont {Neat}},
  \bibinfo {author} {\bibfnamefont {K.}~\bibnamefont {Okawa}}, \bibinfo
  {author} {\bibfnamefont {L.}~\bibnamefont {Bawden}}, \bibinfo {author}
  {\bibfnamefont {I.}~\bibnamefont {Markovi{\'c}}}, \bibinfo {author}
  {\bibfnamefont {F.}~\bibnamefont {Mazzola}}, \bibinfo {author} {\bibfnamefont
  {J.}~\bibnamefont {Feng}}, \bibinfo {author} {\bibfnamefont {V.}~\bibnamefont
  {Sunko}}, \bibinfo {author} {\bibfnamefont {J.}~\bibnamefont {Riley}},
  \bibinfo {author} {\bibfnamefont {W.}~\bibnamefont {Meevasana}}, \emph
  {et~al.},\ }\bibfield  {title} {\bibinfo {title} {Fermiology and
  superconductivity of topological surface states in {P}d{T}e$_{2}$},\
  }\href@noop {} {\bibfield  {journal} {\bibinfo  {journal} {Phys. Rev. Lett.}\
  }\textbf {\bibinfo {volume} {120}},\ \bibinfo {pages} {156401} (\bibinfo
  {year} {2018})}\BibitemShut {NoStop}%
\bibitem [{\citenamefont {Mukherjee}\ \emph {et~al.}(2019)\citenamefont
  {Mukherjee}, \citenamefont {Jung}, \citenamefont {Weber}, \citenamefont {Xu},
  \citenamefont {Qian}, \citenamefont {Xu}, \citenamefont {Biswas},
  \citenamefont {Kim}, \citenamefont {Chapon}, \citenamefont {Watson} \emph
  {et~al.}}]{109-mukherjee2019fermi}%
  \BibitemOpen
  \bibfield  {author} {\bibinfo {author} {\bibfnamefont {S.}~\bibnamefont
  {Mukherjee}}, \bibinfo {author} {\bibfnamefont {S.~W.}\ \bibnamefont {Jung}},
  \bibinfo {author} {\bibfnamefont {S.~F.}\ \bibnamefont {Weber}}, \bibinfo
  {author} {\bibfnamefont {C.}~\bibnamefont {Xu}}, \bibinfo {author}
  {\bibfnamefont {D.}~\bibnamefont {Qian}}, \bibinfo {author} {\bibfnamefont
  {X.}~\bibnamefont {Xu}}, \bibinfo {author} {\bibfnamefont {P.~K.}\
  \bibnamefont {Biswas}}, \bibinfo {author} {\bibfnamefont {T.~K.}\
  \bibnamefont {Kim}}, \bibinfo {author} {\bibfnamefont {L.~C.}\ \bibnamefont
  {Chapon}}, \bibinfo {author} {\bibfnamefont {M.~D.}\ \bibnamefont {Watson}},
  \emph {et~al.},\ }\bibfield  {title} {\bibinfo {title} {Fermi-crossing
  type-{II} {D}irac fermions and topological surface states in
  {N}i{T}e$_{2}$},\ }\href@noop {} {\bibfield  {journal} {\bibinfo  {journal}
  {arXiv:1912.08535}\ } (\bibinfo {year} {2019})}\BibitemShut {NoStop}%
\bibitem [{\citenamefont {Wang}\ \emph
  {et~al.}(2012{\natexlab{b}})\citenamefont {Wang}, \citenamefont {Richard},
  \citenamefont {Huang}, \citenamefont {Miao}, \citenamefont {Cevey},
  \citenamefont {Xu}, \citenamefont {Sun}, \citenamefont {Qian}, \citenamefont
  {Xu}, \citenamefont {Shi} \emph {et~al.}}]{110-wang2012orbital}%
  \BibitemOpen
  \bibfield  {author} {\bibinfo {author} {\bibfnamefont {X.-P.}\ \bibnamefont
  {Wang}}, \bibinfo {author} {\bibfnamefont {P.}~\bibnamefont {Richard}},
  \bibinfo {author} {\bibfnamefont {Y.-B.}\ \bibnamefont {Huang}}, \bibinfo
  {author} {\bibfnamefont {H.}~\bibnamefont {Miao}}, \bibinfo {author}
  {\bibfnamefont {L.}~\bibnamefont {Cevey}}, \bibinfo {author} {\bibfnamefont
  {N.}~\bibnamefont {Xu}}, \bibinfo {author} {\bibfnamefont {Y.-J.}\
  \bibnamefont {Sun}}, \bibinfo {author} {\bibfnamefont {T.}~\bibnamefont
  {Qian}}, \bibinfo {author} {\bibfnamefont {Y.-M.}\ \bibnamefont {Xu}},
  \bibinfo {author} {\bibfnamefont {M.}~\bibnamefont {Shi}}, \emph {et~al.},\
  }\bibfield  {title} {\bibinfo {title} {Orbital characters determined from
  {F}ermi surface intensity patterns using angle-resolved photoemission
  spectroscopy},\ }\href@noop {} {\bibfield  {journal} {\bibinfo  {journal}
  {Phys. Rev. B}\ }\textbf {\bibinfo {volume} {85}},\ \bibinfo {pages} {214518}
  (\bibinfo {year} {2012}{\natexlab{b}})}\BibitemShut {NoStop}%
\bibitem [{\citenamefont {Lv}\ \emph {et~al.}(2019)\citenamefont {Lv},
  \citenamefont {Qian},\ and\ \citenamefont {Ding}}]{111-lv2019angle}%
  \BibitemOpen
  \bibfield  {author} {\bibinfo {author} {\bibfnamefont {B.}~\bibnamefont
  {Lv}}, \bibinfo {author} {\bibfnamefont {T.}~\bibnamefont {Qian}},\ and\
  \bibinfo {author} {\bibfnamefont {H.}~\bibnamefont {Ding}},\ }\bibfield
  {title} {\bibinfo {title} {Angle-resolved photoemission spectroscopy and its
  application to topological materials},\ }\href@noop {} {\bibfield  {journal}
  {\bibinfo  {journal} {Nat. Rev. Phys.}\ }\textbf {\bibinfo {volume} {1}},\
  \bibinfo {pages} {609} (\bibinfo {year} {2019})}\BibitemShut {NoStop}%
\bibitem [{\citenamefont {H{\"u}fner}(2013)}]{112-hufner2013photoelectron}%
  \BibitemOpen
  \bibfield  {author} {\bibinfo {author} {\bibfnamefont {S.}~\bibnamefont
  {H{\"u}fner}},\ }\href@noop {} {\emph {\bibinfo {title} {Photoelectron
  spectroscopy: principles and applications}}}\ (\bibinfo  {publisher}
  {Springer Science \& Business Media},\ \bibinfo {year} {2013})\BibitemShut
  {NoStop}%
\bibitem [{\citenamefont {Damascelli}(2004)}]{113-damascelli2004probing}%
  \BibitemOpen
  \bibfield  {author} {\bibinfo {author} {\bibfnamefont {A.}~\bibnamefont
  {Damascelli}},\ }\bibfield  {title} {\bibinfo {title} {Probing the electronic
  structure of complex systems by {ARPES}},\ }\href@noop {} {\bibfield
  {journal} {\bibinfo  {journal} {Physica Scripta}\ }\textbf {\bibinfo {volume}
  {2004}},\ \bibinfo {pages} {61} (\bibinfo {year} {2004})}\BibitemShut
  {NoStop}%
\bibitem [{\citenamefont {Chen}\ \emph
  {et~al.}(2019{\natexlab{a}})\citenamefont {Chen}, \citenamefont {Tang},
  \citenamefont {Chen}, \citenamefont {Gu}, \citenamefont {Yang}, \citenamefont
  {Du}, \citenamefont {Zhu}, \citenamefont {Wang}, \citenamefont {Wang},\ and\
  \citenamefont {Wen}}]{114-chen2019direct}%
  \BibitemOpen
  \bibfield  {author} {\bibinfo {author} {\bibfnamefont {M.}~\bibnamefont
  {Chen}}, \bibinfo {author} {\bibfnamefont {Q.}~\bibnamefont {Tang}}, \bibinfo
  {author} {\bibfnamefont {X.}~\bibnamefont {Chen}}, \bibinfo {author}
  {\bibfnamefont {Q.}~\bibnamefont {Gu}}, \bibinfo {author} {\bibfnamefont
  {H.}~\bibnamefont {Yang}}, \bibinfo {author} {\bibfnamefont {Z.}~\bibnamefont
  {Du}}, \bibinfo {author} {\bibfnamefont {X.}~\bibnamefont {Zhu}}, \bibinfo
  {author} {\bibfnamefont {E.}~\bibnamefont {Wang}}, \bibinfo {author}
  {\bibfnamefont {Q.-H.}\ \bibnamefont {Wang}},\ and\ \bibinfo {author}
  {\bibfnamefont {H.-H.}\ \bibnamefont {Wen}},\ }\bibfield  {title} {\bibinfo
  {title} {Direct visualization of sign-reversal s$\pm$ superconducting gaps in
  {F}e{T}e$_{0.55}${S}e$_{0.45}$},\ }\href@noop {} {\bibfield  {journal}
  {\bibinfo  {journal} {Phys. Rev. B}\ }\textbf {\bibinfo {volume} {99}},\
  \bibinfo {pages} {014507} (\bibinfo {year} {2019}{\natexlab{a}})}\BibitemShut
  {NoStop}%
\bibitem [{\citenamefont {Miao}\ \emph {et~al.}(2012)\citenamefont {Miao},
  \citenamefont {Richard}, \citenamefont {Tanaka}, \citenamefont {Nakayama},
  \citenamefont {Qian}, \citenamefont {Umezawa}, \citenamefont {Sato},
  \citenamefont {Xu}, \citenamefont {Shi}, \citenamefont {Xu} \emph
  {et~al.}}]{115-miao2012isotropic}%
  \BibitemOpen
  \bibfield  {author} {\bibinfo {author} {\bibfnamefont {H.}~\bibnamefont
  {Miao}}, \bibinfo {author} {\bibfnamefont {P.}~\bibnamefont {Richard}},
  \bibinfo {author} {\bibfnamefont {Y.}~\bibnamefont {Tanaka}}, \bibinfo
  {author} {\bibfnamefont {K.}~\bibnamefont {Nakayama}}, \bibinfo {author}
  {\bibfnamefont {T.}~\bibnamefont {Qian}}, \bibinfo {author} {\bibfnamefont
  {K.}~\bibnamefont {Umezawa}}, \bibinfo {author} {\bibfnamefont
  {T.}~\bibnamefont {Sato}}, \bibinfo {author} {\bibfnamefont {Y.-M.}\
  \bibnamefont {Xu}}, \bibinfo {author} {\bibfnamefont {Y.}~\bibnamefont
  {Shi}}, \bibinfo {author} {\bibfnamefont {N.}~\bibnamefont {Xu}}, \emph
  {et~al.},\ }\bibfield  {title} {\bibinfo {title} {Isotropic superconducting
  gaps with enhanced pairing on electron {F}ermi surfaces in
  {F}e{T}e$_{0.55}${S}e$_{0.45}$},\ }\href@noop {} {\bibfield  {journal}
  {\bibinfo  {journal} {Phys. Rev. B}\ }\textbf {\bibinfo {volume} {85}},\
  \bibinfo {pages} {094506} (\bibinfo {year} {2012})}\BibitemShut {NoStop}%
\bibitem [{\citenamefont {Lipscombe}\ \emph {et~al.}(2011)\citenamefont
  {Lipscombe}, \citenamefont {Chen}, \citenamefont {Fang}, \citenamefont
  {Perring}, \citenamefont {Abernathy}, \citenamefont {Christianson},
  \citenamefont {Egami}, \citenamefont {Wang}, \citenamefont {Hu},\ and\
  \citenamefont {Dai}}]{116-lipscombe2011spin}%
  \BibitemOpen
  \bibfield  {author} {\bibinfo {author} {\bibfnamefont {O.}~\bibnamefont
  {Lipscombe}}, \bibinfo {author} {\bibfnamefont {G.}~\bibnamefont {Chen}},
  \bibinfo {author} {\bibfnamefont {C.}~\bibnamefont {Fang}}, \bibinfo {author}
  {\bibfnamefont {T.}~\bibnamefont {Perring}}, \bibinfo {author} {\bibfnamefont
  {D.}~\bibnamefont {Abernathy}}, \bibinfo {author} {\bibfnamefont
  {A.}~\bibnamefont {Christianson}}, \bibinfo {author} {\bibfnamefont
  {T.}~\bibnamefont {Egami}}, \bibinfo {author} {\bibfnamefont
  {N.}~\bibnamefont {Wang}}, \bibinfo {author} {\bibfnamefont {J.}~\bibnamefont
  {Hu}},\ and\ \bibinfo {author} {\bibfnamefont {P.}~\bibnamefont {Dai}},\
  }\bibfield  {title} {\bibinfo {title} {Spin waves in the ($\pi$, 0)
  magnetically ordered iron chalcogenide {F}e$_{1.05}${T}e},\ }\href@noop {}
  {\bibfield  {journal} {\bibinfo  {journal} {Phys. Rev. Lett.}\ }\textbf
  {\bibinfo {volume} {106}},\ \bibinfo {pages} {057004} (\bibinfo {year}
  {2011})}\BibitemShut {NoStop}%
\bibitem [{\citenamefont {Homes}\ \emph {et~al.}(2010)\citenamefont {Homes},
  \citenamefont {Akrap}, \citenamefont {Wen}, \citenamefont {Xu}, \citenamefont
  {Lin}, \citenamefont {Li},\ and\ \citenamefont
  {Gu}}]{117-homes2010electronic}%
  \BibitemOpen
  \bibfield  {author} {\bibinfo {author} {\bibfnamefont {C.}~\bibnamefont
  {Homes}}, \bibinfo {author} {\bibfnamefont {A.}~\bibnamefont {Akrap}},
  \bibinfo {author} {\bibfnamefont {J.}~\bibnamefont {Wen}}, \bibinfo {author}
  {\bibfnamefont {Z.}~\bibnamefont {Xu}}, \bibinfo {author} {\bibfnamefont
  {Z.}~\bibnamefont {Lin}}, \bibinfo {author} {\bibfnamefont {Q.}~\bibnamefont
  {Li}},\ and\ \bibinfo {author} {\bibfnamefont {G.}~\bibnamefont {Gu}},\
  }\bibfield  {title} {\bibinfo {title} {Electronic correlations and unusual
  superconducting response in the optical properties of the iron chalcogenide
  {F}e{T}e$_{0.55}${S}e$_{0.45}$},\ }\href@noop {} {\bibfield  {journal}
  {\bibinfo  {journal} {Phys. Rev. B}\ }\textbf {\bibinfo {volume} {81}},\
  \bibinfo {pages} {180508} (\bibinfo {year} {2010})}\BibitemShut {NoStop}%
\bibitem [{\citenamefont {Escudero}\ and\ \citenamefont
  {L{\'o}pez-Romero}(2015)}]{118-escudero2015energy}%
  \BibitemOpen
  \bibfield  {author} {\bibinfo {author} {\bibfnamefont {R.}~\bibnamefont
  {Escudero}}\ and\ \bibinfo {author} {\bibfnamefont {R.~E.}\ \bibnamefont
  {L{\'o}pez-Romero}},\ }\bibfield  {title} {\bibinfo {title} {The energy gap
  of the compound {F}e{S}e$_{0.5}${T}e$_{0.5}$ determined by specific heat and
  point contact spectroscopy},\ }\href@noop {} {\bibfield  {journal} {\bibinfo
  {journal} {Solid State Commun.}\ }\textbf {\bibinfo {volume} {220}},\
  \bibinfo {pages} {21} (\bibinfo {year} {2015})}\BibitemShut {NoStop}%
\bibitem [{\citenamefont {Wu}\ \emph {et~al.}(2020{\natexlab{a}})\citenamefont
  {Wu}, \citenamefont {Almoalem}, \citenamefont {Feldman}, \citenamefont {Lee},
  \citenamefont {Kanigel},\ and\ \citenamefont
  {Blumberg}}]{119-wu2020superconductivity}%
  \BibitemOpen
  \bibfield  {author} {\bibinfo {author} {\bibfnamefont {S.-F.}\ \bibnamefont
  {Wu}}, \bibinfo {author} {\bibfnamefont {A.}~\bibnamefont {Almoalem}},
  \bibinfo {author} {\bibfnamefont {I.}~\bibnamefont {Feldman}}, \bibinfo
  {author} {\bibfnamefont {A.}~\bibnamefont {Lee}}, \bibinfo {author}
  {\bibfnamefont {A.}~\bibnamefont {Kanigel}},\ and\ \bibinfo {author}
  {\bibfnamefont {G.}~\bibnamefont {Blumberg}},\ }\bibfield  {title} {\bibinfo
  {title} {Superconductivity and phonon self-energy effects in
  {F}e$_{1+y}${T}e$_{0.6}${S}e$_{0.4}$},\ }\href@noop {} {\bibfield  {journal}
  {\bibinfo  {journal} {Phys. Rev. Research}\ }\textbf {\bibinfo {volume}
  {2}},\ \bibinfo {pages} {013373} (\bibinfo {year}
  {2020}{\natexlab{a}})}\BibitemShut {NoStop}%
\bibitem [{\citenamefont {Lee}(2019)}]{120-Lee2019spontan}%
  \BibitemOpen
  \bibfield  {author} {\bibinfo {author} {\bibfnamefont {P.~A.}\ \bibnamefont
  {Lee}},\ }\href@noop {} {\bibinfo {title} {Spontaneous vortex formation and
  {M}ajorana zero mode in iron based superconductor}},\ \bibinfo {howpublished}
  {\url{https://doi.org/10.36471/JCCM_December_2018_03}} (\bibinfo {year}
  {2019})\BibitemShut {NoStop}%
\bibitem [{\citenamefont {Yang}\ and\ \citenamefont
  {Nagaosa}(2014)}]{121-yang2014classification}%
  \BibitemOpen
  \bibfield  {author} {\bibinfo {author} {\bibfnamefont {B.-J.}\ \bibnamefont
  {Yang}}\ and\ \bibinfo {author} {\bibfnamefont {N.}~\bibnamefont {Nagaosa}},\
  }\bibfield  {title} {\bibinfo {title} {Classification of stable
  three-dimensional {D}irac semimetals with nontrivial topology},\ }\href@noop
  {} {\bibfield  {journal} {\bibinfo  {journal} {Nat. Commun.}\ }\textbf
  {\bibinfo {volume} {5}},\ \bibinfo {pages} {4898} (\bibinfo {year}
  {2014})}\BibitemShut {NoStop}%
\bibitem [{\citenamefont {Xu}\ \emph {et~al.}(2015{\natexlab{b}})\citenamefont
  {Xu}, \citenamefont {Neupane}, \citenamefont {Belopolski}, \citenamefont
  {Liu}, \citenamefont {Alidoust}, \citenamefont {Bian}, \citenamefont {Jia},
  \citenamefont {Landolt}, \citenamefont {Slomski}, \citenamefont {Dil} \emph
  {et~al.}}]{122-xu2015unconventional}%
  \BibitemOpen
  \bibfield  {author} {\bibinfo {author} {\bibfnamefont {S.-Y.}\ \bibnamefont
  {Xu}}, \bibinfo {author} {\bibfnamefont {M.}~\bibnamefont {Neupane}},
  \bibinfo {author} {\bibfnamefont {I.}~\bibnamefont {Belopolski}}, \bibinfo
  {author} {\bibfnamefont {C.}~\bibnamefont {Liu}}, \bibinfo {author}
  {\bibfnamefont {N.}~\bibnamefont {Alidoust}}, \bibinfo {author}
  {\bibfnamefont {G.}~\bibnamefont {Bian}}, \bibinfo {author} {\bibfnamefont
  {S.}~\bibnamefont {Jia}}, \bibinfo {author} {\bibfnamefont {G.}~\bibnamefont
  {Landolt}}, \bibinfo {author} {\bibfnamefont {B.}~\bibnamefont {Slomski}},
  \bibinfo {author} {\bibfnamefont {J.~H.}\ \bibnamefont {Dil}}, \emph
  {et~al.},\ }\bibfield  {title} {\bibinfo {title} {Unconventional
  transformation of spin {D}irac phase across a topological quantum phase
  transition},\ }\href@noop {} {\bibfield  {journal} {\bibinfo  {journal} {Nat.
  Commun.}\ }\textbf {\bibinfo {volume} {6}},\ \bibinfo {pages} {5297}
  (\bibinfo {year} {2015}{\natexlab{b}})}\BibitemShut {NoStop}%
\bibitem [{\citenamefont {Jozwiak}\ \emph {et~al.}(2016)\citenamefont
  {Jozwiak}, \citenamefont {Sobota}, \citenamefont {Gotlieb}, \citenamefont
  {Kemper}, \citenamefont {Rotundu}, \citenamefont {Birgeneau}, \citenamefont
  {Hussain}, \citenamefont {Lee}, \citenamefont {Shen},\ and\ \citenamefont
  {Lanzara}}]{123-jozwiak2016spin}%
  \BibitemOpen
  \bibfield  {author} {\bibinfo {author} {\bibfnamefont {C.}~\bibnamefont
  {Jozwiak}}, \bibinfo {author} {\bibfnamefont {J.~A.}\ \bibnamefont {Sobota}},
  \bibinfo {author} {\bibfnamefont {K.}~\bibnamefont {Gotlieb}}, \bibinfo
  {author} {\bibfnamefont {A.~F.}\ \bibnamefont {Kemper}}, \bibinfo {author}
  {\bibfnamefont {C.~R.}\ \bibnamefont {Rotundu}}, \bibinfo {author}
  {\bibfnamefont {R.~J.}\ \bibnamefont {Birgeneau}}, \bibinfo {author}
  {\bibfnamefont {Z.}~\bibnamefont {Hussain}}, \bibinfo {author} {\bibfnamefont
  {D.-H.}\ \bibnamefont {Lee}}, \bibinfo {author} {\bibfnamefont {Z.-X.}\
  \bibnamefont {Shen}},\ and\ \bibinfo {author} {\bibfnamefont
  {A.}~\bibnamefont {Lanzara}},\ }\bibfield  {title} {\bibinfo {title}
  {Spin-polarized surface resonances accompanying topological surface state
  formation},\ }\href@noop {} {\bibfield  {journal} {\bibinfo  {journal} {Nat.
  Commun.}\ }\textbf {\bibinfo {volume} {7}},\ \bibinfo {pages} {13143}
  (\bibinfo {year} {2016})}\BibitemShut {NoStop}%
\bibitem [{\citenamefont {Neupane}\ \emph {et~al.}(2015)\citenamefont
  {Neupane}, \citenamefont {Xu}, \citenamefont {Alidoust}, \citenamefont
  {Sankar}, \citenamefont {Belopolski}, \citenamefont {Sanchez}, \citenamefont
  {Bian}, \citenamefont {Liu}, \citenamefont {Chang}, \citenamefont {Jeng}
  \emph {et~al.}}]{124-neupane2015surface}%
  \BibitemOpen
  \bibfield  {author} {\bibinfo {author} {\bibfnamefont {M.}~\bibnamefont
  {Neupane}}, \bibinfo {author} {\bibfnamefont {S.-Y.}\ \bibnamefont {Xu}},
  \bibinfo {author} {\bibfnamefont {N.}~\bibnamefont {Alidoust}}, \bibinfo
  {author} {\bibfnamefont {R.}~\bibnamefont {Sankar}}, \bibinfo {author}
  {\bibfnamefont {I.}~\bibnamefont {Belopolski}}, \bibinfo {author}
  {\bibfnamefont {D.~S.}\ \bibnamefont {Sanchez}}, \bibinfo {author}
  {\bibfnamefont {G.}~\bibnamefont {Bian}}, \bibinfo {author} {\bibfnamefont
  {C.}~\bibnamefont {Liu}}, \bibinfo {author} {\bibfnamefont {T.-R.}\
  \bibnamefont {Chang}}, \bibinfo {author} {\bibfnamefont {H.-T.}\ \bibnamefont
  {Jeng}}, \emph {et~al.},\ }\bibfield  {title} {\bibinfo {title} {Surface
  versus bulk {D}irac state tuning in a three-dimensional topological {D}irac
  semimetal},\ }\href@noop {} {\bibfield  {journal} {\bibinfo  {journal} {Phys.
  Rev. B}\ }\textbf {\bibinfo {volume} {91}},\ \bibinfo {pages} {241114}
  (\bibinfo {year} {2015})}\BibitemShut {NoStop}%
\bibitem [{\citenamefont {Abrikosov}(1998)}]{125-abrikosov1998quantum}%
  \BibitemOpen
  \bibfield  {author} {\bibinfo {author} {\bibfnamefont {A.}~\bibnamefont
  {Abrikosov}},\ }\bibfield  {title} {\bibinfo {title} {Quantum
  magnetoresistance},\ }\href@noop {} {\bibfield  {journal} {\bibinfo
  {journal} {Phys. Rev. B}\ }\textbf {\bibinfo {volume} {58}},\ \bibinfo
  {pages} {2788} (\bibinfo {year} {1998})}\BibitemShut {NoStop}%
\bibitem [{\citenamefont {Parish}\ and\ \citenamefont
  {Littlewood}(2003)}]{126-parish2003non}%
  \BibitemOpen
  \bibfield  {author} {\bibinfo {author} {\bibfnamefont {M.}~\bibnamefont
  {Parish}}\ and\ \bibinfo {author} {\bibfnamefont {P.}~\bibnamefont
  {Littlewood}},\ }\bibfield  {title} {\bibinfo {title} {Non-saturating
  magnetoresistance in heavily disordered semiconductors},\ }\href@noop {}
  {\bibfield  {journal} {\bibinfo  {journal} {Nature}\ }\textbf {\bibinfo
  {volume} {426}},\ \bibinfo {pages} {162} (\bibinfo {year}
  {2003})}\BibitemShut {NoStop}%
\bibitem [{\citenamefont {Abrikosov}(2000)}]{127-abrikosov2000quantum}%
  \BibitemOpen
  \bibfield  {author} {\bibinfo {author} {\bibfnamefont {A.}~\bibnamefont
  {Abrikosov}},\ }\bibfield  {title} {\bibinfo {title} {Quantum linear
  magnetoresistance},\ }\href@noop {} {\bibfield  {journal} {\bibinfo
  {journal} {EPL (Europhysics Letters)}\ }\textbf {\bibinfo {volume} {49}},\
  \bibinfo {pages} {789} (\bibinfo {year} {2000})}\BibitemShut {NoStop}%
\bibitem [{\citenamefont {Sun}\ \emph {et~al.}(2014)\citenamefont {Sun},
  \citenamefont {Taen}, \citenamefont {Yamada}, \citenamefont {Pyon},
  \citenamefont {Nishizaki}, \citenamefont {Shi},\ and\ \citenamefont
  {Tamegai}}]{128-sun2014multiband}%
  \BibitemOpen
  \bibfield  {author} {\bibinfo {author} {\bibfnamefont {Y.}~\bibnamefont
  {Sun}}, \bibinfo {author} {\bibfnamefont {T.}~\bibnamefont {Taen}}, \bibinfo
  {author} {\bibfnamefont {T.}~\bibnamefont {Yamada}}, \bibinfo {author}
  {\bibfnamefont {S.}~\bibnamefont {Pyon}}, \bibinfo {author} {\bibfnamefont
  {T.}~\bibnamefont {Nishizaki}}, \bibinfo {author} {\bibfnamefont
  {Z.}~\bibnamefont {Shi}},\ and\ \bibinfo {author} {\bibfnamefont
  {T.}~\bibnamefont {Tamegai}},\ }\bibfield  {title} {\bibinfo {title}
  {Multiband effects and possible {D}irac fermions in
  {F}e$_{1+y}${T}e$_{0.6}${S}e$_{0.4}$},\ }\href@noop {} {\bibfield  {journal}
  {\bibinfo  {journal} {Phys. Rev. B}\ }\textbf {\bibinfo {volume} {89}},\
  \bibinfo {pages} {144512} (\bibinfo {year} {2014})}\BibitemShut {NoStop}%
\bibitem [{\citenamefont {Rameau}\ \emph {et~al.}(2019)\citenamefont {Rameau},
  \citenamefont {Zaki}, \citenamefont {Gu}, \citenamefont {Johnson},\ and\
  \citenamefont {Weinert}}]{129-rameau2019interplay}%
  \BibitemOpen
  \bibfield  {author} {\bibinfo {author} {\bibfnamefont {J.}~\bibnamefont
  {Rameau}}, \bibinfo {author} {\bibfnamefont {N.}~\bibnamefont {Zaki}},
  \bibinfo {author} {\bibfnamefont {G.}~\bibnamefont {Gu}}, \bibinfo {author}
  {\bibfnamefont {P.}~\bibnamefont {Johnson}},\ and\ \bibinfo {author}
  {\bibfnamefont {M.}~\bibnamefont {Weinert}},\ }\bibfield  {title} {\bibinfo
  {title} {Interplay of paramagnetism and topology in the {F}e-chalcogenide
  high-{T}$_{c}$ superconductors},\ }\href@noop {} {\bibfield  {journal}
  {\bibinfo  {journal} {Phys. Rev. B}\ }\textbf {\bibinfo {volume} {99}},\
  \bibinfo {pages} {205117} (\bibinfo {year} {2019})}\BibitemShut {NoStop}%
\bibitem [{\citenamefont {Lohani}\ \emph {et~al.}(2020)\citenamefont {Lohani},
  \citenamefont {Hazra}, \citenamefont {Ribak}, \citenamefont {Nitzav},
  \citenamefont {Fu}, \citenamefont {Yan}, \citenamefont {Randeria},\ and\
  \citenamefont {Kanigel}}]{130-lohani2020band}%
  \BibitemOpen
  \bibfield  {author} {\bibinfo {author} {\bibfnamefont {H.}~\bibnamefont
  {Lohani}}, \bibinfo {author} {\bibfnamefont {T.}~\bibnamefont {Hazra}},
  \bibinfo {author} {\bibfnamefont {A.}~\bibnamefont {Ribak}}, \bibinfo
  {author} {\bibfnamefont {Y.}~\bibnamefont {Nitzav}}, \bibinfo {author}
  {\bibfnamefont {H.}~\bibnamefont {Fu}}, \bibinfo {author} {\bibfnamefont
  {B.}~\bibnamefont {Yan}}, \bibinfo {author} {\bibfnamefont {M.}~\bibnamefont
  {Randeria}},\ and\ \bibinfo {author} {\bibfnamefont {A.}~\bibnamefont
  {Kanigel}},\ }\bibfield  {title} {\bibinfo {title} {Band inversion and
  topology of the bulk electronic structure in
  {F}e{S}e$_{0.45}${T}e$_{0.55}$},\ }\href@noop {} {\bibfield  {journal}
  {\bibinfo  {journal} {Phys. Rev. B}\ }\textbf {\bibinfo {volume} {101}},\
  \bibinfo {pages} {245146} (\bibinfo {year} {2020})}\BibitemShut {NoStop}%
\bibitem [{\citenamefont {Zhu}\ \emph {et~al.}(2017)\citenamefont {Zhu},
  \citenamefont {Cui}, \citenamefont {Lei}, \citenamefont {Wang}, \citenamefont
  {Shang}, \citenamefont {Meng}, \citenamefont {Ma}, \citenamefont {Luo},
  \citenamefont {Wu}, \citenamefont {Sun} \emph {et~al.}}]{131-zhu2017tuning}%
  \BibitemOpen
  \bibfield  {author} {\bibinfo {author} {\bibfnamefont {C.}~\bibnamefont
  {Zhu}}, \bibinfo {author} {\bibfnamefont {J.}~\bibnamefont {Cui}}, \bibinfo
  {author} {\bibfnamefont {B.}~\bibnamefont {Lei}}, \bibinfo {author}
  {\bibfnamefont {N.}~\bibnamefont {Wang}}, \bibinfo {author} {\bibfnamefont
  {C.}~\bibnamefont {Shang}}, \bibinfo {author} {\bibfnamefont
  {F.}~\bibnamefont {Meng}}, \bibinfo {author} {\bibfnamefont {L.}~\bibnamefont
  {Ma}}, \bibinfo {author} {\bibfnamefont {X.}~\bibnamefont {Luo}}, \bibinfo
  {author} {\bibfnamefont {T.}~\bibnamefont {Wu}}, \bibinfo {author}
  {\bibfnamefont {Z.}~\bibnamefont {Sun}}, \emph {et~al.},\ }\bibfield  {title}
  {\bibinfo {title} {Tuning electronic properties of
  {F}e{S}e$_{0.5}${T}e$_{0.5}$ thin flakes using a solid ion conductor
  field-effect transistor},\ }\href@noop {} {\bibfield  {journal} {\bibinfo
  {journal} {Phys. Rev. B}\ }\textbf {\bibinfo {volume} {95}},\ \bibinfo
  {pages} {174513} (\bibinfo {year} {2017})}\BibitemShut {NoStop}%
\bibitem [{\citenamefont {Binnig}\ and\ \citenamefont
  {Rohrer}(1987)}]{132-binnig1987scanning}%
  \BibitemOpen
  \bibfield  {author} {\bibinfo {author} {\bibfnamefont {G.}~\bibnamefont
  {Binnig}}\ and\ \bibinfo {author} {\bibfnamefont {H.}~\bibnamefont
  {Rohrer}},\ }\bibfield  {title} {\bibinfo {title} {Scanning tunneling
  microscopy—from birth to adolescence},\ }\href@noop {} {\bibfield
  {journal} {\bibinfo  {journal} {Rev. Mod. Phys.}\ }\textbf {\bibinfo {volume}
  {59}},\ \bibinfo {pages} {615} (\bibinfo {year} {1987})}\BibitemShut
  {NoStop}%
\bibitem [{\citenamefont {Chen}(1993)}]{133-chen1993introduction}%
  \BibitemOpen
  \bibfield  {author} {\bibinfo {author} {\bibfnamefont {C.~J.}\ \bibnamefont
  {Chen}},\ }\href@noop {} {\emph {\bibinfo {title} {Introduction to scanning
  tunneling microscopy}}},\ Vol.~\bibinfo {volume} {4}\ (\bibinfo  {publisher}
  {Oxford University Press on Demand},\ \bibinfo {year} {1993})\BibitemShut
  {NoStop}%
\bibitem [{\citenamefont {Antipov}\ \emph {et~al.}(2018)\citenamefont
  {Antipov}, \citenamefont {Bargerbos}, \citenamefont {Winkler}, \citenamefont
  {Bauer}, \citenamefont {Rossi},\ and\ \citenamefont
  {Lutchyn}}]{134-antipov2018effects}%
  \BibitemOpen
  \bibfield  {author} {\bibinfo {author} {\bibfnamefont {A.~E.}\ \bibnamefont
  {Antipov}}, \bibinfo {author} {\bibfnamefont {A.}~\bibnamefont {Bargerbos}},
  \bibinfo {author} {\bibfnamefont {G.~W.}\ \bibnamefont {Winkler}}, \bibinfo
  {author} {\bibfnamefont {B.}~\bibnamefont {Bauer}}, \bibinfo {author}
  {\bibfnamefont {E.}~\bibnamefont {Rossi}},\ and\ \bibinfo {author}
  {\bibfnamefont {R.~M.}\ \bibnamefont {Lutchyn}},\ }\bibfield  {title}
  {\bibinfo {title} {Effects of gate-induced electric fields on semiconductor
  {M}ajorana nanowires},\ }\href@noop {} {\bibfield  {journal} {\bibinfo
  {journal} {Phys. Rev. X}\ }\textbf {\bibinfo {volume} {8}},\ \bibinfo {pages}
  {031041} (\bibinfo {year} {2018})}\BibitemShut {NoStop}%
\bibitem [{\citenamefont {Lee}\ \emph {et~al.}(2014)\citenamefont {Lee},
  \citenamefont {Jiang}, \citenamefont {Houzet}, \citenamefont {Aguado},
  \citenamefont {Lieber},\ and\ \citenamefont
  {De~Franceschi}}]{135-lee2014spin}%
  \BibitemOpen
  \bibfield  {author} {\bibinfo {author} {\bibfnamefont {E.~J.}\ \bibnamefont
  {Lee}}, \bibinfo {author} {\bibfnamefont {X.}~\bibnamefont {Jiang}}, \bibinfo
  {author} {\bibfnamefont {M.}~\bibnamefont {Houzet}}, \bibinfo {author}
  {\bibfnamefont {R.}~\bibnamefont {Aguado}}, \bibinfo {author} {\bibfnamefont
  {C.~M.}\ \bibnamefont {Lieber}},\ and\ \bibinfo {author} {\bibfnamefont
  {S.}~\bibnamefont {De~Franceschi}},\ }\bibfield  {title} {\bibinfo {title}
  {Spin-resolved {A}ndreev levels and parity crossings in hybrid
  superconductor--semiconductor nanostructures},\ }\href@noop {} {\bibfield
  {journal} {\bibinfo  {journal} {Nat. Nanotechnol.}\ }\textbf {\bibinfo
  {volume} {9}},\ \bibinfo {pages} {79} (\bibinfo {year} {2014})}\BibitemShut
  {NoStop}%
\bibitem [{\citenamefont {Van~Wees}\ \emph {et~al.}(1992)\citenamefont
  {Van~Wees}, \citenamefont {De~Vries}, \citenamefont {Magn{\'e}e},\ and\
  \citenamefont {Klapwijk}}]{136-van1992excess}%
  \BibitemOpen
  \bibfield  {author} {\bibinfo {author} {\bibfnamefont {B.}~\bibnamefont
  {Van~Wees}}, \bibinfo {author} {\bibfnamefont {P.}~\bibnamefont {De~Vries}},
  \bibinfo {author} {\bibfnamefont {P.}~\bibnamefont {Magn{\'e}e}},\ and\
  \bibinfo {author} {\bibfnamefont {T.}~\bibnamefont {Klapwijk}},\ }\bibfield
  {title} {\bibinfo {title} {Excess conductance of superconductor-semiconductor
  interfaces due to phase conjugation between electrons and holes},\
  }\href@noop {} {\bibfield  {journal} {\bibinfo  {journal} {Phys. Rev. Lett.}\
  }\textbf {\bibinfo {volume} {69}},\ \bibinfo {pages} {510} (\bibinfo {year}
  {1992})}\BibitemShut {NoStop}%
\bibitem [{\citenamefont {Marmorkos}\ \emph {et~al.}(1993)\citenamefont
  {Marmorkos}, \citenamefont {Beenakker},\ and\ \citenamefont
  {Jalabert}}]{137-marmorkos1993three}%
  \BibitemOpen
  \bibfield  {author} {\bibinfo {author} {\bibfnamefont {I.}~\bibnamefont
  {Marmorkos}}, \bibinfo {author} {\bibfnamefont {C.}~\bibnamefont
  {Beenakker}},\ and\ \bibinfo {author} {\bibfnamefont {R.}~\bibnamefont
  {Jalabert}},\ }\bibfield  {title} {\bibinfo {title} {Three signatures of
  phase-coherent {A}ndreev reflection},\ }\href@noop {} {\bibfield  {journal}
  {\bibinfo  {journal} {Phys. Rev. B}\ }\textbf {\bibinfo {volume} {48}},\
  \bibinfo {pages} {2811} (\bibinfo {year} {1993})}\BibitemShut {NoStop}%
\bibitem [{\citenamefont {Kim}\ \emph {et~al.}(2018)\citenamefont {Kim},
  \citenamefont {Shin}, \citenamefont {Kim}, \citenamefont {Song},\ and\
  \citenamefont {Doh}}]{138-kim2018zero}%
  \BibitemOpen
  \bibfield  {author} {\bibinfo {author} {\bibfnamefont {N.-H.}\ \bibnamefont
  {Kim}}, \bibinfo {author} {\bibfnamefont {Y.-S.}\ \bibnamefont {Shin}},
  \bibinfo {author} {\bibfnamefont {H.-S.}\ \bibnamefont {Kim}}, \bibinfo
  {author} {\bibfnamefont {J.-D.}\ \bibnamefont {Song}},\ and\ \bibinfo
  {author} {\bibfnamefont {Y.-J.}\ \bibnamefont {Doh}},\ }\bibfield  {title}
  {\bibinfo {title} {Zero bias conductance peak in {I}n{A}s nanowire coupled to
  superconducting electrodes},\ }\href@noop {} {\bibfield  {journal} {\bibinfo
  {journal} {Curr. Appl. Phys.}\ }\textbf {\bibinfo {volume} {18}},\ \bibinfo
  {pages} {384} (\bibinfo {year} {2018})}\BibitemShut {NoStop}%
\bibitem [{\citenamefont {Nguyen}\ \emph {et~al.}(1992)\citenamefont {Nguyen},
  \citenamefont {Kroemer},\ and\ \citenamefont {Hu}}]{139-nguyen1992anomalous}%
  \BibitemOpen
  \bibfield  {author} {\bibinfo {author} {\bibfnamefont {C.}~\bibnamefont
  {Nguyen}}, \bibinfo {author} {\bibfnamefont {H.}~\bibnamefont {Kroemer}},\
  and\ \bibinfo {author} {\bibfnamefont {E.~L.}\ \bibnamefont {Hu}},\
  }\bibfield  {title} {\bibinfo {title} {Anomalous {A}ndreev conductance in
  {I}n{A}s-{A}l{S}b quantum well structures with nb electrodes},\ }\href@noop
  {} {\bibfield  {journal} {\bibinfo  {journal} {Phys. Rev. Lett.}\ }\textbf
  {\bibinfo {volume} {69}},\ \bibinfo {pages} {2847} (\bibinfo {year}
  {1992})}\BibitemShut {NoStop}%
\bibitem [{\citenamefont {Xiong}\ \emph {et~al.}(1993)\citenamefont {Xiong},
  \citenamefont {Xiao},\ and\ \citenamefont {Laibowitz}}]{140-xiong1993subgap}%
  \BibitemOpen
  \bibfield  {author} {\bibinfo {author} {\bibfnamefont {P.}~\bibnamefont
  {Xiong}}, \bibinfo {author} {\bibfnamefont {G.}~\bibnamefont {Xiao}},\ and\
  \bibinfo {author} {\bibfnamefont {R.}~\bibnamefont {Laibowitz}},\ }\bibfield
  {title} {\bibinfo {title} {Subgap and above-gap differential resistance
  anomalies in superconductor-normal-metal microjunctions},\ }\href@noop {}
  {\bibfield  {journal} {\bibinfo  {journal} {Phys. Rev. Lett.}\ }\textbf
  {\bibinfo {volume} {71}},\ \bibinfo {pages} {1907} (\bibinfo {year}
  {1993})}\BibitemShut {NoStop}%
\bibitem [{\citenamefont {Kouwenhoven}\ and\ \citenamefont
  {Glazman}(2001)}]{141-kouwenhoven2001revival}%
  \BibitemOpen
  \bibfield  {author} {\bibinfo {author} {\bibfnamefont {L.}~\bibnamefont
  {Kouwenhoven}}\ and\ \bibinfo {author} {\bibfnamefont {L.}~\bibnamefont
  {Glazman}},\ }\bibfield  {title} {\bibinfo {title} {Revival of the kondo
  effect},\ }\href@noop {} {\bibfield  {journal} {\bibinfo  {journal} {Physics
  World}\ }\textbf {\bibinfo {volume} {14}},\ \bibinfo {pages} {33} (\bibinfo
  {year} {2001})}\BibitemShut {NoStop}%
\bibitem [{\citenamefont {Lee}\ \emph {et~al.}(2012)\citenamefont {Lee},
  \citenamefont {Jiang}, \citenamefont {Aguado}, \citenamefont {Katsaros},
  \citenamefont {Lieber},\ and\ \citenamefont
  {De~Franceschi}}]{142-lee2012zero}%
  \BibitemOpen
  \bibfield  {author} {\bibinfo {author} {\bibfnamefont {E.~J.}\ \bibnamefont
  {Lee}}, \bibinfo {author} {\bibfnamefont {X.}~\bibnamefont {Jiang}}, \bibinfo
  {author} {\bibfnamefont {R.}~\bibnamefont {Aguado}}, \bibinfo {author}
  {\bibfnamefont {G.}~\bibnamefont {Katsaros}}, \bibinfo {author}
  {\bibfnamefont {C.~M.}\ \bibnamefont {Lieber}},\ and\ \bibinfo {author}
  {\bibfnamefont {S.}~\bibnamefont {De~Franceschi}},\ }\bibfield  {title}
  {\bibinfo {title} {Zero-bias anomaly in a nanowire quantum dot coupled to
  superconductors},\ }\href@noop {} {\bibfield  {journal} {\bibinfo  {journal}
  {Phys. Rev. Lett.}\ }\textbf {\bibinfo {volume} {109}},\ \bibinfo {pages}
  {186802} (\bibinfo {year} {2012})}\BibitemShut {NoStop}%
\bibitem [{\citenamefont {Churchill}\ \emph {et~al.}(2013)\citenamefont
  {Churchill}, \citenamefont {Fatemi}, \citenamefont {Grove-Rasmussen},
  \citenamefont {Deng}, \citenamefont {Caroff}, \citenamefont {Xu},\ and\
  \citenamefont {Marcus}}]{143-churchill2013superconductor}%
  \BibitemOpen
  \bibfield  {author} {\bibinfo {author} {\bibfnamefont {H.}~\bibnamefont
  {Churchill}}, \bibinfo {author} {\bibfnamefont {V.}~\bibnamefont {Fatemi}},
  \bibinfo {author} {\bibfnamefont {K.}~\bibnamefont {Grove-Rasmussen}},
  \bibinfo {author} {\bibfnamefont {M.}~\bibnamefont {Deng}}, \bibinfo {author}
  {\bibfnamefont {P.}~\bibnamefont {Caroff}}, \bibinfo {author} {\bibfnamefont
  {H.}~\bibnamefont {Xu}},\ and\ \bibinfo {author} {\bibfnamefont {C.~M.}\
  \bibnamefont {Marcus}},\ }\bibfield  {title} {\bibinfo {title}
  {Superconductor-nanowire devices from tunneling to the multichannel regime:
  Zero-bias oscillations and magnetoconductance crossover},\ }\href@noop {}
  {\bibfield  {journal} {\bibinfo  {journal} {Phys. Rev. B}\ }\textbf {\bibinfo
  {volume} {87}},\ \bibinfo {pages} {241401} (\bibinfo {year}
  {2013})}\BibitemShut {NoStop}%
\bibitem [{\citenamefont {Shen}\ and\ \citenamefont
  {Rowell}(1968)}]{144-shen1968zero}%
  \BibitemOpen
  \bibfield  {author} {\bibinfo {author} {\bibfnamefont {L.}~\bibnamefont
  {Shen}}\ and\ \bibinfo {author} {\bibfnamefont {J.}~\bibnamefont {Rowell}},\
  }\bibfield  {title} {\bibinfo {title} {Zero-bias tunneling
  anomalies—temperature, voltage, and magnetic field dependence},\
  }\href@noop {} {\bibfield  {journal} {\bibinfo  {journal} {Phys. Rev.}\
  }\textbf {\bibinfo {volume} {165}},\ \bibinfo {pages} {566} (\bibinfo {year}
  {1968})}\BibitemShut {NoStop}%
\bibitem [{\citenamefont {Ternes}\ \emph {et~al.}(2006)\citenamefont {Ternes},
  \citenamefont {Schneider}, \citenamefont {Cuevas}, \citenamefont {Lutz},
  \citenamefont {Hirjibehedin},\ and\ \citenamefont
  {Heinrich}}]{145-ternes2006subgap}%
  \BibitemOpen
  \bibfield  {author} {\bibinfo {author} {\bibfnamefont {M.}~\bibnamefont
  {Ternes}}, \bibinfo {author} {\bibfnamefont {W.-D.}\ \bibnamefont
  {Schneider}}, \bibinfo {author} {\bibfnamefont {J.-C.}\ \bibnamefont
  {Cuevas}}, \bibinfo {author} {\bibfnamefont {C.~P.}\ \bibnamefont {Lutz}},
  \bibinfo {author} {\bibfnamefont {C.~F.}\ \bibnamefont {Hirjibehedin}},\ and\
  \bibinfo {author} {\bibfnamefont {A.~J.}\ \bibnamefont {Heinrich}},\
  }\bibfield  {title} {\bibinfo {title} {Subgap structure in asymmetric
  superconducting tunnel junctions},\ }\href@noop {} {\bibfield  {journal}
  {\bibinfo  {journal} {Phys. Rev. B}\ }\textbf {\bibinfo {volume} {74}},\
  \bibinfo {pages} {132501} (\bibinfo {year} {2006})}\BibitemShut {NoStop}%
\bibitem [{\citenamefont {Levy}\ \emph {et~al.}(2013)\citenamefont {Levy},
  \citenamefont {Zhang}, \citenamefont {Ha}, \citenamefont {Sharifi},
  \citenamefont {Talin}, \citenamefont {Kuk},\ and\ \citenamefont
  {Stroscio}}]{146-levy2013experimental}%
  \BibitemOpen
  \bibfield  {author} {\bibinfo {author} {\bibfnamefont {N.}~\bibnamefont
  {Levy}}, \bibinfo {author} {\bibfnamefont {T.}~\bibnamefont {Zhang}},
  \bibinfo {author} {\bibfnamefont {J.}~\bibnamefont {Ha}}, \bibinfo {author}
  {\bibfnamefont {F.}~\bibnamefont {Sharifi}}, \bibinfo {author} {\bibfnamefont
  {A.~A.}\ \bibnamefont {Talin}}, \bibinfo {author} {\bibfnamefont
  {Y.}~\bibnamefont {Kuk}},\ and\ \bibinfo {author} {\bibfnamefont {J.~A.}\
  \bibnamefont {Stroscio}},\ }\bibfield  {title} {\bibinfo {title}
  {Experimental evidence for s-wave pairing symmetry in superconducting
  {C}u$_{x}${B}i$_{2}${S}e$_{3}$ single crystals using a scanning tunneling
  microscope},\ }\href@noop {} {\bibfield  {journal} {\bibinfo  {journal}
  {Phys. Rev. Lett.}\ }\textbf {\bibinfo {volume} {110}},\ \bibinfo {pages}
  {117001} (\bibinfo {year} {2013})}\BibitemShut {NoStop}%
\bibitem [{\citenamefont {Naaman}\ \emph {et~al.}(2001)\citenamefont {Naaman},
  \citenamefont {Teizer},\ and\ \citenamefont
  {Dynes}}]{147-naaman2001fluctuation}%
  \BibitemOpen
  \bibfield  {author} {\bibinfo {author} {\bibfnamefont {O.}~\bibnamefont
  {Naaman}}, \bibinfo {author} {\bibfnamefont {W.}~\bibnamefont {Teizer}},\
  and\ \bibinfo {author} {\bibfnamefont {R.}~\bibnamefont {Dynes}},\ }\bibfield
   {title} {\bibinfo {title} {Fluctuation dominated {J}osephson tunneling with
  a scanning tunneling microscope},\ }\href@noop {} {\bibfield  {journal}
  {\bibinfo  {journal} {Phys. Rev. Lett.}\ }\textbf {\bibinfo {volume} {87}},\
  \bibinfo {pages} {097004} (\bibinfo {year} {2001})}\BibitemShut {NoStop}%
\bibitem [{\citenamefont {Naaman}\ and\ \citenamefont
  {Dynes}(2004)}]{148-naaman2004subharmonic}%
  \BibitemOpen
  \bibfield  {author} {\bibinfo {author} {\bibfnamefont {O.}~\bibnamefont
  {Naaman}}\ and\ \bibinfo {author} {\bibfnamefont {R.}~\bibnamefont {Dynes}},\
  }\bibfield  {title} {\bibinfo {title} {Subharmonic gap structure in
  superconducting scanning tunneling microscope junctions},\ }\href@noop {}
  {\bibfield  {journal} {\bibinfo  {journal} {Solid State Commun.}\ }\textbf
  {\bibinfo {volume} {129}},\ \bibinfo {pages} {299} (\bibinfo {year}
  {2004})}\BibitemShut {NoStop}%
\bibitem [{\citenamefont {Hikami}\ \emph {et~al.}(1980)\citenamefont {Hikami},
  \citenamefont {Larkin},\ and\ \citenamefont {Nagaoka}}]{149-hikami1980spin}%
  \BibitemOpen
  \bibfield  {author} {\bibinfo {author} {\bibfnamefont {S.}~\bibnamefont
  {Hikami}}, \bibinfo {author} {\bibfnamefont {A.~I.}\ \bibnamefont {Larkin}},\
  and\ \bibinfo {author} {\bibfnamefont {Y.}~\bibnamefont {Nagaoka}},\
  }\bibfield  {title} {\bibinfo {title} {Spin-orbit interaction and
  magnetoresistance in the two dimensional random system},\ }\href@noop {}
  {\bibfield  {journal} {\bibinfo  {journal} {Prog. Theor. Phys.}\ }\textbf
  {\bibinfo {volume} {63}},\ \bibinfo {pages} {707} (\bibinfo {year}
  {1980})}\BibitemShut {NoStop}%
\bibitem [{\citenamefont {Pikulin}\ \emph {et~al.}(2012)\citenamefont
  {Pikulin}, \citenamefont {Dahlhaus}, \citenamefont {Wimmer}, \citenamefont
  {Schomerus},\ and\ \citenamefont {Beenakker}}]{150-pikulin2012zero}%
  \BibitemOpen
  \bibfield  {author} {\bibinfo {author} {\bibfnamefont {D.~I.}\ \bibnamefont
  {Pikulin}}, \bibinfo {author} {\bibfnamefont {J.}~\bibnamefont {Dahlhaus}},
  \bibinfo {author} {\bibfnamefont {M.}~\bibnamefont {Wimmer}}, \bibinfo
  {author} {\bibfnamefont {H.}~\bibnamefont {Schomerus}},\ and\ \bibinfo
  {author} {\bibfnamefont {C.}~\bibnamefont {Beenakker}},\ }\bibfield  {title}
  {\bibinfo {title} {A zero-voltage conductance peak from weak antilocalization
  in a {M}ajorana nanowire},\ }\href@noop {} {\bibfield  {journal} {\bibinfo
  {journal} {New J Phys.}\ }\textbf {\bibinfo {volume} {14}},\ \bibinfo {pages}
  {125011} (\bibinfo {year} {2012})}\BibitemShut {NoStop}%
\bibitem [{\citenamefont {Bagrets}\ and\ \citenamefont
  {Altland}(2012)}]{151-bagrets2012class}%
  \BibitemOpen
  \bibfield  {author} {\bibinfo {author} {\bibfnamefont {D.}~\bibnamefont
  {Bagrets}}\ and\ \bibinfo {author} {\bibfnamefont {A.}~\bibnamefont
  {Altland}},\ }\bibfield  {title} {\bibinfo {title} {Class-{D} spectral peak
  in {M}ajorana quantum wires},\ }\href@noop {} {\bibfield  {journal} {\bibinfo
   {journal} {Phys. Rev. Lett.}\ }\textbf {\bibinfo {volume} {109}},\ \bibinfo
  {pages} {227005} (\bibinfo {year} {2012})}\BibitemShut {NoStop}%
\bibitem [{\citenamefont {Liu}\ \emph {et~al.}(2012)\citenamefont {Liu},
  \citenamefont {Potter}, \citenamefont {Law},\ and\ \citenamefont
  {Lee}}]{152-liu2012zero}%
  \BibitemOpen
  \bibfield  {author} {\bibinfo {author} {\bibfnamefont {J.}~\bibnamefont
  {Liu}}, \bibinfo {author} {\bibfnamefont {A.~C.}\ \bibnamefont {Potter}},
  \bibinfo {author} {\bibfnamefont {K.~T.}\ \bibnamefont {Law}},\ and\ \bibinfo
  {author} {\bibfnamefont {P.~A.}\ \bibnamefont {Lee}},\ }\bibfield  {title}
  {\bibinfo {title} {Zero-bias peaks in the tunneling conductance of
  spin-orbit-coupled superconducting wires with and without {M}ajorana
  end-states},\ }\href@noop {} {\bibfield  {journal} {\bibinfo  {journal}
  {Phys. Rev. Lett.}\ }\textbf {\bibinfo {volume} {109}},\ \bibinfo {pages}
  {267002} (\bibinfo {year} {2012})}\BibitemShut {NoStop}%
\bibitem [{\citenamefont {Pan}\ \emph {et~al.}(2020)\citenamefont {Pan},
  \citenamefont {Cole}, \citenamefont {Sau},\ and\ \citenamefont
  {Sarma}}]{153-pan2020generic}%
  \BibitemOpen
  \bibfield  {author} {\bibinfo {author} {\bibfnamefont {H.}~\bibnamefont
  {Pan}}, \bibinfo {author} {\bibfnamefont {W.~S.}\ \bibnamefont {Cole}},
  \bibinfo {author} {\bibfnamefont {J.~D.}\ \bibnamefont {Sau}},\ and\ \bibinfo
  {author} {\bibfnamefont {S.~D.}\ \bibnamefont {Sarma}},\ }\bibfield  {title}
  {\bibinfo {title} {Generic quantized zero-bias conductance peaks in
  superconductor-semiconductor hybrid structures},\ }\href@noop {} {\bibfield
  {journal} {\bibinfo  {journal} {Phys. Rev. B}\ }\textbf {\bibinfo {volume}
  {101}},\ \bibinfo {pages} {024506} (\bibinfo {year} {2020})}\BibitemShut
  {NoStop}%
\bibitem [{\citenamefont {Anderson}(1958)}]{154-anderson1958absence}%
  \BibitemOpen
  \bibfield  {author} {\bibinfo {author} {\bibfnamefont {P.~W.}\ \bibnamefont
  {Anderson}},\ }\bibfield  {title} {\bibinfo {title} {Absence of diffusion in
  certain random lattices},\ }\href@noop {} {\bibfield  {journal} {\bibinfo
  {journal} {Phys. Rev.}\ }\textbf {\bibinfo {volume} {109}},\ \bibinfo {pages}
  {1492} (\bibinfo {year} {1958})}\BibitemShut {NoStop}%
\bibitem [{\citenamefont {Checkelsky}\ \emph {et~al.}(2009)\citenamefont
  {Checkelsky}, \citenamefont {Hor}, \citenamefont {Liu}, \citenamefont {Qu},
  \citenamefont {Cava},\ and\ \citenamefont {Ong}}]{155-checkelsky2009quantum}%
  \BibitemOpen
  \bibfield  {author} {\bibinfo {author} {\bibfnamefont {J.~G.}\ \bibnamefont
  {Checkelsky}}, \bibinfo {author} {\bibfnamefont {Y.~S.}\ \bibnamefont {Hor}},
  \bibinfo {author} {\bibfnamefont {M.-H.}\ \bibnamefont {Liu}}, \bibinfo
  {author} {\bibfnamefont {D.-X.}\ \bibnamefont {Qu}}, \bibinfo {author}
  {\bibfnamefont {R.~J.}\ \bibnamefont {Cava}},\ and\ \bibinfo {author}
  {\bibfnamefont {N.}~\bibnamefont {Ong}},\ }\bibfield  {title} {\bibinfo
  {title} {Quantum interference in macroscopic crystals of nonmetallic
  {B}i$_{2}${S}e$_{3}$},\ }\href@noop {} {\bibfield  {journal} {\bibinfo
  {journal} {Phys. Rev. Lett.}\ }\textbf {\bibinfo {volume} {103}},\ \bibinfo
  {pages} {246601} (\bibinfo {year} {2009})}\BibitemShut {NoStop}%
\bibitem [{\citenamefont {Kells}\ \emph {et~al.}(2012)\citenamefont {Kells},
  \citenamefont {Meidan},\ and\ \citenamefont {Brouwer}}]{156-kells2012near}%
  \BibitemOpen
  \bibfield  {author} {\bibinfo {author} {\bibfnamefont {G.}~\bibnamefont
  {Kells}}, \bibinfo {author} {\bibfnamefont {D.}~\bibnamefont {Meidan}},\ and\
  \bibinfo {author} {\bibfnamefont {P.}~\bibnamefont {Brouwer}},\ }\bibfield
  {title} {\bibinfo {title} {Near-zero-energy end states in topologically
  trivial spin-orbit coupled superconducting nanowires with a smooth
  confinement},\ }\href@noop {} {\bibfield  {journal} {\bibinfo  {journal}
  {Phys. Rev. B}\ }\textbf {\bibinfo {volume} {86}},\ \bibinfo {pages} {100503}
  (\bibinfo {year} {2012})}\BibitemShut {NoStop}%
\bibitem [{\citenamefont {Liu}\ \emph {et~al.}(2017)\citenamefont {Liu},
  \citenamefont {Sau}, \citenamefont {Stanescu},\ and\ \citenamefont
  {Sarma}}]{157-liu2017andreev}%
  \BibitemOpen
  \bibfield  {author} {\bibinfo {author} {\bibfnamefont {C.-X.}\ \bibnamefont
  {Liu}}, \bibinfo {author} {\bibfnamefont {J.~D.}\ \bibnamefont {Sau}},
  \bibinfo {author} {\bibfnamefont {T.~D.}\ \bibnamefont {Stanescu}},\ and\
  \bibinfo {author} {\bibfnamefont {S.~D.}\ \bibnamefont {Sarma}},\ }\bibfield
  {title} {\bibinfo {title} {Andreev bound states versus {M}ajorana bound
  states in quantum dot-nanowire-superconductor hybrid structures: Trivial
  versus topological zero-bias conductance peaks},\ }\href@noop {} {\bibfield
  {journal} {\bibinfo  {journal} {Phys. Rev. B}\ }\textbf {\bibinfo {volume}
  {96}},\ \bibinfo {pages} {075161} (\bibinfo {year} {2017})}\BibitemShut
  {NoStop}%
\bibitem [{\citenamefont {Moore}\ \emph
  {et~al.}(2018{\natexlab{a}})\citenamefont {Moore}, \citenamefont {Stanescu},\
  and\ \citenamefont {Tewari}}]{158-moore2018two}%
  \BibitemOpen
  \bibfield  {author} {\bibinfo {author} {\bibfnamefont {C.}~\bibnamefont
  {Moore}}, \bibinfo {author} {\bibfnamefont {T.~D.}\ \bibnamefont
  {Stanescu}},\ and\ \bibinfo {author} {\bibfnamefont {S.}~\bibnamefont
  {Tewari}},\ }\bibfield  {title} {\bibinfo {title} {Two-terminal charge
  tunneling: Disentangling {M}ajorana zero modes from partially separated
  {A}ndreev bound states in semiconductor-superconductor heterostructures},\
  }\href@noop {} {\bibfield  {journal} {\bibinfo  {journal} {Phys. Rev. B}\
  }\textbf {\bibinfo {volume} {97}},\ \bibinfo {pages} {165302} (\bibinfo
  {year} {2018}{\natexlab{a}})}\BibitemShut {NoStop}%
\bibitem [{\citenamefont {Moore}\ \emph
  {et~al.}(2018{\natexlab{b}})\citenamefont {Moore}, \citenamefont {Zeng},
  \citenamefont {Stanescu},\ and\ \citenamefont
  {Tewari}}]{159-moore2018quantized}%
  \BibitemOpen
  \bibfield  {author} {\bibinfo {author} {\bibfnamefont {C.}~\bibnamefont
  {Moore}}, \bibinfo {author} {\bibfnamefont {C.}~\bibnamefont {Zeng}},
  \bibinfo {author} {\bibfnamefont {T.~D.}\ \bibnamefont {Stanescu}},\ and\
  \bibinfo {author} {\bibfnamefont {S.}~\bibnamefont {Tewari}},\ }\bibfield
  {title} {\bibinfo {title} {Quantized zero-bias conductance plateau in
  semiconductor-superconductor heterostructures without topological {M}ajorana
  zero modes},\ }\href@noop {} {\bibfield  {journal} {\bibinfo  {journal}
  {Phys. Rev. B}\ }\textbf {\bibinfo {volume} {98}},\ \bibinfo {pages} {155314}
  (\bibinfo {year} {2018}{\natexlab{b}})}\BibitemShut {NoStop}%
\bibitem [{\citenamefont {Deng}\ \emph {et~al.}(2016)\citenamefont {Deng},
  \citenamefont {Vaitiek{\.e}nas}, \citenamefont {Hansen}, \citenamefont
  {Danon}, \citenamefont {Leijnse}, \citenamefont {Flensberg}, \citenamefont
  {Nyg{\aa}rd}, \citenamefont {Krogstrup},\ and\ \citenamefont
  {Marcus}}]{160-deng2016majorana}%
  \BibitemOpen
  \bibfield  {author} {\bibinfo {author} {\bibfnamefont {M.}~\bibnamefont
  {Deng}}, \bibinfo {author} {\bibfnamefont {S.}~\bibnamefont
  {Vaitiek{\.e}nas}}, \bibinfo {author} {\bibfnamefont {E.~B.}\ \bibnamefont
  {Hansen}}, \bibinfo {author} {\bibfnamefont {J.}~\bibnamefont {Danon}},
  \bibinfo {author} {\bibfnamefont {M.}~\bibnamefont {Leijnse}}, \bibinfo
  {author} {\bibfnamefont {K.}~\bibnamefont {Flensberg}}, \bibinfo {author}
  {\bibfnamefont {J.}~\bibnamefont {Nyg{\aa}rd}}, \bibinfo {author}
  {\bibfnamefont {P.}~\bibnamefont {Krogstrup}},\ and\ \bibinfo {author}
  {\bibfnamefont {C.~M.}\ \bibnamefont {Marcus}},\ }\bibfield  {title}
  {\bibinfo {title} {Majorana bound state in a coupled quantum-dot
  hybrid-nanowire system},\ }\href@noop {} {\bibfield  {journal} {\bibinfo
  {journal} {Science}\ }\textbf {\bibinfo {volume} {354}},\ \bibinfo {pages}
  {1557} (\bibinfo {year} {2016})}\BibitemShut {NoStop}%
\bibitem [{\citenamefont {Heinrich}\ \emph {et~al.}(2018)\citenamefont
  {Heinrich}, \citenamefont {Pascual},\ and\ \citenamefont
  {Franke}}]{161-heinrich2018single}%
  \BibitemOpen
  \bibfield  {author} {\bibinfo {author} {\bibfnamefont {B.~W.}\ \bibnamefont
  {Heinrich}}, \bibinfo {author} {\bibfnamefont {J.~I.}\ \bibnamefont
  {Pascual}},\ and\ \bibinfo {author} {\bibfnamefont {K.~J.}\ \bibnamefont
  {Franke}},\ }\bibfield  {title} {\bibinfo {title} {Single magnetic adsorbates
  on s-wave superconductors},\ }\href@noop {} {\bibfield  {journal} {\bibinfo
  {journal} {Prog. Surf. Sci.}\ }\textbf {\bibinfo {volume} {93}},\ \bibinfo
  {pages} {1} (\bibinfo {year} {2018})}\BibitemShut {NoStop}%
\bibitem [{\citenamefont {Hatter}\ \emph {et~al.}(2015)\citenamefont {Hatter},
  \citenamefont {Heinrich}, \citenamefont {Ruby}, \citenamefont {Pascual},\
  and\ \citenamefont {Franke}}]{162-hatter2015magnetic}%
  \BibitemOpen
  \bibfield  {author} {\bibinfo {author} {\bibfnamefont {N.}~\bibnamefont
  {Hatter}}, \bibinfo {author} {\bibfnamefont {B.~W.}\ \bibnamefont
  {Heinrich}}, \bibinfo {author} {\bibfnamefont {M.}~\bibnamefont {Ruby}},
  \bibinfo {author} {\bibfnamefont {J.~I.}\ \bibnamefont {Pascual}},\ and\
  \bibinfo {author} {\bibfnamefont {K.~J.}\ \bibnamefont {Franke}},\ }\bibfield
   {title} {\bibinfo {title} {Magnetic anisotropy in {S}hiba bound states
  across a quantum phase transition},\ }\href@noop {} {\bibfield  {journal}
  {\bibinfo  {journal} {Nat. Commun.}\ }\textbf {\bibinfo {volume} {6}},\
  \bibinfo {pages} {8988} (\bibinfo {year} {2015})}\BibitemShut {NoStop}%
\bibitem [{\citenamefont {Kashiwaya}\ and\ \citenamefont
  {Tanaka}(2000)}]{163-kashiwaya2000tunnelling}%
  \BibitemOpen
  \bibfield  {author} {\bibinfo {author} {\bibfnamefont {S.}~\bibnamefont
  {Kashiwaya}}\ and\ \bibinfo {author} {\bibfnamefont {Y.}~\bibnamefont
  {Tanaka}},\ }\bibfield  {title} {\bibinfo {title} {Tunnelling effects on
  surface bound states in unconventional superconductors},\ }\href@noop {}
  {\bibfield  {journal} {\bibinfo  {journal} {Rep. Prog. Phys.}\ }\textbf
  {\bibinfo {volume} {63}},\ \bibinfo {pages} {1641} (\bibinfo {year}
  {2000})}\BibitemShut {NoStop}%
\bibitem [{\citenamefont {Kashiwaya}\ \emph {et~al.}(1996)\citenamefont
  {Kashiwaya}, \citenamefont {Tanaka}, \citenamefont {Koyanagi},\ and\
  \citenamefont {Kajimura}}]{164-kashiwaya1996theory}%
  \BibitemOpen
  \bibfield  {author} {\bibinfo {author} {\bibfnamefont {S.}~\bibnamefont
  {Kashiwaya}}, \bibinfo {author} {\bibfnamefont {Y.}~\bibnamefont {Tanaka}},
  \bibinfo {author} {\bibfnamefont {M.}~\bibnamefont {Koyanagi}},\ and\
  \bibinfo {author} {\bibfnamefont {K.}~\bibnamefont {Kajimura}},\ }\bibfield
  {title} {\bibinfo {title} {Theory for tunneling spectroscopy of anisotropic
  superconductors},\ }\href@noop {} {\bibfield  {journal} {\bibinfo  {journal}
  {Phys. Rev. B}\ }\textbf {\bibinfo {volume} {53}},\ \bibinfo {pages} {2667}
  (\bibinfo {year} {1996})}\BibitemShut {NoStop}%
\bibitem [{\citenamefont {L{\"o}fwander}\ \emph {et~al.}(2001)\citenamefont
  {L{\"o}fwander}, \citenamefont {Shumeiko},\ and\ \citenamefont
  {Wendin}}]{165-lofwander2001andreev}%
  \BibitemOpen
  \bibfield  {author} {\bibinfo {author} {\bibfnamefont {T.}~\bibnamefont
  {L{\"o}fwander}}, \bibinfo {author} {\bibfnamefont {V.}~\bibnamefont
  {Shumeiko}},\ and\ \bibinfo {author} {\bibfnamefont {G.}~\bibnamefont
  {Wendin}},\ }\bibfield  {title} {\bibinfo {title} {Andreev bound states in
  high-{T}$_{c}$ superconducting junctions},\ }\href@noop {} {\bibfield
  {journal} {\bibinfo  {journal} {Supercond. Sci. Tech.}\ }\textbf {\bibinfo
  {volume} {14}},\ \bibinfo {pages} {R53} (\bibinfo {year} {2001})}\BibitemShut
  {NoStop}%
\bibitem [{\citenamefont {Hu}(1994)}]{166-hu1994midgap}%
  \BibitemOpen
  \bibfield  {author} {\bibinfo {author} {\bibfnamefont {C.-R.}\ \bibnamefont
  {Hu}},\ }\bibfield  {title} {\bibinfo {title} {Midgap surface states as a
  novel signature for d$_{x_{a}^{2}-x_{b}^{2}}$-wave superconductivity},\
  }\href@noop {} {\bibfield  {journal} {\bibinfo  {journal} {Phys. Rev. Lett.}\
  }\textbf {\bibinfo {volume} {72}},\ \bibinfo {pages} {1526} (\bibinfo {year}
  {1994})}\BibitemShut {NoStop}%
\bibitem [{\citenamefont {Tanaka}\ \emph {et~al.}(2002)\citenamefont {Tanaka},
  \citenamefont {Tanuma}, \citenamefont {Kuroki},\ and\ \citenamefont
  {Kashiwaya}}]{167-tanaka2002theory}%
  \BibitemOpen
  \bibfield  {author} {\bibinfo {author} {\bibfnamefont {Y.}~\bibnamefont
  {Tanaka}}, \bibinfo {author} {\bibfnamefont {Y.}~\bibnamefont {Tanuma}},
  \bibinfo {author} {\bibfnamefont {K.}~\bibnamefont {Kuroki}},\ and\ \bibinfo
  {author} {\bibfnamefont {S.}~\bibnamefont {Kashiwaya}},\ }\bibfield  {title}
  {\bibinfo {title} {Theory of magnetotunneling spectroscopy in spin triplet
  p-wave superconductors},\ }\href@noop {} {\bibfield  {journal} {\bibinfo
  {journal} {J Phys. Soc. Jpn.}\ }\textbf {\bibinfo {volume} {71}},\ \bibinfo
  {pages} {2102} (\bibinfo {year} {2002})}\BibitemShut {NoStop}%
\bibitem [{\citenamefont {Tanaka}\ and\ \citenamefont
  {Tamura}(2018)}]{168-tanaka2018surface}%
  \BibitemOpen
  \bibfield  {author} {\bibinfo {author} {\bibfnamefont {Y.}~\bibnamefont
  {Tanaka}}\ and\ \bibinfo {author} {\bibfnamefont {S.}~\bibnamefont
  {Tamura}},\ }\bibfield  {title} {\bibinfo {title} {Surface {A}ndreev bound
  states and odd-frequency pairing in topological superconductor junctions},\
  }\href@noop {} {\bibfield  {journal} {\bibinfo  {journal} {J Low Temp.
  Phys.}\ }\textbf {\bibinfo {volume} {191}},\ \bibinfo {pages} {61} (\bibinfo
  {year} {2018})}\BibitemShut {NoStop}%
\bibitem [{\citenamefont {Kobayashi}\ \emph {et~al.}(2015)\citenamefont
  {Kobayashi}, \citenamefont {Tanaka},\ and\ \citenamefont
  {Sato}}]{169-kobayashi2015fragile}%
  \BibitemOpen
  \bibfield  {author} {\bibinfo {author} {\bibfnamefont {S.}~\bibnamefont
  {Kobayashi}}, \bibinfo {author} {\bibfnamefont {Y.}~\bibnamefont {Tanaka}},\
  and\ \bibinfo {author} {\bibfnamefont {M.}~\bibnamefont {Sato}},\ }\bibfield
  {title} {\bibinfo {title} {Fragile surface zero-energy flat bands in
  three-dimensional chiral superconductors},\ }\href@noop {} {\bibfield
  {journal} {\bibinfo  {journal} {Phys. Rev. B}\ }\textbf {\bibinfo {volume}
  {92}},\ \bibinfo {pages} {214514} (\bibinfo {year} {2015})}\BibitemShut
  {NoStop}%
\bibitem [{\citenamefont {Tamura}\ \emph {et~al.}(2017)\citenamefont {Tamura},
  \citenamefont {Kobayashi}, \citenamefont {Bo},\ and\ \citenamefont
  {Tanaka}}]{170-tamura2017theory}%
  \BibitemOpen
  \bibfield  {author} {\bibinfo {author} {\bibfnamefont {S.}~\bibnamefont
  {Tamura}}, \bibinfo {author} {\bibfnamefont {S.}~\bibnamefont {Kobayashi}},
  \bibinfo {author} {\bibfnamefont {L.}~\bibnamefont {Bo}},\ and\ \bibinfo
  {author} {\bibfnamefont {Y.}~\bibnamefont {Tanaka}},\ }\bibfield  {title}
  {\bibinfo {title} {Theory of surface {A}ndreev bound states and tunneling
  spectroscopy in three-dimensional chiral superconductors},\ }\href@noop {}
  {\bibfield  {journal} {\bibinfo  {journal} {Phys. Rev. B}\ }\textbf {\bibinfo
  {volume} {95}},\ \bibinfo {pages} {104511} (\bibinfo {year}
  {2017})}\BibitemShut {NoStop}%
\bibitem [{\citenamefont {Hsieh}\ and\ \citenamefont
  {Fu}(2012)}]{171-hsieh2012majorana}%
  \BibitemOpen
  \bibfield  {author} {\bibinfo {author} {\bibfnamefont {T.~H.}\ \bibnamefont
  {Hsieh}}\ and\ \bibinfo {author} {\bibfnamefont {L.}~\bibnamefont {Fu}},\
  }\bibfield  {title} {\bibinfo {title} {Majorana fermions and exotic surface
  {A}ndreev bound states in topological superconductors: application to
  {C}u$_{x}${B}i$_{2}${S}e$_{3}$},\ }\href@noop {} {\bibfield  {journal}
  {\bibinfo  {journal} {Phys. Rev. Lett.}\ }\textbf {\bibinfo {volume} {108}},\
  \bibinfo {pages} {107005} (\bibinfo {year} {2012})}\BibitemShut {NoStop}%
\bibitem [{\citenamefont {Wei}\ \emph {et~al.}(1998)\citenamefont {Wei},
  \citenamefont {Yeh}, \citenamefont {Garrigus},\ and\ \citenamefont
  {Strasik}}]{172-wei1998directional}%
  \BibitemOpen
  \bibfield  {author} {\bibinfo {author} {\bibfnamefont {J.}~\bibnamefont
  {Wei}}, \bibinfo {author} {\bibfnamefont {N.-C.}\ \bibnamefont {Yeh}},
  \bibinfo {author} {\bibfnamefont {D.}~\bibnamefont {Garrigus}},\ and\
  \bibinfo {author} {\bibfnamefont {M.}~\bibnamefont {Strasik}},\ }\bibfield
  {title} {\bibinfo {title} {Directional tunneling and {A}ndreev reflection on
  {Y}{B}a$_{2}${C}u$_{3}${O}$_{7-x}$ single crystals: predominance of d-wave
  pairing symmetry verified with the generalized {B}londer, {T}inkham, and
  {K}lapwijk theory},\ }\href@noop {} {\bibfield  {journal} {\bibinfo
  {journal} {Phys. Rev. Lett.}\ }\textbf {\bibinfo {volume} {81}},\ \bibinfo
  {pages} {2542} (\bibinfo {year} {1998})}\BibitemShut {NoStop}%
\bibitem [{\citenamefont {Kashiwaya}\ \emph {et~al.}(2011)\citenamefont
  {Kashiwaya}, \citenamefont {Kashiwaya}, \citenamefont {Kambara},
  \citenamefont {Furuta}, \citenamefont {Yaguchi}, \citenamefont {Tanaka},\
  and\ \citenamefont {Maeno}}]{173-kashiwaya2011edge}%
  \BibitemOpen
  \bibfield  {author} {\bibinfo {author} {\bibfnamefont {S.}~\bibnamefont
  {Kashiwaya}}, \bibinfo {author} {\bibfnamefont {H.}~\bibnamefont
  {Kashiwaya}}, \bibinfo {author} {\bibfnamefont {H.}~\bibnamefont {Kambara}},
  \bibinfo {author} {\bibfnamefont {T.}~\bibnamefont {Furuta}}, \bibinfo
  {author} {\bibfnamefont {H.}~\bibnamefont {Yaguchi}}, \bibinfo {author}
  {\bibfnamefont {Y.}~\bibnamefont {Tanaka}},\ and\ \bibinfo {author}
  {\bibfnamefont {Y.}~\bibnamefont {Maeno}},\ }\bibfield  {title} {\bibinfo
  {title} {Edge states of {S}r$_{2}${R}u{O}$_{4}$ detected by in-plane
  tunneling spectroscopy},\ }\href@noop {} {\bibfield  {journal} {\bibinfo
  {journal} {Phys. Rev. Lett.}\ }\textbf {\bibinfo {volume} {107}},\ \bibinfo
  {pages} {077003} (\bibinfo {year} {2011})}\BibitemShut {NoStop}%
\bibitem [{\citenamefont {Sasaki}\ \emph {et~al.}(2011)\citenamefont {Sasaki},
  \citenamefont {Kriener}, \citenamefont {Segawa}, \citenamefont {Yada},
  \citenamefont {Tanaka}, \citenamefont {Sato},\ and\ \citenamefont
  {Ando}}]{174-sasaki2011topological}%
  \BibitemOpen
  \bibfield  {author} {\bibinfo {author} {\bibfnamefont {S.}~\bibnamefont
  {Sasaki}}, \bibinfo {author} {\bibfnamefont {M.}~\bibnamefont {Kriener}},
  \bibinfo {author} {\bibfnamefont {K.}~\bibnamefont {Segawa}}, \bibinfo
  {author} {\bibfnamefont {K.}~\bibnamefont {Yada}}, \bibinfo {author}
  {\bibfnamefont {Y.}~\bibnamefont {Tanaka}}, \bibinfo {author} {\bibfnamefont
  {M.}~\bibnamefont {Sato}},\ and\ \bibinfo {author} {\bibfnamefont
  {Y.}~\bibnamefont {Ando}},\ }\bibfield  {title} {\bibinfo {title}
  {Topological superconductivity in {C}u$_{x}${B}i$_{2}${S}e$_{3}$},\
  }\href@noop {} {\bibfield  {journal} {\bibinfo  {journal} {Phys. Rev. Lett.}\
  }\textbf {\bibinfo {volume} {107}},\ \bibinfo {pages} {217001} (\bibinfo
  {year} {2011})}\BibitemShut {NoStop}%
\bibitem [{\citenamefont {Hess}\ \emph {et~al.}(1990)\citenamefont {Hess},
  \citenamefont {Robinson},\ and\ \citenamefont
  {Waszczak}}]{175-hess1990vortex}%
  \BibitemOpen
  \bibfield  {author} {\bibinfo {author} {\bibfnamefont {H.}~\bibnamefont
  {Hess}}, \bibinfo {author} {\bibfnamefont {R.}~\bibnamefont {Robinson}},\
  and\ \bibinfo {author} {\bibfnamefont {J.}~\bibnamefont {Waszczak}},\
  }\bibfield  {title} {\bibinfo {title} {Vortex-core structure observed with a
  scanning tunneling microscope},\ }\href@noop {} {\bibfield  {journal}
  {\bibinfo  {journal} {Phys. Rev. Lett.}\ }\textbf {\bibinfo {volume} {64}},\
  \bibinfo {pages} {2711} (\bibinfo {year} {1990})}\BibitemShut {NoStop}%
\bibitem [{\citenamefont {Hosur}\ \emph {et~al.}(2011)\citenamefont {Hosur},
  \citenamefont {Ghaemi}, \citenamefont {Mong},\ and\ \citenamefont
  {Vishwanath}}]{176-hosur2011majorana}%
  \BibitemOpen
  \bibfield  {author} {\bibinfo {author} {\bibfnamefont {P.}~\bibnamefont
  {Hosur}}, \bibinfo {author} {\bibfnamefont {P.}~\bibnamefont {Ghaemi}},
  \bibinfo {author} {\bibfnamefont {R.~S.}\ \bibnamefont {Mong}},\ and\
  \bibinfo {author} {\bibfnamefont {A.}~\bibnamefont {Vishwanath}},\ }\bibfield
   {title} {\bibinfo {title} {Majorana modes at the ends of superconductor
  vortices in doped topological insulators},\ }\href@noop {} {\bibfield
  {journal} {\bibinfo  {journal} {Phys. Rev. Lett.}\ }\textbf {\bibinfo
  {volume} {107}},\ \bibinfo {pages} {097001} (\bibinfo {year}
  {2011})}\BibitemShut {NoStop}%
\bibitem [{\citenamefont {Chiu}\ \emph {et~al.}(2011)\citenamefont {Chiu},
  \citenamefont {Gilbert},\ and\ \citenamefont {Hughes}}]{177-chiu2011vortex}%
  \BibitemOpen
  \bibfield  {author} {\bibinfo {author} {\bibfnamefont {C.-K.}\ \bibnamefont
  {Chiu}}, \bibinfo {author} {\bibfnamefont {M.~J.}\ \bibnamefont {Gilbert}},\
  and\ \bibinfo {author} {\bibfnamefont {T.~L.}\ \bibnamefont {Hughes}},\
  }\bibfield  {title} {\bibinfo {title} {Vortex lines in topological
  insulator-superconductor heterostructures},\ }\href@noop {} {\bibfield
  {journal} {\bibinfo  {journal} {Phys. Rev. B}\ }\textbf {\bibinfo {volume}
  {84}},\ \bibinfo {pages} {144507} (\bibinfo {year} {2011})}\BibitemShut
  {NoStop}%
\bibitem [{\citenamefont {Hung}\ \emph {et~al.}(2013)\citenamefont {Hung},
  \citenamefont {Ghaemi}, \citenamefont {Hughes},\ and\ \citenamefont
  {Gilbert}}]{178-hung2013vortex}%
  \BibitemOpen
  \bibfield  {author} {\bibinfo {author} {\bibfnamefont {H.-H.}\ \bibnamefont
  {Hung}}, \bibinfo {author} {\bibfnamefont {P.}~\bibnamefont {Ghaemi}},
  \bibinfo {author} {\bibfnamefont {T.~L.}\ \bibnamefont {Hughes}},\ and\
  \bibinfo {author} {\bibfnamefont {M.~J.}\ \bibnamefont {Gilbert}},\
  }\bibfield  {title} {\bibinfo {title} {Vortex lattices in the superconducting
  phases of doped topological insulators and heterostructures},\ }\href@noop {}
  {\bibfield  {journal} {\bibinfo  {journal} {Phys. Rev. B}\ }\textbf {\bibinfo
  {volume} {87}},\ \bibinfo {pages} {035401} (\bibinfo {year}
  {2013})}\BibitemShut {NoStop}%
\bibitem [{\citenamefont {Wang}\ and\ \citenamefont
  {Fu}(2017)}]{179-wang2017topological}%
  \BibitemOpen
  \bibfield  {author} {\bibinfo {author} {\bibfnamefont {Y.}~\bibnamefont
  {Wang}}\ and\ \bibinfo {author} {\bibfnamefont {L.}~\bibnamefont {Fu}},\
  }\bibfield  {title} {\bibinfo {title} {Topological phase transitions in
  multicomponent superconductors},\ }\href@noop {} {\bibfield  {journal}
  {\bibinfo  {journal} {Phys. Rev. Lett.}\ }\textbf {\bibinfo {volume} {119}},\
  \bibinfo {pages} {187003} (\bibinfo {year} {2017})}\BibitemShut {NoStop}%
\bibitem [{\citenamefont {Principi}\ \emph {et~al.}(2012)\citenamefont
  {Principi}, \citenamefont {Polini}, \citenamefont {Asgari},\ and\
  \citenamefont {MacDonald}}]{180-principi2012tunneling}%
  \BibitemOpen
  \bibfield  {author} {\bibinfo {author} {\bibfnamefont {A.}~\bibnamefont
  {Principi}}, \bibinfo {author} {\bibfnamefont {M.}~\bibnamefont {Polini}},
  \bibinfo {author} {\bibfnamefont {R.}~\bibnamefont {Asgari}},\ and\ \bibinfo
  {author} {\bibfnamefont {A.}~\bibnamefont {MacDonald}},\ }\bibfield  {title}
  {\bibinfo {title} {The tunneling density-of-states of interacting massless
  {D}irac fermions},\ }\href@noop {} {\bibfield  {journal} {\bibinfo  {journal}
  {Solid State Commun.}\ }\textbf {\bibinfo {volume} {152}},\ \bibinfo {pages}
  {1456} (\bibinfo {year} {2012})}\BibitemShut {NoStop}%
\bibitem [{\citenamefont {De~Gennes}(2018)}]{181-de2018superconductivity}%
  \BibitemOpen
  \bibfield  {author} {\bibinfo {author} {\bibfnamefont {P.-G.}\ \bibnamefont
  {De~Gennes}},\ }\href@noop {} {\emph {\bibinfo {title} {Superconductivity of
  metals and alloys}}}\ (\bibinfo  {publisher} {CRC Press},\ \bibinfo {year}
  {2018})\BibitemShut {NoStop}%
\bibitem [{\citenamefont {Joyez}\ \emph {et~al.}(1994)\citenamefont {Joyez},
  \citenamefont {Lafarge}, \citenamefont {Filipe}, \citenamefont {Esteve},\
  and\ \citenamefont {Devoret}}]{182-joyez1994observation}%
  \BibitemOpen
  \bibfield  {author} {\bibinfo {author} {\bibfnamefont {P.}~\bibnamefont
  {Joyez}}, \bibinfo {author} {\bibfnamefont {P.}~\bibnamefont {Lafarge}},
  \bibinfo {author} {\bibfnamefont {A.}~\bibnamefont {Filipe}}, \bibinfo
  {author} {\bibfnamefont {D.}~\bibnamefont {Esteve}},\ and\ \bibinfo {author}
  {\bibfnamefont {M.}~\bibnamefont {Devoret}},\ }\bibfield  {title} {\bibinfo
  {title} {Observation of parity-induced suppression of {J}osephson tunneling
  in the superconducting single electron transistor},\ }\href@noop {}
  {\bibfield  {journal} {\bibinfo  {journal} {Phys. Rev. Lett.}\ }\textbf
  {\bibinfo {volume} {72}},\ \bibinfo {pages} {2458} (\bibinfo {year}
  {1994})}\BibitemShut {NoStop}%
\bibitem [{\citenamefont {Aumentado}\ \emph {et~al.}(2004)\citenamefont
  {Aumentado}, \citenamefont {Keller}, \citenamefont {Martinis},\ and\
  \citenamefont {Devoret}}]{183-aumentado2004nonequilibrium}%
  \BibitemOpen
  \bibfield  {author} {\bibinfo {author} {\bibfnamefont {J.}~\bibnamefont
  {Aumentado}}, \bibinfo {author} {\bibfnamefont {M.~W.}\ \bibnamefont
  {Keller}}, \bibinfo {author} {\bibfnamefont {J.~M.}\ \bibnamefont
  {Martinis}},\ and\ \bibinfo {author} {\bibfnamefont {M.~H.}\ \bibnamefont
  {Devoret}},\ }\bibfield  {title} {\bibinfo {title} {Nonequilibrium
  quasiparticles and 2e periodicity in single-{c}ooper-pair transistors},\
  }\href@noop {} {\bibfield  {journal} {\bibinfo  {journal} {Phys. Rev. Lett.}\
  }\textbf {\bibinfo {volume} {92}},\ \bibinfo {pages} {066802} (\bibinfo
  {year} {2004})}\BibitemShut {NoStop}%
\bibitem [{\citenamefont {Cheng}\ \emph {et~al.}(2012)\citenamefont {Cheng},
  \citenamefont {Lutchyn},\ and\ \citenamefont
  {Sarma}}]{184-cheng2012topological}%
  \BibitemOpen
  \bibfield  {author} {\bibinfo {author} {\bibfnamefont {M.}~\bibnamefont
  {Cheng}}, \bibinfo {author} {\bibfnamefont {R.~M.}\ \bibnamefont {Lutchyn}},\
  and\ \bibinfo {author} {\bibfnamefont {S.~D.}\ \bibnamefont {Sarma}},\
  }\bibfield  {title} {\bibinfo {title} {Topological protection of {M}ajorana
  qubits},\ }\href@noop {} {\bibfield  {journal} {\bibinfo  {journal} {Phys.
  Rev. B}\ }\textbf {\bibinfo {volume} {85}},\ \bibinfo {pages} {165124}
  (\bibinfo {year} {2012})}\BibitemShut {NoStop}%
\bibitem [{\citenamefont {Rainis}\ and\ \citenamefont
  {Loss}(2012)}]{185-rainis2012majorana}%
  \BibitemOpen
  \bibfield  {author} {\bibinfo {author} {\bibfnamefont {D.}~\bibnamefont
  {Rainis}}\ and\ \bibinfo {author} {\bibfnamefont {D.}~\bibnamefont {Loss}},\
  }\bibfield  {title} {\bibinfo {title} {Majorana qubit decoherence by
  quasiparticle poisoning},\ }\href@noop {} {\bibfield  {journal} {\bibinfo
  {journal} {Phys. Rev. B}\ }\textbf {\bibinfo {volume} {85}},\ \bibinfo
  {pages} {174533} (\bibinfo {year} {2012})}\BibitemShut {NoStop}%
\bibitem [{\citenamefont {Colbert}\ and\ \citenamefont
  {Lee}(2014)}]{186-colbert2014proposal}%
  \BibitemOpen
  \bibfield  {author} {\bibinfo {author} {\bibfnamefont {J.~R.}\ \bibnamefont
  {Colbert}}\ and\ \bibinfo {author} {\bibfnamefont {P.~A.}\ \bibnamefont
  {Lee}},\ }\bibfield  {title} {\bibinfo {title} {Proposal to measure the
  quasiparticle poisoning time of {M}ajorana bound states},\ }\href@noop {}
  {\bibfield  {journal} {\bibinfo  {journal} {Phys. Rev. B}\ }\textbf {\bibinfo
  {volume} {89}},\ \bibinfo {pages} {140505} (\bibinfo {year}
  {2014})}\BibitemShut {NoStop}%
\bibitem [{\citenamefont {Sarma}\ \emph {et~al.}(2016)\citenamefont {Sarma},
  \citenamefont {Nag},\ and\ \citenamefont {Sau}}]{187-sarma2016infer}%
  \BibitemOpen
  \bibfield  {author} {\bibinfo {author} {\bibfnamefont {S.~D.}\ \bibnamefont
  {Sarma}}, \bibinfo {author} {\bibfnamefont {A.}~\bibnamefont {Nag}},\ and\
  \bibinfo {author} {\bibfnamefont {J.~D.}\ \bibnamefont {Sau}},\ }\bibfield
  {title} {\bibinfo {title} {How to infer non-{A}belian statistics and
  topological visibility from tunneling conductance properties of realistic
  {M}ajorana nanowires},\ }\href@noop {} {\bibfield  {journal} {\bibinfo
  {journal} {Phys. Rev. B}\ }\textbf {\bibinfo {volume} {94}},\ \bibinfo
  {pages} {035143} (\bibinfo {year} {2016})}\BibitemShut {NoStop}%
\bibitem [{\citenamefont {Van~Woerkom}\ \emph {et~al.}(2015)\citenamefont
  {Van~Woerkom}, \citenamefont {Geresdi},\ and\ \citenamefont
  {Kouwenhoven}}]{188-van2015one}%
  \BibitemOpen
  \bibfield  {author} {\bibinfo {author} {\bibfnamefont {D.~J.}\ \bibnamefont
  {Van~Woerkom}}, \bibinfo {author} {\bibfnamefont {A.}~\bibnamefont
  {Geresdi}},\ and\ \bibinfo {author} {\bibfnamefont {L.~P.}\ \bibnamefont
  {Kouwenhoven}},\ }\bibfield  {title} {\bibinfo {title} {One minute parity
  lifetime of a {N}b{T}i{N} {C}ooper-pair transistor},\ }\href@noop {}
  {\bibfield  {journal} {\bibinfo  {journal} {Nat. Phys.}\ }\textbf {\bibinfo
  {volume} {11}},\ \bibinfo {pages} {547} (\bibinfo {year} {2015})}\BibitemShut
  {NoStop}%
\bibitem [{\citenamefont {Higginbotham}\ \emph {et~al.}(2015)\citenamefont
  {Higginbotham}, \citenamefont {Albrecht}, \citenamefont {Kir{\v{s}}anskas},
  \citenamefont {Chang}, \citenamefont {Kuemmeth}, \citenamefont {Krogstrup},
  \citenamefont {Jespersen}, \citenamefont {Nyg{\aa}rd}, \citenamefont
  {Flensberg},\ and\ \citenamefont {Marcus}}]{189-higginbotham2015parity}%
  \BibitemOpen
  \bibfield  {author} {\bibinfo {author} {\bibfnamefont {A.~P.}\ \bibnamefont
  {Higginbotham}}, \bibinfo {author} {\bibfnamefont {S.~M.}\ \bibnamefont
  {Albrecht}}, \bibinfo {author} {\bibfnamefont {G.}~\bibnamefont
  {Kir{\v{s}}anskas}}, \bibinfo {author} {\bibfnamefont {W.}~\bibnamefont
  {Chang}}, \bibinfo {author} {\bibfnamefont {F.}~\bibnamefont {Kuemmeth}},
  \bibinfo {author} {\bibfnamefont {P.}~\bibnamefont {Krogstrup}}, \bibinfo
  {author} {\bibfnamefont {T.~S.}\ \bibnamefont {Jespersen}}, \bibinfo {author}
  {\bibfnamefont {J.}~\bibnamefont {Nyg{\aa}rd}}, \bibinfo {author}
  {\bibfnamefont {K.}~\bibnamefont {Flensberg}},\ and\ \bibinfo {author}
  {\bibfnamefont {C.~M.}\ \bibnamefont {Marcus}},\ }\bibfield  {title}
  {\bibinfo {title} {Parity lifetime of bound states in a proximitized
  semiconductor nanowire},\ }\href@noop {} {\bibfield  {journal} {\bibinfo
  {journal} {Nat. Phys.}\ }\textbf {\bibinfo {volume} {11}},\ \bibinfo {pages}
  {1017} (\bibinfo {year} {2015})}\BibitemShut {NoStop}%
\bibitem [{\citenamefont {Albrecht}\ \emph {et~al.}(2017)\citenamefont
  {Albrecht}, \citenamefont {Hansen}, \citenamefont {Higginbotham},
  \citenamefont {Kuemmeth}, \citenamefont {Jespersen}, \citenamefont
  {Nyg{\aa}rd}, \citenamefont {Krogstrup}, \citenamefont {Danon}, \citenamefont
  {Flensberg},\ and\ \citenamefont {Marcus}}]{190-albrecht2017transport}%
  \BibitemOpen
  \bibfield  {author} {\bibinfo {author} {\bibfnamefont {S.}~\bibnamefont
  {Albrecht}}, \bibinfo {author} {\bibfnamefont {E.}~\bibnamefont {Hansen}},
  \bibinfo {author} {\bibfnamefont {A.~P.}\ \bibnamefont {Higginbotham}},
  \bibinfo {author} {\bibfnamefont {F.}~\bibnamefont {Kuemmeth}}, \bibinfo
  {author} {\bibfnamefont {T.}~\bibnamefont {Jespersen}}, \bibinfo {author}
  {\bibfnamefont {J.}~\bibnamefont {Nyg{\aa}rd}}, \bibinfo {author}
  {\bibfnamefont {P.}~\bibnamefont {Krogstrup}}, \bibinfo {author}
  {\bibfnamefont {J.}~\bibnamefont {Danon}}, \bibinfo {author} {\bibfnamefont
  {K.}~\bibnamefont {Flensberg}},\ and\ \bibinfo {author} {\bibfnamefont
  {C.}~\bibnamefont {Marcus}},\ }\bibfield  {title} {\bibinfo {title}
  {Transport signatures of quasiparticle poisoning in a {M}ajorana island},\
  }\href@noop {} {\bibfield  {journal} {\bibinfo  {journal} {Phys. Rev. Lett.}\
  }\textbf {\bibinfo {volume} {118}},\ \bibinfo {pages} {137701} (\bibinfo
  {year} {2017})}\BibitemShut {NoStop}%
\bibitem [{\citenamefont {Klein}\ \emph {et~al.}(2014)\citenamefont {Klein},
  \citenamefont {Grasland}, \citenamefont {Cercellier}, \citenamefont
  {Toulemonde},\ and\ \citenamefont {Marcenat}}]{191-klein2014vortex}%
  \BibitemOpen
  \bibfield  {author} {\bibinfo {author} {\bibfnamefont {T.}~\bibnamefont
  {Klein}}, \bibinfo {author} {\bibfnamefont {H.}~\bibnamefont {Grasland}},
  \bibinfo {author} {\bibfnamefont {H.}~\bibnamefont {Cercellier}}, \bibinfo
  {author} {\bibfnamefont {P.}~\bibnamefont {Toulemonde}},\ and\ \bibinfo
  {author} {\bibfnamefont {C.}~\bibnamefont {Marcenat}},\ }\bibfield  {title}
  {\bibinfo {title} {Vortex creep down to 0.3 {K} in superconducting
  {F}e({T}e,{S}e) single crystals},\ }\href@noop {} {\bibfield  {journal}
  {\bibinfo  {journal} {Phys. Rev. B}\ }\textbf {\bibinfo {volume} {89}},\
  \bibinfo {pages} {014514} (\bibinfo {year} {2014})}\BibitemShut {NoStop}%
\bibitem [{\citenamefont {Eley}\ \emph {et~al.}(2017)\citenamefont {Eley},
  \citenamefont {Miura}, \citenamefont {Maiorov},\ and\ \citenamefont
  {Civale}}]{192-eley2017universal}%
  \BibitemOpen
  \bibfield  {author} {\bibinfo {author} {\bibfnamefont {S.}~\bibnamefont
  {Eley}}, \bibinfo {author} {\bibfnamefont {M.}~\bibnamefont {Miura}},
  \bibinfo {author} {\bibfnamefont {B.}~\bibnamefont {Maiorov}},\ and\ \bibinfo
  {author} {\bibfnamefont {L.}~\bibnamefont {Civale}},\ }\bibfield  {title}
  {\bibinfo {title} {Universal lower limit on vortex creep in
  superconductors},\ }\href@noop {} {\bibfield  {journal} {\bibinfo  {journal}
  {Nat. Mater.}\ }\textbf {\bibinfo {volume} {16}},\ \bibinfo {pages} {409}
  (\bibinfo {year} {2017})}\BibitemShut {NoStop}%
\bibitem [{\citenamefont {Wray}\ \emph {et~al.}(2010)\citenamefont {Wray},
  \citenamefont {Xu}, \citenamefont {Xia}, \citenamefont {San~Hor},
  \citenamefont {Qian}, \citenamefont {Fedorov}, \citenamefont {Lin},
  \citenamefont {Bansil}, \citenamefont {Cava},\ and\ \citenamefont
  {Hasan}}]{193-wray2010observation}%
  \BibitemOpen
  \bibfield  {author} {\bibinfo {author} {\bibfnamefont {L.~A.}\ \bibnamefont
  {Wray}}, \bibinfo {author} {\bibfnamefont {S.-Y.}\ \bibnamefont {Xu}},
  \bibinfo {author} {\bibfnamefont {Y.}~\bibnamefont {Xia}}, \bibinfo {author}
  {\bibfnamefont {Y.}~\bibnamefont {San~Hor}}, \bibinfo {author} {\bibfnamefont
  {D.}~\bibnamefont {Qian}}, \bibinfo {author} {\bibfnamefont {A.~V.}\
  \bibnamefont {Fedorov}}, \bibinfo {author} {\bibfnamefont {H.}~\bibnamefont
  {Lin}}, \bibinfo {author} {\bibfnamefont {A.}~\bibnamefont {Bansil}},
  \bibinfo {author} {\bibfnamefont {R.~J.}\ \bibnamefont {Cava}},\ and\
  \bibinfo {author} {\bibfnamefont {M.~Z.}\ \bibnamefont {Hasan}},\ }\bibfield
  {title} {\bibinfo {title} {Observation of topological order in a
  superconducting doped topological insulator},\ }\href@noop {} {\bibfield
  {journal} {\bibinfo  {journal} {Nat. Phys.}\ }\textbf {\bibinfo {volume}
  {6}},\ \bibinfo {pages} {855} (\bibinfo {year} {2010})}\BibitemShut {NoStop}%
\bibitem [{\citenamefont {Altland}\ and\ \citenamefont
  {Zirnbauer}(1997)}]{194-altland1997nonstandard}%
  \BibitemOpen
  \bibfield  {author} {\bibinfo {author} {\bibfnamefont {A.}~\bibnamefont
  {Altland}}\ and\ \bibinfo {author} {\bibfnamefont {M.~R.}\ \bibnamefont
  {Zirnbauer}},\ }\bibfield  {title} {\bibinfo {title} {Nonstandard symmetry
  classes in mesoscopic normal-superconducting hybrid structures},\ }\href@noop
  {} {\bibfield  {journal} {\bibinfo  {journal} {Phys. Rev. B}\ }\textbf
  {\bibinfo {volume} {55}},\ \bibinfo {pages} {1142} (\bibinfo {year}
  {1997})}\BibitemShut {NoStop}%
\bibitem [{\citenamefont {Schnyder}\ \emph {et~al.}(2008)\citenamefont
  {Schnyder}, \citenamefont {Ryu}, \citenamefont {Furusaki},\ and\
  \citenamefont {Ludwig}}]{195-schnyder2008classification}%
  \BibitemOpen
  \bibfield  {author} {\bibinfo {author} {\bibfnamefont {A.~P.}\ \bibnamefont
  {Schnyder}}, \bibinfo {author} {\bibfnamefont {S.}~\bibnamefont {Ryu}},
  \bibinfo {author} {\bibfnamefont {A.}~\bibnamefont {Furusaki}},\ and\
  \bibinfo {author} {\bibfnamefont {A.~W.}\ \bibnamefont {Ludwig}},\ }\bibfield
   {title} {\bibinfo {title} {Classification of topological insulators and
  superconductors in three spatial dimensions},\ }\href@noop {} {\bibfield
  {journal} {\bibinfo  {journal} {Phys. Rev. B}\ }\textbf {\bibinfo {volume}
  {78}},\ \bibinfo {pages} {195125} (\bibinfo {year} {2008})}\BibitemShut
  {NoStop}%
\bibitem [{\citenamefont {Teo}\ and\ \citenamefont
  {Kane}(2010)}]{196-teo2010topological}%
  \BibitemOpen
  \bibfield  {author} {\bibinfo {author} {\bibfnamefont {J.~C.}\ \bibnamefont
  {Teo}}\ and\ \bibinfo {author} {\bibfnamefont {C.~L.}\ \bibnamefont {Kane}},\
  }\bibfield  {title} {\bibinfo {title} {Topological defects and gapless modes
  in insulators and superconductors},\ }\href@noop {} {\bibfield  {journal}
  {\bibinfo  {journal} {Phys. Rev. B}\ }\textbf {\bibinfo {volume} {82}},\
  \bibinfo {pages} {115120} (\bibinfo {year} {2010})}\BibitemShut {NoStop}%
\bibitem [{\citenamefont {Chiu}\ \emph {et~al.}(2016)\citenamefont {Chiu},
  \citenamefont {Teo}, \citenamefont {Schnyder},\ and\ \citenamefont
  {Ryu}}]{197-chiu2016classification}%
  \BibitemOpen
  \bibfield  {author} {\bibinfo {author} {\bibfnamefont {C.-K.}\ \bibnamefont
  {Chiu}}, \bibinfo {author} {\bibfnamefont {J.~C.}\ \bibnamefont {Teo}},
  \bibinfo {author} {\bibfnamefont {A.~P.}\ \bibnamefont {Schnyder}},\ and\
  \bibinfo {author} {\bibfnamefont {S.}~\bibnamefont {Ryu}},\ }\bibfield
  {title} {\bibinfo {title} {Classification of topological quantum matter with
  symmetries},\ }\href@noop {} {\bibfield  {journal} {\bibinfo  {journal} {Rev.
  Mod. Phys.}\ }\textbf {\bibinfo {volume} {88}},\ \bibinfo {pages} {035005}
  (\bibinfo {year} {2016})}\BibitemShut {NoStop}%
\bibitem [{\citenamefont {Xu}\ \emph {et~al.}(2016)\citenamefont {Xu},
  \citenamefont {Lian}, \citenamefont {Tang}, \citenamefont {Qi},\ and\
  \citenamefont {Zhang}}]{198-xu2016topological}%
  \BibitemOpen
  \bibfield  {author} {\bibinfo {author} {\bibfnamefont {G.}~\bibnamefont
  {Xu}}, \bibinfo {author} {\bibfnamefont {B.}~\bibnamefont {Lian}}, \bibinfo
  {author} {\bibfnamefont {P.}~\bibnamefont {Tang}}, \bibinfo {author}
  {\bibfnamefont {X.-L.}\ \bibnamefont {Qi}},\ and\ \bibinfo {author}
  {\bibfnamefont {S.-C.}\ \bibnamefont {Zhang}},\ }\bibfield  {title} {\bibinfo
  {title} {Topological superconductivity on the surface of {F}e-based
  superconductors},\ }\href@noop {} {\bibfield  {journal} {\bibinfo  {journal}
  {Phys. Rev. Lett.}\ }\textbf {\bibinfo {volume} {117}},\ \bibinfo {pages}
  {047001} (\bibinfo {year} {2016})}\BibitemShut {NoStop}%
\bibitem [{\citenamefont {Chiu}\ \emph {et~al.}(2012)\citenamefont {Chiu},
  \citenamefont {Ghaemi},\ and\ \citenamefont
  {Hughes}}]{199-chiu2012stabilization}%
  \BibitemOpen
  \bibfield  {author} {\bibinfo {author} {\bibfnamefont {C.-K.}\ \bibnamefont
  {Chiu}}, \bibinfo {author} {\bibfnamefont {P.}~\bibnamefont {Ghaemi}},\ and\
  \bibinfo {author} {\bibfnamefont {T.~L.}\ \bibnamefont {Hughes}},\ }\bibfield
   {title} {\bibinfo {title} {Stabilization of {M}ajorana modes in magnetic
  vortices in the superconducting phase of topological insulators using
  topologically trivial bands},\ }\href@noop {} {\bibfield  {journal} {\bibinfo
   {journal} {Phys. Rev. Lett.}\ }\textbf {\bibinfo {volume} {109}},\ \bibinfo
  {pages} {237009} (\bibinfo {year} {2012})}\BibitemShut {NoStop}%
\bibitem [{\citenamefont {Ghazaryan}\ \emph {et~al.}(2020)\citenamefont
  {Ghazaryan}, \citenamefont {Lopes}, \citenamefont {Hosur}, \citenamefont
  {Gilbert},\ and\ \citenamefont {Ghaemi}}]{200-ghazaryan2020effect}%
  \BibitemOpen
  \bibfield  {author} {\bibinfo {author} {\bibfnamefont {A.}~\bibnamefont
  {Ghazaryan}}, \bibinfo {author} {\bibfnamefont {P.~L.}\ \bibnamefont
  {Lopes}}, \bibinfo {author} {\bibfnamefont {P.}~\bibnamefont {Hosur}},
  \bibinfo {author} {\bibfnamefont {M.~J.}\ \bibnamefont {Gilbert}},\ and\
  \bibinfo {author} {\bibfnamefont {P.}~\bibnamefont {Ghaemi}},\ }\bibfield
  {title} {\bibinfo {title} {Effect of zeeman coupling on the {M}ajorana vortex
  modes in iron-based topological superconductors},\ }\href@noop {} {\bibfield
  {journal} {\bibinfo  {journal} {Phys. Rev. B}\ }\textbf {\bibinfo {volume}
  {101}},\ \bibinfo {pages} {020504} (\bibinfo {year} {2020})}\BibitemShut
  {NoStop}%
\bibitem [{\citenamefont {Qin}\ \emph {et~al.}(2019{\natexlab{a}})\citenamefont
  {Qin}, \citenamefont {Hu}, \citenamefont {Wu}, \citenamefont {Dai},
  \citenamefont {Fang}, \citenamefont {Zhang},\ and\ \citenamefont
  {Hu}}]{201-qin2019topological}%
  \BibitemOpen
  \bibfield  {author} {\bibinfo {author} {\bibfnamefont {S.}~\bibnamefont
  {Qin}}, \bibinfo {author} {\bibfnamefont {L.}~\bibnamefont {Hu}}, \bibinfo
  {author} {\bibfnamefont {X.}~\bibnamefont {Wu}}, \bibinfo {author}
  {\bibfnamefont {X.}~\bibnamefont {Dai}}, \bibinfo {author} {\bibfnamefont
  {C.}~\bibnamefont {Fang}}, \bibinfo {author} {\bibfnamefont {F.-C.}\
  \bibnamefont {Zhang}},\ and\ \bibinfo {author} {\bibfnamefont
  {J.}~\bibnamefont {Hu}},\ }\bibfield  {title} {\bibinfo {title} {Topological
  vortex phase transitions in iron-based superconductors},\ }\href@noop {}
  {\bibfield  {journal} {\bibinfo  {journal} {Sci. Bull.}\ }\textbf {\bibinfo
  {volume} {64}},\ \bibinfo {pages} {1207} (\bibinfo {year}
  {2019}{\natexlab{a}})}\BibitemShut {NoStop}%
\bibitem [{\citenamefont {K{\"o}nig}\ and\ \citenamefont
  {Coleman}(2019)}]{202-konig2019crystalline}%
  \BibitemOpen
  \bibfield  {author} {\bibinfo {author} {\bibfnamefont {E.~J.}\ \bibnamefont
  {K{\"o}nig}}\ and\ \bibinfo {author} {\bibfnamefont {P.}~\bibnamefont
  {Coleman}},\ }\bibfield  {title} {\bibinfo {title}
  {Crystalline-symmetry-protected helical {M}ajorana modes in the iron
  pnictides},\ }\href@noop {} {\bibfield  {journal} {\bibinfo  {journal} {Phys.
  Rev. Lett.}\ }\textbf {\bibinfo {volume} {122}},\ \bibinfo {pages} {207001}
  (\bibinfo {year} {2019})}\BibitemShut {NoStop}%
\bibitem [{\citenamefont {Qin}\ \emph {et~al.}(2019{\natexlab{b}})\citenamefont
  {Qin}, \citenamefont {Hu}, \citenamefont {Le}, \citenamefont {Zeng},
  \citenamefont {Zhang}, \citenamefont {Fang},\ and\ \citenamefont
  {Hu}}]{203-qin2019quasi}%
  \BibitemOpen
  \bibfield  {author} {\bibinfo {author} {\bibfnamefont {S.}~\bibnamefont
  {Qin}}, \bibinfo {author} {\bibfnamefont {L.}~\bibnamefont {Hu}}, \bibinfo
  {author} {\bibfnamefont {C.}~\bibnamefont {Le}}, \bibinfo {author}
  {\bibfnamefont {J.}~\bibnamefont {Zeng}}, \bibinfo {author} {\bibfnamefont
  {F.-C.}\ \bibnamefont {Zhang}}, \bibinfo {author} {\bibfnamefont
  {C.}~\bibnamefont {Fang}},\ and\ \bibinfo {author} {\bibfnamefont
  {J.}~\bibnamefont {Hu}},\ }\bibfield  {title} {\bibinfo {title} {Quasi-1{D}
  topological nodal vortex line phase in doped superconducting 3{D} {D}irac
  semimetals},\ }\href@noop {} {\bibfield  {journal} {\bibinfo  {journal}
  {Phys. Rev. Lett.}\ }\textbf {\bibinfo {volume} {123}},\ \bibinfo {pages}
  {027003} (\bibinfo {year} {2019}{\natexlab{b}})}\BibitemShut {NoStop}%
\bibitem [{\citenamefont {Law}\ \emph {et~al.}(2009)\citenamefont {Law},
  \citenamefont {Lee},\ and\ \citenamefont {Ng}}]{204-law2009majorana}%
  \BibitemOpen
  \bibfield  {author} {\bibinfo {author} {\bibfnamefont {K.~T.}\ \bibnamefont
  {Law}}, \bibinfo {author} {\bibfnamefont {P.~A.}\ \bibnamefont {Lee}},\ and\
  \bibinfo {author} {\bibfnamefont {T.~K.}\ \bibnamefont {Ng}},\ }\bibfield
  {title} {\bibinfo {title} {Majorana fermion induced resonant {A}ndreev
  reflection},\ }\href@noop {} {\bibfield  {journal} {\bibinfo  {journal}
  {Phys. Rev. Lett.}\ }\textbf {\bibinfo {volume} {103}},\ \bibinfo {pages}
  {237001} (\bibinfo {year} {2009})}\BibitemShut {NoStop}%
\bibitem [{\citenamefont {Chang}\ \emph {et~al.}(2012)\citenamefont {Chang},
  \citenamefont {Mendez},\ and\ \citenamefont
  {Tejedor}}]{207-chang2012resonant}%
  \BibitemOpen
  \bibfield  {author} {\bibinfo {author} {\bibfnamefont {L.~L.}\ \bibnamefont
  {Chang}}, \bibinfo {author} {\bibfnamefont {E.}~\bibnamefont {Mendez}},\ and\
  \bibinfo {author} {\bibfnamefont {C.}~\bibnamefont {Tejedor}},\ }\href@noop
  {} {\emph {\bibinfo {title} {Resonant Tunneling in Semiconductors: Physics
  and Applications}}},\ Vol.\ \bibinfo {volume} {277}\ (\bibinfo  {publisher}
  {Springer Science \& Business Media},\ \bibinfo {year} {2012})\BibitemShut
  {NoStop}%
\bibitem [{\citenamefont {Tsu}\ and\ \citenamefont
  {Esaki}(1973)}]{205-tsu1973tunneling}%
  \BibitemOpen
  \bibfield  {author} {\bibinfo {author} {\bibfnamefont {R.}~\bibnamefont
  {Tsu}}\ and\ \bibinfo {author} {\bibfnamefont {L.}~\bibnamefont {Esaki}},\
  }\bibfield  {title} {\bibinfo {title} {Tunneling in a finite superlattice},\
  }\href@noop {} {\bibfield  {journal} {\bibinfo  {journal} {Appl. Phys.
  Lett.}\ }\textbf {\bibinfo {volume} {22}},\ \bibinfo {pages} {562} (\bibinfo
  {year} {1973})}\BibitemShut {NoStop}%
\bibitem [{\citenamefont {Chang}\ \emph {et~al.}(1974)\citenamefont {Chang},
  \citenamefont {Esaki},\ and\ \citenamefont {Tsu}}]{206-chang1974resonant}%
  \BibitemOpen
  \bibfield  {author} {\bibinfo {author} {\bibfnamefont {L.}~\bibnamefont
  {Chang}}, \bibinfo {author} {\bibfnamefont {L.}~\bibnamefont {Esaki}},\ and\
  \bibinfo {author} {\bibfnamefont {R.}~\bibnamefont {Tsu}},\ }\bibfield
  {title} {\bibinfo {title} {Resonant tunneling in semiconductor double
  barriers},\ }\href@noop {} {\bibfield  {journal} {\bibinfo  {journal} {Appl.
  Phys. Lett.}\ }\textbf {\bibinfo {volume} {24}},\ \bibinfo {pages} {593}
  (\bibinfo {year} {1974})}\BibitemShut {NoStop}%
\bibitem [{\citenamefont {He}\ \emph {et~al.}(2014)\citenamefont {He},
  \citenamefont {Ng}, \citenamefont {Lee},\ and\ \citenamefont
  {Law}}]{208-he2014selective}%
  \BibitemOpen
  \bibfield  {author} {\bibinfo {author} {\bibfnamefont {J.~J.}\ \bibnamefont
  {He}}, \bibinfo {author} {\bibfnamefont {T.~K.}\ \bibnamefont {Ng}}, \bibinfo
  {author} {\bibfnamefont {P.~A.}\ \bibnamefont {Lee}},\ and\ \bibinfo {author}
  {\bibfnamefont {K.~T.}\ \bibnamefont {Law}},\ }\bibfield  {title} {\bibinfo
  {title} {Selective equal-spin {A}ndreev reflections induced by {M}ajorana
  fermions},\ }\href@noop {} {\bibfield  {journal} {\bibinfo  {journal} {Phys.
  Rev. Lett.}\ }\textbf {\bibinfo {volume} {112}},\ \bibinfo {pages} {037001}
  (\bibinfo {year} {2014})}\BibitemShut {NoStop}%
\bibitem [{\citenamefont {Haim}\ \emph {et~al.}(2015)\citenamefont {Haim},
  \citenamefont {Berg}, \citenamefont {von Oppen},\ and\ \citenamefont
  {Oreg}}]{209-haim2015signatures}%
  \BibitemOpen
  \bibfield  {author} {\bibinfo {author} {\bibfnamefont {A.}~\bibnamefont
  {Haim}}, \bibinfo {author} {\bibfnamefont {E.}~\bibnamefont {Berg}}, \bibinfo
  {author} {\bibfnamefont {F.}~\bibnamefont {von Oppen}},\ and\ \bibinfo
  {author} {\bibfnamefont {Y.}~\bibnamefont {Oreg}},\ }\bibfield  {title}
  {\bibinfo {title} {Signatures of {M}ajorana zero modes in spin-resolved
  current correlations},\ }\href@noop {} {\bibfield  {journal} {\bibinfo
  {journal} {Phys. Rev. Lett.}\ }\textbf {\bibinfo {volume} {114}},\ \bibinfo
  {pages} {166406} (\bibinfo {year} {2015})}\BibitemShut {NoStop}%
\bibitem [{\citenamefont {Kawakami}\ and\ \citenamefont
  {Hu}(2015)}]{210-kawakami2015evolution}%
  \BibitemOpen
  \bibfield  {author} {\bibinfo {author} {\bibfnamefont {T.}~\bibnamefont
  {Kawakami}}\ and\ \bibinfo {author} {\bibfnamefont {X.}~\bibnamefont {Hu}},\
  }\bibfield  {title} {\bibinfo {title} {Evolution of density of states and a
  spin-resolved checkerboard-type pattern associated with the {M}ajorana bound
  state},\ }\href@noop {} {\bibfield  {journal} {\bibinfo  {journal} {Phys.
  Rev. Lett.}\ }\textbf {\bibinfo {volume} {115}},\ \bibinfo {pages} {177001}
  (\bibinfo {year} {2015})}\BibitemShut {NoStop}%
\bibitem [{\citenamefont {Hu}\ \emph {et~al.}(2016)\citenamefont {Hu},
  \citenamefont {Li}, \citenamefont {Xu}, \citenamefont {Zhou},\ and\
  \citenamefont {Zhang}}]{211-hu2016theory}%
  \BibitemOpen
  \bibfield  {author} {\bibinfo {author} {\bibfnamefont {L.-H.}\ \bibnamefont
  {Hu}}, \bibinfo {author} {\bibfnamefont {C.}~\bibnamefont {Li}}, \bibinfo
  {author} {\bibfnamefont {D.-H.}\ \bibnamefont {Xu}}, \bibinfo {author}
  {\bibfnamefont {Y.}~\bibnamefont {Zhou}},\ and\ \bibinfo {author}
  {\bibfnamefont {F.-C.}\ \bibnamefont {Zhang}},\ }\bibfield  {title} {\bibinfo
  {title} {Theory of spin-selective {A}ndreev reflection in the vortex core of
  a topological superconductor},\ }\href@noop {} {\bibfield  {journal}
  {\bibinfo  {journal} {Phys. Rev. B}\ }\textbf {\bibinfo {volume} {94}},\
  \bibinfo {pages} {224501} (\bibinfo {year} {2016})}\BibitemShut {NoStop}%
\bibitem [{\citenamefont {Flensberg}(2010)}]{212-flensberg2010tunneling}%
  \BibitemOpen
  \bibfield  {author} {\bibinfo {author} {\bibfnamefont {K.}~\bibnamefont
  {Flensberg}},\ }\bibfield  {title} {\bibinfo {title} {Tunneling
  characteristics of a chain of {M}ajorana bound states},\ }\href@noop {}
  {\bibfield  {journal} {\bibinfo  {journal} {Phys. Rev. B}\ }\textbf {\bibinfo
  {volume} {82}},\ \bibinfo {pages} {180516} (\bibinfo {year}
  {2010})}\BibitemShut {NoStop}%
\bibitem [{\citenamefont {Wimmer}\ \emph {et~al.}(2011)\citenamefont {Wimmer},
  \citenamefont {Akhmerov}, \citenamefont {Dahlhaus},\ and\ \citenamefont
  {Beenakker}}]{213-wimmer2011quantum}%
  \BibitemOpen
  \bibfield  {author} {\bibinfo {author} {\bibfnamefont {M.}~\bibnamefont
  {Wimmer}}, \bibinfo {author} {\bibfnamefont {A.}~\bibnamefont {Akhmerov}},
  \bibinfo {author} {\bibfnamefont {J.}~\bibnamefont {Dahlhaus}},\ and\
  \bibinfo {author} {\bibfnamefont {C.}~\bibnamefont {Beenakker}},\ }\bibfield
  {title} {\bibinfo {title} {Quantum point contact as a probe of a topological
  superconductor},\ }\href@noop {} {\bibfield  {journal} {\bibinfo  {journal}
  {New J Phys.}\ }\textbf {\bibinfo {volume} {13}},\ \bibinfo {pages} {053016}
  (\bibinfo {year} {2011})}\BibitemShut {NoStop}%
\bibitem [{\citenamefont {Fidkowski}\ \emph {et~al.}(2012)\citenamefont
  {Fidkowski}, \citenamefont {Alicea}, \citenamefont {Lindner}, \citenamefont
  {Lutchyn},\ and\ \citenamefont {Fisher}}]{214-fidkowski2012universal}%
  \BibitemOpen
  \bibfield  {author} {\bibinfo {author} {\bibfnamefont {L.}~\bibnamefont
  {Fidkowski}}, \bibinfo {author} {\bibfnamefont {J.}~\bibnamefont {Alicea}},
  \bibinfo {author} {\bibfnamefont {N.~H.}\ \bibnamefont {Lindner}}, \bibinfo
  {author} {\bibfnamefont {R.~M.}\ \bibnamefont {Lutchyn}},\ and\ \bibinfo
  {author} {\bibfnamefont {M.~P.}\ \bibnamefont {Fisher}},\ }\bibfield  {title}
  {\bibinfo {title} {Universal transport signatures of {M}ajorana fermions in
  superconductor-{L}uttinger liquid junctions},\ }\href@noop {} {\bibfield
  {journal} {\bibinfo  {journal} {Phys. Rev. B}\ }\textbf {\bibinfo {volume}
  {85}},\ \bibinfo {pages} {245121} (\bibinfo {year} {2012})}\BibitemShut
  {NoStop}%
\bibitem [{\citenamefont {Sengupta}\ \emph {et~al.}(2001)\citenamefont
  {Sengupta}, \citenamefont {{\v{Z}}uti{\'c}}, \citenamefont {Kwon},
  \citenamefont {Yakovenko},\ and\ \citenamefont
  {Sarma}}]{215-sengupta2001midgap}%
  \BibitemOpen
  \bibfield  {author} {\bibinfo {author} {\bibfnamefont {K.}~\bibnamefont
  {Sengupta}}, \bibinfo {author} {\bibfnamefont {I.}~\bibnamefont
  {{\v{Z}}uti{\'c}}}, \bibinfo {author} {\bibfnamefont {H.-J.}\ \bibnamefont
  {Kwon}}, \bibinfo {author} {\bibfnamefont {V.~M.}\ \bibnamefont
  {Yakovenko}},\ and\ \bibinfo {author} {\bibfnamefont {S.~D.}\ \bibnamefont
  {Sarma}},\ }\bibfield  {title} {\bibinfo {title} {Midgap edge states and
  pairing symmetry of quasi-one-dimensional organic superconductors},\
  }\href@noop {} {\bibfield  {journal} {\bibinfo  {journal} {Phys. Rev. B}\
  }\textbf {\bibinfo {volume} {63}},\ \bibinfo {pages} {144531} (\bibinfo
  {year} {2001})}\BibitemShut {NoStop}%
\bibitem [{\citenamefont {Nichele}\ \emph {et~al.}(2017)\citenamefont
  {Nichele}, \citenamefont {Drachmann}, \citenamefont {Whiticar}, \citenamefont
  {O’Farrell}, \citenamefont {Suominen}, \citenamefont {Fornieri},
  \citenamefont {Wang}, \citenamefont {Gardner}, \citenamefont {Thomas},
  \citenamefont {Hatke} \emph {et~al.}}]{216-nichele2017scaling}%
  \BibitemOpen
  \bibfield  {author} {\bibinfo {author} {\bibfnamefont {F.}~\bibnamefont
  {Nichele}}, \bibinfo {author} {\bibfnamefont {A.~C.}\ \bibnamefont
  {Drachmann}}, \bibinfo {author} {\bibfnamefont {A.~M.}\ \bibnamefont
  {Whiticar}}, \bibinfo {author} {\bibfnamefont {E.~C.}\ \bibnamefont
  {O’Farrell}}, \bibinfo {author} {\bibfnamefont {H.~J.}\ \bibnamefont
  {Suominen}}, \bibinfo {author} {\bibfnamefont {A.}~\bibnamefont {Fornieri}},
  \bibinfo {author} {\bibfnamefont {T.}~\bibnamefont {Wang}}, \bibinfo {author}
  {\bibfnamefont {G.~C.}\ \bibnamefont {Gardner}}, \bibinfo {author}
  {\bibfnamefont {C.}~\bibnamefont {Thomas}}, \bibinfo {author} {\bibfnamefont
  {A.~T.}\ \bibnamefont {Hatke}}, \emph {et~al.},\ }\bibfield  {title}
  {\bibinfo {title} {Scaling of {M}ajorana zero-bias conductance peaks},\
  }\href@noop {} {\bibfield  {journal} {\bibinfo  {journal} {Phys. Rev. Lett.}\
  }\textbf {\bibinfo {volume} {119}},\ \bibinfo {pages} {136803} (\bibinfo
  {year} {2017})}\BibitemShut {NoStop}%
\bibitem [{\citenamefont {Setiawan}\ \emph {et~al.}(2017)\citenamefont
  {Setiawan}, \citenamefont {Liu}, \citenamefont {Sau},\ and\ \citenamefont
  {Sarma}}]{217-setiawan2017electron}%
  \BibitemOpen
  \bibfield  {author} {\bibinfo {author} {\bibfnamefont {F.}~\bibnamefont
  {Setiawan}}, \bibinfo {author} {\bibfnamefont {C.-X.}\ \bibnamefont {Liu}},
  \bibinfo {author} {\bibfnamefont {J.~D.}\ \bibnamefont {Sau}},\ and\ \bibinfo
  {author} {\bibfnamefont {S.~D.}\ \bibnamefont {Sarma}},\ }\bibfield  {title}
  {\bibinfo {title} {Electron temperature and tunnel coupling dependence of
  zero-bias and almost-zero-bias conductance peaks in {M}ajorana nanowires},\
  }\href@noop {} {\bibfield  {journal} {\bibinfo  {journal} {Phys. Rev. B}\
  }\textbf {\bibinfo {volume} {96}},\ \bibinfo {pages} {184520} (\bibinfo
  {year} {2017})}\BibitemShut {NoStop}%
\bibitem [{\citenamefont {Pientka}\ \emph {et~al.}(2012)\citenamefont
  {Pientka}, \citenamefont {Kells}, \citenamefont {Romito}, \citenamefont
  {Brouwer},\ and\ \citenamefont {Von~Oppen}}]{218-pientka2012enhanced}%
  \BibitemOpen
  \bibfield  {author} {\bibinfo {author} {\bibfnamefont {F.}~\bibnamefont
  {Pientka}}, \bibinfo {author} {\bibfnamefont {G.}~\bibnamefont {Kells}},
  \bibinfo {author} {\bibfnamefont {A.}~\bibnamefont {Romito}}, \bibinfo
  {author} {\bibfnamefont {P.~W.}\ \bibnamefont {Brouwer}},\ and\ \bibinfo
  {author} {\bibfnamefont {F.}~\bibnamefont {Von~Oppen}},\ }\bibfield  {title}
  {\bibinfo {title} {Enhanced zero-bias {M}ajorana peak in the differential
  tunneling conductance of disordered multisubband quantum-wire/superconductor
  junctions},\ }\href@noop {} {\bibfield  {journal} {\bibinfo  {journal} {Phys.
  Rev. Lett.}\ }\textbf {\bibinfo {volume} {109}},\ \bibinfo {pages} {227006}
  (\bibinfo {year} {2012})}\BibitemShut {NoStop}%
\bibitem [{\citenamefont {Van~Wees}\ \emph {et~al.}(1988)\citenamefont
  {Van~Wees}, \citenamefont {Van~Houten}, \citenamefont {Beenakker},
  \citenamefont {Williamson}, \citenamefont {Kouwenhoven}, \citenamefont
  {Van~der Marel},\ and\ \citenamefont {Foxon}}]{219-van1988quantized}%
  \BibitemOpen
  \bibfield  {author} {\bibinfo {author} {\bibfnamefont {B.}~\bibnamefont
  {Van~Wees}}, \bibinfo {author} {\bibfnamefont {H.}~\bibnamefont
  {Van~Houten}}, \bibinfo {author} {\bibfnamefont {C.}~\bibnamefont
  {Beenakker}}, \bibinfo {author} {\bibfnamefont {J.~G.}\ \bibnamefont
  {Williamson}}, \bibinfo {author} {\bibfnamefont {L.}~\bibnamefont
  {Kouwenhoven}}, \bibinfo {author} {\bibfnamefont {D.}~\bibnamefont {Van~der
  Marel}},\ and\ \bibinfo {author} {\bibfnamefont {C.}~\bibnamefont {Foxon}},\
  }\bibfield  {title} {\bibinfo {title} {Quantized conductance of point
  contacts in a two-dimensional electron gas},\ }\href@noop {} {\bibfield
  {journal} {\bibinfo  {journal} {Phys. Rev. Lett.}\ }\textbf {\bibinfo
  {volume} {60}},\ \bibinfo {pages} {848} (\bibinfo {year} {1988})}\BibitemShut
  {NoStop}%
\bibitem [{\citenamefont {Kammhuber}\ \emph {et~al.}(2016)\citenamefont
  {Kammhuber}, \citenamefont {Cassidy}, \citenamefont {Zhang}, \citenamefont
  {G{\"u}l}, \citenamefont {Pei}, \citenamefont {de~Moor}, \citenamefont
  {Nijholt}, \citenamefont {Watanabe}, \citenamefont {Taniguchi}, \citenamefont
  {Car} \emph {et~al.}}]{220-kammhuber2016conductance}%
  \BibitemOpen
  \bibfield  {author} {\bibinfo {author} {\bibfnamefont {J.}~\bibnamefont
  {Kammhuber}}, \bibinfo {author} {\bibfnamefont {M.~C.}\ \bibnamefont
  {Cassidy}}, \bibinfo {author} {\bibfnamefont {H.}~\bibnamefont {Zhang}},
  \bibinfo {author} {\bibfnamefont {{\"O}.}~\bibnamefont {G{\"u}l}}, \bibinfo
  {author} {\bibfnamefont {F.}~\bibnamefont {Pei}}, \bibinfo {author}
  {\bibfnamefont {M.~W.}\ \bibnamefont {de~Moor}}, \bibinfo {author}
  {\bibfnamefont {B.}~\bibnamefont {Nijholt}}, \bibinfo {author} {\bibfnamefont
  {K.}~\bibnamefont {Watanabe}}, \bibinfo {author} {\bibfnamefont
  {T.}~\bibnamefont {Taniguchi}}, \bibinfo {author} {\bibfnamefont
  {D.}~\bibnamefont {Car}}, \emph {et~al.},\ }\bibfield  {title} {\bibinfo
  {title} {Conductance quantization at zero magnetic field in {I}n{S}b
  nanowires},\ }\href@noop {} {\bibfield  {journal} {\bibinfo  {journal} {Nano
  Lett.}\ }\textbf {\bibinfo {volume} {16}},\ \bibinfo {pages} {3482} (\bibinfo
  {year} {2016})}\BibitemShut {NoStop}%
\bibitem [{\citenamefont {Beenakker}(1992)}]{221-beenakker1992quantum}%
  \BibitemOpen
  \bibfield  {author} {\bibinfo {author} {\bibfnamefont {C.}~\bibnamefont
  {Beenakker}},\ }\bibfield  {title} {\bibinfo {title} {Quantum transport in
  semiconductor-superconductor microjunctions},\ }\href@noop {} {\bibfield
  {journal} {\bibinfo  {journal} {Phys. Rev. B}\ }\textbf {\bibinfo {volume}
  {46}},\ \bibinfo {pages} {12841} (\bibinfo {year} {1992})}\BibitemShut
  {NoStop}%
\bibitem [{\citenamefont {Kj{\ae}rgaard}\ \emph {et~al.}(2016)\citenamefont
  {Kj{\ae}rgaard}, \citenamefont {Nichele}, \citenamefont {Suominen},
  \citenamefont {Nowak}, \citenamefont {Wimmer}, \citenamefont {Akhmerov},
  \citenamefont {Folk}, \citenamefont {Flensberg}, \citenamefont {Shabani},
  \citenamefont {Palmstr{\o}m} \emph {et~al.}}]{222-kjaergaard2016quantized}%
  \BibitemOpen
  \bibfield  {author} {\bibinfo {author} {\bibfnamefont {M.}~\bibnamefont
  {Kj{\ae}rgaard}}, \bibinfo {author} {\bibfnamefont {F.}~\bibnamefont
  {Nichele}}, \bibinfo {author} {\bibfnamefont {H.~J.}\ \bibnamefont
  {Suominen}}, \bibinfo {author} {\bibfnamefont {M.}~\bibnamefont {Nowak}},
  \bibinfo {author} {\bibfnamefont {M.}~\bibnamefont {Wimmer}}, \bibinfo
  {author} {\bibfnamefont {A.}~\bibnamefont {Akhmerov}}, \bibinfo {author}
  {\bibfnamefont {J.}~\bibnamefont {Folk}}, \bibinfo {author} {\bibfnamefont
  {K.}~\bibnamefont {Flensberg}}, \bibinfo {author} {\bibfnamefont
  {J.}~\bibnamefont {Shabani}}, \bibinfo {author} {\bibfnamefont {w.~C.}\
  \bibnamefont {Palmstr{\o}m}}, \emph {et~al.},\ }\bibfield  {title} {\bibinfo
  {title} {Quantized conductance doubling and hard gap in a two-dimensional
  semiconductor--superconductor heterostructure},\ }\href@noop {} {\bibfield
  {journal} {\bibinfo  {journal} {Nat. Commun.}\ }\textbf {\bibinfo {volume}
  {7}},\ \bibinfo {pages} {12841} (\bibinfo {year} {2016})}\BibitemShut
  {NoStop}%
\bibitem [{\citenamefont {Zhang}\ \emph {et~al.}(2017)\citenamefont {Zhang},
  \citenamefont {G{\"u}l}, \citenamefont {Conesa-Boj}, \citenamefont {Nowak},
  \citenamefont {Wimmer}, \citenamefont {Zuo}, \citenamefont {Mourik},
  \citenamefont {De~Vries}, \citenamefont {Van~Veen}, \citenamefont {De~Moor}
  \emph {et~al.}}]{223-zhang2017ballistic}%
  \BibitemOpen
  \bibfield  {author} {\bibinfo {author} {\bibfnamefont {H.}~\bibnamefont
  {Zhang}}, \bibinfo {author} {\bibfnamefont {{\"O}.}~\bibnamefont {G{\"u}l}},
  \bibinfo {author} {\bibfnamefont {S.}~\bibnamefont {Conesa-Boj}}, \bibinfo
  {author} {\bibfnamefont {M.~P.}\ \bibnamefont {Nowak}}, \bibinfo {author}
  {\bibfnamefont {M.}~\bibnamefont {Wimmer}}, \bibinfo {author} {\bibfnamefont
  {K.}~\bibnamefont {Zuo}}, \bibinfo {author} {\bibfnamefont {V.}~\bibnamefont
  {Mourik}}, \bibinfo {author} {\bibfnamefont {F.~K.}\ \bibnamefont
  {De~Vries}}, \bibinfo {author} {\bibfnamefont {J.}~\bibnamefont {Van~Veen}},
  \bibinfo {author} {\bibfnamefont {M.~W.}\ \bibnamefont {De~Moor}}, \emph
  {et~al.},\ }\bibfield  {title} {\bibinfo {title} {Ballistic superconductivity
  in semiconductor nanowires},\ }\href@noop {} {\bibfield  {journal} {\bibinfo
  {journal} {Nat. Commun.}\ }\textbf {\bibinfo {volume} {8}},\ \bibinfo {pages}
  {16025} (\bibinfo {year} {2017})}\BibitemShut {NoStop}%
\bibitem [{\citenamefont {G{\"u}l}\ \emph {et~al.}(2018)\citenamefont
  {G{\"u}l}, \citenamefont {Zhang}, \citenamefont {Bommer}, \citenamefont
  {de~Moor}, \citenamefont {Car}, \citenamefont {Plissard}, \citenamefont
  {Bakkers}, \citenamefont {Geresdi}, \citenamefont {Watanabe}, \citenamefont
  {Taniguchi} \emph {et~al.}}]{224-gul2018ballistic}%
  \BibitemOpen
  \bibfield  {author} {\bibinfo {author} {\bibfnamefont {{\"O}.}~\bibnamefont
  {G{\"u}l}}, \bibinfo {author} {\bibfnamefont {H.}~\bibnamefont {Zhang}},
  \bibinfo {author} {\bibfnamefont {J.~D.}\ \bibnamefont {Bommer}}, \bibinfo
  {author} {\bibfnamefont {M.~W.}\ \bibnamefont {de~Moor}}, \bibinfo {author}
  {\bibfnamefont {D.}~\bibnamefont {Car}}, \bibinfo {author} {\bibfnamefont
  {S.~R.}\ \bibnamefont {Plissard}}, \bibinfo {author} {\bibfnamefont {E.~P.}\
  \bibnamefont {Bakkers}}, \bibinfo {author} {\bibfnamefont {A.}~\bibnamefont
  {Geresdi}}, \bibinfo {author} {\bibfnamefont {K.}~\bibnamefont {Watanabe}},
  \bibinfo {author} {\bibfnamefont {T.}~\bibnamefont {Taniguchi}}, \emph
  {et~al.},\ }\bibfield  {title} {\bibinfo {title} {Ballistic {M}ajorana
  nanowire devices},\ }\href@noop {} {\bibfield  {journal} {\bibinfo  {journal}
  {Nat. Nanotechnol.}\ }\textbf {\bibinfo {volume} {13}},\ \bibinfo {pages}
  {192} (\bibinfo {year} {2018})}\BibitemShut {NoStop}%
\bibitem [{\citenamefont {Kitaev}(2001)}]{225-kitaev2001unpaired}%
  \BibitemOpen
  \bibfield  {author} {\bibinfo {author} {\bibfnamefont {A.~Y.}\ \bibnamefont
  {Kitaev}},\ }\bibfield  {title} {\bibinfo {title} {Unpaired {M}ajorana
  fermions in quantum wires},\ }\href@noop {} {\bibfield  {journal} {\bibinfo
  {journal} {Phys. Uspekhi}\ }\textbf {\bibinfo {volume} {44}},\ \bibinfo
  {pages} {131} (\bibinfo {year} {2001})}\BibitemShut {NoStop}%
\bibitem [{\citenamefont {Tsui}\ \emph {et~al.}(2019)\citenamefont {Tsui},
  \citenamefont {Li}, \citenamefont {Huang}, \citenamefont {Louie},\ and\
  \citenamefont {Lee}}]{226-tsui2019classification}%
  \BibitemOpen
  \bibfield  {author} {\bibinfo {author} {\bibfnamefont {L.}~\bibnamefont
  {Tsui}}, \bibinfo {author} {\bibfnamefont {Z.-X.}\ \bibnamefont {Li}},
  \bibinfo {author} {\bibfnamefont {Y.-T.}\ \bibnamefont {Huang}}, \bibinfo
  {author} {\bibfnamefont {S.~G.}\ \bibnamefont {Louie}},\ and\ \bibinfo
  {author} {\bibfnamefont {D.-H.}\ \bibnamefont {Lee}},\ }\bibfield  {title}
  {\bibinfo {title} {Classification of topological trivial matter with
  non-trivial defects},\ }\href@noop {} {\bibfield  {journal} {\bibinfo
  {journal} {Sci. Bull.}\ }\textbf {\bibinfo {volume} {64}},\ \bibinfo {pages}
  {575} (\bibinfo {year} {2019})}\BibitemShut {NoStop}%
\bibitem [{\citenamefont {Chan}\ \emph {et~al.}(2017)\citenamefont {Chan},
  \citenamefont {Zhang}, \citenamefont {Poon}, \citenamefont {He},
  \citenamefont {Wang},\ and\ \citenamefont {Liu}}]{227-chan2017generic}%
  \BibitemOpen
  \bibfield  {author} {\bibinfo {author} {\bibfnamefont {C.}~\bibnamefont
  {Chan}}, \bibinfo {author} {\bibfnamefont {L.}~\bibnamefont {Zhang}},
  \bibinfo {author} {\bibfnamefont {T.~F.~J.}\ \bibnamefont {Poon}}, \bibinfo
  {author} {\bibfnamefont {Y.-P.}\ \bibnamefont {He}}, \bibinfo {author}
  {\bibfnamefont {Y.-Q.}\ \bibnamefont {Wang}},\ and\ \bibinfo {author}
  {\bibfnamefont {X.-J.}\ \bibnamefont {Liu}},\ }\bibfield  {title} {\bibinfo
  {title} {Generic theory for {M}ajorana zero modes in 2{D} superconductors},\
  }\href@noop {} {\bibfield  {journal} {\bibinfo  {journal} {Phys. Rev. Lett.}\
  }\textbf {\bibinfo {volume} {119}},\ \bibinfo {pages} {047001} (\bibinfo
  {year} {2017})}\BibitemShut {NoStop}%
\bibitem [{\citenamefont {Qi}\ \emph {et~al.}(2010{\natexlab{b}})\citenamefont
  {Qi}, \citenamefont {Hughes},\ and\ \citenamefont
  {Zhang}}]{228-qi2010topological}%
  \BibitemOpen
  \bibfield  {author} {\bibinfo {author} {\bibfnamefont {X.-L.}\ \bibnamefont
  {Qi}}, \bibinfo {author} {\bibfnamefont {T.~L.}\ \bibnamefont {Hughes}},\
  and\ \bibinfo {author} {\bibfnamefont {S.-C.}\ \bibnamefont {Zhang}},\
  }\bibfield  {title} {\bibinfo {title} {Topological invariants for the {F}ermi
  surface of a time-reversal-invariant superconductor},\ }\href@noop {}
  {\bibfield  {journal} {\bibinfo  {journal} {Phys. Rev. B}\ }\textbf {\bibinfo
  {volume} {81}},\ \bibinfo {pages} {134508} (\bibinfo {year}
  {2010}{\natexlab{b}})}\BibitemShut {NoStop}%
\bibitem [{\citenamefont {Yan}\ \emph {et~al.}(2017)\citenamefont {Yan},
  \citenamefont {Bi},\ and\ \citenamefont {Wang}}]{229-yan2017majorana}%
  \BibitemOpen
  \bibfield  {author} {\bibinfo {author} {\bibfnamefont {Z.}~\bibnamefont
  {Yan}}, \bibinfo {author} {\bibfnamefont {R.}~\bibnamefont {Bi}},\ and\
  \bibinfo {author} {\bibfnamefont {Z.}~\bibnamefont {Wang}},\ }\bibfield
  {title} {\bibinfo {title} {Majorana zero modes protected by a hopf invariant
  in topologically trivial superconductors},\ }\href@noop {} {\bibfield
  {journal} {\bibinfo  {journal} {Phys. Rev. Lett.}\ }\textbf {\bibinfo
  {volume} {118}},\ \bibinfo {pages} {147003} (\bibinfo {year}
  {2017})}\BibitemShut {NoStop}%
\bibitem [{\citenamefont {Agterberg}\ and\ \citenamefont
  {Tsunetsugu}(2008)}]{230-agterberg2008dislocations}%
  \BibitemOpen
  \bibfield  {author} {\bibinfo {author} {\bibfnamefont {D.}~\bibnamefont
  {Agterberg}}\ and\ \bibinfo {author} {\bibfnamefont {H.}~\bibnamefont
  {Tsunetsugu}},\ }\bibfield  {title} {\bibinfo {title} {Dislocations and
  vortices in pair-density-wave superconductors},\ }\href@noop {} {\bibfield
  {journal} {\bibinfo  {journal} {Nat. Phys.}\ }\textbf {\bibinfo {volume}
  {4}},\ \bibinfo {pages} {639} (\bibinfo {year} {2008})}\BibitemShut {NoStop}%
\bibitem [{\citenamefont {Khurana}(1990)}]{231-khurana1990stm}%
  \BibitemOpen
  \bibfield  {author} {\bibinfo {author} {\bibfnamefont {A.}~\bibnamefont
  {Khurana}},\ }\bibfield  {title} {\bibinfo {title} {{STM} unravels the vortex
  core in type-{II} superconductors},\ }\href@noop {} {\bibfield  {journal}
  {\bibinfo  {journal} {Phys. Today}\ }\textbf {\bibinfo {volume} {43}},\
  \bibinfo {pages} {17} (\bibinfo {year} {1990})}\BibitemShut {NoStop}%
\bibitem [{\citenamefont {Kopnin}\ and\ \citenamefont
  {Volovik}(1997)}]{232-kopnin1997flux}%
  \BibitemOpen
  \bibfield  {author} {\bibinfo {author} {\bibfnamefont {N.}~\bibnamefont
  {Kopnin}}\ and\ \bibinfo {author} {\bibfnamefont {G.}~\bibnamefont
  {Volovik}},\ }\bibfield  {title} {\bibinfo {title} {Flux flow in d-wave
  superconductors: Low temperature universality and scaling},\ }\href@noop {}
  {\bibfield  {journal} {\bibinfo  {journal} {Phys. Rev. Lett.}\ }\textbf
  {\bibinfo {volume} {79}},\ \bibinfo {pages} {1377} (\bibinfo {year}
  {1997})}\BibitemShut {NoStop}%
\bibitem [{\citenamefont {Franz}\ and\ \citenamefont
  {Te{\v{s}}anovi{\'c}}(1998)}]{233-franz1998self}%
  \BibitemOpen
  \bibfield  {author} {\bibinfo {author} {\bibfnamefont {M.}~\bibnamefont
  {Franz}}\ and\ \bibinfo {author} {\bibfnamefont {Z.}~\bibnamefont
  {Te{\v{s}}anovi{\'c}}},\ }\bibfield  {title} {\bibinfo {title}
  {Self-consistent electronic structure of a d$_{x^{2}-y^{2}}$ and a
  d$_{x^{2}-y^{2}}$+id$_{xy}$ vortex},\ }\href@noop {} {\bibfield  {journal}
  {\bibinfo  {journal} {Phys. Rev. Lett.}\ }\textbf {\bibinfo {volume} {80}},\
  \bibinfo {pages} {4763} (\bibinfo {year} {1998})}\BibitemShut {NoStop}%
\bibitem [{\citenamefont {Berthod}\ \emph {et~al.}(2017)\citenamefont
  {Berthod}, \citenamefont {Maggio-Aprile}, \citenamefont {Bru{\'e}r},
  \citenamefont {Erb},\ and\ \citenamefont
  {Renner}}]{234-berthod2017observation}%
  \BibitemOpen
  \bibfield  {author} {\bibinfo {author} {\bibfnamefont {C.}~\bibnamefont
  {Berthod}}, \bibinfo {author} {\bibfnamefont {I.}~\bibnamefont
  {Maggio-Aprile}}, \bibinfo {author} {\bibfnamefont {J.}~\bibnamefont
  {Bru{\'e}r}}, \bibinfo {author} {\bibfnamefont {A.}~\bibnamefont {Erb}},\
  and\ \bibinfo {author} {\bibfnamefont {C.}~\bibnamefont {Renner}},\
  }\bibfield  {title} {\bibinfo {title} {Observation of {C}aroli--de
  {G}ennes--{M}atricon vortex states in {Y}{B}a$_{2}${C}u$_{3}${O}$_{7-x}$},\
  }\href@noop {} {\bibfield  {journal} {\bibinfo  {journal} {Phys. Rev. Lett.}\
  }\textbf {\bibinfo {volume} {119}},\ \bibinfo {pages} {237001} (\bibinfo
  {year} {2017})}\BibitemShut {NoStop}%
\bibitem [{\citenamefont {Berthod}(2016)}]{235-berthod2016vortex}%
  \BibitemOpen
  \bibfield  {author} {\bibinfo {author} {\bibfnamefont {C.}~\bibnamefont
  {Berthod}},\ }\bibfield  {title} {\bibinfo {title} {Vortex spectroscopy in
  the vortex glass: A real-space numerical approach},\ }\href@noop {}
  {\bibfield  {journal} {\bibinfo  {journal} {Phys. Rev. B}\ }\textbf {\bibinfo
  {volume} {94}},\ \bibinfo {pages} {184510} (\bibinfo {year}
  {2016})}\BibitemShut {NoStop}%
\bibitem [{\citenamefont {Hayashi}\ \emph {et~al.}(1998)\citenamefont
  {Hayashi}, \citenamefont {Isoshima}, \citenamefont {Ichioka},\ and\
  \citenamefont {Machida}}]{236-hayashi1998low}%
  \BibitemOpen
  \bibfield  {author} {\bibinfo {author} {\bibfnamefont {N.}~\bibnamefont
  {Hayashi}}, \bibinfo {author} {\bibfnamefont {T.}~\bibnamefont {Isoshima}},
  \bibinfo {author} {\bibfnamefont {M.}~\bibnamefont {Ichioka}},\ and\ \bibinfo
  {author} {\bibfnamefont {K.}~\bibnamefont {Machida}},\ }\bibfield  {title}
  {\bibinfo {title} {Low-lying quasiparticle excitations around a vortex core
  in quantum limit},\ }\href@noop {} {\bibfield  {journal} {\bibinfo  {journal}
  {Phys. Rev. Lett.}\ }\textbf {\bibinfo {volume} {80}},\ \bibinfo {pages}
  {2921} (\bibinfo {year} {1998})}\BibitemShut {NoStop}%
\bibitem [{\citenamefont {Shan}\ \emph {et~al.}(2011)\citenamefont {Shan},
  \citenamefont {Wang}, \citenamefont {Shen}, \citenamefont {Zeng},
  \citenamefont {Huang}, \citenamefont {Li}, \citenamefont {Wang},
  \citenamefont {Yang}, \citenamefont {Ren}, \citenamefont {Wang} \emph
  {et~al.}}]{237-shan2011observation}%
  \BibitemOpen
  \bibfield  {author} {\bibinfo {author} {\bibfnamefont {L.}~\bibnamefont
  {Shan}}, \bibinfo {author} {\bibfnamefont {Y.-L.}\ \bibnamefont {Wang}},
  \bibinfo {author} {\bibfnamefont {B.}~\bibnamefont {Shen}}, \bibinfo {author}
  {\bibfnamefont {B.}~\bibnamefont {Zeng}}, \bibinfo {author} {\bibfnamefont
  {Y.}~\bibnamefont {Huang}}, \bibinfo {author} {\bibfnamefont
  {A.}~\bibnamefont {Li}}, \bibinfo {author} {\bibfnamefont {D.}~\bibnamefont
  {Wang}}, \bibinfo {author} {\bibfnamefont {H.}~\bibnamefont {Yang}}, \bibinfo
  {author} {\bibfnamefont {C.}~\bibnamefont {Ren}}, \bibinfo {author}
  {\bibfnamefont {Q.-H.}\ \bibnamefont {Wang}}, \emph {et~al.},\ }\bibfield
  {title} {\bibinfo {title} {Observation of ordered vortices with {A}ndreev
  bound states in {B}a$_{0.6}${K}$_{0.4}${F}e$_{2}${A}s$_{2}$},\ }\href@noop {}
  {\bibfield  {journal} {\bibinfo  {journal} {Nat. Phys.}\ }\textbf {\bibinfo
  {volume} {7}},\ \bibinfo {pages} {325} (\bibinfo {year} {2011})}\BibitemShut
  {NoStop}%
\bibitem [{\citenamefont {Hanaguri}\ \emph {et~al.}(2012)\citenamefont
  {Hanaguri}, \citenamefont {Kitagawa}, \citenamefont {Matsubayashi},
  \citenamefont {Mazaki}, \citenamefont {Uwatoko},\ and\ \citenamefont
  {Takagi}}]{238-hanaguri2012scanning}%
  \BibitemOpen
  \bibfield  {author} {\bibinfo {author} {\bibfnamefont {T.}~\bibnamefont
  {Hanaguri}}, \bibinfo {author} {\bibfnamefont {K.}~\bibnamefont {Kitagawa}},
  \bibinfo {author} {\bibfnamefont {K.}~\bibnamefont {Matsubayashi}}, \bibinfo
  {author} {\bibfnamefont {Y.}~\bibnamefont {Mazaki}}, \bibinfo {author}
  {\bibfnamefont {Y.}~\bibnamefont {Uwatoko}},\ and\ \bibinfo {author}
  {\bibfnamefont {H.}~\bibnamefont {Takagi}},\ }\bibfield  {title} {\bibinfo
  {title} {Scanning tunneling microscopy/spectroscopy of vortices in
  {L}i{F}e{A}s},\ }\href@noop {} {\bibfield  {journal} {\bibinfo  {journal}
  {Phys. Rev. B}\ }\textbf {\bibinfo {volume} {85}},\ \bibinfo {pages} {214505}
  (\bibinfo {year} {2012})}\BibitemShut {NoStop}%
\bibitem [{\citenamefont {Chen}\ \emph
  {et~al.}(2018{\natexlab{b}})\citenamefont {Chen}, \citenamefont {Chen},
  \citenamefont {Yang}, \citenamefont {Du}, \citenamefont {Zhu}, \citenamefont
  {Wang},\ and\ \citenamefont {Wen}}]{239-chen2018discrete}%
  \BibitemOpen
  \bibfield  {author} {\bibinfo {author} {\bibfnamefont {M.}~\bibnamefont
  {Chen}}, \bibinfo {author} {\bibfnamefont {X.}~\bibnamefont {Chen}}, \bibinfo
  {author} {\bibfnamefont {H.}~\bibnamefont {Yang}}, \bibinfo {author}
  {\bibfnamefont {Z.}~\bibnamefont {Du}}, \bibinfo {author} {\bibfnamefont
  {X.}~\bibnamefont {Zhu}}, \bibinfo {author} {\bibfnamefont {E.}~\bibnamefont
  {Wang}},\ and\ \bibinfo {author} {\bibfnamefont {H.-H.}\ \bibnamefont
  {Wen}},\ }\bibfield  {title} {\bibinfo {title} {Discrete energy levels of
  {C}aroli-de {G}ennes-{M}atricon states in quantum limit in
  {F}e{T}e$_{0.55}${S}e$_{0.45}$},\ }\href@noop {} {\bibfield  {journal}
  {\bibinfo  {journal} {Nat. Commun.}\ }\textbf {\bibinfo {volume} {9}},\
  \bibinfo {pages} {970} (\bibinfo {year} {2018}{\natexlab{b}})}\BibitemShut
  {NoStop}%
\bibitem [{\citenamefont {Berthod}(2018)}]{240-berthod2018signatures}%
  \BibitemOpen
  \bibfield  {author} {\bibinfo {author} {\bibfnamefont {C.}~\bibnamefont
  {Berthod}},\ }\bibfield  {title} {\bibinfo {title} {Signatures of nodeless
  multiband superconductivity and particle-hole crossover in the vortex cores
  of {F}e{T}e$_{0.55}${S}e$_{0.45}$},\ }\href@noop {} {\bibfield  {journal}
  {\bibinfo  {journal} {Phys. Rev. B}\ }\textbf {\bibinfo {volume} {98}},\
  \bibinfo {pages} {144519} (\bibinfo {year} {2018})}\BibitemShut {NoStop}%
\bibitem [{\citenamefont {Hanaguri}\ \emph {et~al.}(2019)\citenamefont
  {Hanaguri}, \citenamefont {Kasahara}, \citenamefont {B{\"o}ker},
  \citenamefont {Eremin}, \citenamefont {Shibauchi},\ and\ \citenamefont
  {Matsuda}}]{241-hanaguri2019quantum}%
  \BibitemOpen
  \bibfield  {author} {\bibinfo {author} {\bibfnamefont {T.}~\bibnamefont
  {Hanaguri}}, \bibinfo {author} {\bibfnamefont {S.}~\bibnamefont {Kasahara}},
  \bibinfo {author} {\bibfnamefont {J.}~\bibnamefont {B{\"o}ker}}, \bibinfo
  {author} {\bibfnamefont {I.}~\bibnamefont {Eremin}}, \bibinfo {author}
  {\bibfnamefont {T.}~\bibnamefont {Shibauchi}},\ and\ \bibinfo {author}
  {\bibfnamefont {Y.}~\bibnamefont {Matsuda}},\ }\bibfield  {title} {\bibinfo
  {title} {Quantum vortex core and missing pseudogap in the multiband
  {BCS}-{BEC} crossover superconductor {F}e{S}e},\ }\href@noop {} {\bibfield
  {journal} {\bibinfo  {journal} {Phys. Rev. Lett.}\ }\textbf {\bibinfo
  {volume} {122}},\ \bibinfo {pages} {077001} (\bibinfo {year}
  {2019})}\BibitemShut {NoStop}%
\bibitem [{\citenamefont {Chen}\ \emph
  {et~al.}(2020{\natexlab{b}})\citenamefont {Chen}, \citenamefont {Liu},
  \citenamefont {Bao}, \citenamefont {Yan}, \citenamefont {Wang}, \citenamefont
  {Zhang},\ and\ \citenamefont {Feng}}]{242-chen2020observation}%
  \BibitemOpen
  \bibfield  {author} {\bibinfo {author} {\bibfnamefont {C.}~\bibnamefont
  {Chen}}, \bibinfo {author} {\bibfnamefont {Q.}~\bibnamefont {Liu}}, \bibinfo
  {author} {\bibfnamefont {W.-C.}\ \bibnamefont {Bao}}, \bibinfo {author}
  {\bibfnamefont {Y.}~\bibnamefont {Yan}}, \bibinfo {author} {\bibfnamefont
  {Q.-H.}\ \bibnamefont {Wang}}, \bibinfo {author} {\bibfnamefont
  {T.}~\bibnamefont {Zhang}},\ and\ \bibinfo {author} {\bibfnamefont
  {D.}~\bibnamefont {Feng}},\ }\bibfield  {title} {\bibinfo {title}
  {Observation of discrete conventional {C}aroli--de {G}ennes--{M}atricon
  states in the vortex core of single-layer {F}e{S}e/{S}r{T}i{O}$_{3}$},\
  }\href@noop {} {\bibfield  {journal} {\bibinfo  {journal} {Phys. Rev. Lett.}\
  }\textbf {\bibinfo {volume} {124}},\ \bibinfo {pages} {097001} (\bibinfo
  {year} {2020}{\natexlab{b}})}\BibitemShut {NoStop}%
\bibitem [{\citenamefont {Chang}\ \emph {et~al.}(2014)\citenamefont {Chang},
  \citenamefont {Matsuura}, \citenamefont {Schnyder},\ and\ \citenamefont
  {Ryu}}]{243-chang2014majorana}%
  \BibitemOpen
  \bibfield  {author} {\bibinfo {author} {\bibfnamefont {P.-Y.}\ \bibnamefont
  {Chang}}, \bibinfo {author} {\bibfnamefont {S.}~\bibnamefont {Matsuura}},
  \bibinfo {author} {\bibfnamefont {A.~P.}\ \bibnamefont {Schnyder}},\ and\
  \bibinfo {author} {\bibfnamefont {S.}~\bibnamefont {Ryu}},\ }\bibfield
  {title} {\bibinfo {title} {Majorana vortex-bound states in three-dimensional
  nodal noncentrosymmetric superconductors},\ }\href@noop {} {\bibfield
  {journal} {\bibinfo  {journal} {Phys. Rev. B}\ }\textbf {\bibinfo {volume}
  {90}},\ \bibinfo {pages} {174504} (\bibinfo {year} {2014})}\BibitemShut
  {NoStop}%
\bibitem [{\citenamefont {Lee}\ and\ \citenamefont
  {Schnyder}(2016)}]{244-lee2016structure}%
  \BibitemOpen
  \bibfield  {author} {\bibinfo {author} {\bibfnamefont {D.}~\bibnamefont
  {Lee}}\ and\ \bibinfo {author} {\bibfnamefont {A.~P.}\ \bibnamefont
  {Schnyder}},\ }\bibfield  {title} {\bibinfo {title} {Structure of
  vortex-bound states in spin-singlet chiral superconductors},\ }\href@noop {}
  {\bibfield  {journal} {\bibinfo  {journal} {Phys. Rev. B}\ }\textbf {\bibinfo
  {volume} {93}},\ \bibinfo {pages} {064522} (\bibinfo {year}
  {2016})}\BibitemShut {NoStop}%
\bibitem [{\citenamefont {Jackiw}\ and\ \citenamefont
  {Rossi}(1981)}]{246-jackiw1981zero}%
  \BibitemOpen
  \bibfield  {author} {\bibinfo {author} {\bibfnamefont {R.}~\bibnamefont
  {Jackiw}}\ and\ \bibinfo {author} {\bibfnamefont {P.}~\bibnamefont {Rossi}},\
  }\bibfield  {title} {\bibinfo {title} {Zero modes of the vortex-fermion
  system},\ }\href@noop {} {\bibfield  {journal} {\bibinfo  {journal} {Nucl.
  Phys. B}\ }\textbf {\bibinfo {volume} {190}},\ \bibinfo {pages} {681}
  (\bibinfo {year} {1981})}\BibitemShut {NoStop}%
\bibitem [{\citenamefont {Ghaemi}\ and\ \citenamefont
  {Wilczek}(2012)}]{247-ghaemi2012near}%
  \BibitemOpen
  \bibfield  {author} {\bibinfo {author} {\bibfnamefont {P.}~\bibnamefont
  {Ghaemi}}\ and\ \bibinfo {author} {\bibfnamefont {F.}~\bibnamefont
  {Wilczek}},\ }\bibfield  {title} {\bibinfo {title} {Near-zero modes in
  superconducting graphene},\ }\href@noop {} {\bibfield  {journal} {\bibinfo
  {journal} {Phys. Scripta}\ }\textbf {\bibinfo {volume} {2012}},\ \bibinfo
  {pages} {014019} (\bibinfo {year} {2012})}\BibitemShut {NoStop}%
\bibitem [{\citenamefont {Khaymovich}\ \emph {et~al.}(2009)\citenamefont
  {Khaymovich}, \citenamefont {Kopnin}, \citenamefont {Mel’Nikov},\ and\
  \citenamefont {Shereshevskii}}]{245-khaymovich2009vortex}%
  \BibitemOpen
  \bibfield  {author} {\bibinfo {author} {\bibfnamefont {I.}~\bibnamefont
  {Khaymovich}}, \bibinfo {author} {\bibfnamefont {N.}~\bibnamefont {Kopnin}},
  \bibinfo {author} {\bibfnamefont {A.}~\bibnamefont {Mel’Nikov}},\ and\
  \bibinfo {author} {\bibfnamefont {I.}~\bibnamefont {Shereshevskii}},\
  }\bibfield  {title} {\bibinfo {title} {Vortex core states in superconducting
  graphene},\ }\href@noop {} {\bibfield  {journal} {\bibinfo  {journal} {Phys.
  Rev. B}\ }\textbf {\bibinfo {volume} {79}},\ \bibinfo {pages} {224506}
  (\bibinfo {year} {2009})}\BibitemShut {NoStop}%
\bibitem [{\citenamefont {Cheng}\ \emph {et~al.}(2010)\citenamefont {Cheng},
  \citenamefont {Lutchyn}, \citenamefont {Galitski},\ and\ \citenamefont
  {Sarma}}]{248-cheng2010tunneling}%
  \BibitemOpen
  \bibfield  {author} {\bibinfo {author} {\bibfnamefont {M.}~\bibnamefont
  {Cheng}}, \bibinfo {author} {\bibfnamefont {R.~M.}\ \bibnamefont {Lutchyn}},
  \bibinfo {author} {\bibfnamefont {V.}~\bibnamefont {Galitski}},\ and\
  \bibinfo {author} {\bibfnamefont {S.~D.}\ \bibnamefont {Sarma}},\ }\bibfield
  {title} {\bibinfo {title} {Tunneling of anyonic {M}ajorana excitations in
  topological superconductors},\ }\href@noop {} {\bibfield  {journal} {\bibinfo
   {journal} {Phys. Rev. B}\ }\textbf {\bibinfo {volume} {82}},\ \bibinfo
  {pages} {094504} (\bibinfo {year} {2010})}\BibitemShut {NoStop}%
\bibitem [{\citenamefont {Cheng}\ \emph {et~al.}(2009)\citenamefont {Cheng},
  \citenamefont {Lutchyn}, \citenamefont {Galitski},\ and\ \citenamefont
  {Sarma}}]{249-cheng2009splitting}%
  \BibitemOpen
  \bibfield  {author} {\bibinfo {author} {\bibfnamefont {M.}~\bibnamefont
  {Cheng}}, \bibinfo {author} {\bibfnamefont {R.~M.}\ \bibnamefont {Lutchyn}},
  \bibinfo {author} {\bibfnamefont {V.}~\bibnamefont {Galitski}},\ and\
  \bibinfo {author} {\bibfnamefont {S.~D.}\ \bibnamefont {Sarma}},\ }\bibfield
  {title} {\bibinfo {title} {Splitting of {M}ajorana-fermion modes due to
  intervortex tunneling in a p$_{x}$+ ip$_{y}$ superconductor},\ }\href@noop {}
  {\bibfield  {journal} {\bibinfo  {journal} {Phys. Rev. Lett.}\ }\textbf
  {\bibinfo {volume} {103}},\ \bibinfo {pages} {107001} (\bibinfo {year}
  {2009})}\BibitemShut {NoStop}%
\bibitem [{\citenamefont {Biswas}(2013)}]{250-biswas2013majorana}%
  \BibitemOpen
  \bibfield  {author} {\bibinfo {author} {\bibfnamefont {R.~R.}\ \bibnamefont
  {Biswas}},\ }\bibfield  {title} {\bibinfo {title} {Majorana fermions in
  vortex lattices},\ }\href@noop {} {\bibfield  {journal} {\bibinfo  {journal}
  {Phys. Rev. Lett.}\ }\textbf {\bibinfo {volume} {111}},\ \bibinfo {pages}
  {136401} (\bibinfo {year} {2013})}\BibitemShut {NoStop}%
\bibitem [{\citenamefont {Chiu}\ \emph {et~al.}(2015)\citenamefont {Chiu},
  \citenamefont {Pikulin},\ and\ \citenamefont {Franz}}]{251-chiu2015strongly}%
  \BibitemOpen
  \bibfield  {author} {\bibinfo {author} {\bibfnamefont {C.-K.}\ \bibnamefont
  {Chiu}}, \bibinfo {author} {\bibfnamefont {D.}~\bibnamefont {Pikulin}},\ and\
  \bibinfo {author} {\bibfnamefont {M.}~\bibnamefont {Franz}},\ }\bibfield
  {title} {\bibinfo {title} {Strongly interacting {M}ajorana fermions},\
  }\href@noop {} {\bibfield  {journal} {\bibinfo  {journal} {Phys. Rev. B}\
  }\textbf {\bibinfo {volume} {91}},\ \bibinfo {pages} {165402} (\bibinfo
  {year} {2015})}\BibitemShut {NoStop}%
\bibitem [{\citenamefont {Liu}\ and\ \citenamefont
  {Franz}(2015)}]{252-liu2015electronic}%
  \BibitemOpen
  \bibfield  {author} {\bibinfo {author} {\bibfnamefont {T.}~\bibnamefont
  {Liu}}\ and\ \bibinfo {author} {\bibfnamefont {M.}~\bibnamefont {Franz}},\
  }\bibfield  {title} {\bibinfo {title} {Electronic structure of topological
  superconductors in the presence of a vortex lattice},\ }\href@noop {}
  {\bibfield  {journal} {\bibinfo  {journal} {Phys. Rev. B}\ }\textbf {\bibinfo
  {volume} {92}},\ \bibinfo {pages} {134519} (\bibinfo {year}
  {2015})}\BibitemShut {NoStop}%
\bibitem [{\citenamefont {Rahmani}\ and\ \citenamefont
  {Franz}(2019)}]{253-rahmani2019interacting}%
  \BibitemOpen
  \bibfield  {author} {\bibinfo {author} {\bibfnamefont {A.}~\bibnamefont
  {Rahmani}}\ and\ \bibinfo {author} {\bibfnamefont {M.}~\bibnamefont
  {Franz}},\ }\bibfield  {title} {\bibinfo {title} {Interacting {M}ajorana
  fermions},\ }\href@noop {} {\bibfield  {journal} {\bibinfo  {journal} {Rep.
  Prog. Phys.}\ }\textbf {\bibinfo {volume} {82}},\ \bibinfo {pages} {084501}
  (\bibinfo {year} {2019})}\BibitemShut {NoStop}%
\bibitem [{\citenamefont {Schubert}\ \emph {et~al.}(2012)\citenamefont
  {Schubert}, \citenamefont {Fehske}, \citenamefont {Fritz},\ and\
  \citenamefont {Vojta}}]{262-schubert2012fate}%
  \BibitemOpen
  \bibfield  {author} {\bibinfo {author} {\bibfnamefont {G.}~\bibnamefont
  {Schubert}}, \bibinfo {author} {\bibfnamefont {H.}~\bibnamefont {Fehske}},
  \bibinfo {author} {\bibfnamefont {L.}~\bibnamefont {Fritz}},\ and\ \bibinfo
  {author} {\bibfnamefont {M.}~\bibnamefont {Vojta}},\ }\bibfield  {title}
  {\bibinfo {title} {Fate of topological-insulator surface states under strong
  disorder},\ }\href@noop {} {\bibfield  {journal} {\bibinfo  {journal} {Phys.
  Rev. B}\ }\textbf {\bibinfo {volume} {85}},\ \bibinfo {pages} {201105}
  (\bibinfo {year} {2012})}\BibitemShut {NoStop}%
\bibitem [{\citenamefont {Beidenkopf}\ \emph {et~al.}(2011)\citenamefont
  {Beidenkopf}, \citenamefont {Roushan}, \citenamefont {Seo}, \citenamefont
  {Gorman}, \citenamefont {Drozdov}, \citenamefont {San~Hor}, \citenamefont
  {Cava},\ and\ \citenamefont {Yazdani}}]{254-beidenkopf2011spatial}%
  \BibitemOpen
  \bibfield  {author} {\bibinfo {author} {\bibfnamefont {H.}~\bibnamefont
  {Beidenkopf}}, \bibinfo {author} {\bibfnamefont {P.}~\bibnamefont {Roushan}},
  \bibinfo {author} {\bibfnamefont {J.}~\bibnamefont {Seo}}, \bibinfo {author}
  {\bibfnamefont {L.}~\bibnamefont {Gorman}}, \bibinfo {author} {\bibfnamefont
  {I.}~\bibnamefont {Drozdov}}, \bibinfo {author} {\bibfnamefont
  {Y.}~\bibnamefont {San~Hor}}, \bibinfo {author} {\bibfnamefont {R.~J.}\
  \bibnamefont {Cava}},\ and\ \bibinfo {author} {\bibfnamefont
  {A.}~\bibnamefont {Yazdani}},\ }\bibfield  {title} {\bibinfo {title} {Spatial
  fluctuations of helical {D}irac fermions on the surface of topological
  insulators},\ }\href@noop {} {\bibfield  {journal} {\bibinfo  {journal} {Nat.
  Phys.}\ }\textbf {\bibinfo {volume} {7}},\ \bibinfo {pages} {939} (\bibinfo
  {year} {2011})}\BibitemShut {NoStop}%
\bibitem [{\citenamefont {Kotta}\ \emph {et~al.}(2020)\citenamefont {Kotta},
  \citenamefont {Miao}, \citenamefont {Xu}, \citenamefont {Breitweiser},
  \citenamefont {Jozwiak}, \citenamefont {Bostwick}, \citenamefont {Rotenberg},
  \citenamefont {Zhang}, \citenamefont {Wu}, \citenamefont {Suzuki} \emph
  {et~al.}}]{255-kotta2020spectromicroscopic}%
  \BibitemOpen
  \bibfield  {author} {\bibinfo {author} {\bibfnamefont {E.}~\bibnamefont
  {Kotta}}, \bibinfo {author} {\bibfnamefont {L.}~\bibnamefont {Miao}},
  \bibinfo {author} {\bibfnamefont {Y.}~\bibnamefont {Xu}}, \bibinfo {author}
  {\bibfnamefont {S.~A.}\ \bibnamefont {Breitweiser}}, \bibinfo {author}
  {\bibfnamefont {C.}~\bibnamefont {Jozwiak}}, \bibinfo {author} {\bibfnamefont
  {A.}~\bibnamefont {Bostwick}}, \bibinfo {author} {\bibfnamefont
  {E.}~\bibnamefont {Rotenberg}}, \bibinfo {author} {\bibfnamefont
  {W.}~\bibnamefont {Zhang}}, \bibinfo {author} {\bibfnamefont
  {W.}~\bibnamefont {Wu}}, \bibinfo {author} {\bibfnamefont {T.}~\bibnamefont
  {Suzuki}}, \emph {et~al.},\ }\bibfield  {title} {\bibinfo {title}
  {Spectromicroscopic measurement of surface and bulk band structure interplay
  in a disordered topological insulator},\ }\href@noop {} {\bibfield  {journal}
  {\bibinfo  {journal} {Nat. Phys.}\ }\textbf {\bibinfo {volume} {16}},\
  \bibinfo {pages} {285} (\bibinfo {year} {2020})}\BibitemShut {NoStop}%
\bibitem [{\citenamefont {Martin}\ \emph {et~al.}(2008)\citenamefont {Martin},
  \citenamefont {Akerman}, \citenamefont {Ulbricht}, \citenamefont {Lohmann},
  \citenamefont {Smet}, \citenamefont {Von~Klitzing},\ and\ \citenamefont
  {Yacoby}}]{256-martin2008observation}%
  \BibitemOpen
  \bibfield  {author} {\bibinfo {author} {\bibfnamefont {J.}~\bibnamefont
  {Martin}}, \bibinfo {author} {\bibfnamefont {N.}~\bibnamefont {Akerman}},
  \bibinfo {author} {\bibfnamefont {G.}~\bibnamefont {Ulbricht}}, \bibinfo
  {author} {\bibfnamefont {T.}~\bibnamefont {Lohmann}}, \bibinfo {author}
  {\bibfnamefont {J.~V.}\ \bibnamefont {Smet}}, \bibinfo {author}
  {\bibfnamefont {K.}~\bibnamefont {Von~Klitzing}},\ and\ \bibinfo {author}
  {\bibfnamefont {A.}~\bibnamefont {Yacoby}},\ }\bibfield  {title} {\bibinfo
  {title} {Observation of electron--hole puddles in graphene using a scanning
  single-electron transistor},\ }\href@noop {} {\bibfield  {journal} {\bibinfo
  {journal} {Nat. Phys.}\ }\textbf {\bibinfo {volume} {4}},\ \bibinfo {pages}
  {144} (\bibinfo {year} {2008})}\BibitemShut {NoStop}%
\bibitem [{\citenamefont {Rhodes}\ \emph {et~al.}(2019)\citenamefont {Rhodes},
  \citenamefont {Chae}, \citenamefont {Ribeiro-Palau},\ and\ \citenamefont
  {Hone}}]{257-rhodes2019disorder}%
  \BibitemOpen
  \bibfield  {author} {\bibinfo {author} {\bibfnamefont {D.}~\bibnamefont
  {Rhodes}}, \bibinfo {author} {\bibfnamefont {S.~H.}\ \bibnamefont {Chae}},
  \bibinfo {author} {\bibfnamefont {R.}~\bibnamefont {Ribeiro-Palau}},\ and\
  \bibinfo {author} {\bibfnamefont {J.}~\bibnamefont {Hone}},\ }\bibfield
  {title} {\bibinfo {title} {Disorder in van der waals heterostructures of 2{D}
  materials},\ }\href@noop {} {\bibfield  {journal} {\bibinfo  {journal} {Nat.
  Mater.}\ }\textbf {\bibinfo {volume} {18}},\ \bibinfo {pages} {541} (\bibinfo
  {year} {2019})}\BibitemShut {NoStop}%
\bibitem [{\citenamefont {He}\ \emph {et~al.}(2011)\citenamefont {He},
  \citenamefont {Li}, \citenamefont {Zhang}, \citenamefont {Karki},
  \citenamefont {Jin}, \citenamefont {Sales}, \citenamefont {Sefat},
  \citenamefont {McGuire}, \citenamefont {Mandrus},\ and\ \citenamefont
  {Plummer}}]{258-he2011nanoscale}%
  \BibitemOpen
  \bibfield  {author} {\bibinfo {author} {\bibfnamefont {X.}~\bibnamefont
  {He}}, \bibinfo {author} {\bibfnamefont {G.}~\bibnamefont {Li}}, \bibinfo
  {author} {\bibfnamefont {J.}~\bibnamefont {Zhang}}, \bibinfo {author}
  {\bibfnamefont {A.}~\bibnamefont {Karki}}, \bibinfo {author} {\bibfnamefont
  {R.}~\bibnamefont {Jin}}, \bibinfo {author} {\bibfnamefont {B.~C.}\
  \bibnamefont {Sales}}, \bibinfo {author} {\bibfnamefont {A.}~\bibnamefont
  {Sefat}}, \bibinfo {author} {\bibfnamefont {M.~A.}\ \bibnamefont {McGuire}},
  \bibinfo {author} {\bibfnamefont {D.}~\bibnamefont {Mandrus}},\ and\ \bibinfo
  {author} {\bibfnamefont {E.}~\bibnamefont {Plummer}},\ }\bibfield  {title}
  {\bibinfo {title} {Nanoscale chemical phase separation in
  {F}e{T}e$_{0.55}${S}e$_{0.45}$ as seen via scanning tunneling spectroscopy},\
  }\href@noop {} {\bibfield  {journal} {\bibinfo  {journal} {Phys. Rev. B}\
  }\textbf {\bibinfo {volume} {83}},\ \bibinfo {pages} {220502} (\bibinfo
  {year} {2011})}\BibitemShut {NoStop}%
\bibitem [{\citenamefont {Lin}\ \emph {et~al.}(2013)\citenamefont {Lin},
  \citenamefont {Li}, \citenamefont {Sales}, \citenamefont {Jesse},
  \citenamefont {Sefat}, \citenamefont {Kalinin},\ and\ \citenamefont
  {Pan}}]{259-lin2013direct}%
  \BibitemOpen
  \bibfield  {author} {\bibinfo {author} {\bibfnamefont {W.}~\bibnamefont
  {Lin}}, \bibinfo {author} {\bibfnamefont {Q.}~\bibnamefont {Li}}, \bibinfo
  {author} {\bibfnamefont {B.~C.}\ \bibnamefont {Sales}}, \bibinfo {author}
  {\bibfnamefont {S.}~\bibnamefont {Jesse}}, \bibinfo {author} {\bibfnamefont
  {A.~S.}\ \bibnamefont {Sefat}}, \bibinfo {author} {\bibfnamefont {S.~V.}\
  \bibnamefont {Kalinin}},\ and\ \bibinfo {author} {\bibfnamefont
  {M.}~\bibnamefont {Pan}},\ }\bibfield  {title} {\bibinfo {title} {Direct
  probe of interplay between local structure and superconductivity in
  {F}e{T}e$_{0.55}${S}e$_{0.45}$},\ }\href@noop {} {\bibfield  {journal}
  {\bibinfo  {journal} {ACS Nano}\ }\textbf {\bibinfo {volume} {7}},\ \bibinfo
  {pages} {2634} (\bibinfo {year} {2013})}\BibitemShut {NoStop}%
\bibitem [{\citenamefont {Singh}\ \emph {et~al.}(2013)\citenamefont {Singh},
  \citenamefont {White}, \citenamefont {Schmaus}, \citenamefont {Tsurkan},
  \citenamefont {Loidl}, \citenamefont {Deisenhofer},\ and\ \citenamefont
  {Wahl}}]{260-singh2013spatial}%
  \BibitemOpen
  \bibfield  {author} {\bibinfo {author} {\bibfnamefont {U.~R.}\ \bibnamefont
  {Singh}}, \bibinfo {author} {\bibfnamefont {S.~C.}\ \bibnamefont {White}},
  \bibinfo {author} {\bibfnamefont {S.}~\bibnamefont {Schmaus}}, \bibinfo
  {author} {\bibfnamefont {V.}~\bibnamefont {Tsurkan}}, \bibinfo {author}
  {\bibfnamefont {A.}~\bibnamefont {Loidl}}, \bibinfo {author} {\bibfnamefont
  {J.}~\bibnamefont {Deisenhofer}},\ and\ \bibinfo {author} {\bibfnamefont
  {P.}~\bibnamefont {Wahl}},\ }\bibfield  {title} {\bibinfo {title} {Spatial
  inhomogeneity of the superconducting gap and order parameter in
  {F}e{S}e$_{0.4}${T}e$_{0.6}$},\ }\href@noop {} {\bibfield  {journal}
  {\bibinfo  {journal} {Phys. Rev. B}\ }\textbf {\bibinfo {volume} {88}},\
  \bibinfo {pages} {155124} (\bibinfo {year} {2013})}\BibitemShut {NoStop}%
\bibitem [{\citenamefont {Massee}\ \emph {et~al.}(2015)\citenamefont {Massee},
  \citenamefont {Sprau}, \citenamefont {Wang}, \citenamefont {Davis},
  \citenamefont {Ghigo}, \citenamefont {Gu},\ and\ \citenamefont
  {Kwok}}]{261-massee2015imaging}%
  \BibitemOpen
  \bibfield  {author} {\bibinfo {author} {\bibfnamefont {F.}~\bibnamefont
  {Massee}}, \bibinfo {author} {\bibfnamefont {P.~O.}\ \bibnamefont {Sprau}},
  \bibinfo {author} {\bibfnamefont {Y.-L.}\ \bibnamefont {Wang}}, \bibinfo
  {author} {\bibfnamefont {J.~S.}\ \bibnamefont {Davis}}, \bibinfo {author}
  {\bibfnamefont {G.}~\bibnamefont {Ghigo}}, \bibinfo {author} {\bibfnamefont
  {G.~D.}\ \bibnamefont {Gu}},\ and\ \bibinfo {author} {\bibfnamefont {W.-K.}\
  \bibnamefont {Kwok}},\ }\bibfield  {title} {\bibinfo {title} {Imaging
  atomic-scale effects of high-energy ion irradiation on superconductivity and
  vortex pinning in {F}e({S}e,{T}e)},\ }\href@noop {} {\bibfield  {journal}
  {\bibinfo  {journal} {Sci. Adv.}\ }\textbf {\bibinfo {volume} {1}},\ \bibinfo
  {pages} {e1500033} (\bibinfo {year} {2015})}\BibitemShut {NoStop}%
\bibitem [{\citenamefont {Sacksteder}\ \emph {et~al.}(2015)\citenamefont
  {Sacksteder}, \citenamefont {Ohtsuki},\ and\ \citenamefont
  {Kobayashi}}]{263-sacksteder2015modification}%
  \BibitemOpen
  \bibfield  {author} {\bibinfo {author} {\bibfnamefont {V.}~\bibnamefont
  {Sacksteder}}, \bibinfo {author} {\bibfnamefont {T.}~\bibnamefont
  {Ohtsuki}},\ and\ \bibinfo {author} {\bibfnamefont {K.}~\bibnamefont
  {Kobayashi}},\ }\bibfield  {title} {\bibinfo {title} {Modification and
  control of topological insulator surface states using surface disorder},\
  }\href@noop {} {\bibfield  {journal} {\bibinfo  {journal} {Phys. Rev.
  Applied}\ }\textbf {\bibinfo {volume} {3}},\ \bibinfo {pages} {064006}
  (\bibinfo {year} {2015})}\BibitemShut {NoStop}%
\bibitem [{\citenamefont {Pan}\ and\ \citenamefont
  {Sarma}(2020)}]{264-pan2020physical}%
  \BibitemOpen
  \bibfield  {author} {\bibinfo {author} {\bibfnamefont {H.}~\bibnamefont
  {Pan}}\ and\ \bibinfo {author} {\bibfnamefont {S.~D.}\ \bibnamefont
  {Sarma}},\ }\bibfield  {title} {\bibinfo {title} {Physical mechanisms for
  zero-bias conductance peaks in {M}ajorana nanowires},\ }\href@noop {}
  {\bibfield  {journal} {\bibinfo  {journal} {Phys. Rev. Research}\ }\textbf
  {\bibinfo {volume} {2}},\ \bibinfo {pages} {013377} (\bibinfo {year}
  {2020})}\BibitemShut {NoStop}%
\bibitem [{\citenamefont {Fu}\ \emph {et~al.}(2007)\citenamefont {Fu},
  \citenamefont {Kane},\ and\ \citenamefont {Mele}}]{265-fu2007topological}%
  \BibitemOpen
  \bibfield  {author} {\bibinfo {author} {\bibfnamefont {L.}~\bibnamefont
  {Fu}}, \bibinfo {author} {\bibfnamefont {C.~L.}\ \bibnamefont {Kane}},\ and\
  \bibinfo {author} {\bibfnamefont {E.~J.}\ \bibnamefont {Mele}},\ }\bibfield
  {title} {\bibinfo {title} {Topological insulators in three dimensions},\
  }\href@noop {} {\bibfield  {journal} {\bibinfo  {journal} {Phys. Rev. Lett.}\
  }\textbf {\bibinfo {volume} {98}},\ \bibinfo {pages} {106803} (\bibinfo
  {year} {2007})}\BibitemShut {NoStop}%
\bibitem [{\citenamefont {Noguchi}\ \emph {et~al.}(2019)\citenamefont
  {Noguchi}, \citenamefont {Takahashi}, \citenamefont {Kuroda}, \citenamefont
  {Ochi}, \citenamefont {Shirasawa}, \citenamefont {Sakano}, \citenamefont
  {Bareille}, \citenamefont {Nakayama}, \citenamefont {Watson}, \citenamefont
  {Yaji} \emph {et~al.}}]{266-noguchi2019weak}%
  \BibitemOpen
  \bibfield  {author} {\bibinfo {author} {\bibfnamefont {R.}~\bibnamefont
  {Noguchi}}, \bibinfo {author} {\bibfnamefont {T.}~\bibnamefont {Takahashi}},
  \bibinfo {author} {\bibfnamefont {K.}~\bibnamefont {Kuroda}}, \bibinfo
  {author} {\bibfnamefont {M.}~\bibnamefont {Ochi}}, \bibinfo {author}
  {\bibfnamefont {T.}~\bibnamefont {Shirasawa}}, \bibinfo {author}
  {\bibfnamefont {M.}~\bibnamefont {Sakano}}, \bibinfo {author} {\bibfnamefont
  {C.}~\bibnamefont {Bareille}}, \bibinfo {author} {\bibfnamefont
  {M.}~\bibnamefont {Nakayama}}, \bibinfo {author} {\bibfnamefont
  {M.}~\bibnamefont {Watson}}, \bibinfo {author} {\bibfnamefont
  {K.}~\bibnamefont {Yaji}}, \emph {et~al.},\ }\bibfield  {title} {\bibinfo
  {title} {A weak topological insulator state in quasi-one-dimensional bismuth
  iodide},\ }\href@noop {} {\bibfield  {journal} {\bibinfo  {journal} {Nature}\
  }\textbf {\bibinfo {volume} {566}},\ \bibinfo {pages} {518} (\bibinfo {year}
  {2019})}\BibitemShut {NoStop}%
\bibitem [{\citenamefont {Shi}\ \emph {et~al.}(2017)\citenamefont {Shi},
  \citenamefont {Han}, \citenamefont {Richard}, \citenamefont {Wu},
  \citenamefont {Peng}, \citenamefont {Qian}, \citenamefont {Wang},
  \citenamefont {Hu}, \citenamefont {Sun},\ and\ \citenamefont
  {Ding}}]{267-shi2017fete1}%
  \BibitemOpen
  \bibfield  {author} {\bibinfo {author} {\bibfnamefont {X.}~\bibnamefont
  {Shi}}, \bibinfo {author} {\bibfnamefont {Z.-Q.}\ \bibnamefont {Han}},
  \bibinfo {author} {\bibfnamefont {P.}~\bibnamefont {Richard}}, \bibinfo
  {author} {\bibfnamefont {X.-X.}\ \bibnamefont {Wu}}, \bibinfo {author}
  {\bibfnamefont {X.-L.}\ \bibnamefont {Peng}}, \bibinfo {author}
  {\bibfnamefont {T.}~\bibnamefont {Qian}}, \bibinfo {author} {\bibfnamefont
  {S.-C.}\ \bibnamefont {Wang}}, \bibinfo {author} {\bibfnamefont {J.-P.}\
  \bibnamefont {Hu}}, \bibinfo {author} {\bibfnamefont {Y.-J.}\ \bibnamefont
  {Sun}},\ and\ \bibinfo {author} {\bibfnamefont {H.}~\bibnamefont {Ding}},\
  }\bibfield  {title} {\bibinfo {title} {{F}e{T}e$_{1-x}${S}e$_{x}$ monolayer
  films: towards the realization of high-temperature connate topological
  superconductivity},\ }\href@noop {} {\bibfield  {journal} {\bibinfo
  {journal} {Sci. Bull.}\ }\textbf {\bibinfo {volume} {62}},\ \bibinfo {pages}
  {503} (\bibinfo {year} {2017})}\BibitemShut {NoStop}%
\bibitem [{\citenamefont {Peng}\ \emph {et~al.}(2019)\citenamefont {Peng},
  \citenamefont {Li}, \citenamefont {Wu}, \citenamefont {Deng}, \citenamefont
  {Shi}, \citenamefont {Fan}, \citenamefont {Li}, \citenamefont {Huang},
  \citenamefont {Qian}, \citenamefont {Richard} \emph
  {et~al.}}]{268-peng2019observation}%
  \BibitemOpen
  \bibfield  {author} {\bibinfo {author} {\bibfnamefont {X.-L.}\ \bibnamefont
  {Peng}}, \bibinfo {author} {\bibfnamefont {Y.}~\bibnamefont {Li}}, \bibinfo
  {author} {\bibfnamefont {X.-X.}\ \bibnamefont {Wu}}, \bibinfo {author}
  {\bibfnamefont {H.-B.}\ \bibnamefont {Deng}}, \bibinfo {author}
  {\bibfnamefont {X.}~\bibnamefont {Shi}}, \bibinfo {author} {\bibfnamefont
  {W.-H.}\ \bibnamefont {Fan}}, \bibinfo {author} {\bibfnamefont
  {M.}~\bibnamefont {Li}}, \bibinfo {author} {\bibfnamefont {Y.-B.}\
  \bibnamefont {Huang}}, \bibinfo {author} {\bibfnamefont {T.}~\bibnamefont
  {Qian}}, \bibinfo {author} {\bibfnamefont {P.}~\bibnamefont {Richard}}, \emph
  {et~al.},\ }\bibfield  {title} {\bibinfo {title} {Observation of topological
  transition in high-{T}$_{c}$ superconducting monolayer
  {F}e{T}e$_{1-x}${S}e$_{x}$ films on {S}r{T}i{O}$_{3}$ (001)},\ }\href@noop {}
  {\bibfield  {journal} {\bibinfo  {journal} {Phys. Rev. B}\ }\textbf {\bibinfo
  {volume} {100}},\ \bibinfo {pages} {155134} (\bibinfo {year}
  {2019})}\BibitemShut {NoStop}%
\bibitem [{\citenamefont {Machida}\ \emph {et~al.}(2019)\citenamefont
  {Machida}, \citenamefont {Sun}, \citenamefont {Pyon}, \citenamefont {Takeda},
  \citenamefont {Kohsaka}, \citenamefont {Hanaguri}, \citenamefont {Sasagawa},\
  and\ \citenamefont {Tamegai}}]{269-machida2019zero}%
  \BibitemOpen
  \bibfield  {author} {\bibinfo {author} {\bibfnamefont {T.}~\bibnamefont
  {Machida}}, \bibinfo {author} {\bibfnamefont {Y.}~\bibnamefont {Sun}},
  \bibinfo {author} {\bibfnamefont {S.}~\bibnamefont {Pyon}}, \bibinfo {author}
  {\bibfnamefont {S.}~\bibnamefont {Takeda}}, \bibinfo {author} {\bibfnamefont
  {Y.}~\bibnamefont {Kohsaka}}, \bibinfo {author} {\bibfnamefont
  {T.}~\bibnamefont {Hanaguri}}, \bibinfo {author} {\bibfnamefont
  {T.}~\bibnamefont {Sasagawa}},\ and\ \bibinfo {author} {\bibfnamefont
  {T.}~\bibnamefont {Tamegai}},\ }\bibfield  {title} {\bibinfo {title}
  {Zero-energy vortex bound state in the superconducting topological surface
  state of {F}e({S}e,{T}e)},\ }\href@noop {} {\bibfield  {journal} {\bibinfo
  {journal} {Nat. Mater.}\ }\textbf {\bibinfo {volume} {18}},\ \bibinfo {pages}
  {811} (\bibinfo {year} {2019})}\BibitemShut {NoStop}%
\bibitem [{\citenamefont {Chen}\ \emph
  {et~al.}(2019{\natexlab{b}})\citenamefont {Chen}, \citenamefont {Chen},
  \citenamefont {Duan}, \citenamefont {Zhu}, \citenamefont {Yang},\ and\
  \citenamefont {Wen}}]{270-chen2019observation}%
  \BibitemOpen
  \bibfield  {author} {\bibinfo {author} {\bibfnamefont {X.}~\bibnamefont
  {Chen}}, \bibinfo {author} {\bibfnamefont {M.}~\bibnamefont {Chen}}, \bibinfo
  {author} {\bibfnamefont {W.}~\bibnamefont {Duan}}, \bibinfo {author}
  {\bibfnamefont {X.}~\bibnamefont {Zhu}}, \bibinfo {author} {\bibfnamefont
  {H.}~\bibnamefont {Yang}},\ and\ \bibinfo {author} {\bibfnamefont {H.-H.}\
  \bibnamefont {Wen}},\ }\bibfield  {title} {\bibinfo {title} {Observation and
  characterization of the zero energy conductance peak in the vortex core state
  of {F}e{T}e$_{0.55}${S}e$_{0.45}$},\ }\href@noop {} {\bibfield  {journal}
  {\bibinfo  {journal} {arXiv:1909.01686}\ } (\bibinfo {year}
  {2019}{\natexlab{b}})}\BibitemShut {NoStop}%
\bibitem [{\citenamefont {Chiu}\ \emph {et~al.}(2020)\citenamefont {Chiu},
  \citenamefont {Machida}, \citenamefont {Huang}, \citenamefont {Hanaguri},\
  and\ \citenamefont {Zhang}}]{271-chiu2020scalable}%
  \BibitemOpen
  \bibfield  {author} {\bibinfo {author} {\bibfnamefont {C.-K.}\ \bibnamefont
  {Chiu}}, \bibinfo {author} {\bibfnamefont {T.}~\bibnamefont {Machida}},
  \bibinfo {author} {\bibfnamefont {Y.}~\bibnamefont {Huang}}, \bibinfo
  {author} {\bibfnamefont {T.}~\bibnamefont {Hanaguri}},\ and\ \bibinfo
  {author} {\bibfnamefont {F.-C.}\ \bibnamefont {Zhang}},\ }\bibfield  {title}
  {\bibinfo {title} {Scalable {M}ajorana vortex modes in iron-based
  superconductors},\ }\href@noop {} {\bibfield  {journal} {\bibinfo  {journal}
  {Sci. Adv.}\ }\textbf {\bibinfo {volume} {6}},\ \bibinfo {pages} {eaay0443}
  (\bibinfo {year} {2020})}\BibitemShut {NoStop}%
\bibitem [{\citenamefont {Wang}\ \emph {et~al.}(2020)\citenamefont {Wang},
  \citenamefont {Rodriguez}, \citenamefont {Jiao}, \citenamefont {Howard},
  \citenamefont {Graham}, \citenamefont {Gu}, \citenamefont {Hughes},
  \citenamefont {Morr},\ and\ \citenamefont {Madhavan}}]{272-wang2020evidence}%
  \BibitemOpen
  \bibfield  {author} {\bibinfo {author} {\bibfnamefont {Z.}~\bibnamefont
  {Wang}}, \bibinfo {author} {\bibfnamefont {J.~O.}\ \bibnamefont {Rodriguez}},
  \bibinfo {author} {\bibfnamefont {L.}~\bibnamefont {Jiao}}, \bibinfo {author}
  {\bibfnamefont {S.}~\bibnamefont {Howard}}, \bibinfo {author} {\bibfnamefont
  {M.}~\bibnamefont {Graham}}, \bibinfo {author} {\bibfnamefont
  {G.}~\bibnamefont {Gu}}, \bibinfo {author} {\bibfnamefont {T.~L.}\
  \bibnamefont {Hughes}}, \bibinfo {author} {\bibfnamefont {D.~K.}\
  \bibnamefont {Morr}},\ and\ \bibinfo {author} {\bibfnamefont
  {V.}~\bibnamefont {Madhavan}},\ }\bibfield  {title} {\bibinfo {title}
  {Evidence for dispersing 1{D} {M}ajorana channels in an iron-based
  superconductor},\ }\href@noop {} {\bibfield  {journal} {\bibinfo  {journal}
  {Science}\ }\textbf {\bibinfo {volume} {367}},\ \bibinfo {pages} {104}
  (\bibinfo {year} {2020})}\BibitemShut {NoStop}%
\bibitem [{\citenamefont {Liu}\ \emph {et~al.}(2018)\citenamefont {Liu},
  \citenamefont {Chen}, \citenamefont {Zhang}, \citenamefont {Peng},
  \citenamefont {Yan}, \citenamefont {Lou}, \citenamefont {Huang},
  \citenamefont {Tian}, \citenamefont {Dong}, \citenamefont {Wang} \emph
  {et~al.}}]{273-liu2018robust}%
  \BibitemOpen
  \bibfield  {author} {\bibinfo {author} {\bibfnamefont {Q.}~\bibnamefont
  {Liu}}, \bibinfo {author} {\bibfnamefont {C.}~\bibnamefont {Chen}}, \bibinfo
  {author} {\bibfnamefont {T.}~\bibnamefont {Zhang}}, \bibinfo {author}
  {\bibfnamefont {R.}~\bibnamefont {Peng}}, \bibinfo {author} {\bibfnamefont
  {Y.-J.}\ \bibnamefont {Yan}}, \bibinfo {author} {\bibfnamefont
  {X.}~\bibnamefont {Lou}}, \bibinfo {author} {\bibfnamefont {Y.-L.}\
  \bibnamefont {Huang}}, \bibinfo {author} {\bibfnamefont {J.-P.}\ \bibnamefont
  {Tian}}, \bibinfo {author} {\bibfnamefont {X.-L.}\ \bibnamefont {Dong}},
  \bibinfo {author} {\bibfnamefont {G.-W.}\ \bibnamefont {Wang}}, \emph
  {et~al.},\ }\bibfield  {title} {\bibinfo {title} {Robust and clean {M}ajorana
  zero mode in the vortex core of high-temperature superconductor
  ({L}i$_{0.84}${F}e$_{0.16}$){OH}{F}e{S}e},\ }\href@noop {} {\bibfield
  {journal} {\bibinfo  {journal} {Phys. Rev. X}\ }\textbf {\bibinfo {volume}
  {8}},\ \bibinfo {pages} {041056} (\bibinfo {year} {2018})}\BibitemShut
  {NoStop}%
\bibitem [{\citenamefont {Chen}\ \emph
  {et~al.}(2019{\natexlab{c}})\citenamefont {Chen}, \citenamefont {Liu},
  \citenamefont {Zhang}, \citenamefont {Li}, \citenamefont {Shen},
  \citenamefont {Dong}, \citenamefont {Zhao}, \citenamefont {Zhang},\ and\
  \citenamefont {Feng}}]{274-chen2019quantized}%
  \BibitemOpen
  \bibfield  {author} {\bibinfo {author} {\bibfnamefont {C.}~\bibnamefont
  {Chen}}, \bibinfo {author} {\bibfnamefont {Q.}~\bibnamefont {Liu}}, \bibinfo
  {author} {\bibfnamefont {T.}~\bibnamefont {Zhang}}, \bibinfo {author}
  {\bibfnamefont {D.}~\bibnamefont {Li}}, \bibinfo {author} {\bibfnamefont
  {P.}~\bibnamefont {Shen}}, \bibinfo {author} {\bibfnamefont {X.}~\bibnamefont
  {Dong}}, \bibinfo {author} {\bibfnamefont {Z.-X.}\ \bibnamefont {Zhao}},
  \bibinfo {author} {\bibfnamefont {T.}~\bibnamefont {Zhang}},\ and\ \bibinfo
  {author} {\bibfnamefont {D.}~\bibnamefont {Feng}},\ }\bibfield  {title}
  {\bibinfo {title} {Quantized conductance of {M}ajorana zero mode in the
  vortex of the topological superconductor
  ({L}i$_{0.84}${F}e$_{0.16}$){OH}{F}e{S}e},\ }\href@noop {} {\bibfield
  {journal} {\bibinfo  {journal} {Chin. Phys. Lett.}\ }\textbf {\bibinfo
  {volume} {36}},\ \bibinfo {pages} {057403} (\bibinfo {year}
  {2019}{\natexlab{c}})}\BibitemShut {NoStop}%
\bibitem [{\citenamefont {Liu}\ \emph {et~al.}(2020)\citenamefont {Liu},
  \citenamefont {Chen}, \citenamefont {Liu}, \citenamefont {Wang},
  \citenamefont {Liu}, \citenamefont {Ye}, \citenamefont {Wang}, \citenamefont
  {Hu},\ and\ \citenamefont {Wang}}]{275-liu2020zero}%
  \BibitemOpen
  \bibfield  {author} {\bibinfo {author} {\bibfnamefont {C.}~\bibnamefont
  {Liu}}, \bibinfo {author} {\bibfnamefont {C.}~\bibnamefont {Chen}}, \bibinfo
  {author} {\bibfnamefont {X.}~\bibnamefont {Liu}}, \bibinfo {author}
  {\bibfnamefont {Z.}~\bibnamefont {Wang}}, \bibinfo {author} {\bibfnamefont
  {Y.}~\bibnamefont {Liu}}, \bibinfo {author} {\bibfnamefont {S.}~\bibnamefont
  {Ye}}, \bibinfo {author} {\bibfnamefont {Z.}~\bibnamefont {Wang}}, \bibinfo
  {author} {\bibfnamefont {J.}~\bibnamefont {Hu}},\ and\ \bibinfo {author}
  {\bibfnamefont {J.}~\bibnamefont {Wang}},\ }\bibfield  {title} {\bibinfo
  {title} {Zero-energy bound states in the high-temperature superconductors at
  the two-dimensional limit},\ }\href@noop {} {\bibfield  {journal} {\bibinfo
  {journal} {Sci. Adv.}\ }\textbf {\bibinfo {volume} {6}},\ \bibinfo {pages}
  {eaax7547} (\bibinfo {year} {2020})}\BibitemShut {NoStop}%
\bibitem [{\citenamefont {Jiang}\ \emph {et~al.}(2019)\citenamefont {Jiang},
  \citenamefont {Dai},\ and\ \citenamefont {Wang}}]{276-jiang2019quantum}%
  \BibitemOpen
  \bibfield  {author} {\bibinfo {author} {\bibfnamefont {K.}~\bibnamefont
  {Jiang}}, \bibinfo {author} {\bibfnamefont {X.}~\bibnamefont {Dai}},\ and\
  \bibinfo {author} {\bibfnamefont {Z.}~\bibnamefont {Wang}},\ }\bibfield
  {title} {\bibinfo {title} {Quantum anomalous vortex and {M}ajorana zero mode
  in iron-based superconductor {F}e({T}e,{S}e)},\ }\href@noop {} {\bibfield
  {journal} {\bibinfo  {journal} {Phys. Rev. X}\ }\textbf {\bibinfo {volume}
  {9}},\ \bibinfo {pages} {011033} (\bibinfo {year} {2019})}\BibitemShut
  {NoStop}%
\bibitem [{\citenamefont {Zhang}\ \emph
  {et~al.}(2020{\natexlab{a}})\citenamefont {Zhang}, \citenamefont {Yin},
  \citenamefont {Dai}, \citenamefont {Zhao}, \citenamefont {Chang},
  \citenamefont {Shumiya}, \citenamefont {Jiang}, \citenamefont {Zheng},
  \citenamefont {Bian}, \citenamefont {Multer} \emph
  {et~al.}}]{277-zhang2020field}%
  \BibitemOpen
  \bibfield  {author} {\bibinfo {author} {\bibfnamefont {S.~S.}\ \bibnamefont
  {Zhang}}, \bibinfo {author} {\bibfnamefont {J.-X.}\ \bibnamefont {Yin}},
  \bibinfo {author} {\bibfnamefont {G.}~\bibnamefont {Dai}}, \bibinfo {author}
  {\bibfnamefont {L.}~\bibnamefont {Zhao}}, \bibinfo {author} {\bibfnamefont
  {T.-R.}\ \bibnamefont {Chang}}, \bibinfo {author} {\bibfnamefont
  {N.}~\bibnamefont {Shumiya}}, \bibinfo {author} {\bibfnamefont
  {K.}~\bibnamefont {Jiang}}, \bibinfo {author} {\bibfnamefont
  {H.}~\bibnamefont {Zheng}}, \bibinfo {author} {\bibfnamefont
  {G.}~\bibnamefont {Bian}}, \bibinfo {author} {\bibfnamefont {D.}~\bibnamefont
  {Multer}}, \emph {et~al.},\ }\bibfield  {title} {\bibinfo {title} {Field-free
  platform for {M}ajorana-like zero mode in superconductors with a topological
  surface state},\ }\href@noop {} {\bibfield  {journal} {\bibinfo  {journal}
  {Phys. Rev. B}\ }\textbf {\bibinfo {volume} {101}},\ \bibinfo {pages}
  {100507} (\bibinfo {year} {2020}{\natexlab{a}})}\BibitemShut {NoStop}%
\bibitem [{\citenamefont {Fan}\ \emph {et~al.}(2020)\citenamefont {Fan},
  \citenamefont {Yang}, \citenamefont {Qian}, \citenamefont {Chen},
  \citenamefont {Zhang}, \citenamefont {Li}, \citenamefont {Huang},
  \citenamefont {Xing}, \citenamefont {Kong}, \citenamefont {Liu} \emph
  {et~al.}}]{278-fan2020reversible}%
  \BibitemOpen
  \bibfield  {author} {\bibinfo {author} {\bibfnamefont {P.}~\bibnamefont
  {Fan}}, \bibinfo {author} {\bibfnamefont {F.}~\bibnamefont {Yang}}, \bibinfo
  {author} {\bibfnamefont {G.}~\bibnamefont {Qian}}, \bibinfo {author}
  {\bibfnamefont {H.}~\bibnamefont {Chen}}, \bibinfo {author} {\bibfnamefont
  {Y.-Y.}\ \bibnamefont {Zhang}}, \bibinfo {author} {\bibfnamefont
  {G.}~\bibnamefont {Li}}, \bibinfo {author} {\bibfnamefont {Z.}~\bibnamefont
  {Huang}}, \bibinfo {author} {\bibfnamefont {Y.}~\bibnamefont {Xing}},
  \bibinfo {author} {\bibfnamefont {L.}~\bibnamefont {Kong}}, \bibinfo {author}
  {\bibfnamefont {W.}~\bibnamefont {Liu}}, \emph {et~al.},\ }\bibfield  {title}
  {\bibinfo {title} {Reversible transition between {Y}u-{S}hiba-{R}usinov state
  and {M}ajorana zero mode by magnetic adatom manipulation in an iron-based
  superconductor},\ }\href@noop {} {\bibfield  {journal} {\bibinfo  {journal}
  {arXiv:2001.07376}\ } (\bibinfo {year} {2020})}\BibitemShut {NoStop}%
\bibitem [{\citenamefont {Gray}\ \emph {et~al.}(2019)\citenamefont {Gray},
  \citenamefont {Freudenstein}, \citenamefont {Zhao}, \citenamefont
  {O’Connor}, \citenamefont {Jenkins}, \citenamefont {Kumar}, \citenamefont
  {Hoek}, \citenamefont {Kopec}, \citenamefont {Huh}, \citenamefont {Taniguchi}
  \emph {et~al.}}]{279-gray2019evidence}%
  \BibitemOpen
  \bibfield  {author} {\bibinfo {author} {\bibfnamefont {M.~J.}\ \bibnamefont
  {Gray}}, \bibinfo {author} {\bibfnamefont {J.}~\bibnamefont {Freudenstein}},
  \bibinfo {author} {\bibfnamefont {S.~Y.~F.}\ \bibnamefont {Zhao}}, \bibinfo
  {author} {\bibfnamefont {R.}~\bibnamefont {O’Connor}}, \bibinfo {author}
  {\bibfnamefont {S.}~\bibnamefont {Jenkins}}, \bibinfo {author} {\bibfnamefont
  {N.}~\bibnamefont {Kumar}}, \bibinfo {author} {\bibfnamefont
  {M.}~\bibnamefont {Hoek}}, \bibinfo {author} {\bibfnamefont {A.}~\bibnamefont
  {Kopec}}, \bibinfo {author} {\bibfnamefont {S.}~\bibnamefont {Huh}}, \bibinfo
  {author} {\bibfnamefont {T.}~\bibnamefont {Taniguchi}}, \emph {et~al.},\
  }\bibfield  {title} {\bibinfo {title} {Evidence for helical hinge zero modes
  in an {F}e-based superconductor},\ }\href@noop {} {\bibfield  {journal}
  {\bibinfo  {journal} {Nano Lett.}\ }\textbf {\bibinfo {volume} {19}},\
  \bibinfo {pages} {4890} (\bibinfo {year} {2019})}\BibitemShut {NoStop}%
\bibitem [{\citenamefont {Chen}\ \emph
  {et~al.}(2020{\natexlab{c}})\citenamefont {Chen}, \citenamefont {Jiang},
  \citenamefont {Zhang}, \citenamefont {Liu}, \citenamefont {Liu},
  \citenamefont {Wang},\ and\ \citenamefont {Wang}}]{280-chen2020atomic}%
  \BibitemOpen
  \bibfield  {author} {\bibinfo {author} {\bibfnamefont {C.}~\bibnamefont
  {Chen}}, \bibinfo {author} {\bibfnamefont {K.}~\bibnamefont {Jiang}},
  \bibinfo {author} {\bibfnamefont {Y.}~\bibnamefont {Zhang}}, \bibinfo
  {author} {\bibfnamefont {C.}~\bibnamefont {Liu}}, \bibinfo {author}
  {\bibfnamefont {Y.}~\bibnamefont {Liu}}, \bibinfo {author} {\bibfnamefont
  {Z.}~\bibnamefont {Wang}},\ and\ \bibinfo {author} {\bibfnamefont
  {J.}~\bibnamefont {Wang}},\ }\bibfield  {title} {\bibinfo {title} {Atomic
  line defects and zero-energy end states in monolayer {F}e({T}e,{S}e)
  high-temperature superconductors},\ }\href@noop {} {\bibfield  {journal}
  {\bibinfo  {journal} {Nat. Phys.}\ }\textbf {\bibinfo {volume} {16}},\
  \bibinfo {pages} {536} (\bibinfo {year} {2020}{\natexlab{c}})}\BibitemShut
  {NoStop}%
\bibitem [{\citenamefont {Wu}\ \emph {et~al.}(2020{\natexlab{b}})\citenamefont
  {Wu}, \citenamefont {Yin}, \citenamefont {Liu},\ and\ \citenamefont
  {Hu}}]{281-wu2020topological}%
  \BibitemOpen
  \bibfield  {author} {\bibinfo {author} {\bibfnamefont {X.}~\bibnamefont
  {Wu}}, \bibinfo {author} {\bibfnamefont {J.-X.}\ \bibnamefont {Yin}},
  \bibinfo {author} {\bibfnamefont {C.-X.}\ \bibnamefont {Liu}},\ and\ \bibinfo
  {author} {\bibfnamefont {J.}~\bibnamefont {Hu}},\ }\bibfield  {title}
  {\bibinfo {title} {Topological magnetic line defects in {F}e({T}e,{S}e)
  high-temperature superconductors},\ }\href@noop {} {\bibfield  {journal}
  {\bibinfo  {journal} {arXiv:2004.05848}\ } (\bibinfo {year}
  {2020}{\natexlab{b}})}\BibitemShut {NoStop}%
\bibitem [{\citenamefont {Zhang}\ \emph
  {et~al.}(2020{\natexlab{b}})\citenamefont {Zhang}, \citenamefont {Jiang},
  \citenamefont {Zhang}, \citenamefont {Wang},\ and\ \citenamefont
  {Wang}}]{282-zhang2020atomic}%
  \BibitemOpen
  \bibfield  {author} {\bibinfo {author} {\bibfnamefont {Y.}~\bibnamefont
  {Zhang}}, \bibinfo {author} {\bibfnamefont {K.}~\bibnamefont {Jiang}},
  \bibinfo {author} {\bibfnamefont {F.}~\bibnamefont {Zhang}}, \bibinfo
  {author} {\bibfnamefont {J.}~\bibnamefont {Wang}},\ and\ \bibinfo {author}
  {\bibfnamefont {Z.}~\bibnamefont {Wang}},\ }\bibfield  {title} {\bibinfo
  {title} {Atomic line defects in unconventional superconductors as a new route
  toward one dimensional topological superconductors},\ }\href@noop {}
  {\bibfield  {journal} {\bibinfo  {journal} {arXiv:2004.05860}\ } (\bibinfo
  {year} {2020}{\natexlab{b}})}\BibitemShut {NoStop}%
\bibitem [{\citenamefont {Zaki}\ \emph {et~al.}(2019)\citenamefont {Zaki},
  \citenamefont {Gu}, \citenamefont {Tsvelik}, \citenamefont {Wu},\ and\
  \citenamefont {Johnson}}]{283-zaki2019time}%
  \BibitemOpen
  \bibfield  {author} {\bibinfo {author} {\bibfnamefont {N.}~\bibnamefont
  {Zaki}}, \bibinfo {author} {\bibfnamefont {G.}~\bibnamefont {Gu}}, \bibinfo
  {author} {\bibfnamefont {A.}~\bibnamefont {Tsvelik}}, \bibinfo {author}
  {\bibfnamefont {C.}~\bibnamefont {Wu}},\ and\ \bibinfo {author}
  {\bibfnamefont {P.}~\bibnamefont {Johnson}},\ }\bibfield  {title} {\bibinfo
  {title} {Time reversal symmetry breaking in the {F}e-chalcogenide
  superconductors},\ }\href@noop {} {\bibfield  {journal} {\bibinfo  {journal}
  {arXiv:1907.11602}\ } (\bibinfo {year} {2019})}\BibitemShut {NoStop}%
\bibitem [{\citenamefont {Hu}\ \emph {et~al.}(2020)\citenamefont {Hu},
  \citenamefont {Johnson},\ and\ \citenamefont {Wu}}]{284-hu2020pairing}%
  \BibitemOpen
  \bibfield  {author} {\bibinfo {author} {\bibfnamefont {L.-H.}\ \bibnamefont
  {Hu}}, \bibinfo {author} {\bibfnamefont {P.}~\bibnamefont {Johnson}},\ and\
  \bibinfo {author} {\bibfnamefont {C.}~\bibnamefont {Wu}},\ }\bibfield
  {title} {\bibinfo {title} {Pairing symmetry and topological surface state in
  iron-chalcogenide superconductors},\ }\href@noop {} {\bibfield  {journal}
  {\bibinfo  {journal} {Phys. Rev. Research}\ }\textbf {\bibinfo {volume}
  {2}},\ \bibinfo {pages} {022021} (\bibinfo {year} {2020})}\BibitemShut
  {NoStop}%
\bibitem [{\citenamefont {Wu}\ \emph {et~al.}(2020{\natexlab{c}})\citenamefont
  {Wu}, \citenamefont {Chung}, \citenamefont {Liu},\ and\ \citenamefont
  {Kim}}]{285-wu2020topological}%
  \BibitemOpen
  \bibfield  {author} {\bibinfo {author} {\bibfnamefont {X.}~\bibnamefont
  {Wu}}, \bibinfo {author} {\bibfnamefont {S.~B.}\ \bibnamefont {Chung}},
  \bibinfo {author} {\bibfnamefont {C.-X.}\ \bibnamefont {Liu}},\ and\ \bibinfo
  {author} {\bibfnamefont {E.-A.}\ \bibnamefont {Kim}},\ }\bibfield  {title}
  {\bibinfo {title} {Topological orders competing for the {D}irac surface state
  in {F}e{S}e{T}e surfaces},\ }\href@noop {} {\bibfield  {journal} {\bibinfo
  {journal} {arXiv:2004.13068}\ } (\bibinfo {year}
  {2020}{\natexlab{c}})}\BibitemShut {NoStop}%
\bibitem [{\citenamefont {Kawakami}\ and\ \citenamefont
  {Sato}(2019)}]{286-kawakami2019topological}%
  \BibitemOpen
  \bibfield  {author} {\bibinfo {author} {\bibfnamefont {T.}~\bibnamefont
  {Kawakami}}\ and\ \bibinfo {author} {\bibfnamefont {M.}~\bibnamefont
  {Sato}},\ }\bibfield  {title} {\bibinfo {title} {Topological crystalline
  superconductivity in {D}irac semimetal phase of iron-based superconductors},\
  }\href@noop {} {\bibfield  {journal} {\bibinfo  {journal} {Phys. Rev. B}\
  }\textbf {\bibinfo {volume} {100}},\ \bibinfo {pages} {094520} (\bibinfo
  {year} {2019})}\BibitemShut {NoStop}%
\bibitem [{\citenamefont {Luo}\ \emph {et~al.}(2020)\citenamefont {Luo},
  \citenamefont {Chen}, \citenamefont {Wang},\ and\ \citenamefont
  {Yu}}]{287-luo2020topological}%
  \BibitemOpen
  \bibfield  {author} {\bibinfo {author} {\bibfnamefont {X.}~\bibnamefont
  {Luo}}, \bibinfo {author} {\bibfnamefont {Y.-G.}\ \bibnamefont {Chen}},
  \bibinfo {author} {\bibfnamefont {Z.}~\bibnamefont {Wang}},\ and\ \bibinfo
  {author} {\bibfnamefont {Y.}~\bibnamefont {Yu}},\ }\bibfield  {title}
  {\bibinfo {title} {Topological superconductor from superconducting
  topological surface states and fault-tolerant quantum computing},\
  }\href@noop {} {\bibfield  {journal} {\bibinfo  {journal} {arXiv:2003.11752}\
  } (\bibinfo {year} {2020})}\BibitemShut {NoStop}%
\bibitem [{\citenamefont {Chen}\ \emph
  {et~al.}(2018{\natexlab{c}})\citenamefont {Chen}, \citenamefont {Chen},
  \citenamefont {Yang}, \citenamefont {Du},\ and\ \citenamefont
  {Wen}}]{288-chen2018superconductivity}%
  \BibitemOpen
  \bibfield  {author} {\bibinfo {author} {\bibfnamefont {M.}~\bibnamefont
  {Chen}}, \bibinfo {author} {\bibfnamefont {X.}~\bibnamefont {Chen}}, \bibinfo
  {author} {\bibfnamefont {H.}~\bibnamefont {Yang}}, \bibinfo {author}
  {\bibfnamefont {Z.}~\bibnamefont {Du}},\ and\ \bibinfo {author}
  {\bibfnamefont {H.-H.}\ \bibnamefont {Wen}},\ }\bibfield  {title} {\bibinfo
  {title} {Superconductivity with twofold symmetry in
  {B}i$_{2}${T}e$_{3}$/{F}e{T}e$_{0.55}${S}e$_{0.45}$ heterostructures},\
  }\href@noop {} {\bibfield  {journal} {\bibinfo  {journal} {Sci. Adv.}\
  }\textbf {\bibinfo {volume} {4}},\ \bibinfo {pages} {eaat1084} (\bibinfo
  {year} {2018}{\natexlab{c}})}\BibitemShut {NoStop}%
\bibitem [{\citenamefont {Zhao}\ \emph {et~al.}(2018)\citenamefont {Zhao},
  \citenamefont {Rachmilowitz}, \citenamefont {Ren}, \citenamefont {Han},
  \citenamefont {Schneeloch}, \citenamefont {Zhong}, \citenamefont {Gu},
  \citenamefont {Wang},\ and\ \citenamefont
  {Zeljkovic}}]{289-zhao2018superconducting}%
  \BibitemOpen
  \bibfield  {author} {\bibinfo {author} {\bibfnamefont {H.}~\bibnamefont
  {Zhao}}, \bibinfo {author} {\bibfnamefont {B.}~\bibnamefont {Rachmilowitz}},
  \bibinfo {author} {\bibfnamefont {Z.}~\bibnamefont {Ren}}, \bibinfo {author}
  {\bibfnamefont {R.}~\bibnamefont {Han}}, \bibinfo {author} {\bibfnamefont
  {J.}~\bibnamefont {Schneeloch}}, \bibinfo {author} {\bibfnamefont
  {R.}~\bibnamefont {Zhong}}, \bibinfo {author} {\bibfnamefont
  {G.}~\bibnamefont {Gu}}, \bibinfo {author} {\bibfnamefont {Z.}~\bibnamefont
  {Wang}},\ and\ \bibinfo {author} {\bibfnamefont {I.}~\bibnamefont
  {Zeljkovic}},\ }\bibfield  {title} {\bibinfo {title} {Superconducting
  proximity effect in a topological insulator using {F}e({T}e,{S}e)},\
  }\href@noop {} {\bibfield  {journal} {\bibinfo  {journal} {Phys. Rev. B}\
  }\textbf {\bibinfo {volume} {97}},\ \bibinfo {pages} {224504} (\bibinfo
  {year} {2018})}\BibitemShut {NoStop}%
\bibitem [{\citenamefont {Bao}\ \emph {et~al.}(2018)\citenamefont {Bao},
  \citenamefont {Tang}, \citenamefont {Lu},\ and\ \citenamefont
  {Wang}}]{290-bao2018visualizing}%
  \BibitemOpen
  \bibfield  {author} {\bibinfo {author} {\bibfnamefont {W.-C.}\ \bibnamefont
  {Bao}}, \bibinfo {author} {\bibfnamefont {Q.-K.}\ \bibnamefont {Tang}},
  \bibinfo {author} {\bibfnamefont {D.-C.}\ \bibnamefont {Lu}},\ and\ \bibinfo
  {author} {\bibfnamefont {Q.-H.}\ \bibnamefont {Wang}},\ }\bibfield  {title}
  {\bibinfo {title} {Visualizing the d vector in a nematic triplet
  superconductor},\ }\href@noop {} {\bibfield  {journal} {\bibinfo  {journal}
  {Phys. Rev. B}\ }\textbf {\bibinfo {volume} {98}},\ \bibinfo {pages} {054502}
  (\bibinfo {year} {2018})}\BibitemShut {NoStop}%
\bibitem [{\citenamefont {Chen}\ \emph
  {et~al.}(2019{\natexlab{d}})\citenamefont {Chen}, \citenamefont {Chen},
  \citenamefont {Duan}, \citenamefont {Yang},\ and\ \citenamefont
  {Wen}}]{291-chen2019zero}%
  \BibitemOpen
  \bibfield  {author} {\bibinfo {author} {\bibfnamefont {X.}~\bibnamefont
  {Chen}}, \bibinfo {author} {\bibfnamefont {M.}~\bibnamefont {Chen}}, \bibinfo
  {author} {\bibfnamefont {W.}~\bibnamefont {Duan}}, \bibinfo {author}
  {\bibfnamefont {H.}~\bibnamefont {Yang}},\ and\ \bibinfo {author}
  {\bibfnamefont {H.-H.}\ \bibnamefont {Wen}},\ }\bibfield  {title} {\bibinfo
  {title} {Zero-energy modes on superconducting bismuth islands deposited on
  {F}e({T}e,{S}e)},\ }\href@noop {} {\bibfield  {journal} {\bibinfo  {journal}
  {arXiv:1905.05735}\ } (\bibinfo {year} {2019}{\natexlab{d}})}\BibitemShut
  {NoStop}%
\bibitem [{\citenamefont {Rachmilowitz}\ \emph {et~al.}(2019)\citenamefont
  {Rachmilowitz}, \citenamefont {Zhao}, \citenamefont {Li}, \citenamefont
  {LaFleur}, \citenamefont {Schneeloch}, \citenamefont {Zhong}, \citenamefont
  {Gu},\ and\ \citenamefont {Zeljkovic}}]{292-rachmilowitz2019proximity}%
  \BibitemOpen
  \bibfield  {author} {\bibinfo {author} {\bibfnamefont {B.}~\bibnamefont
  {Rachmilowitz}}, \bibinfo {author} {\bibfnamefont {H.}~\bibnamefont {Zhao}},
  \bibinfo {author} {\bibfnamefont {H.}~\bibnamefont {Li}}, \bibinfo {author}
  {\bibfnamefont {A.}~\bibnamefont {LaFleur}}, \bibinfo {author} {\bibfnamefont
  {J.}~\bibnamefont {Schneeloch}}, \bibinfo {author} {\bibfnamefont
  {R.}~\bibnamefont {Zhong}}, \bibinfo {author} {\bibfnamefont
  {G.}~\bibnamefont {Gu}},\ and\ \bibinfo {author} {\bibfnamefont
  {I.}~\bibnamefont {Zeljkovic}},\ }\bibfield  {title} {\bibinfo {title}
  {Proximity-induced superconductivity in a topological crystalline
  insulator},\ }\href@noop {} {\bibfield  {journal} {\bibinfo  {journal} {Phys.
  Rev. B}\ }\textbf {\bibinfo {volume} {100}},\ \bibinfo {pages} {241402}
  (\bibinfo {year} {2019})}\BibitemShut {NoStop}%
\bibitem [{\citenamefont {Dong}\ \emph {et~al.}(2019)\citenamefont {Dong},
  \citenamefont {Zhao}, \citenamefont {Zeljkovic}, \citenamefont {Wilson},\
  and\ \citenamefont {Harter}}]{293-dong2019bulk}%
  \BibitemOpen
  \bibfield  {author} {\bibinfo {author} {\bibfnamefont {L.}~\bibnamefont
  {Dong}}, \bibinfo {author} {\bibfnamefont {H.}~\bibnamefont {Zhao}}, \bibinfo
  {author} {\bibfnamefont {I.}~\bibnamefont {Zeljkovic}}, \bibinfo {author}
  {\bibfnamefont {S.~D.}\ \bibnamefont {Wilson}},\ and\ \bibinfo {author}
  {\bibfnamefont {J.~W.}\ \bibnamefont {Harter}},\ }\bibfield  {title}
  {\bibinfo {title} {Bulk superconductivity in {F}e{T}e$_{1-x}${S}e$_{x}$ via
  physicochemical pumping of excess iron},\ }\href@noop {} {\bibfield
  {journal} {\bibinfo  {journal} {Phys. Rev. Materials}\ }\textbf {\bibinfo
  {volume} {3}},\ \bibinfo {pages} {114801} (\bibinfo {year}
  {2019})}\BibitemShut {NoStop}%
\bibitem [{\citenamefont {Zhang}\ \emph
  {et~al.}(2019{\natexlab{b}})\citenamefont {Zhang}, \citenamefont {Cole},\
  and\ \citenamefont {Sarma}}]{294-zhang2019helical}%
  \BibitemOpen
  \bibfield  {author} {\bibinfo {author} {\bibfnamefont {R.-X.}\ \bibnamefont
  {Zhang}}, \bibinfo {author} {\bibfnamefont {W.~S.}\ \bibnamefont {Cole}},\
  and\ \bibinfo {author} {\bibfnamefont {S.~D.}\ \bibnamefont {Sarma}},\
  }\bibfield  {title} {\bibinfo {title} {Helical hinge {M}ajorana modes in
  iron-based superconductors},\ }\href@noop {} {\bibfield  {journal} {\bibinfo
  {journal} {Phys. Rev. Lett.}\ }\textbf {\bibinfo {volume} {122}},\ \bibinfo
  {pages} {187001} (\bibinfo {year} {2019}{\natexlab{b}})}\BibitemShut
  {NoStop}%
\bibitem [{\citenamefont {Wu}\ \emph {et~al.}(2020{\natexlab{d}})\citenamefont
  {Wu}, \citenamefont {Benalcazar}, \citenamefont {Li}, \citenamefont
  {Thomale}, \citenamefont {Liu},\ and\ \citenamefont
  {Hu}}]{295-wu2020boundary}%
  \BibitemOpen
  \bibfield  {author} {\bibinfo {author} {\bibfnamefont {X.}~\bibnamefont
  {Wu}}, \bibinfo {author} {\bibfnamefont {W.~A.}\ \bibnamefont {Benalcazar}},
  \bibinfo {author} {\bibfnamefont {Y.}~\bibnamefont {Li}}, \bibinfo {author}
  {\bibfnamefont {R.}~\bibnamefont {Thomale}}, \bibinfo {author} {\bibfnamefont
  {C.-X.}\ \bibnamefont {Liu}},\ and\ \bibinfo {author} {\bibfnamefont
  {J.}~\bibnamefont {Hu}},\ }\bibfield  {title} {\bibinfo {title}
  {Boundary-obstructed topological high-{T}$_{c}$ superconductivity in iron
  pnictides},\ }\href@noop {} {\bibfield  {journal} {\bibinfo  {journal}
  {arXiv:2003.12204}\ } (\bibinfo {year} {2020}{\natexlab{d}})}\BibitemShut
  {NoStop}%
\bibitem [{\citenamefont {Zhang}\ \emph
  {et~al.}(2019{\natexlab{c}})\citenamefont {Zhang}, \citenamefont {Cole},
  \citenamefont {Wu},\ and\ \citenamefont {Sarma}}]{296-zhang2019higher}%
  \BibitemOpen
  \bibfield  {author} {\bibinfo {author} {\bibfnamefont {R.-X.}\ \bibnamefont
  {Zhang}}, \bibinfo {author} {\bibfnamefont {W.~S.}\ \bibnamefont {Cole}},
  \bibinfo {author} {\bibfnamefont {X.}~\bibnamefont {Wu}},\ and\ \bibinfo
  {author} {\bibfnamefont {S.~D.}\ \bibnamefont {Sarma}},\ }\bibfield  {title}
  {\bibinfo {title} {Higher-order topology and nodal topological
  superconductivity in {F}e({S}e,{T}e) heterostructures},\ }\href@noop {}
  {\bibfield  {journal} {\bibinfo  {journal} {Phys. Rev. Lett.}\ }\textbf
  {\bibinfo {volume} {123}},\ \bibinfo {pages} {167001} (\bibinfo {year}
  {2019}{\natexlab{c}})}\BibitemShut {NoStop}%
\bibitem [{\citenamefont {Wu}\ \emph {et~al.}(2019)\citenamefont {Wu},
  \citenamefont {Liu}, \citenamefont {Thomale},\ and\ \citenamefont
  {Liu}}]{297-wu2019high}%
  \BibitemOpen
  \bibfield  {author} {\bibinfo {author} {\bibfnamefont {X.}~\bibnamefont
  {Wu}}, \bibinfo {author} {\bibfnamefont {X.}~\bibnamefont {Liu}}, \bibinfo
  {author} {\bibfnamefont {R.}~\bibnamefont {Thomale}},\ and\ \bibinfo {author}
  {\bibfnamefont {C.-X.}\ \bibnamefont {Liu}},\ }\bibfield  {title} {\bibinfo
  {title} {High-{T}$_{c}$ superconductor {F}e({S}e,{T}e) monolayer: an
  intrinsic, scalable and electrically-tunable {M}ajorana platform},\
  }\href@noop {} {\bibfield  {journal} {\bibinfo  {journal} {arXiv:1905.10648}\
  } (\bibinfo {year} {2019})}\BibitemShut {NoStop}%
\bibitem [{\citenamefont {Chen}\ \emph
  {et~al.}(2019{\natexlab{e}})\citenamefont {Chen}, \citenamefont {Liu},
  \citenamefont {Xu},\ and\ \citenamefont {Liu}}]{298-chen2019lattice}%
  \BibitemOpen
  \bibfield  {author} {\bibinfo {author} {\bibfnamefont {L.}~\bibnamefont
  {Chen}}, \bibinfo {author} {\bibfnamefont {B.}~\bibnamefont {Liu}}, \bibinfo
  {author} {\bibfnamefont {G.}~\bibnamefont {Xu}},\ and\ \bibinfo {author}
  {\bibfnamefont {X.}~\bibnamefont {Liu}},\ }\bibfield  {title} {\bibinfo
  {title} {Lattice distortion induced first and second order topological phase
  transition in rectangular high-{T}$_{c}$ superconducting monolayer},\
  }\href@noop {} {\bibfield  {journal} {\bibinfo  {journal} {arXiv:1909.10402}\
  } (\bibinfo {year} {2019}{\natexlab{e}})}\BibitemShut {NoStop}%
\bibitem [{\citenamefont {Wu}\ \emph {et~al.}(2020{\natexlab{e}})\citenamefont
  {Wu}, \citenamefont {Zhang}, \citenamefont {Xu}, \citenamefont {Hu},\ and\
  \citenamefont {Liu}}]{299-wu2020pursuit}%
  \BibitemOpen
  \bibfield  {author} {\bibinfo {author} {\bibfnamefont {X.}~\bibnamefont
  {Wu}}, \bibinfo {author} {\bibfnamefont {R.-X.}\ \bibnamefont {Zhang}},
  \bibinfo {author} {\bibfnamefont {G.}~\bibnamefont {Xu}}, \bibinfo {author}
  {\bibfnamefont {J.}~\bibnamefont {Hu}},\ and\ \bibinfo {author}
  {\bibfnamefont {C.-X.}\ \bibnamefont {Liu}},\ }\bibfield  {title} {\bibinfo
  {title} {In the pursuit of {M}ajorana modes in iron-based high-{T}$_{c}$
  superconductors},\ }\href@noop {} {\bibfield  {journal} {\bibinfo  {journal}
  {arXiv:2005.03603}\ } (\bibinfo {year} {2020}{\natexlab{e}})}\BibitemShut
  {NoStop}%
\bibitem [{\citenamefont {Liu}\ \emph {et~al.}(2019{\natexlab{d}})\citenamefont
  {Liu}, \citenamefont {Liu}, \citenamefont {Zhang},\ and\ \citenamefont
  {Chiu}}]{300-liu2019protocol}%
  \BibitemOpen
  \bibfield  {author} {\bibinfo {author} {\bibfnamefont {C.-X.}\ \bibnamefont
  {Liu}}, \bibinfo {author} {\bibfnamefont {D.~E.}\ \bibnamefont {Liu}},
  \bibinfo {author} {\bibfnamefont {F.-C.}\ \bibnamefont {Zhang}},\ and\
  \bibinfo {author} {\bibfnamefont {C.-K.}\ \bibnamefont {Chiu}},\ }\bibfield
  {title} {\bibinfo {title} {Protocol for reading out {M}ajorana vortex qubits
  and testing non-{A}belian statistics},\ }\href@noop {} {\bibfield  {journal}
  {\bibinfo  {journal} {Phys. Rev. Applied}\ }\textbf {\bibinfo {volume}
  {12}},\ \bibinfo {pages} {054035} (\bibinfo {year}
  {2019}{\natexlab{d}})}\BibitemShut {NoStop}%
\bibitem [{\citenamefont {November}\ \emph {et~al.}(2019)\citenamefont
  {November}, \citenamefont {Sau}, \citenamefont {Williams},\ and\
  \citenamefont {Hoffman}}]{301-november2019scheme}%
  \BibitemOpen
  \bibfield  {author} {\bibinfo {author} {\bibfnamefont {B.~H.}\ \bibnamefont
  {November}}, \bibinfo {author} {\bibfnamefont {J.~D.}\ \bibnamefont {Sau}},
  \bibinfo {author} {\bibfnamefont {J.~R.}\ \bibnamefont {Williams}},\ and\
  \bibinfo {author} {\bibfnamefont {J.~E.}\ \bibnamefont {Hoffman}},\
  }\bibfield  {title} {\bibinfo {title} {Scheme for {M}ajorana manipulation
  using magnetic force microscopy},\ }\href@noop {} {\bibfield  {journal}
  {\bibinfo  {journal} {arXiv:1905.09792}\ } (\bibinfo {year}
  {2019})}\BibitemShut {NoStop}%
\bibitem [{\citenamefont {Posske}\ \emph {et~al.}(2020)\citenamefont {Posske},
  \citenamefont {Chiu},\ and\ \citenamefont {Thorwart}}]{302-posske2020vortex}%
  \BibitemOpen
  \bibfield  {author} {\bibinfo {author} {\bibfnamefont {T.}~\bibnamefont
  {Posske}}, \bibinfo {author} {\bibfnamefont {C.-K.}\ \bibnamefont {Chiu}},\
  and\ \bibinfo {author} {\bibfnamefont {M.}~\bibnamefont {Thorwart}},\
  }\bibfield  {title} {\bibinfo {title} {Vortex {M}ajorana braiding in a finite
  time},\ }\href@noop {} {\bibfield  {journal} {\bibinfo  {journal} {Phys. Rev.
  Research}\ }\textbf {\bibinfo {volume} {2}},\ \bibinfo {pages} {023205}
  (\bibinfo {year} {2020})}\BibitemShut {NoStop}%
\bibitem [{\citenamefont {Yuan}\ \emph {et~al.}(2019)\citenamefont {Yuan},
  \citenamefont {Pan}, \citenamefont {Wang}, \citenamefont {Fang},
  \citenamefont {Song}, \citenamefont {Wang}, \citenamefont {He}, \citenamefont
  {Ma}, \citenamefont {Zhang}, \citenamefont {Huang} \emph
  {et~al.}}]{303-yuan2019evidence}%
  \BibitemOpen
  \bibfield  {author} {\bibinfo {author} {\bibfnamefont {Y.}~\bibnamefont
  {Yuan}}, \bibinfo {author} {\bibfnamefont {J.}~\bibnamefont {Pan}}, \bibinfo
  {author} {\bibfnamefont {X.}~\bibnamefont {Wang}}, \bibinfo {author}
  {\bibfnamefont {Y.}~\bibnamefont {Fang}}, \bibinfo {author} {\bibfnamefont
  {C.}~\bibnamefont {Song}}, \bibinfo {author} {\bibfnamefont {L.}~\bibnamefont
  {Wang}}, \bibinfo {author} {\bibfnamefont {K.}~\bibnamefont {He}}, \bibinfo
  {author} {\bibfnamefont {X.}~\bibnamefont {Ma}}, \bibinfo {author}
  {\bibfnamefont {H.}~\bibnamefont {Zhang}}, \bibinfo {author} {\bibfnamefont
  {F.}~\bibnamefont {Huang}}, \emph {et~al.},\ }\bibfield  {title} {\bibinfo
  {title} {Evidence of anisotropic {M}ajorana bound states in
  2{M}-{WS}$_{2}$},\ }\href@noop {} {\bibfield  {journal} {\bibinfo  {journal}
  {Nat. Phys.}\ }\textbf {\bibinfo {volume} {15}},\ \bibinfo {pages} {1046}
  (\bibinfo {year} {2019})}\BibitemShut {NoStop}%
\bibitem [{\citenamefont {Xia}\ \emph {et~al.}(2020)\citenamefont {Xia},
  \citenamefont {Shi}, \citenamefont {Zhang}, \citenamefont {Su}, \citenamefont
  {Wang}, \citenamefont {Ding}, \citenamefont {Chen}, \citenamefont {Wang},
  \citenamefont {Zou}, \citenamefont {Yu} \emph {et~al.}}]{304-xia2020bulk}%
  \BibitemOpen
  \bibfield  {author} {\bibinfo {author} {\bibfnamefont {W.}~\bibnamefont
  {Xia}}, \bibinfo {author} {\bibfnamefont {X.}~\bibnamefont {Shi}}, \bibinfo
  {author} {\bibfnamefont {Y.}~\bibnamefont {Zhang}}, \bibinfo {author}
  {\bibfnamefont {H.}~\bibnamefont {Su}}, \bibinfo {author} {\bibfnamefont
  {Q.}~\bibnamefont {Wang}}, \bibinfo {author} {\bibfnamefont {L.}~\bibnamefont
  {Ding}}, \bibinfo {author} {\bibfnamefont {L.}~\bibnamefont {Chen}}, \bibinfo
  {author} {\bibfnamefont {X.}~\bibnamefont {Wang}}, \bibinfo {author}
  {\bibfnamefont {Z.}~\bibnamefont {Zou}}, \bibinfo {author} {\bibfnamefont
  {N.}~\bibnamefont {Yu}}, \emph {et~al.},\ }\bibfield  {title} {\bibinfo
  {title} {Bulk {F}ermi surface of the layered superconductor {T}a{S}e$_{3}$
  with three-dimensional strong topological state},\ }\href@noop {} {\bibfield
  {journal} {\bibinfo  {journal} {Phys. Rev. B}\ }\textbf {\bibinfo {volume}
  {101}},\ \bibinfo {pages} {155117} (\bibinfo {year} {2020})}\BibitemShut
  {NoStop}%
\bibitem [{\citenamefont {Zhang}\ \emph
  {et~al.}(2020{\natexlab{c}})\citenamefont {Zhang}, \citenamefont {Shi},
  \citenamefont {Liu}, \citenamefont {Xia}, \citenamefont {Su}, \citenamefont
  {Chen}, \citenamefont {Wang}, \citenamefont {Yu}, \citenamefont {Zou},
  \citenamefont {Zhao} \emph {et~al.}}]{305-zhang2020hass}%
  \BibitemOpen
  \bibfield  {author} {\bibinfo {author} {\bibfnamefont {G.}~\bibnamefont
  {Zhang}}, \bibinfo {author} {\bibfnamefont {X.}~\bibnamefont {Shi}}, \bibinfo
  {author} {\bibfnamefont {X.}~\bibnamefont {Liu}}, \bibinfo {author}
  {\bibfnamefont {W.}~\bibnamefont {Xia}}, \bibinfo {author} {\bibfnamefont
  {H.}~\bibnamefont {Su}}, \bibinfo {author} {\bibfnamefont {L.}~\bibnamefont
  {Chen}}, \bibinfo {author} {\bibfnamefont {X.}~\bibnamefont {Wang}}, \bibinfo
  {author} {\bibfnamefont {N.}~\bibnamefont {Yu}}, \bibinfo {author}
  {\bibfnamefont {Z.}~\bibnamefont {Zou}}, \bibinfo {author} {\bibfnamefont
  {W.}~\bibnamefont {Zhao}}, \emph {et~al.},\ }\bibfield  {title} {\bibinfo
  {title} {The de hass-van alphen quantum oscillations in {B}a{S}n$_{3}$
  superconductor with multiple {D}irac fermions},\ }\href@noop {} {\bibfield
  {journal} {\bibinfo  {journal} {arXiv:2001.08359}\ } (\bibinfo {year}
  {2020}{\natexlab{c}})}\BibitemShut {NoStop}%
\bibitem [{\citenamefont {Beenakker}(2019)}]{306-beenakker2019search}%
  \BibitemOpen
  \bibfield  {author} {\bibinfo {author} {\bibfnamefont {C.}~\bibnamefont
  {Beenakker}},\ }\bibfield  {title} {\bibinfo {title} {Search for
  non-{A}belian {M}ajorana braiding statistics in superconductors},\
  }\href@noop {} {\bibfield  {journal} {\bibinfo  {journal} {arXiv:1907.06497}\
  } (\bibinfo {year} {2019})}\BibitemShut {NoStop}%
\end{thebibliography}%

\end{document}